\documentclass[a4paper,10pt,reqno,final]{amsart}
\usepackage{epsfig}
\usepackage{psfrag}

\usepackage{amsmath}
\usepackage{amsfonts}
\usepackage{amsthm}

\usepackage[T1]{fontenc}

\usepackage[latin1]{inputenc}

\usepackage[french,english]{babel}

\usepackage[centerlast,small]{subfigure}

\usepackage{hyperref}
\hypersetup{
breaklinks=true,
}

\newcommand{\erf}[1]{\operatorname{erf}#1}
\newcommand{\Rlogo}{\protect\includegraphics[height=1.8ex,keepaspectratio]{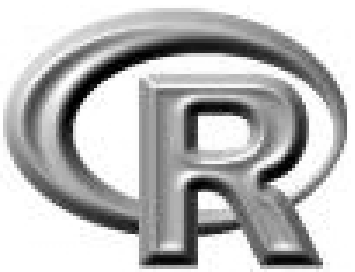}}
\newcommand{\nombresujetttp}{9}
\newcommand{\nombrettp}{304}

\newtheorem{theorem}{Theorem}[section]

\theoremstyle{definition}

\theoremstyle{remark}
\newtheorem{remark}[theorem]{Remark}

\numberwithin{equation}{section}

\newcommand{\Er}{\mathbb{R}}

\newcommand{\Erps}{\mathbb{R_{\, +} ^{\, *}}}

\newcommand{\ve}{\varepsilon}

\makeatletter
\renewcommand{\@seccntformat}[1]{\large{\csname the#1\endcsname}.
\hspace{0.5em}}
\renewcommand{\section}{\@startsection {section}{1}{0mm}%
                                   {-\baselineskip}%
                                   {0.5\baselineskip}%
                                   {\sffamily\large\upshape\bfseries}}
\renewcommand{\subsection}{\@startsection {subsection}{2}{0mm}%
                                   {-0.5\baselineskip}%
                                   {0.5\baselineskip}%
                                   {\sffamily\normalsize\upshape
                                   \bfseries}}
\makeatother

\begin{document}
\selectlanguage{english}

%%%%%%%%%%%%%%%%%%%%%%%%%%%%
% TITLE (for amsart class)
%%%%%%%%%%%%%%%%%%%%%%%%%%%%

%%%%%%%%% Begin of title

\title%
[Model of  joint displacement using sigmoid function]%
{Model of  joint displacement using sigmoid function. Experimental approach for planar pointing task and squat jump}

\author{Thomas Creveaux}
\author{J\'er\^ome Bastien}
\author{Cl\'ement Villars}
\author{Pierre Legreneur}

\date{\today}

\selectlanguage{french}

\address{
Université de Lyon\\
Centre de Recherche et d'Innovation sur le Sport\\
   U.F.R.S.T.A.P.S.\\
   Université Claude Bernard - Lyon 1\\
   27-29, Bd du 11 Novembre 1918\\
   69622 Villeurbanne Cedex\\
France}

% pour utilisation avec hyperref
% \email{\href{mailto:thomas.creveaux@univ-lyon1.fr}{\nolinkurl{thomas.creveaux@univ-lyon1.fr}}}
% \email{\href{mailto:jerome.bastien@univ-lyon1.fr}{\nolinkurl{jerome.bastien@univ-lyon1.fr}}}
% \email{\href{mailto:clement.villars@univ-lyon1.fr}{\nolinkurl{clement.villars@univ-lyon1.fr}}}
% \email{\href{mailto:pierre.legreneur@univ-lyon1.fr}{\nolinkurl{pierre.legreneur@univ-lyon1.fr}}}
\email{thomas.creveaux@univ-lyon1.fr}
\email{jerome.bastien@univ-lyon1.fr}
\email{clement.villars@univ-lyon1.fr}
\email{pierre.legreneur@univ-lyon1.fr}

\selectlanguage{english}


\begin{abstract}
Using an experimental optimization approach, this study investigated whether two human movements, pointing tasks  and squat-jumps, could be modelled with a reduced set of kinematic parameters. Three sigmoid models were proposed to model the evolution of joint angles. The models parameters were optimized to fit the 2D position of the joints obtained from 
\nombrettp\ pointing tasks
and 
120 squat-jumps.
The models were accurate for both movements. This study provides a new framework to model planar movements with a small number of meaningful kinematic parameters, allowing a continuous description of both kinematics and kinetics. Further researches should investigate the implication of the control parameters in relation to motor control and validate this approach for three dimensional movements.
\end{abstract}

%%%%%%%%% End of title

\maketitle

%  optionnel : si plus de trois auteurs.
\markboth{Thomas CREVEAUX et al.}
{\MakeUppercase{%
Model of  joint displacement using sigmoid function. Experimental approach}}

%%%%%%%%%%%%%%%%%%%%%%%%%%%%%%%%%%%%%%%%%%%%%%%%%%%%%%%
%          begin of text
%%%%%%%%%%%%%%%%%%%%%%%%%%%%%%%%%%%%%%%%%%%%%%%%%%%%%%%
%% main tex
%%%%%%%%%%%%%%%%%%%%%%%%%%%%%%%%%%%%%%%%%%%%%%%%%%%%%%%%%%%%%%%%%%%%
%%%%%%%%%%%%%%%%%%%%%%%%%%%%%%%%%%%%%%%%%%%%%%%%%%%%%%%%%%%%%%%%%%%%

%%%%%%%%%%%%%%%%%%%%%%%%%%%%%%%%%%%%%%%%%%%%%%%%%%%%%%%%%%%%%%%%%%%%
%%%%%%%%%%%%%%%%%%%%%%%%%%%%%%%%%%%%%%%%%%%%%%%%%%%%%%%%%%%%%%%%%%%%
\section{Introduction}
\label{intro}

Quantitative analysis of human movement usually relies on the time history of reflective markers fixed to anatomical landmarks obtained from optical systems. These raw data are further used to compute relevant parameters such as velocities, accelerations, moments or powers. During the recent years, the performance of acquisition systems greatly increased, especially consdiering acquisition rate and accuracy. However, raw data still remain noisy, due to the movement of the skin with regard to the bones and finite accuracy of such systems. Furthermore, the effect of noise increases as the data is derived with respect to time, which is a very common task in movement analysis.

To overcome the aforementionned issues, raw data are quite always smoothed or filtered, resulting in well-known decrease of movement amplitude. Specific filtering methods accounting for properties of the skeletal system such as constant length of the limbs have been used but such approaches still suffer from the motion of the markers relatively to the skeletal system. An interesting feature of human motion is the necessity for decelerating the joint displacement before its maximal amplitude (anatomical constraint) in order to protect this joint from any damage \cite{vanIngenShenau1989}. Regarding to kinematics, the anatomical constraint implies that joint angular time history should match an asymmetric sigmoid shape \cite{Zelaznik1986} 
and thus an asymmetric bell-shaped velocity profile \cite{SoechtingLacquantini1981}, which accounts for synergistic actuators' activations at a joint, i.e. agonist and antagonist muscle-tendon systems. In the field of human movement analysis, Plamondon proposed an asymetric model of asymetric sigmoid \cite{plamondon95a,plamondon95b,plamondon98,plamondonchunchua03} but the velocity is not null at the end of the movement so that the anatomical constraint is not satisfied.

Therefore, this study aimed at modelling two different movements, i.e. a pointing task  and an explosive movement, the squat-jump, using a generic model of sigmoidal joint displacement based on meaningfull kinematic parameters which accounts for the anatomical constraint. Three submodels were used to achieve best fitting of experimental data obtained from both movements.

%%%%%%%%%%%%%%%%%%%%%%%%%%%%%%%%%%%%%%%%%%%%%%%%%%%%%%%%%%%%%%%%%%%%%%%%%%%
%%%%%%%%%%%%%%%%%%%%%%%%%%%%%%%%%%%%%%%%%%%%%%%%%%%%%%%%%%%%%%%%%%%%%%%%%%%
\section{Methods}
\label{method}

%%%%%%%%%%%%%%%%%%%%%%%%%%%%%%%%%%%%%%%%%%%%%%%%%%%%%%%%%%%%%%%%%%%%%%%%%%%
\subsection{Model of joint displacement}

\subsubsection{General model}\

\begin{figure}[ht]
\psfrag{timx}{time}
\psfrag{an}{angle}
\psfrag{vit}{velocity}
\psfrag{acc}{acceleration}
\psfrag{ti}{$t_{\text{b}}$}
\psfrag{tm}{$t_0$}
\psfrag{tf}{$t_{\text{e}}$}
\psfrag{C}{$\theta_{\text{b}}$}
\psfrag{B}{$\theta_0$}
\psfrag{A}{$\theta_{\text{e}}$}
\psfrag{K}{$K$}
\begin{center}
\epsfig{file=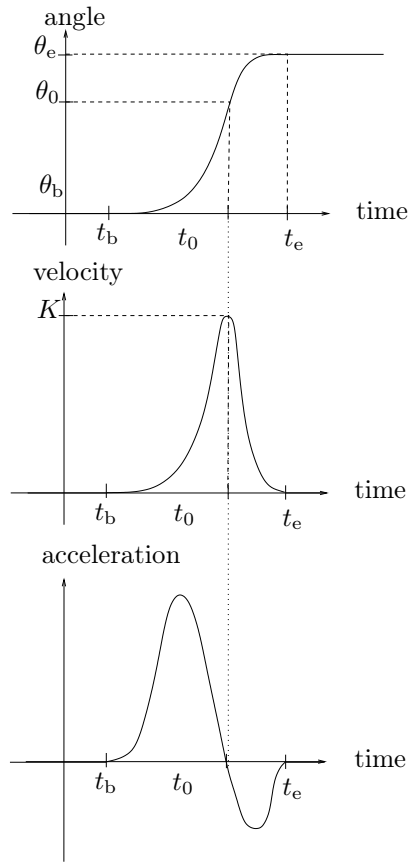, width=5  cm}
\end{center}
\caption{\label{fig10}Shape of used sigmoid: angle, velocity and acceleration versus time (for the increasing case).}
\end{figure}

Accounting for a monotone evolution of a given angle and considering the anatomical constraint requirements,
it is assumed that each angle $\theta$ is characterized by the following properties (figure \ref{fig10}):
\begin{itemize}
\item
at the beginning and at the end of the movement, the velocity and the acceleration are equal to zero;
\item
the angle increases (respectively decreases) throu\-ghout the whole movement;
\item
during the movement, the velocity increases (respectively decreases) until it reaches its maximum (respectively minimum), then decreases (respectively increases). 
\end{itemize}

More precisely, we try to determine a function  $\theta$ from  $[0,T]$ to $\Er$ of  class $C^2$. Let  
$t_{\text{b}},t_0,t_{\text{e}}$ be three instants such that 
\begin{equation}
\label{eqan01newA}
0\leq t_{\text{b}}<t_0<t_{\text{e}}\leq T.
\end{equation}
Let  
$\theta_{\text{b}},\theta_0,\theta_{\text{e}}$ be three real numbers such that 
\begin{equation}
\label{eqan01newAbis}
\theta_{\text{b}}<\theta_0<\theta_{\text{e}}\text{ or }
\theta_{\text{b}}>\theta_0>\theta_{\text{e}}
\end{equation}
We assume that 
\begin{itemize}
\item
$\theta$ is constant and equals $\theta_{\text{b}}$ on $[0,t_{\text{b}}]$ ;
\item
$\theta$ is constant and equals $\theta_{\text{e}}$ on $[t_{\text{e}},T]$ ;
\item
there exists $\ve\in \{-1,1\}$ such that  $\ve\theta$ is strictly 
increasing  on  $[t_{\text{b}},t_{\text{e}} ]$ ; 
\item
$\ve  \theta$ is strictly convex on  
$(t_{\text{b}},t_0)$ ;
\item
$\ve \theta$  is strictly concave on  
$(t_0,t_{\text{e}})$.
\end{itemize}

We set 
\begin{align}
\label{eq100}
& \ve=
\text{Sign}\,\left(\theta_{\text{e}}-\theta_{\text{b}}\right)\in \{-1,1\}.\\
\intertext{Let $K$ be the number defined by } 
\label{eq101}
&
K=
\begin{cases}
\displaystyle{\max_{t\in [t_{\text{b}},t_{\text{e}}]} \theta'(t)}&\text{if $\ve=1$},\\
\displaystyle{\min_{t\in [t_{\text{b}},t_{\text{e}}]} \theta'(t)}&\text{if $\ve=-1$}.
\end{cases}
\end{align}
Since $\theta$ is of class $C^2$, we have 
\begin{subequations}
\label{eqan01tot}
\begin{align}
\label{eqan01a}
&
\theta(t_{\text{b}})=\theta_{\text{b}},\quad 
\theta'(t_{\text{b}})=0,\quad 
\theta''(t_{\text{b}})=0,\\
\label{eqan01b}
&
\theta(t_{\text{e}})=\theta_{\text{e}},\quad 
\theta'(t_{\text{e}})=0,\quad 
\theta''(t_{\text{e}})=0,\\
\label{eqan01c}
&\theta(t_0)=\theta_0,\quad 
\theta'(t_0)=K,\quad
\theta''(t_0)=0,\\
\label{eqan01d}
&\forall t\in(t_{\text{b}},t_0),\quad  \ve\theta''(t)>0,\\
\label{eqan01e}
&\forall t\in(t_0,t_{\text{e}}),\quad  \ve\theta''(t)<0.
\end{align}
\end{subequations}

We consider $\alpha,\beta\in (0,1)$ and $k\in \Er$ defined by 
\begin{equation}
\label{eq109tot}
\alpha=\frac{t_0-t_{\text{b}}}{t_{\text{e}}-t_{\text{b}}} ,\quad 
\beta=\frac{\theta_0-\theta_{\text{b}}}{\theta_{\text{e}}-\theta_{\text{b}}},\quad
k=K\frac{t_{\text{e}}-t_{\text{b}}}{\theta_{\text{e}}-\theta_{\text{b}}}.
\end{equation}
Applying the following change of scale, 
\begin{subequations}
\label{eqan30tot}
\begin{align}
\label{eqan30a}
&\forall t \in \left[t_{\text{b}},t_{\text{e}}\right],\quad u=\frac{t-t_{\text{b}}}{t_{\text{e}}-t_{\text{b}}}\in[0,1],\\
\label{eqan30b}
& \forall u\in [0,1],\quad
g(u)=\frac{\theta\bigl( (t_{\text{e}}-t_{\text{b}})u +t_{\text{b}}\bigr)-\theta_{\text{b}}}{\theta_{\text{e}}-\theta_{\text{b}}}.
\end{align}
\end{subequations}
the problem can be reformulated as follows: 
we look for a function $g$ of class $C^2$
defined on $[0,1]$
satisfying:
\begin{subequations}
\label{eqan40tot}
\begin{align}
\label{eqan40a}
&
g(0)=0,\quad g'(0)=0,\quad g''(0)=0,\\
\label{eqan40b}
&
g(1)=1,\quad g'(1)=0,\quad g''(1)=0,\\
\label{eqan40c}
&
g(\alpha)=\beta,\quad g'(\alpha)=k,\quad g''(\alpha)=0,\\
\label{eqan40d}
&\forall u\in(0,\alpha),\quad  g''(t)>0,\\
\label{eqan40e}
&\forall u\in(\alpha,1),\quad  g''(t)<0.
\end{align}
\end{subequations}

\begin{remark}
\label{remalpha}
Under the assumptions given in \eqref{eqan40tot} 
we have necessarily (see \cite{creveaux09,bastiencreveauxencours12})
\begin{equation}
\label{eqan42}
k\geq \max\left(\frac{\beta}{\alpha},\frac{1-\beta}{1-\alpha}\right)>1.
\end{equation}
\end{remark}

Finally, the function $\theta$ is defined
for all $t \in [0,T]$, by
\begin{equation}
\label{eqan50}
 \theta(t)=
\begin{cases}
\displaystyle{%
\theta_{\text{b}}},&\text{if $t\leq t_{\text{b}}$},\\
\displaystyle{\left(\theta_{\text{e}}-\theta_{\text{b}}\right)
g\left( \frac{t-t_{\text{b}}}{t_{\text{e}}-t_{\text{b}}}\right)
+\theta_{\text{b}}}, &\text{if $ t_{\text{b}}<t< t_{\text{e}}$},\\
\displaystyle{\theta_{\text{e}}},&\text{if $t\geq t_{\text{e}}$}.
\end{cases}
\end{equation}
This function is defined by 7 independent parameters: 
\begin{itemize}
\item
2 time scale parameters ($t_{\text{b}}$ and $t_{\text{e}}$),
\item
2 angle scale parameters ($\theta_{\text{b}}$ and $\theta_{\text{e}}$),
\item
and 3 shape parameters ($\alpha$, $\beta$, $k$).
\end{itemize}
Thus, $\theta$ can be written under the form $\theta_{t_{\text{b}},t_{\text{e}},\theta_{\text{b}},\theta_{\text{e}},\alpha,\beta,k}$.

In the literature, there exist many sigmoidal functions.
However, these models can not be used to solve \eqref{eqan40tot} because the sigmoids
\begin{itemize}
        \item
        are symmetric \cite{menon96,kumazawa00,MR1440301,MR2212485,MR1976213,MR1400774},
        \item
        are defined on $\Er$ and can not be used on a bounded interval \cite{MR1365401,MR2004649,lindenmann1963,bradbury1970,pearson1991},
        \item
        are characterized by not enough parameters \cite{MR0048770,debouche79}.
\end{itemize}

To our knowledge, there is not in the literature, a general construction of non symmetric sigmoid 
satisfying \eqref{eqan40tot}, of class  $C^2$ or $C^\infty$.

In the field of movement analysis, R. Plamondon \cite{plamondon95a,plamondon95b,plamondon98,plamondonchunchua03}
used a log-normal function
\begin{equation}
\label{eqan51}
\Lambda_{t_0,\mu,\sigma^2}(t)=\frac{1}{\sigma \sqrt{2\pi}(t-t_0)}\exp\left({\displaystyle{-\frac{\left(\ln(t-t_0)-\mu\right)^2}{2\sigma ^2}}}\right)
\end{equation}
to describe general movements and applied it to the 
Fitts task \cite{Fitts54,FittsPeterson64}.
By using the Central Limit Theorem, he proved
that, for a large number of agonist and antagonist muscles acting, the proposed function can model the behavior of the system.
However, this work can not be applied to solve \eqref{eqan40tot}.
The major concern with Plamondon's function remains in its asymptotic behavior at the end of the movement.
Especially, it should be observed that the velocity tends to zero as $t$ approaches $+\infty$ and then, the end of movement is not clearly defined. Other works related to the log-normal law \cite{lindenmann1963,bradbury1970,pearson1991} do not solve this problem.
Moreover, Plamandon's model contains not enough parameters to allow the solving of \eqref{eqan40tot}. 

The simplest idea to solve \eqref{eqan40tot} would be to apply Hermite's polynomial interpolation
(e.g. \cite{boor}), but it can be showed that
this method can not be used (see \cite{bastiencreveauxencours12}).

To allow the solving of the system \eqref{eqan40tot}, the model has to include three control parameters which have to be determined in relation to $\alpha$, $\beta$ and $k$.
In the next section, three sigmoid models, SYM, NORM and INVEXP, previously described in \cite{creveaux09}, meeting the mentioned requirements are presented.
These models were successfully used for both pointing tasks \cite{villars08,cvjbplkm2008ISCA,legreneuretali11} and squat-jump \cite{creveaux09,TCJBPLACAPS09ACAPS}. Exhaustive theoretical description of the models will be given in a future paper \cite{bastiencreveauxencours12}.

%%%%%%%%%%%%%%%%%%%%%%%%%%%%%%%%%%%%%%%%%%%%%%%%%%%%%%%%%%%%%%%%%%%%%%%%%%%
\subsubsection{The SYM model}\
\label{fonction_sigmoideC2}

The SYM model was built using a pseudo-symmetry approach. Its function $g$ is defined by the three parameters
$\alpha, \beta \in (0,1)$ and $k>1$.

Let 
$g_{\alpha,\beta,k}$ be a function of class $C^2$ from $[0,\alpha]$ to $\Er$
satisfying \eqref{eqan40a},\eqref{eqan40c},  and \eqref{eqan40d}.
If the function $g$ is defined from $[0,1]$ to $\Er$ by,
\begin{equation}
\label{eqan60}
g(u)=
\begin{cases}
g_{\alpha,\beta,k}(u), &\text{if $u\leq \alpha$},\\
1-g_{1-\alpha,1-\beta,k}(1-u), &\text{if   $u> \alpha$},
\end{cases}
\end{equation}
then, $g$ is of class $C^2$ on  $[0,1]$ and 
\eqref{eqan40tot} holds.
Considering the function $H^{(a,b,\kappa)}$ defined on $[0,\alpha]$ for all $a,b>0$ and $\kappa >2$ by 
\begin{equation}
\label{eqan100}
H^{(a,b,\kappa)}(u)=a\left(1-e^{-bu^\kappa }\right),
\end{equation}
$a$, $b$ and $\kappa$ have to be determined so that 
\eqref{eqan40a}, \eqref{eqan40c} and \eqref{eqan40d}
hold.
We set 
\begin{equation}
\label{defr0}
r_0=\frac{1}{e^{1/2}-1}\approx 1.54
\end{equation}
For all $(\alpha,\beta) \in (0,1)^2$, for all $k$ 
such that 
$k >  r_0 \beta/\alpha$,
there exist 
$(a,b,\kappa)\in \Erps^2\times (2,\infty)$
such that \eqref{eqan40a}, \eqref{eqan40c} and \eqref{eqan40d}
hold for function  $H^{(a,b,\kappa)}$.  
$a$, $b$ and $\kappa$ still need to be defined.
We set  
\begin{subequations}
\label{eqan110tot}
\begin{equation}
\label{eqan110a}
\gamma=\frac{\beta}{k\alpha}\in\left(0,e^{\frac{1}{2}}-1\right).
\end{equation}
It exists a unique  $X\in (1/2,1)$ such that  
\begin{equation}
\label{eqan110b}
\left(e^X-1\right)\frac{1-X}{X}=\gamma,
\end{equation}
and it follows 
\begin{equation}
\label{eqan110c}
\kappa=\frac{1}{1-X},\quad
a=\frac{\beta}{1-e^{-X}},\quad
b=\frac{X}{\alpha^\kappa}.
\end{equation}
\end{subequations}
By setting $(a,b,\kappa)=\mathcal{G}(\alpha, \beta,k)$, the function $g$ is defined
for all $u\in [0,1]$ by
\begin{equation}
\label{eqan61}
g(u)=
\begin{cases}
H^{\mathcal{G}(\alpha, \beta,k)}(u), &\text{if $u\leq \alpha$},\\
1-H^{\mathcal{G}(1-\alpha, 1-\beta,k)}(1-u), &\text{if   $u> \alpha$},
\end{cases}
\end{equation}

%%%%%%%%%%%%%%%%%%%%%%%%%%%%%%%%%%%%%%%%%%%%%%%%%%%%%%%%%%%%%%%%%%%%
\subsubsection{The NORM model}\
\label{fonction_sigmoideCinf}

The NORM model (named from its relation to the normal law) function $g$ is defined by three parameters
$a\in (0,1)$, $p>0$ and $s>0$.

We recall that the  density function of the the normal (or Gaussian) distribution
with mean $m$ and variance $s^2$ is given by:
\begin{equation}
\label{eqan510}
\forall x \in \Er,\quad
f(x)=\frac{1}{s\sqrt{2 \pi}} 
\exp{\left(-\frac{1}{2}{\left(\frac{x-m}{s}\right)}^2\right)}.
\end{equation}
Considering the $\erf$ function defined by
\begin{equation}
\label{eqan500}
\forall x \in \Er,\quad
\erf(t)=\frac{2}{\sqrt{\pi}} \int_0 ^x e^{-t^2}dt, 
\end{equation}
the cumulative distribution function of the normal law is given by 
\begin{equation}
\label{eqan515}
\forall x \in \Er,\quad
\Phi(x)=
\frac{1}{2}\erf\left(\frac{x-m}{\sqrt 2 s}\right)+\frac{1}{2}.
\end{equation}
For all $p>0$, we define the bijection $G$ from $(0,1)$ to $\Er$ by 
\begin{equation}
\label{eqan520}
\forall u\in (0,1),\quad
G(u)=\ln\left(\frac{u^p}{1-u^p}\right).
\end{equation}
Finally, the function $g$ is defined by 
\begin{subequations}
\label{eqan530tot}
\begin{align}
\label{eqan530a}
&\forall t\in (0,1),\quad
g(t)=\Phi(G(t)),\\
\label{eqan530b}
&
g(0)=0,\\
\label{eqan530c}
&
g(1)=1.
\end{align}
\end{subequations}
with $a=G^{-1}(m)$.
%%%%%%%%%%%%%%%%%%%%%%%%%%%%%%%%%%%%%%%%%%%%%%%%%%%%%%%%%%%%%%%%%%%%%%%%%%%

%%%%%%%%%%%%%%%%%%%%%%%%%%%%%%%%%%%%%%%%%%%%%%%%%%%%%%%%%%%%%%%%%%%%
\subsubsection{The INVEXP model}\
\label{fonction_sigmoideCinfTC}

The INVEXP model (derived from the inverse exponential) function $g$ is defined by three parameters
$\lambda,\mu>0$ and $a\in \Er$.
For all $a$, for all $\lambda,\mu$, we set 
\begin{equation}
\label{eqan540}
\alpha=\frac{\lambda}{\lambda+\mu}\in (0,1),
\end{equation}
and 
we consider function $g_{a,\alpha}$ defined by 
if $a=0$
\begin{subequations}
\label{eqan550}
\begin{align}
&g_{a,\alpha}=1,
\intertext{and if  $a>0$}
&
\begin{cases}
&\forall y \in [0,\alpha),\quad \displaystyle{g_{a,\alpha}(y)=1-\exp\left(\frac{t}{a(t-\alpha)}\right)},\\
\\
&\forall y \in [\alpha,1],\quad g_{a,\alpha}(y)=1.
\end{cases}
\end{align}
\end{subequations}
For all $a\in \Er$  and for all $\alpha\in (0,1)$, we consider the function $G_{a,\alpha}$ defined by 
\begin{equation}
\label{eqan560}
\begin{cases}
&\text{if } a\geq 0,\quad G_{a,\alpha}=g_{a,\alpha},\\
&\text{if } a<0,\quad G_{a,\alpha}=g_{-a,1-\alpha}(1-.).
\end{cases}
\end{equation}
For all 
$\lambda,\mu>0$, $f_{\lambda,\mu}$ is defined by 
\begin{subequations}
\label{eqan570tot}
\begin{align}
\label{eqan5700a}
&\forall t\in (0,1),\quad
f_{\lambda,\mu}(t)=\exp\left(-\frac{1}{t^\lambda(1-t)^\mu}\right),\\
\label{eqan570b}
&
f_{\lambda,\mu}(0)=0,\\
\label{eqan570c}
&
f_{\lambda,\mu}(1)=1.
\end{align}
\end{subequations}
For all $a\in \Er$
$\lambda,\mu>0$, $h_{\lambda,\mu,a}$ is defined by 
\begin{equation}
\label{eqan580}
h_{\lambda,\mu,a}=   f_{\lambda,\mu}G_{a,\lambda/(\lambda+\mu)}
\end{equation}
and finally the function $g$ is defined by 
\begin{equation}
\label{eqan590}
\forall t\in [0,1],\quad
g(t)=\frac{\displaystyle{\int_0^t h_{\lambda,\mu,a}(u)du}}{\displaystyle{\int_0^1 h_{\lambda,\mu,a}(u)du}}.
\end{equation}

\subsubsection{Definition domains of the sigmoid models}\
\label{fonction_sigmoide_conclusion}

\begin{figure}
\centering
%% sous figure 1
\subfigure[\label{cc1}]
{\epsfig{file=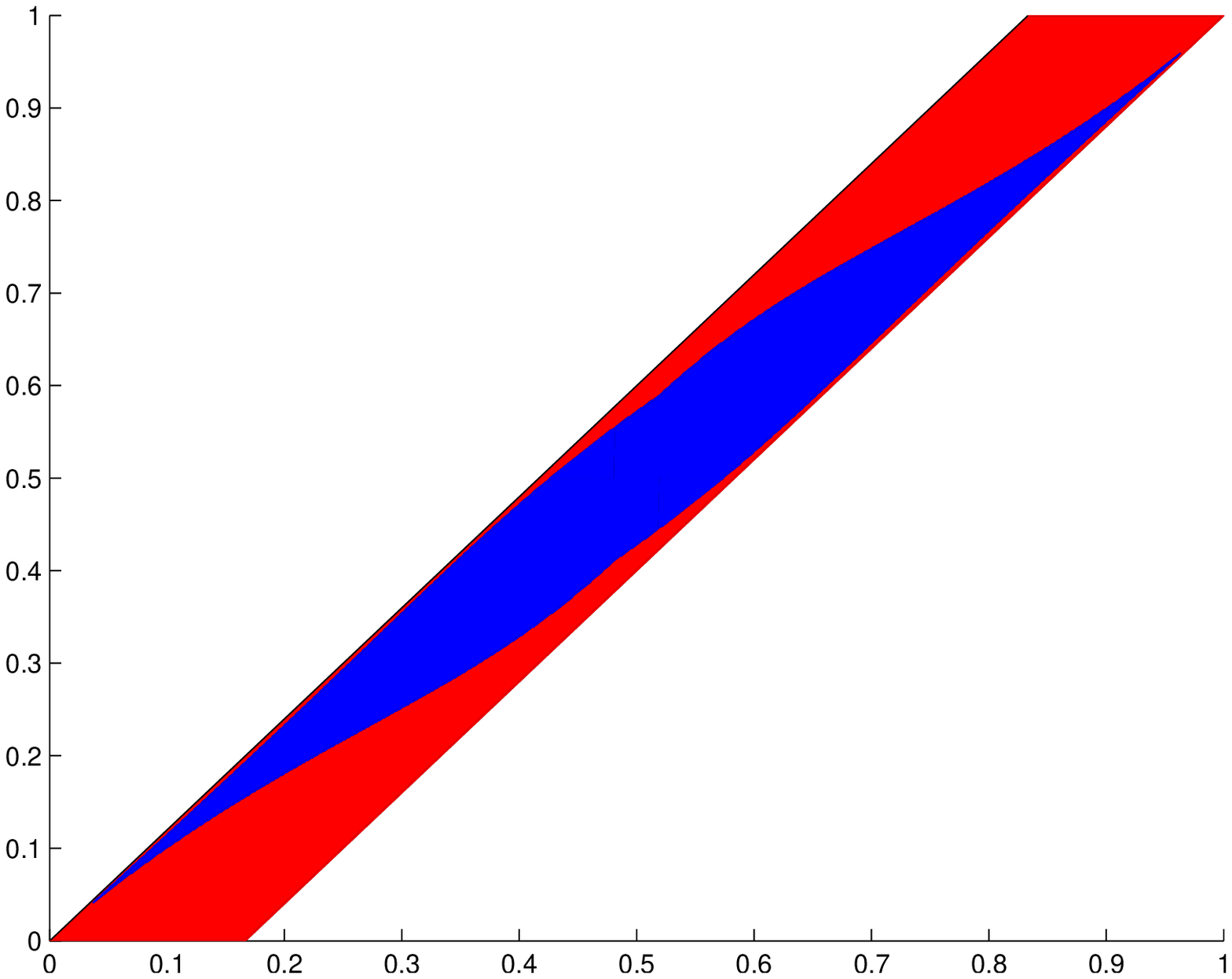, width=6 cm}}
%% sous figure 2
\subfigure[\label{cc2}]
{\epsfig{file=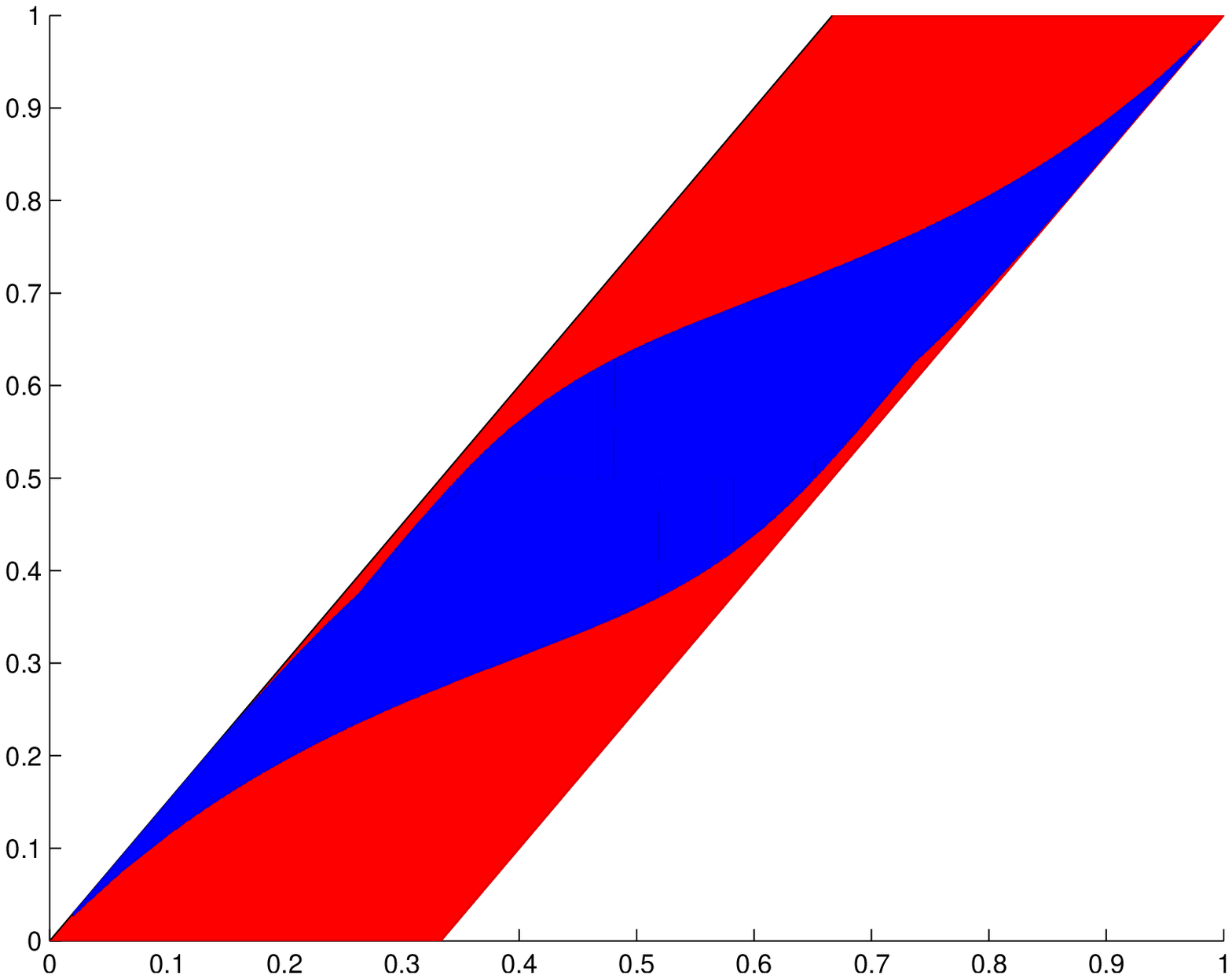, width=6 cm}}
%% sous figure 3
\subfigure[\label{cc3}]
{\epsfig{file=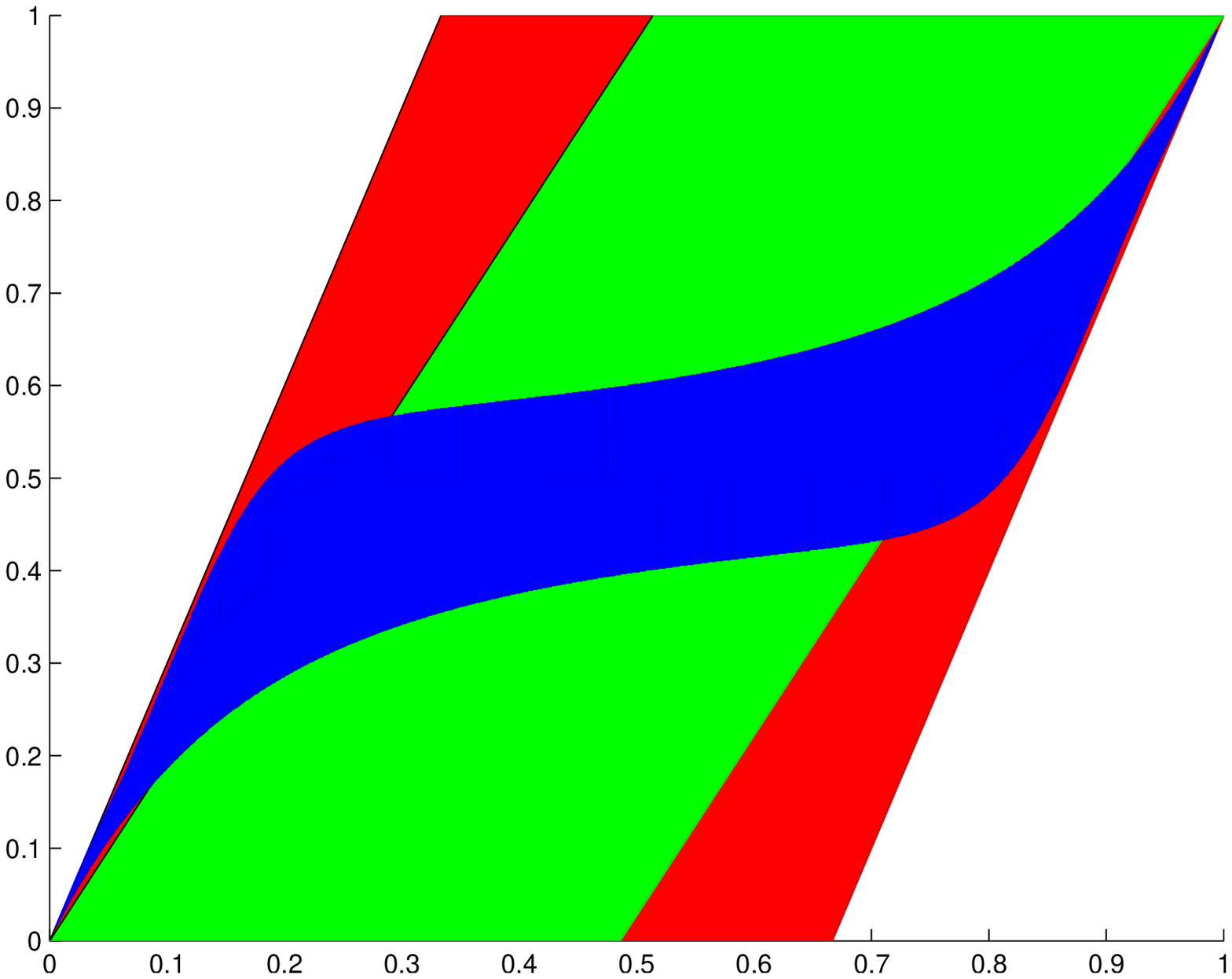, width=6 cm}}
%% sous figure 4
\subfigure[\label{cc4}]
{\epsfig{file=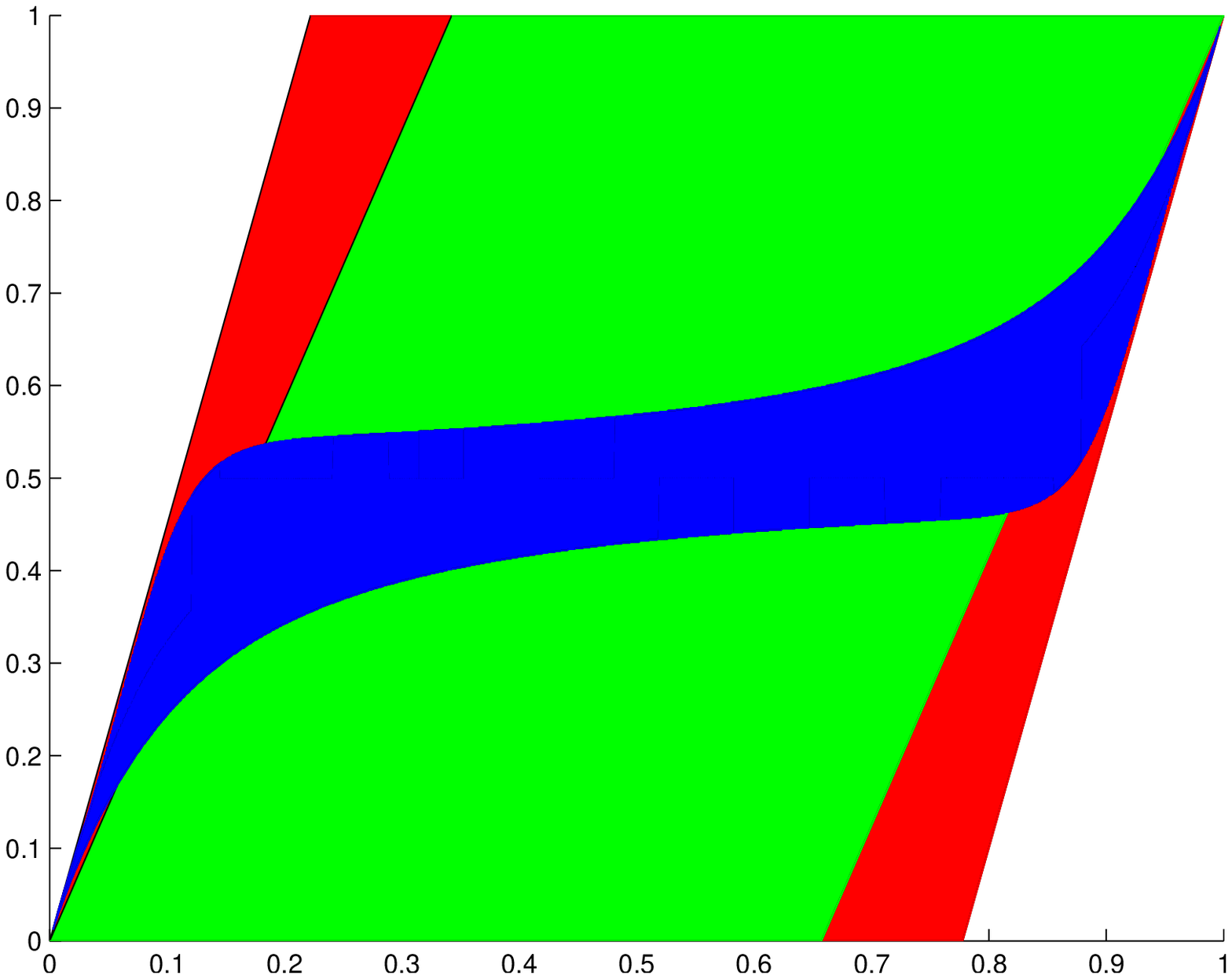, width=6 cm}}
\caption{%
\label{cctot}%
Domain $S_K$ of  the three sigmoids   for $k=1.2$ (a), $k=1.5$ (b), 
for $k=3$ (c), and for  $k=4.5$ (d).
INVEXP, NORM and SYM domains are  plotted in red, blue and green respectively.
According to \eqref{eqan592}, SYM domain is empty for  $k=1.2$ and  $k=1.5$.}
\end{figure}

Each of the three functions is defined by three parameters. We will prove in \cite{bastiencreveauxencours12}
that for all $k>1$, there exist a part $S_k$ of $(0,1)^2$ such that for all 
$(\alpha,\beta) \in S_k$, there exist at least one sigmoid of kind $g$ satisfying 
\eqref{eqan40tot} whose parameters can be determined by splitting \eqref{eqan40tot} in three non-linear equations which can be solved with a numerical solver.
This part $S_k$ is different for the three sigmoid models. 
The bigger part is obtained with the INVEXP model and is given by 
\begin{equation}
\label{eqan591}
(\alpha,\beta)\in S_k^{\text{INVEXP}} \Longleftrightarrow \text{Eq. \eqref{eqan42} holds.}
\end{equation}
This domain is a polygonal part of $[0,1]^2$. 
The domain of a function satisfying \eqref{eqan40tot}
can not be bigger thanks to \eqref{eqan42}.
The  domain $S_k^{\text{SYM}}$ of SYM sigmoid, which is also 
a polygonal part of $[0,1]^2$ is given by 
\begin{equation}
\label{eqan592}
(\alpha,\beta)\in S_k^{\text{SYM}} \Longleftrightarrow 
k\geq r_0\max\left(\frac{\beta}{\alpha},\frac{1-\beta}{1-\alpha}\right)>r_0,
\end{equation}
where 
$r_0$ is defined by \eqref{defr0}.
The domain $S_k^{\text{NORM}}$ of NORM sigmoid can be determined numerically. 
It should be noticed that these three domains are symmetric according to the point $(1/2,1/2)$ and this point
belongs to the three domains (Fig \ref{cctot}).

%%%%% 4 figures faites avec 
%%\programmes\matlab\recherche\CRIS\sigmoide\Cinf\balayage_Sk_erf_fill(1e-3,1.2,1,0.029,0,'domSk1p2',1);
%%\programmes\matlab\recherche\CRIS\sigmoide\Cinf\balayage_Sk_erf_fill(1e-3,1.5,1,0.02,0,'domSk1p5',1);
%%\programmes\matlab\recherche\CRIS\sigmoide\Cinf\balayage_Sk_erf_fill(1e-3,3,1,6e-3,0,'domSk3',1);
%%\programmes\matlab\recherche\CRIS\sigmoide\Cinf\balayage_Sk_erf_fill(1e-3,4.5,1,3e-3,0,'domSk4p5',1);

%%%%% 3 figures faites avec fonction de 
% \programmes\matlab\recherche\CRIS\sigmoide\commun
% Taper 
% ep=1e-2;
% trace_plusieurs_sigmoide(1,20,ep,1/6+2/3*ep,1-ep,5/6-2/3*ep,3,'exemplesigmoNORM',1e-3,6e-3,1);
% trace_plusieurs_sigmoide(2,20,ep,1/6+2/3*ep,1-ep,5/6-2/3*ep,3,'exemplesigmoSYM',1e-3);
% trace_plusieurs_sigmoide(3,20,ep,1/6+2/3*ep,1-ep,5/6-2/3*ep,3,'exemplesigmoINVEXP',1e-3);

\begin{figure}
\centering
%% sous figure 1
\subfigure[\label{exemplesigmoa}]
{\epsfig{file=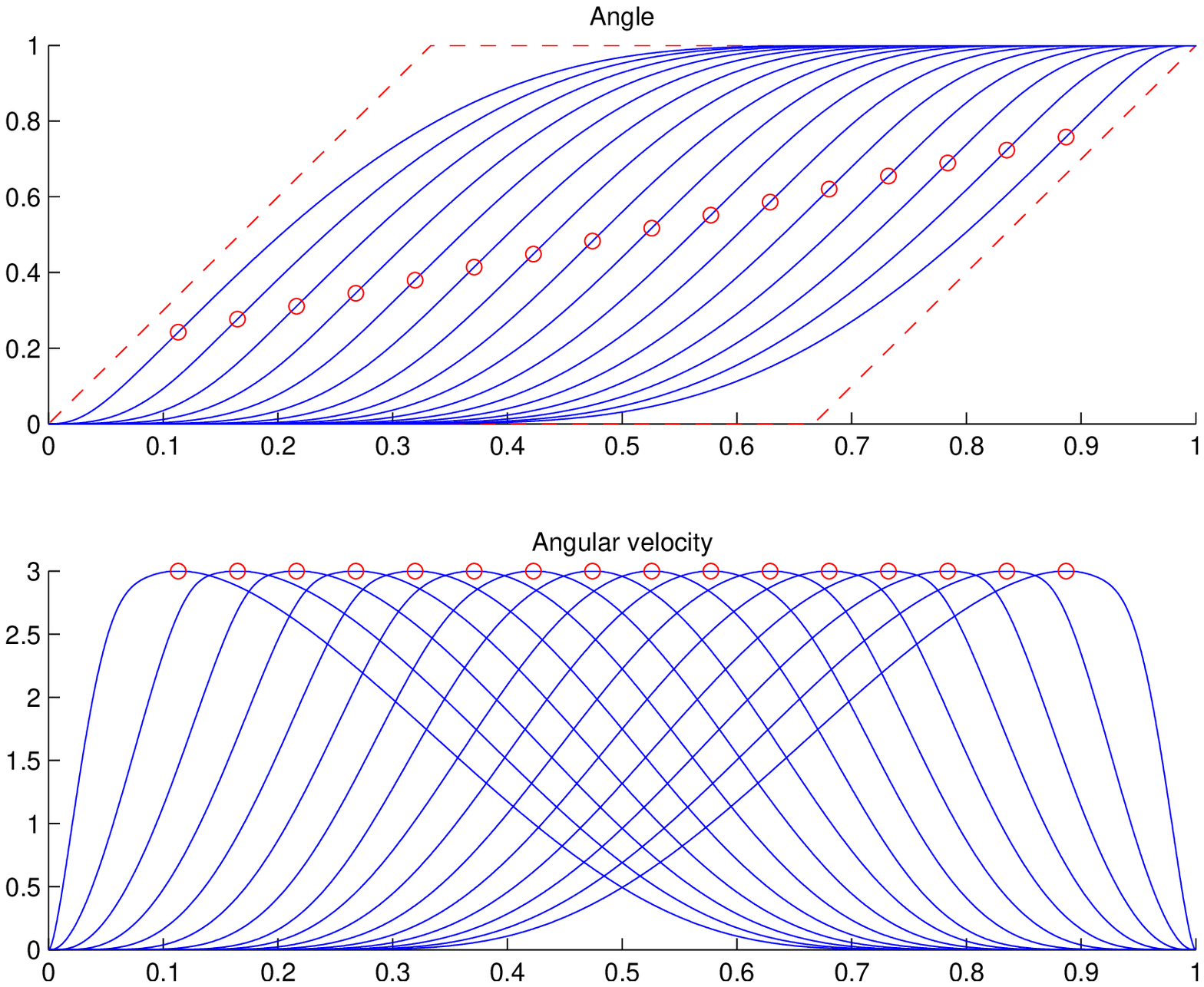, width=6.8 cm}}
%% sous figure 2
\subfigure[\label{exemplesigmob}]
{\epsfig{file=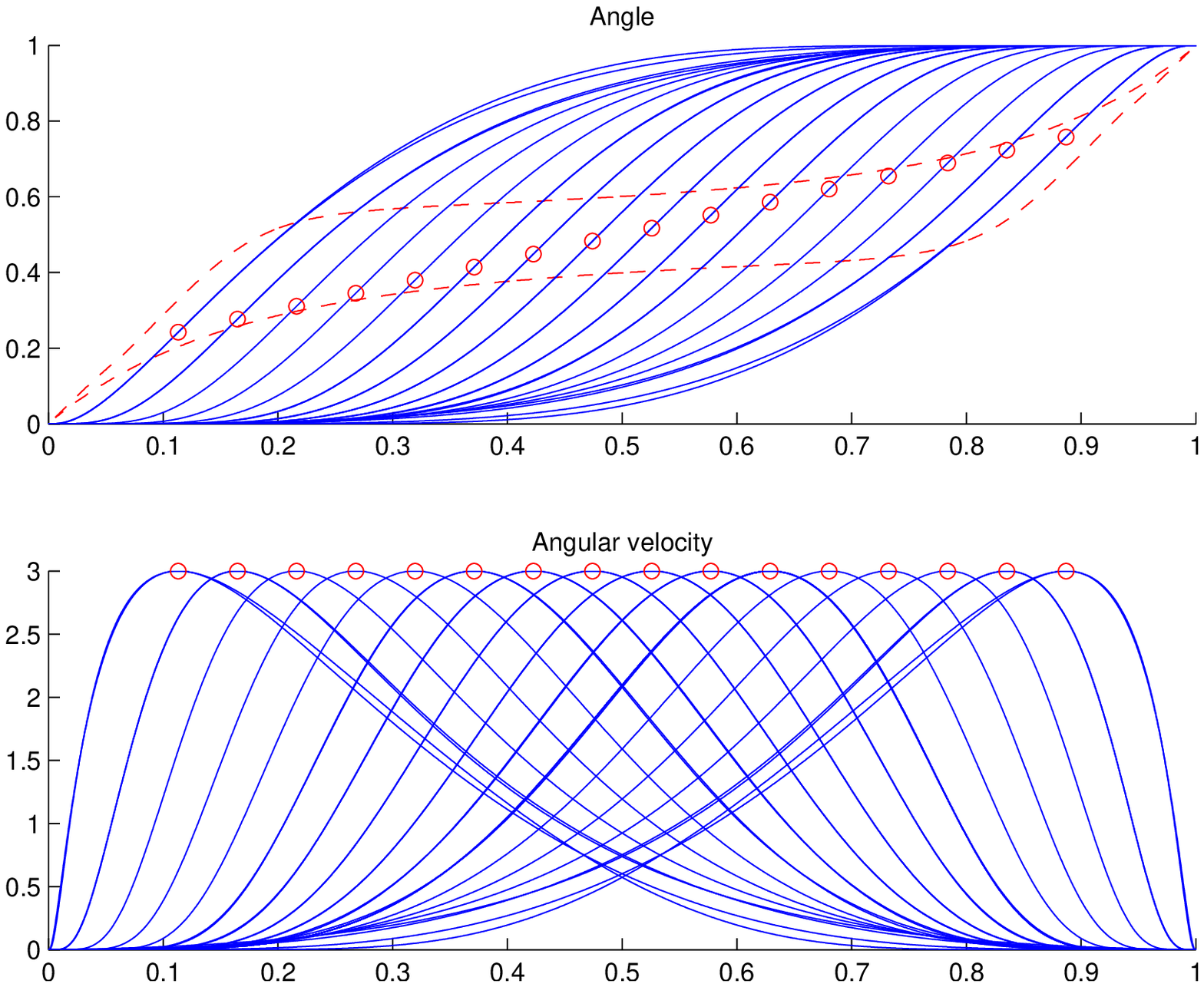, width=6.8 cm}}
%% sous figure 3
\subfigure[\label{exemplesigmoc}]
{\epsfig{file=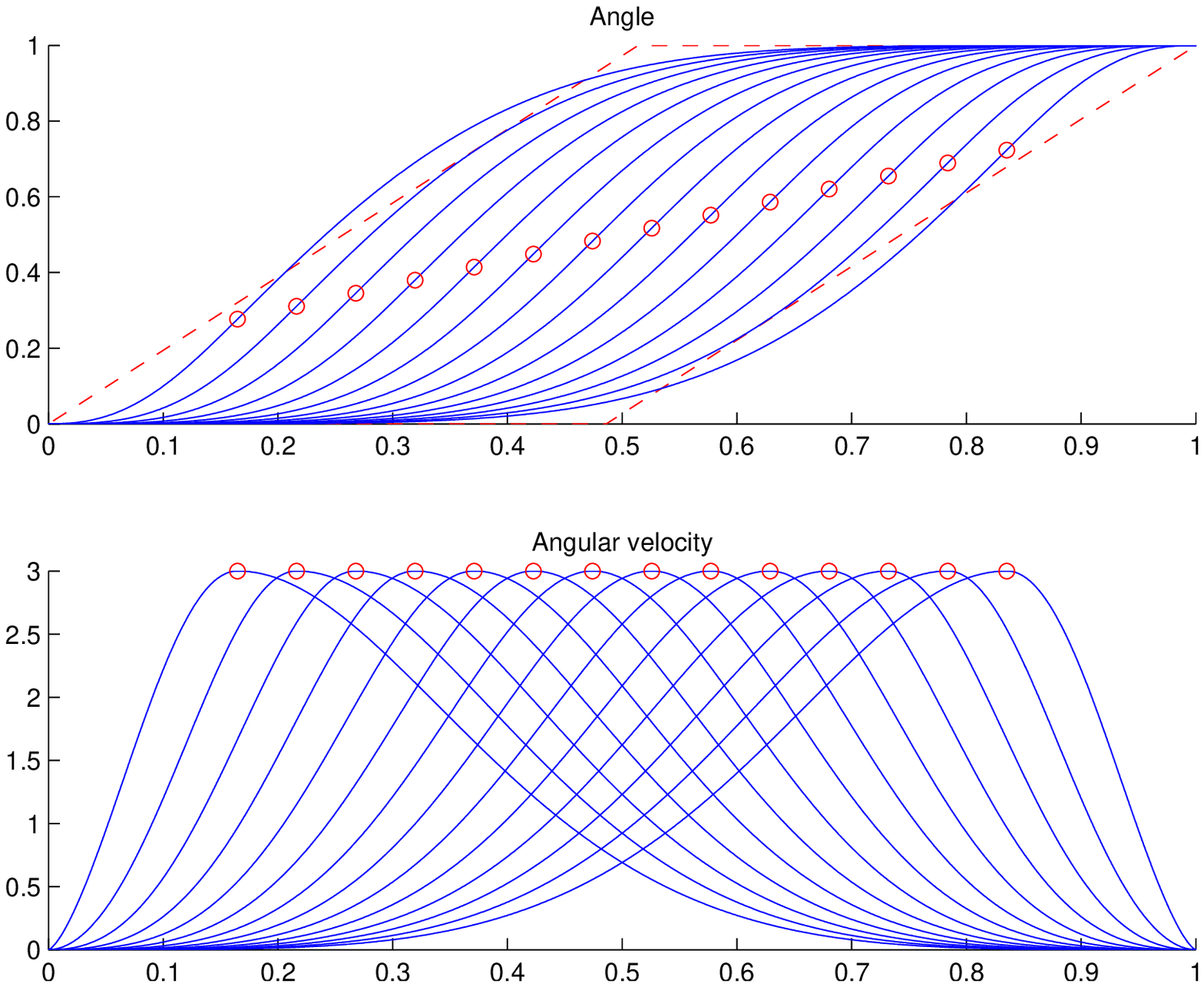, width=6.8  cm}}
\caption{%
\label{exemplesigmotot}%
Examples of curves for angle  and angular velocity,
INVEXP model (a),
NORM model  (b), and
SYM model (c).
The boundaries of domain are plotted in dashed red line. 
Points $(\alpha_i,\beta_i)$ are indicated by red circles.}
\end{figure}

Examples of position and velocity curves obtained from the three models are provided in figure \ref{exemplesigmotot}.
Each sigmoid is defined by $(\alpha_i,\beta_i)$, belonging to a fixed straight line and $k=2$.

%%%%%%%%%%%%%%%%%%%%%%%%%%%%%%%%%%%%%%%%%%%%%%%%%%%%%%%%%%%%%%%%%%%%%%%%%%%
\subsubsection{Specific properties of the sigmoid models}\

\begin{itemize}
\item
The model SYM is very simple and fast to calculate; however, the class of this model is only $C^2$ versus $C^\infty$ for the two other models. 
\item
Since the function $\erf$ is directly implemented in most numerical softwares, the model NORM is fast to calculate. However, 
the domain of this model has been obtained by symmetrization, and for some values of $(\alpha,\beta)$,
two different sigmoids can be obtained. Moreover, the determination of the definition domain is not trivial and the function does not meet the concavity and convexity requirements outside of it.
\item
The part $S_k$ of model INVEXP is the biggest, but this model is harder to calculate, because a numerical 
method of integration has to be used. Since efficient numerical methods exist, this problem can be overcomed and computation time remains reasonable.
\end{itemize}

%%%%%%%%%%%%%%%%%%%%%%%%%%%%%%%%%%%%%%%%%%%%%%%%%%%%%%%%%%%%%%%%
\subsection{Experimental procedures}

\begin{figure}[h]
\centering
%%% sous figure 1
\subfigure[\label{fig100a} : Upper view of the task environment]
{\epsfig{file=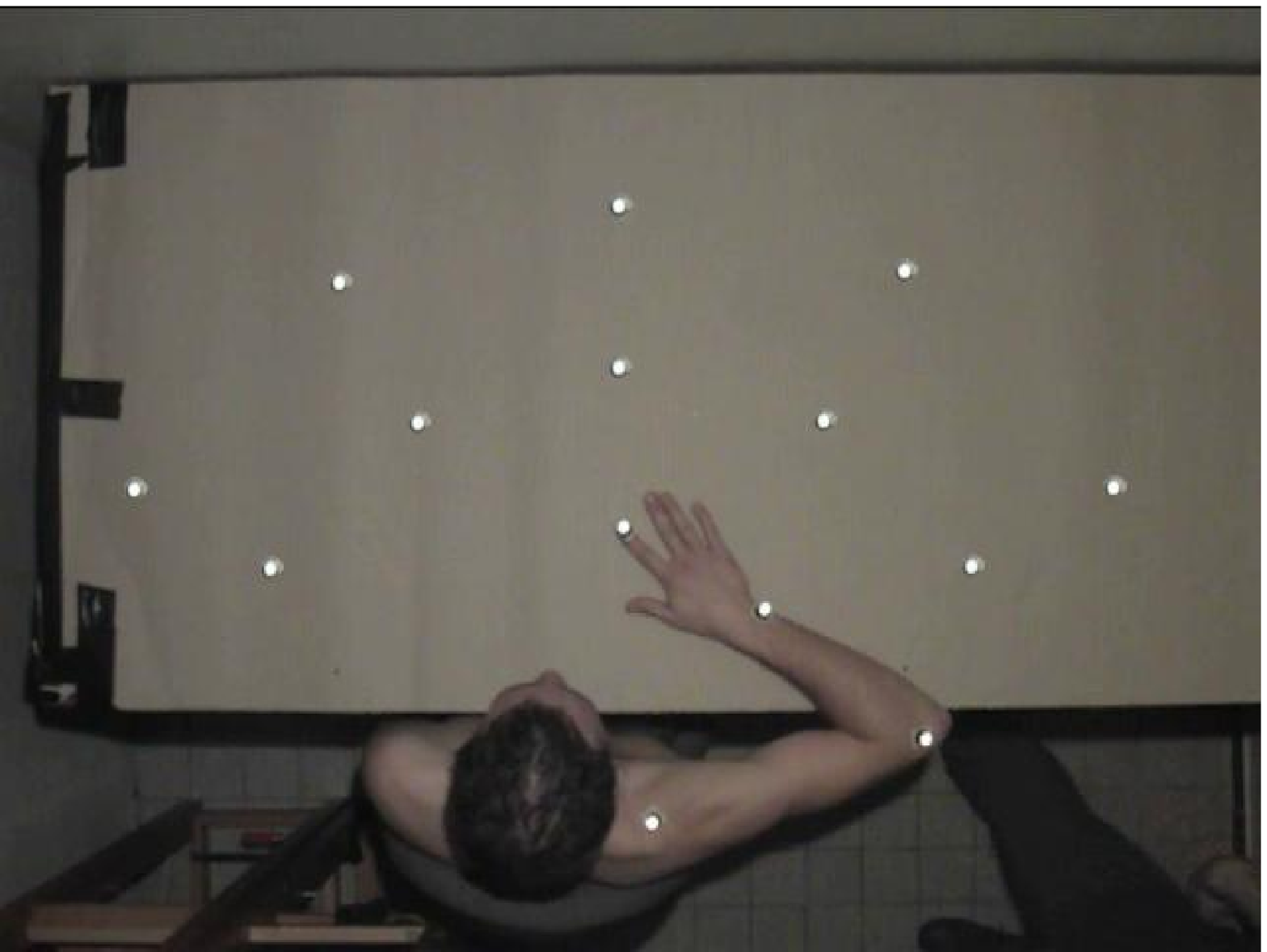, width=5.41 cm}}
%\subfigure[\label{fig100a} : Upper view of the task environment]
%{\includegraphics[bb=10 10 600 600,height=5 cm,width=5 cm]{photobras.jpg}}
%\qquad
%%% sous figure 2
\subfigure[\label{fig100b} : Targets (continuous line) and initial arm position (dashed line)]
{\epsfig{file=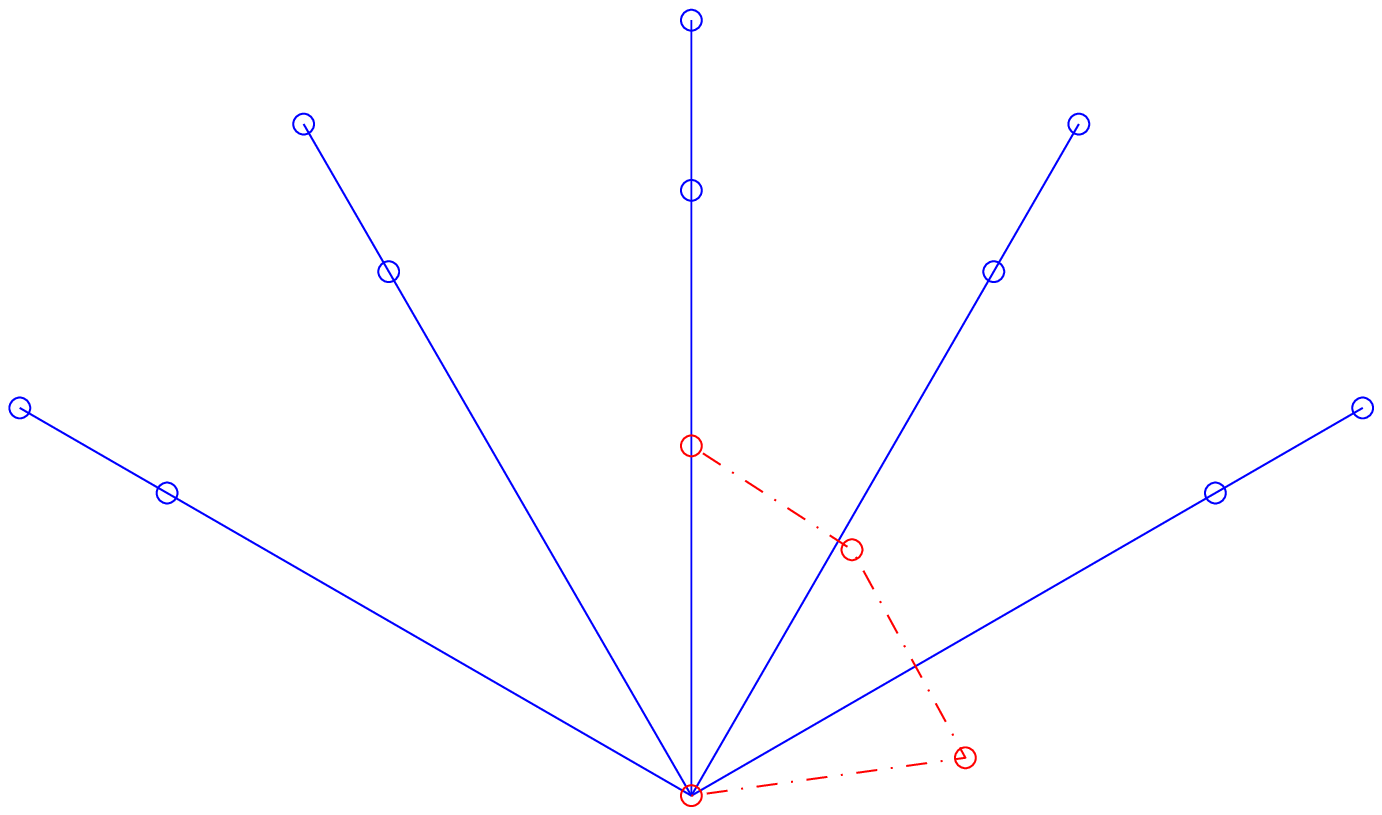, width=5 cm}}
\caption{\label{fig100} Pointing task experience.}
\end{figure}

\begin{figure}
\begin{center}
\epsfig{file=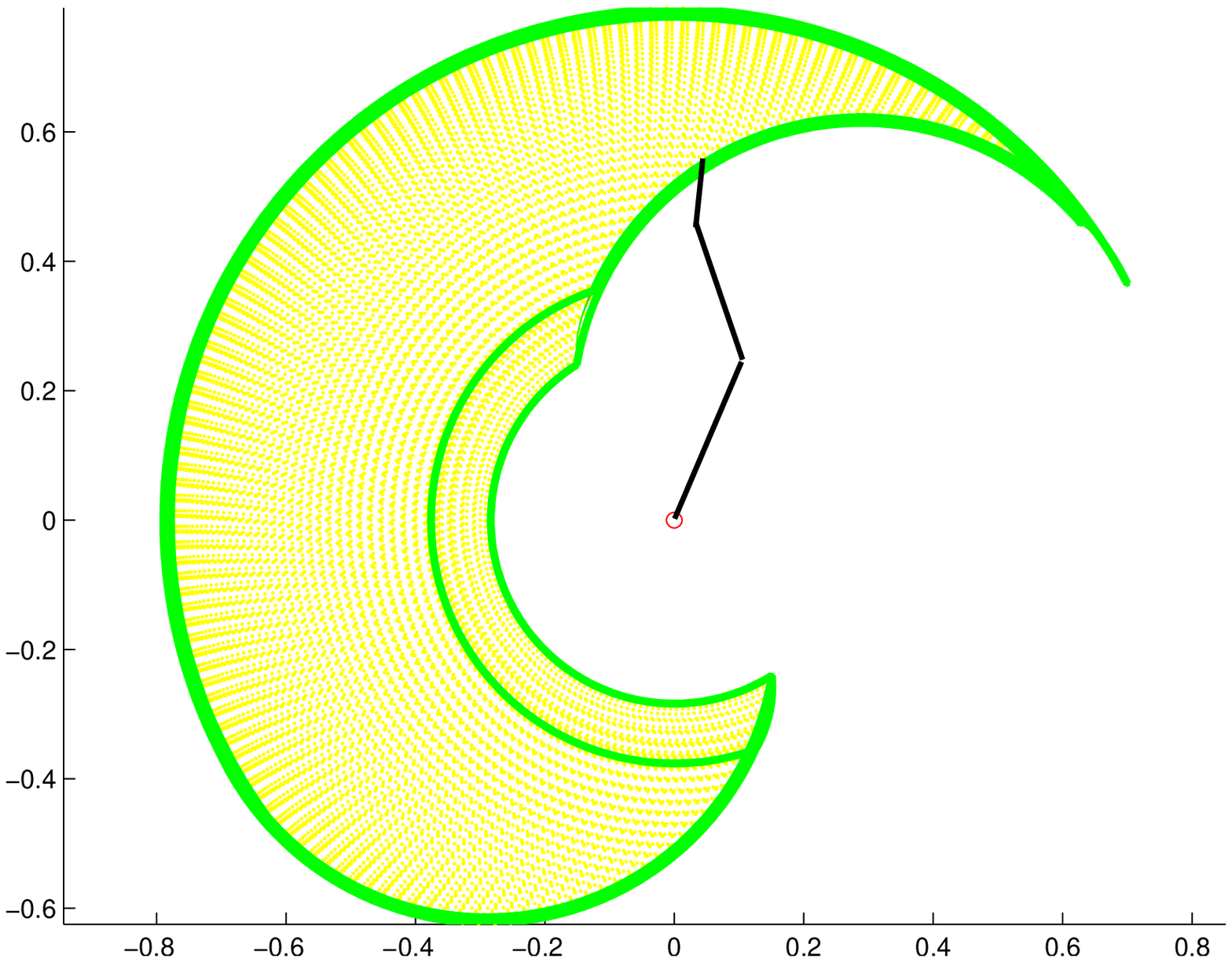, width= 8 cm}
\end{center}
\caption{\label{fig101}Upper limb workspace \cite{jbplkm2008}} 
\end{figure}

%%%%%%%%%%%%%%%%%%%%%%%%%%%%%%%%%%%%%%%%%%%%%%%%%%%%%%%%%%%%%%%%%%%%%%%%%%%
\subsubsection{Pointing task}\

\nombresujetttp\ right-handed male subjects (age: $24.9 \pm 2.42$, height: $177.6 \pm 5.83$ cm and mass: $68.8 \pm 8.18$ kg) were asked to perform pointing tasks in the horizontal plane. 
The total number of pointing tasks was \nombrettp.
Movements were performed for five directions and two distances (Fig \ref{fig100}). For each direction, two spherical targets were placed on a table at 60 and 80 cm from the shoulder. Directions of pointing task ranged regularly from $30^\text{o}$ to $150^\text{o}$ including pointing along the antero-posterior axis. The described position of the targets ensures that each of them is located inside the subjects workspace (Fig \ref{fig101}) when considering a 80 cm upper limb length and anthropometric data presented in \cite{jbplkm2008}. At the beginning of the movement, subjects had to position their arm so that the forefinger is located at 40 cm of the soulder in the antero-posterior direction. During the experiment, subjects sat on a chair whose height was adjusted so that the upper limb remained in the horizontal plane while moving over the table from starting point to targets and the trunk was immobilized by using straps. In order to ensure that the upper limb remained in the horizontal plane, the subjects were instructed to keep the upper limb lying on the table during the movements. Video reflective markers were placed on the subjects at the shoulder (acromion), elbow (olecrane), wrist (middle of radial and ulnar styloid processes) and forefinger extremity to allow further modeling of the upper limb. For each target, subjects performed three movements which were filmed at 25 Hz with a numeric camera JVC \copyright Everio placed above the subjects and oriented vertically. Raw experimental data, i.e. the position of the joints throughout the movement, were extracted from videographic recordings.

%%%%%%%%%%%%%%%%%%%%%%%%%%%%%%%%%%%%%%%%%%%%%%%%%%%%%%%%%%%%%%%%%%%%%%%%%%%
\subsubsection{Squat jump}\

The squat-jump data was obtained from a previous work \cite{SJ_JBYBLM_12}. Each of 13 other subjects performed 10 vertical jum\-ps. Instructions were given for keeping the hands on the hips during the movement to limit the contribution of the upper limbs to the performance. Furthermore, subjects were asked to do no countermovement. The jumps which did not meet both of these requirements were excluded from the study. In order to model the skeleton in a 4 rigid segments system, landmarks were placed on the left fifth metatarsophalangeal, lateral malleolus, lateral femoral epicondyle, greater trochanter and acromion. These landmarks define the foot, the shank, the thigh and the upper body (Head, Arms and Trunk: HAT). The subjects were filmed orthogonally to the sagittal plane at 100 Hz and the ground reaction force was recorded at 1000 Hz from an OR6-7-2000 AMTI force plate.
The center of mass (CoM) position of limbs was computed using anthropometric data \cite{Winter1990}. The whole body CoM (Center of Mass)
position was determined on the one hand from kinematic data and on the other hand from force plate measurements using a double numerical integration procedure. For the latter, subject mass, initial body CoM position and velocity had to be set. These values were computed so that the difference between CoM path obtained from kinetic and kinematic data was minimized in a least square sense. This optimization step was also used to synchronize both recording sources.

%%%%%%%%%%%%%%%%%%%%%%%%%%%%%%%%%%%%%%%%%%%%%%%%%%%%%%%%%%%%%%%%%%%%%%%%%%%
\subsection{Skeletal model}
\label{modele_articulaire}

\begin{figure}[ht]
\psfrag{i}{$\mathbf{i}$}
\psfrag{j}{$\mathbf{j}$}
\psfrag{O}[][l]{$A_0$}
\psfrag{l0}{$l_0$}
\psfrag{l1}{$l_1$}
\psfrag{l2}{$l_2$}
\psfrag{l3}{$l_{p-1}$}
\psfrag{A1}{$A_1$}
\psfrag{A2}{$A_2$}
\psfrag{A3}{$A_3$}
\psfrag{A4}{$A_{p}$}
\psfrag{p0}{$\theta_0$}
\psfrag{p1}{$\theta_1$}
\psfrag{t2}{$\theta_{2}$}
\psfrag{t3}{$\theta_{p-1}$}
\begin{center}
\epsfig{file=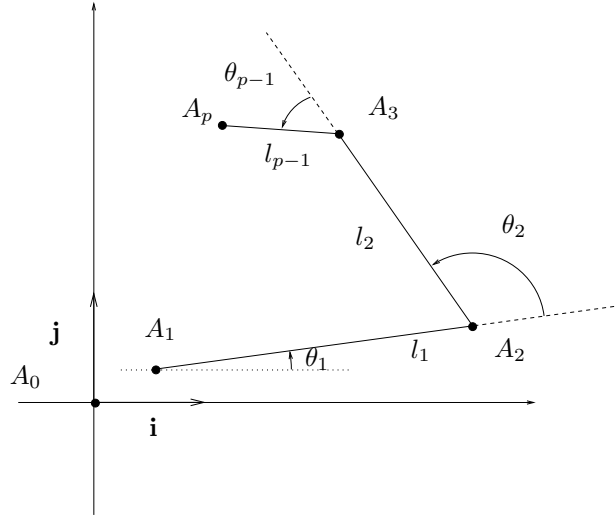, width=8 cm}
\end{center}
\caption{\label{fig111} Skeletal model.}
\end{figure}

For both tasks, the studied limbs were modeled as rigid bodies rotating around frictionless hinge joints. Given $p$ limbs, the joint positions are defined by the points $A_j \left(x_j,y_j \right)$ ($j \in \{1..,\hdots,p\}$) with $p=3$ and $p=4$ for pointing task and squat-jump respectively. From this definition, the position of the joints in the direct orthonormal reference frame $\left(O,\vec{i},\vec{j}\right)$ are related in the complex sense to limb lengths $l_j$ and angles $\theta_j$ (Fig \ref{fig111}) and the affix of $A_j$ is given by
\begin{equation}
\label{anglegeneral}
z_{A_j}=z_{A_1}+\sum_{n=1}^{j-1} l_n \exp\left(i\sum_{k=1}^n \theta_k\right)
\end{equation}
for all $j\in \{2,\hdots,p\}$, where $i$ is the imaginary unit and $z_{A_1}$ is the affix of $A_1$.

It should be noticed that $\theta_1$ and $\theta_j (j\in \{2,\hdots,p\})$ are segmental and joint angles respectively (Figure \ref{fig111}). According to these definitions, the velocities and other derivatives of the joint positions with respect to time can be computed from corresponding derivatives of $\theta_j$ and $A_1$.

\begin{remark}
\label{rem01}
The determination of the positions, velocities and accelerations of 
the points $A_2,\dots, A_p$ require that 
the coordinates and further derivatives of $A_1$ are known.
Three cases should be considered to calculate $x_1$ and  $y_1$:
\begin{itemize}
\item
the point $A_1$ is fixed; 
\item
the point  $A_1$ belongs to a simple curve such as a circle or a parabola;
\item
the center of mass of the subject and angle $\theta_j$, $j\in{1,..,p}$ are known.
\end{itemize}
\end{remark}

%%%%%%%%%%%%%%%%%%%%%%%%%%%%%%%%%%%%%%%%%%%%%%%%%%%%%%%%%%%%%%%%%%%%%%%%%%%
\subsection{Movement model}
\label{fonction_sigmoide}

Considering limbs as rigid bodies, the relative positions of joints are directly related to the angles $\theta_j$ according to \eqref{anglegeneral}.

The evolution of the angles throughout the movement was modeled using three types of sigmoid shaped curves.
The parameters of the sigmoids were computed using an optimization procedure so that the experimental and modeled markers trajectories are closest from each other in the least square sense.
Theoretical results of this section 
have been already partially given in \cite{creveaux09}
and will be presented extensively in a future work
\cite{bastiencreveauxencours12}. We recall that 
to allow dynamic continuity, the required solutions should be models of class $C^2$ at least,
defined on $[0,1]$ and satisfying \eqref{eqan40tot}.

%%%%%%%%%%%%%%%%%%%%%%%%%%%%%%%%%%%%%%%%%%%%%%%%%%%%%%%%%%%%%%%%%%%%%%%%%%%
\subsection{Data processing}

Experimental pointing task and jumping data were modeled using the sigmoid models.
Specific procedures are described below.

%%%%%%%%%%%%%%%%%%%%%%%%%%%%%%%%%%%%%%%%%%%%%%%%%%%%%%%%%%%%%%%%%%%%%%%%%%%
\subsubsection{Shoulder path}\

% toutes figures préparées avec
% dans \programmes\matlab\recherche\CRIS\tache pointage plane\utilisation C2 Cinf et Cinf TC
% lancer figure_pour_article_tpp('S8P2S2','résultats','resfond01','parametrefinaux01');

\begin{figure}[h]
\psfrag{nuage de point}{{\tiny Experimental}}
\psfrag{cercle}{{\tiny Fitted circle}}
\begin{center}
\epsfig{file=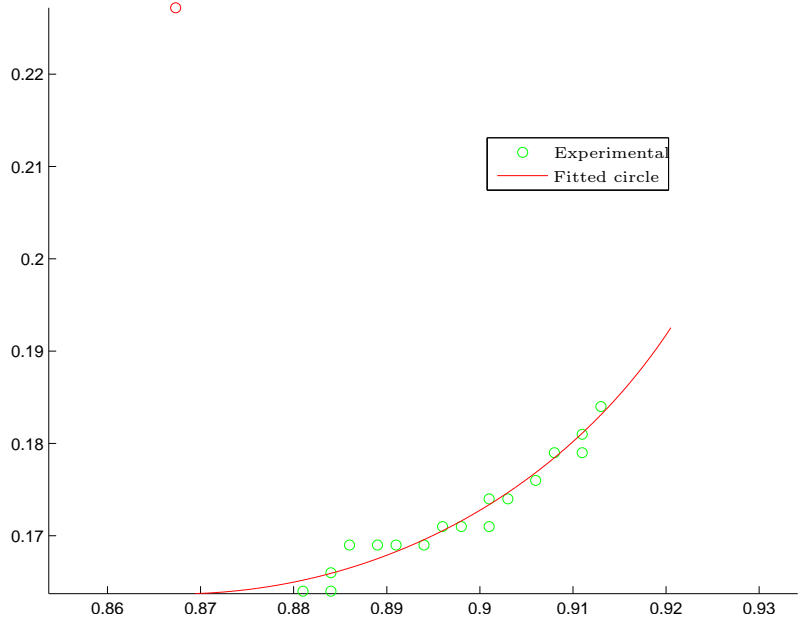, width=10.5  cm}
\end{center}
\caption{\label{fig200} Shoulder trajectory: Experimental points and fitted circle.}
\end{figure}

Experimental data obtained from pointing tasks show that the shoulder joint path is well fitted by an arc of circle for all subjects (Fig. \ref{fig200}).
This movement was observed in spite of trunk immobilization.
Since shoulder is a patella type articulation, this result is not surprising.
The characteristics of the circle can be determined using a least squares method.
The center $\Omega$ and radius $R$ of the circle minimizing the distance sum of squares between experimental data  $(x_k,y_k)$ and theoretical data  $(X_k,Y_k)$ were researched. Firstly, a direct method was used to minimize the sum of 
\begin{equation*}
S={\left((x_k-X_k)^2+(y_k-Y_k)^2-R^2\right)}^2.
\end{equation*}
Secondly, the sum 
\begin{equation*}
S'=\sqrt{(x_k-X_k)^2+(y_k-Y_k)^2}
\end{equation*}
was minimized by using an iterative method. For this method, the results of the direct optimization were used as initial values for $\Omega$ and $R$. This final optimization was performed with the library \emph{M\MakeLowercase{ATLAB LEAST SQUARES GEOMETRIC ELEMENT SOFTWARE}},
available at 
{\footnotesize{\url{http://www.eurometros.org/gen_report.php?category=distributions&pkey=14}}}.
%\path|http://www.eurometros.org/gen_report.php?category=distributions&pkey=14|.
%\path|http://www.eurometros.org/gen_report.php?category=|\linebreak\path|distributions&pkey=14|.
For more details, see 
\cite{razet97,razet98,Ahnetali01,Endoetali07,MR1099915,MR1132365,MR1011392}.

We set $A_0=\Omega$, $l_0=R$ and 
we consider then angle $\theta_0$  defined by:
\begin{equation}
\label{sjeq10}
\theta_0=\widehat{\left(\mathbf{i}, \overrightarrow{A_0A_{1}}\right)} \in (-\pi,\pi].
\end{equation}
We assume then that $l_0$ and $A_0$ are constant and we obtain 
\begin{equation}
\label{sjeq20tot}
z_{A_1}=z_{A_0}+  l_0 \exp(i\theta_0).
\end{equation}
where $x_0$, $y_0$ and $l_0$ are known.
We add then Eq. \eqref{anglegeneral}; thus, assumption of remark \ref{rem01} holds.

The mechanical system is plotted on Fig. \ref{fig111}; the coordinates of $A_0$ and the lengths $l_0$, $l_1$, $l_2$ and $l_3$ are constant and the experimental data are then angle $\theta_0$, $\theta_1$, $\theta_2$ and $\theta_3$.

%%%%%%%%%%%%%%%%%%%%%%%%%%%%%%%%%%%%%%%%%%%%%%%%%%%%%%%%%%%%%%%%%%%%%%%%%%%
\subsubsection{Determination of sigmoid parameters}\
\label{determsigmoparam}

Sigmoid parameters were obtained from a multi-stage optimization procedure.
First, $t_{\text{b}},t_{\text{e}},\theta_{\text{b}},\theta_{\text{e}},\alpha,\beta$ and $\kappa$ were estimated from experimental data.
The scale parameters were defined so that the absolute peak angle velocity occurs between $t_{\text{b}}$ and $t_{\text{e}}$ and the angle velocity sign changes at the endpoints of this interval.
Considering $\theta_{\text{b}}$ and $\theta_{\text{e}}$ as the angle values at $t_{\text{b}}$ and $t_{\text{e}}$, the shape parameters $\alpha,\beta$ and $\kappa$ were determined thanks to Eq. \eqref{eq109tot}.

First optimization consisted in minimizing the sum of square of differences between experimental angles $\theta^i$ and those obtained from the sigmoid models $\sigma^i$ for each of $n$ instants:
\begin{equation*}
\mathcal{S}=\sum_{i=1}^n {\left(\theta^i-\sigma^i\right)}^2,
\end{equation*}
The optimization was achieved for each sigmoid with the \path|lsqcurvefit| function provided in Matlab software.
Initial values of parameters were set from estimations of experimental data described previously. This optimization stage will be further referred to as local optimization.

Secondly, differences between experimental and mo\-del reconstructed joint positions were minimized in a least square sense.
The objective can thus be written as 
\begin{equation*}
\mathcal{S}'=\sum_{j=1}^p \sum_{i=1}^n {\left(X_j^i-x_j^i\right)}^2+{\left(Y_j^i-y_j^i\right)}^2.
\end{equation*}

Compared to the previous stage, this optimization can be considered as global since for the latter, the parameters of the $p$ sigmoids were determined simultaneously.
Computation of model-based joint positions implies the lengths of the limbs to be provided.
For the pointing tasks, the optimization was performed using (i) mean experimental limb lengths (semi-global optimization) and (ii) limb lengths as model parameters (global optimization).

%%%%%%%%%%%%%%%%%%%%%%%%%%%%%%%%%%%%%%%%%%%%%%%%%%%%%%%%%%%%%%%%%%%%%%%%%%%
\subsubsection{Squat-jump specific procedure}\

The modeling of the jump focused on the position of the joints in a reference frame located at the distal extremity of the foot. Thus, the optimization consisted in fitting the experimental joint positions of ankle, knee, hip and shoulder with the model parameters in this reference frame. Since joints do not remain fully extended after the take-off, differences were not taken into account during the whole movement. This prevented the model from underestimating the necessary amplitude of joint extensions. Therefore, differences between experimental and model-based data were considered during the intervals corresponding to increase of vertical joint coordinates in the given reference frame (e.g. the error at the ankle joint was only taken into account while the vertical distance between the knee and the foot extremity increased).
The objective $\mathcal{S}'$ was used to achieve this optimization stage, which will be further related as kinematic.

Second stage of optimization included non-linear constraints on position, velocity and acceleration of the body CoM computed from sigmoid model. It was imposed that the body CoM position computed from both the sigmoid model and the force plate data were similar at the instant $t_1$ for which the marker located on the distal extremity of the foot started to move upward. At this instant, equality for the coordinates of both velocity and acceleration of body CoM obtained from kinetic and kinematic data was also required. Finally, body CoM vertical acceleration was constrained to be greater than -9.81 m.s$^{-2}$ before $t_1$ ensuring that take-off occurs necessarily after $t_1$. From $t_1$ to the end of the jump, the movement of $A_0$ was set so that kinetic and kinematic-based movement of the CoM were similar. This results in a continuous characterization of the movement position, velocity and acceleration. It should be noticed that using similar constraints for jerk and further derivatives could have led to description of class $C^3$ and higher.

%%%%%%%%%%%%%%%%%%%%%%%%%%%%%%%%%%%%%%%%%%%%%%%%%%%%%%%%%%%%%%%%%%%%%%%%%%%
\subsection{Modeling accuracy}

For each instant $i$ and joint $j$, the optimization accuracy can be quantified by the difference between experimental data ($x_j^i$ and $y_j^i$) and sigmoid modeled data ($X_j^i$ and $Y_j^i$):
\begin{subequations}
\begin{equation} 
\label{erreurcoordind} 
\varepsilon_{i,j}=\sqrt{\left(X_j^i-x_j^i\right)^2+\left(Y_j^i-y_j^i\right)^2}.
\end{equation}
In further analysis, maximal $\varepsilon_{\max}$ and mean $\varepsilon_{\text{mean}}$ values of these differences are used to account for the fitting accuracy of the modeling procedures:
\begin{align}
\label{erreurcoordmax}
& \varepsilon_{\max}=\max_{i,j} \varepsilon_{i,j},\\
\label{erreurcoordmean}
& \varepsilon_{\text{mean}}=\underset{{i,j}}{{\text{mean}}} \,\varepsilon_{i,j}.
\end{align}
\end{subequations}

%%%%%%%%%%%%%%%%%%%%%%%%%%%%%%%%%%%%%%%%%%%%%%%%%%%%%%%%%%%%%%%%%%%%%%%%%%%
\subsection{Statistical analysis}

For both maximal and mean errors given in \eqref{erreurcoordmax} and \eqref{erreurcoordmean}, the Shapiro-Wilk test reported unnormal distributions.
Thus, statistical tests were realized on normally distributed $\log_{10}$ of observations (i.e., errors and computation time).
Firstly, anovas for repeated measures were performed for errors and computation time.
When anovas reported significant results, post-hoc tests were performed to check for differences between the sigmoid models and the optimization procedures.
All the tests were realized with \Rlogo\ \cite{R} and statistical significance was set at $95 \%$ confidence level, i.e. $p<0.05$.

%%%%%%%%%%%%%%%%%%%%%%%%%%%%%%%%%%%%%%%%%%%%%%%%%%%%%%%%%%%%%%%%%%%%%%%%%%%
%%%%%%%%%%%%%%%%%%%%%%%%%%%%%%%%%%%%%%%%%%%%%%%%%%%%%%%%%%%%%%%%%%%%%%%%%%%
\section{Results}

\begin{figure}[ht]
\psfrag{Angle 0 (en degrés) de S8P2S2}{}
\psfrag{Angle 1 (en degrés) de S8P2S2}{}
\psfrag{Angle 2 (en degrés) de S8P2S2}{}
\psfrag{Angle 3 (en degrés) de S8P2S2}{}
\centering
%%% sous figure 1
\subfigure[\label{220a}: Angle 0]
{\epsfig{file=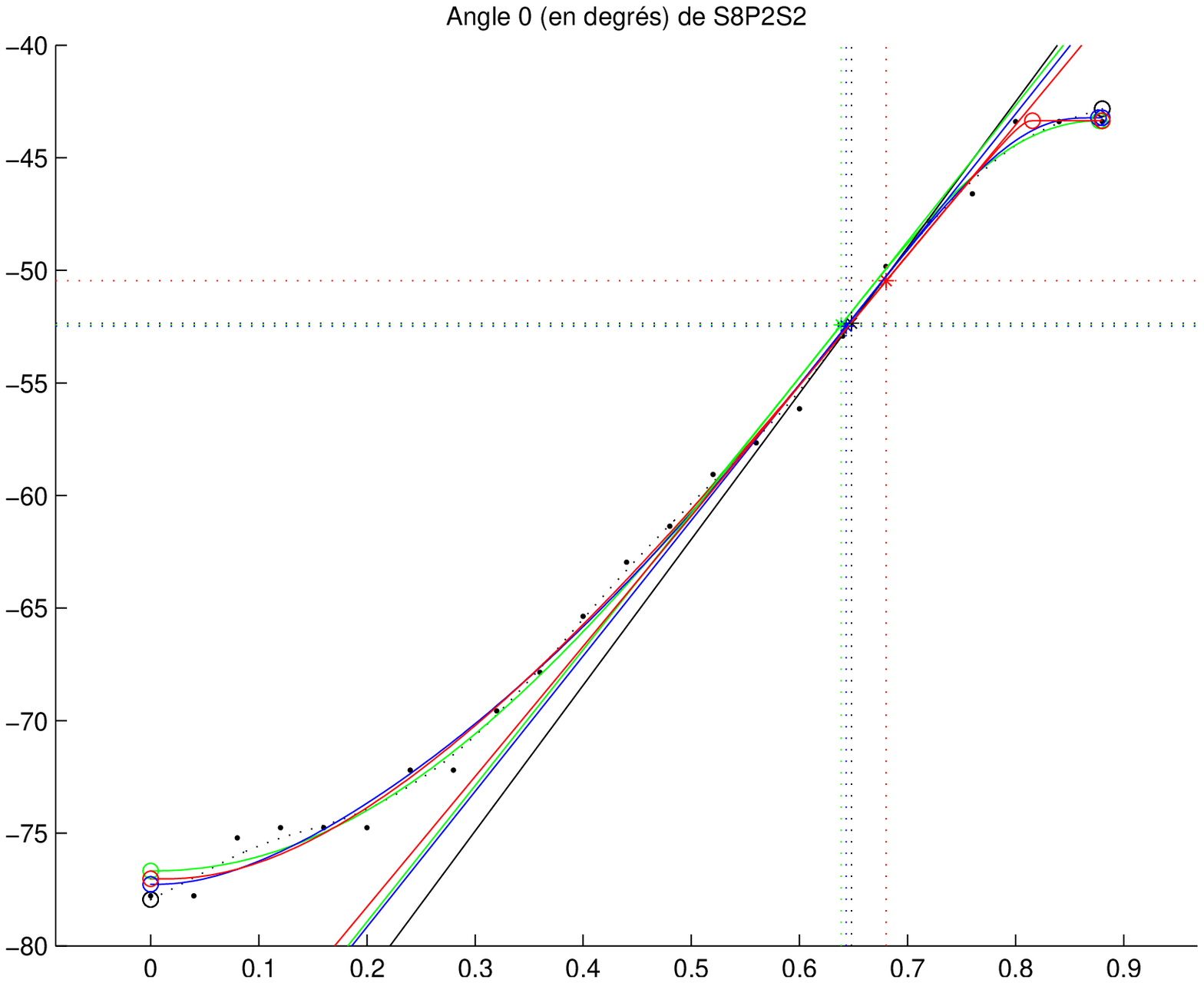, width=6 cm}}
%%% sous figure 2
\subfigure[\label{220b}: Angle 1]
{\epsfig{file=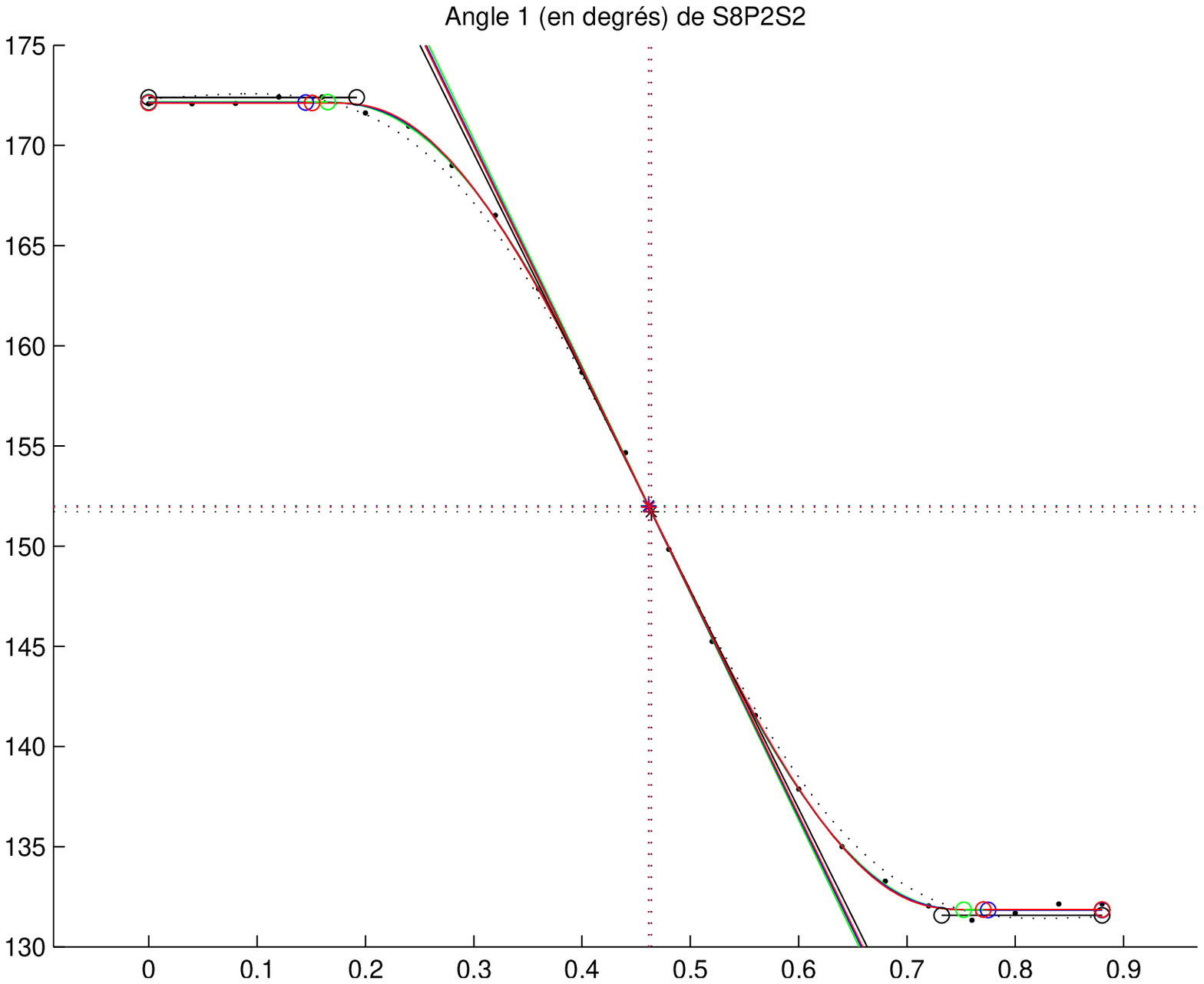, width=6 cm}}
%%% sous figure 3
\subfigure[\label{220c}: Angle 2]
{\epsfig{file=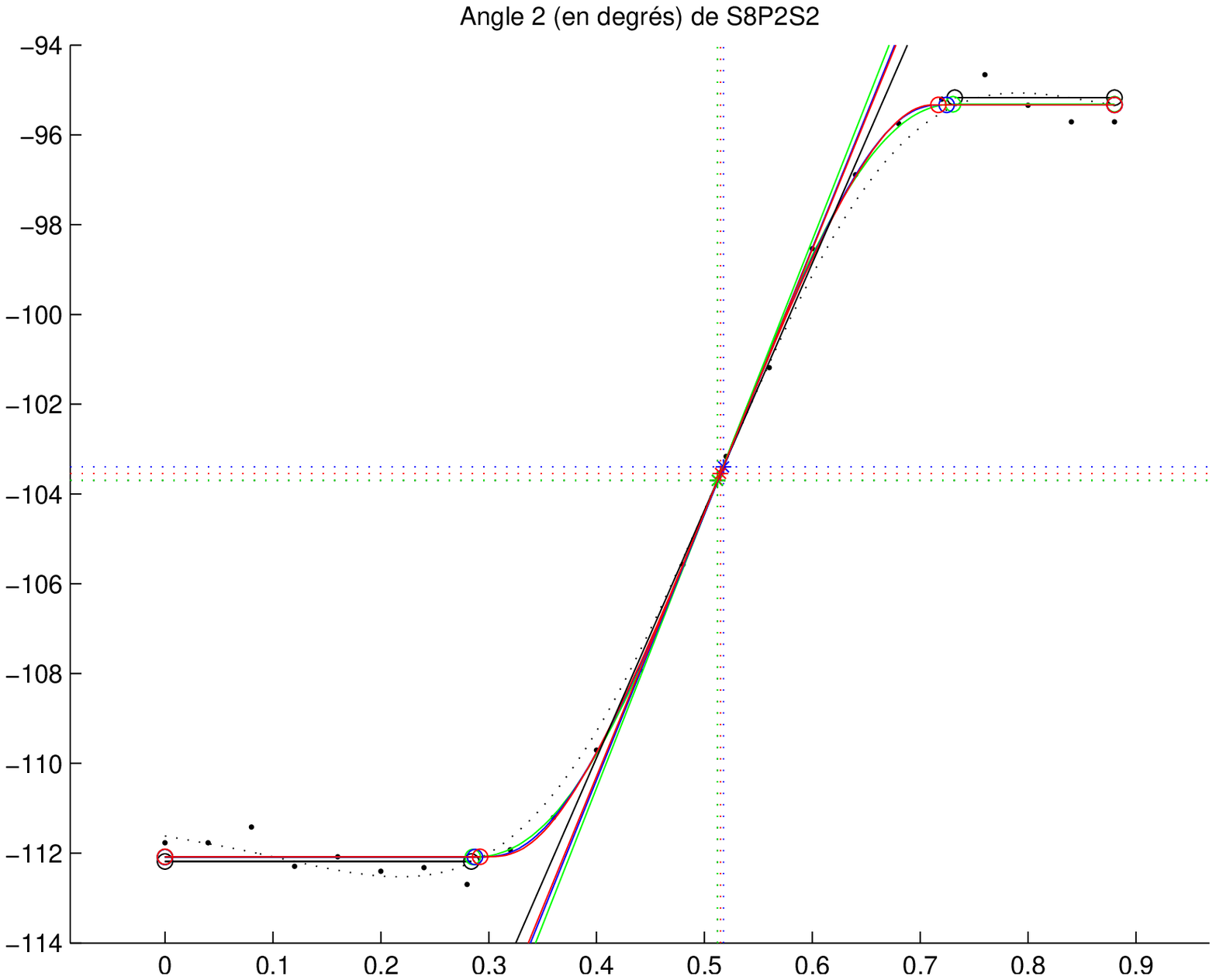, width=6 cm}}
%%% sous figure 4
\subfigure[\label{220d}: Angle 3]
{\epsfig{file=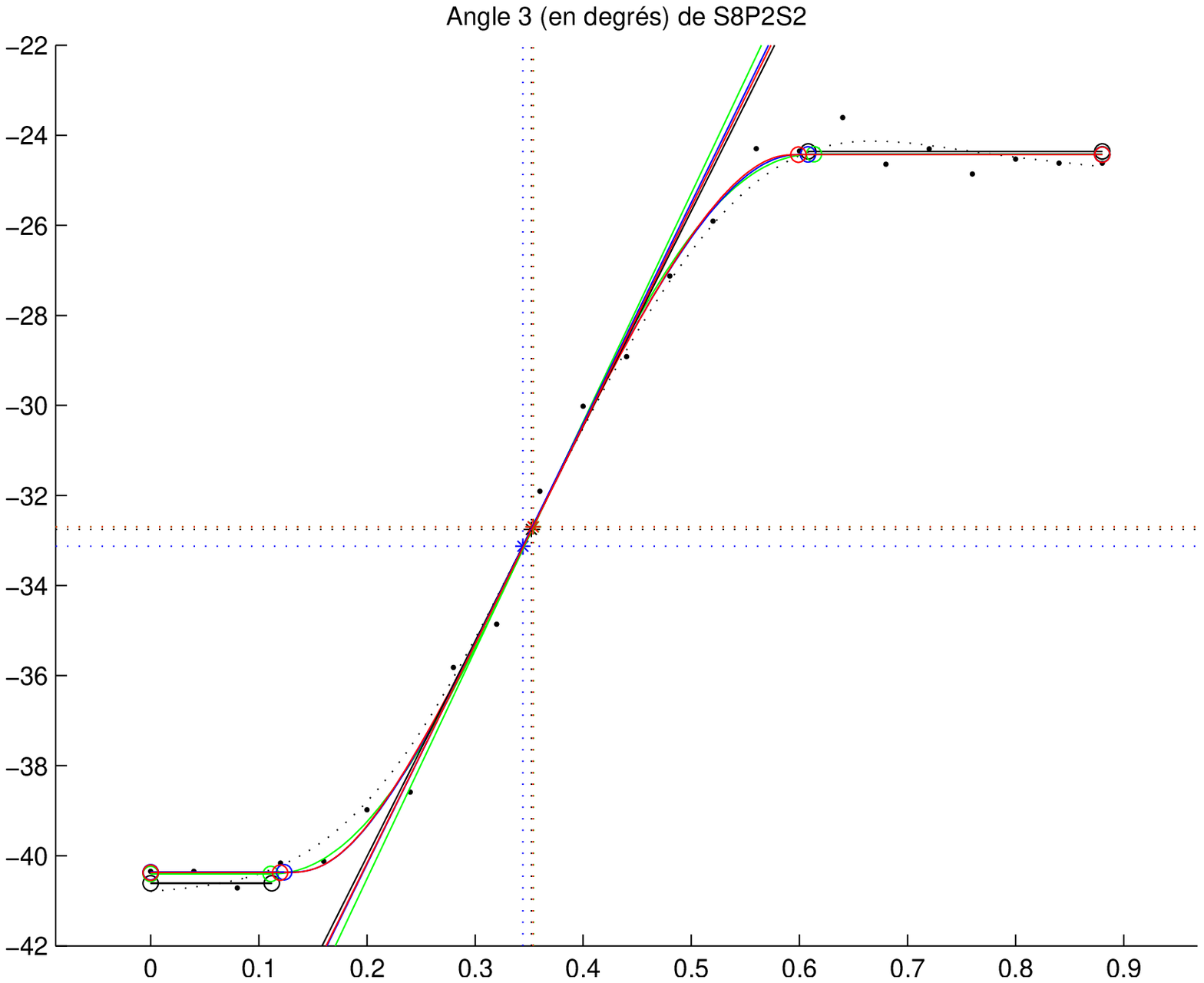, width=6 cm}}
\caption{\label{220tot} Angles (in degrees)  according time. Experimental data $\theta_j^i$ are plotted by black points,
INVEXP sigmoid model is  plotted in red continuous line, 
NORM sigmoid model is plotted in blue continuous line, 
SYM sigmoid model is plotted in green continuous line. 
On the figure, each points
of coordinates $\left(t_{\text{b}}^{(j)},\theta_{\text{b}}^{(j)}\right)$ and $\left(t_{\text{e}}^{(j)},\theta_{\text{e}}^{(j)}\right)$
are plotted by a circle. 
We add also the tangent of the curves in $\theta_{0}^{(j)}$.}
\end{figure}

\begin{figure}[ht]
\centering
%%% sous figure 1
\subfigure[\label{230a}: Angle 0]
{\epsfig{file=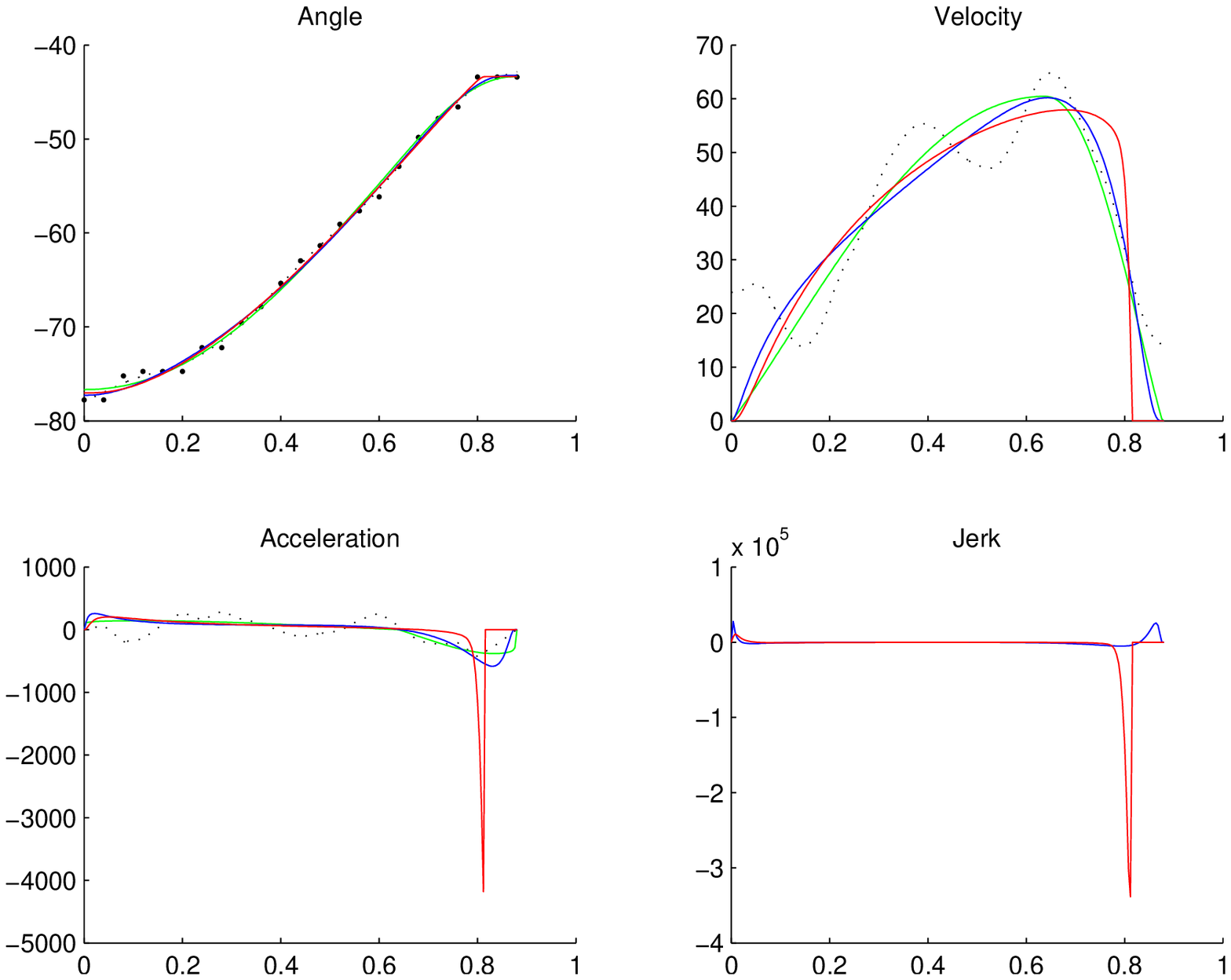, width=6 cm}}
%%% sous figure 2
\subfigure[\label{230b}: Angle 1]
{\epsfig{file=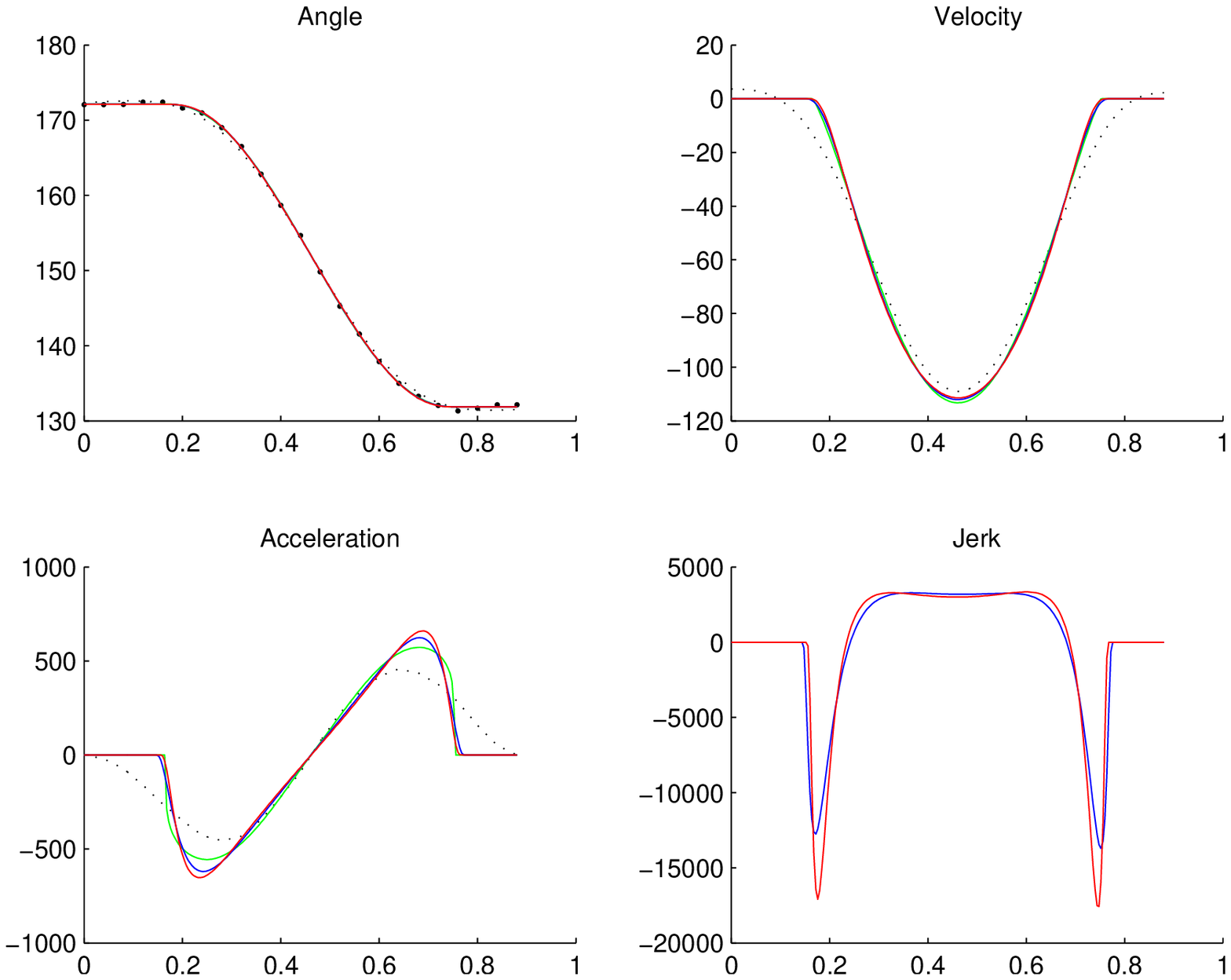, width=6 cm}}
%%% sous figure 3
\subfigure[\label{230c}: Angle 2]
{\epsfig{file=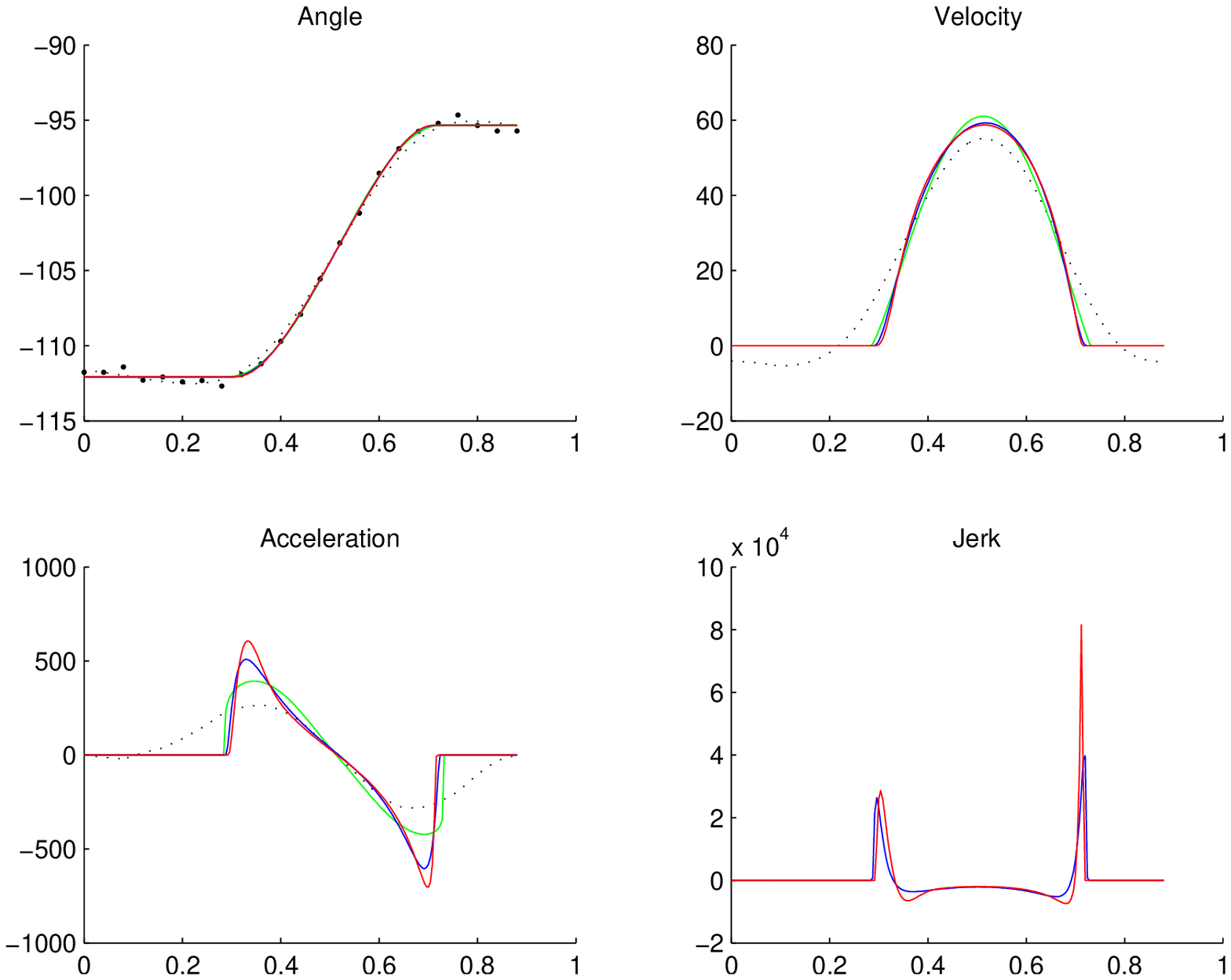, width=6 cm}}
%%% sous figure 4
\subfigure[\label{230d}: Angle 3]
{\epsfig{file=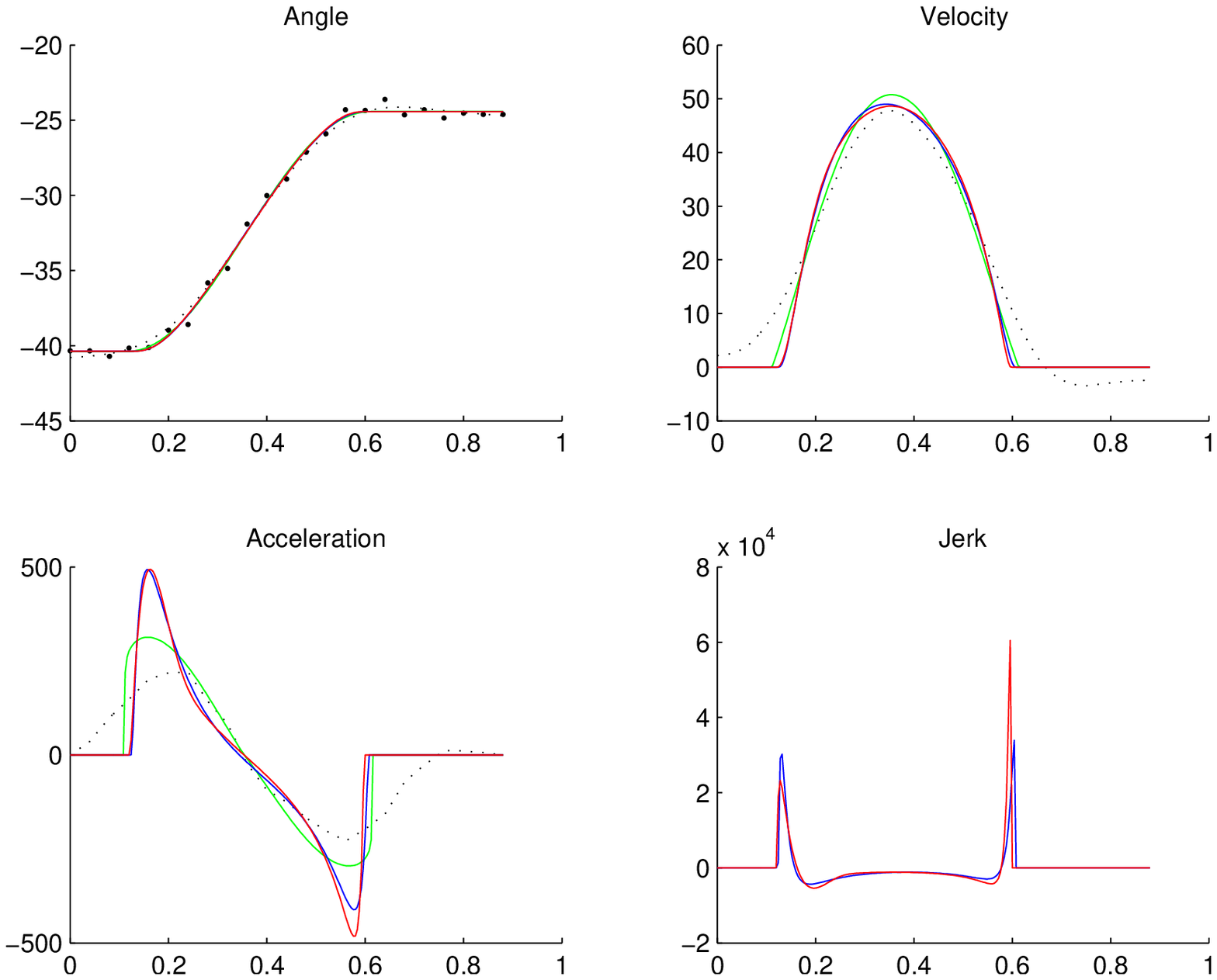, width=6 cm}}
\caption{\label{230tot} Angles (in degrees)  and angular velocity, acceleration and jerk, for optimization for each angle.
Derivatives of order 1 to 3 have been calculated by using analytical expression of displacements expressed with sigmoid laws.
Experimental data $\theta_j^i $ are plotted by black points,
Smoothing data $\widehat \theta_j$ are plotted by dashed black line, 
INVEXP sigmoid model is  plotted in red continuous line, 
NORM sigmoid model is plotted in blue continuous line, 
SYM sigmoid model is plotted in green continuous line.}
\end{figure}

\begin{figure}[ht]
\centering
%%% sous figure 1
\subfigure[\label{230aglob}: Angle 0]
{\epsfig{file=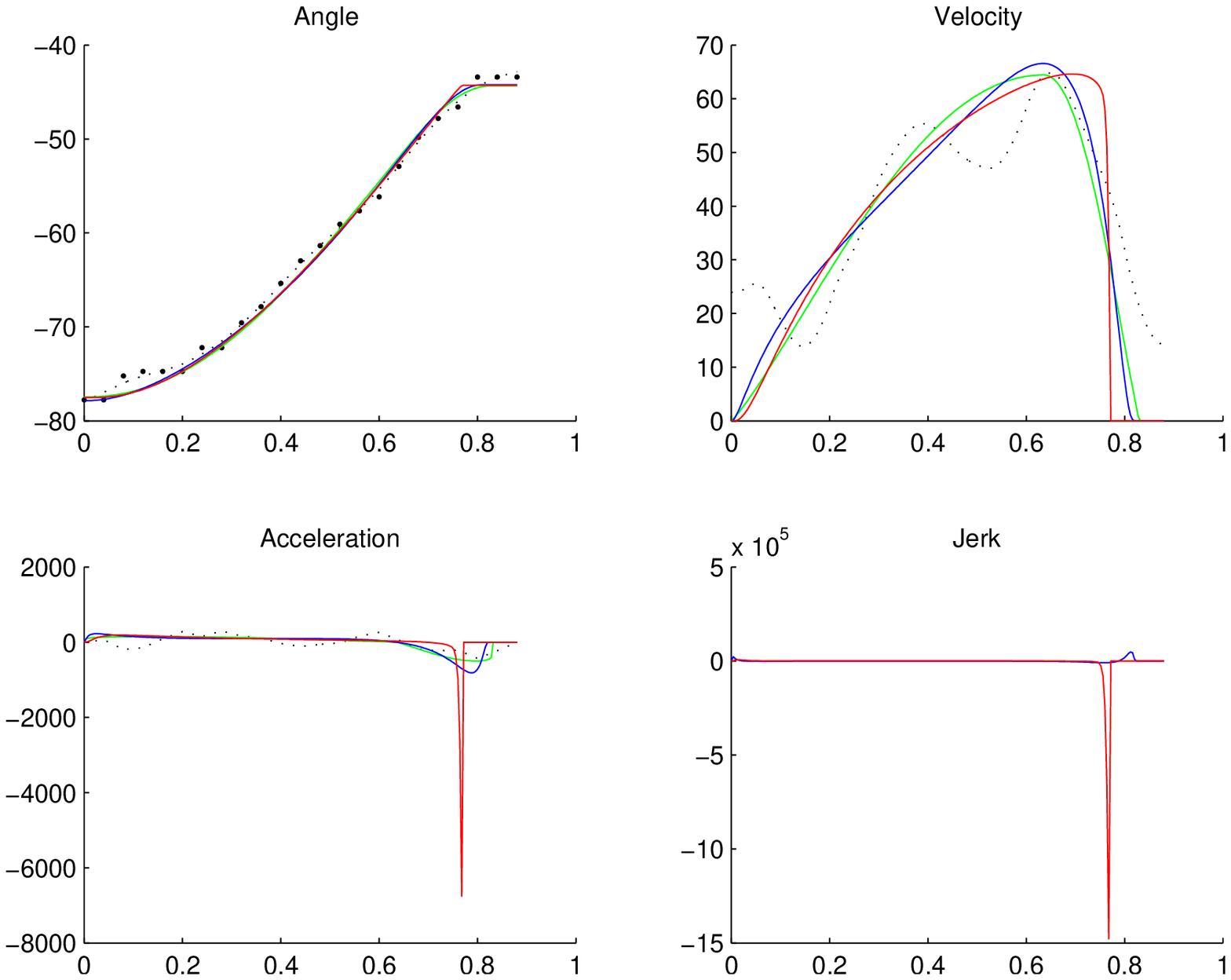, width=6 cm}}
%%% sous figure 2
\subfigure[\label{230bglob}: Angle 1]
{\epsfig{file=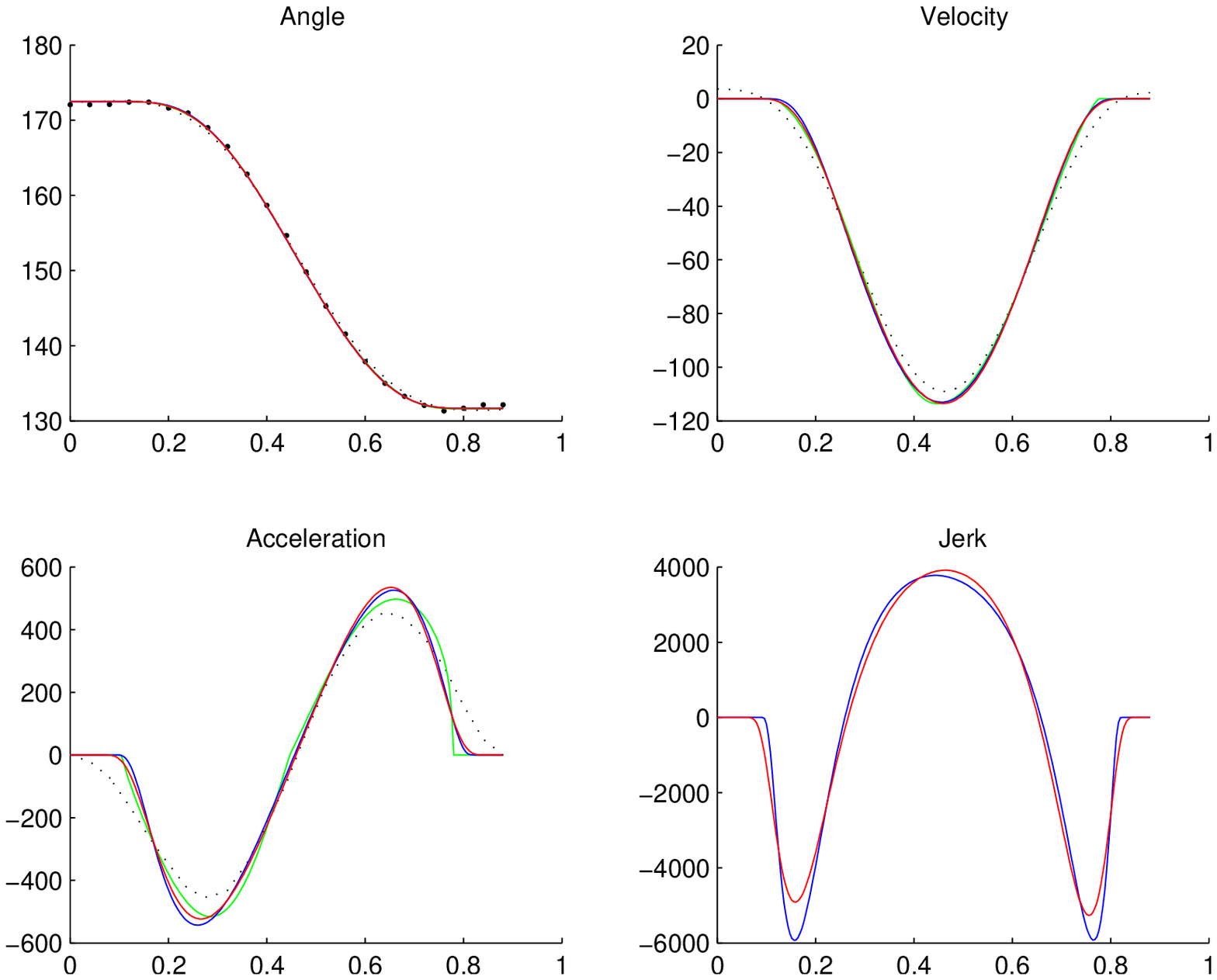, width=6 cm}}
%%% sous figure 3
\subfigure[\label{230cglob}: Angle 2]
{\epsfig{file=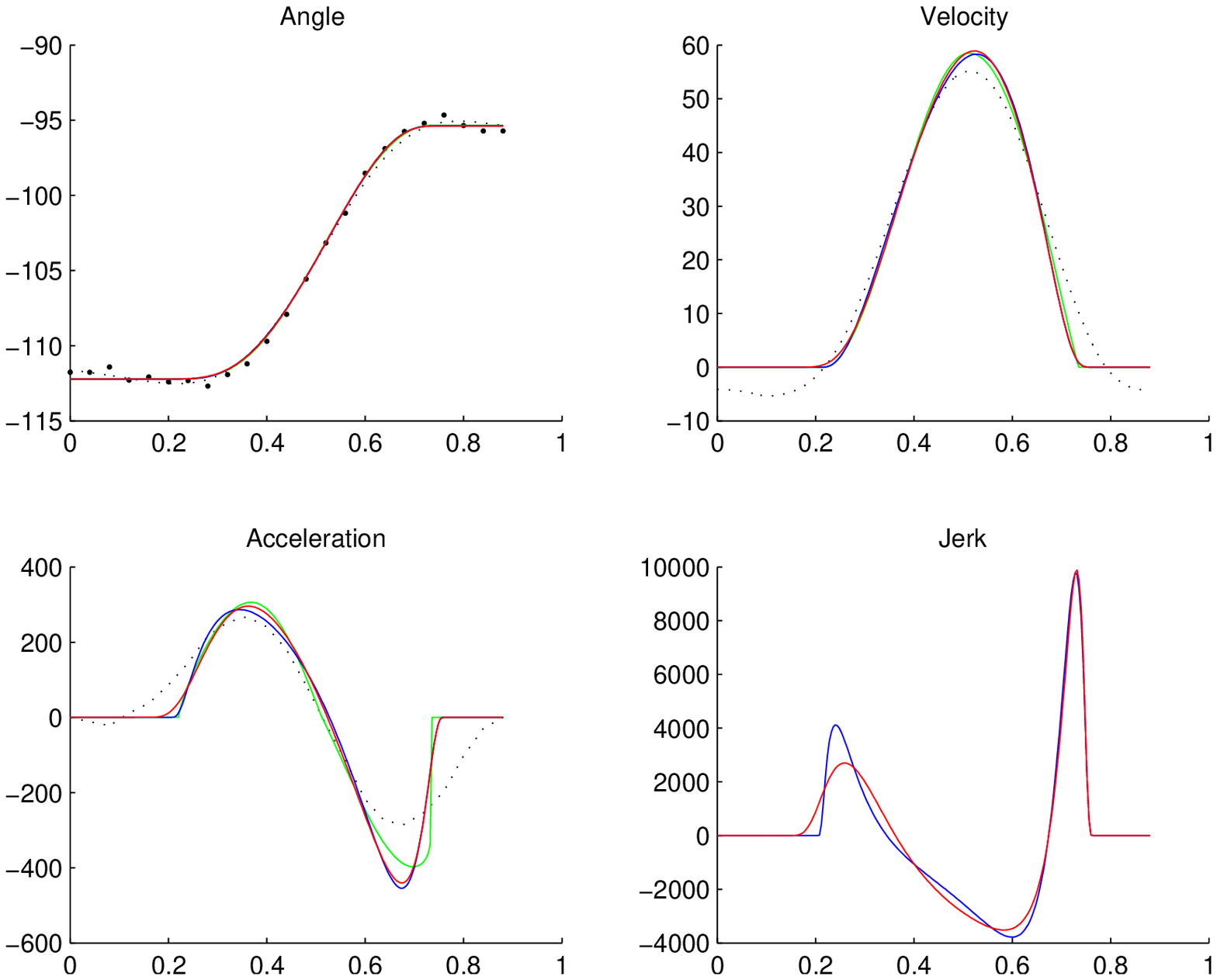, width=6 cm}}
%%% sous figure 4
\subfigure[\label{230dglob}: Angle 3]
{\epsfig{file=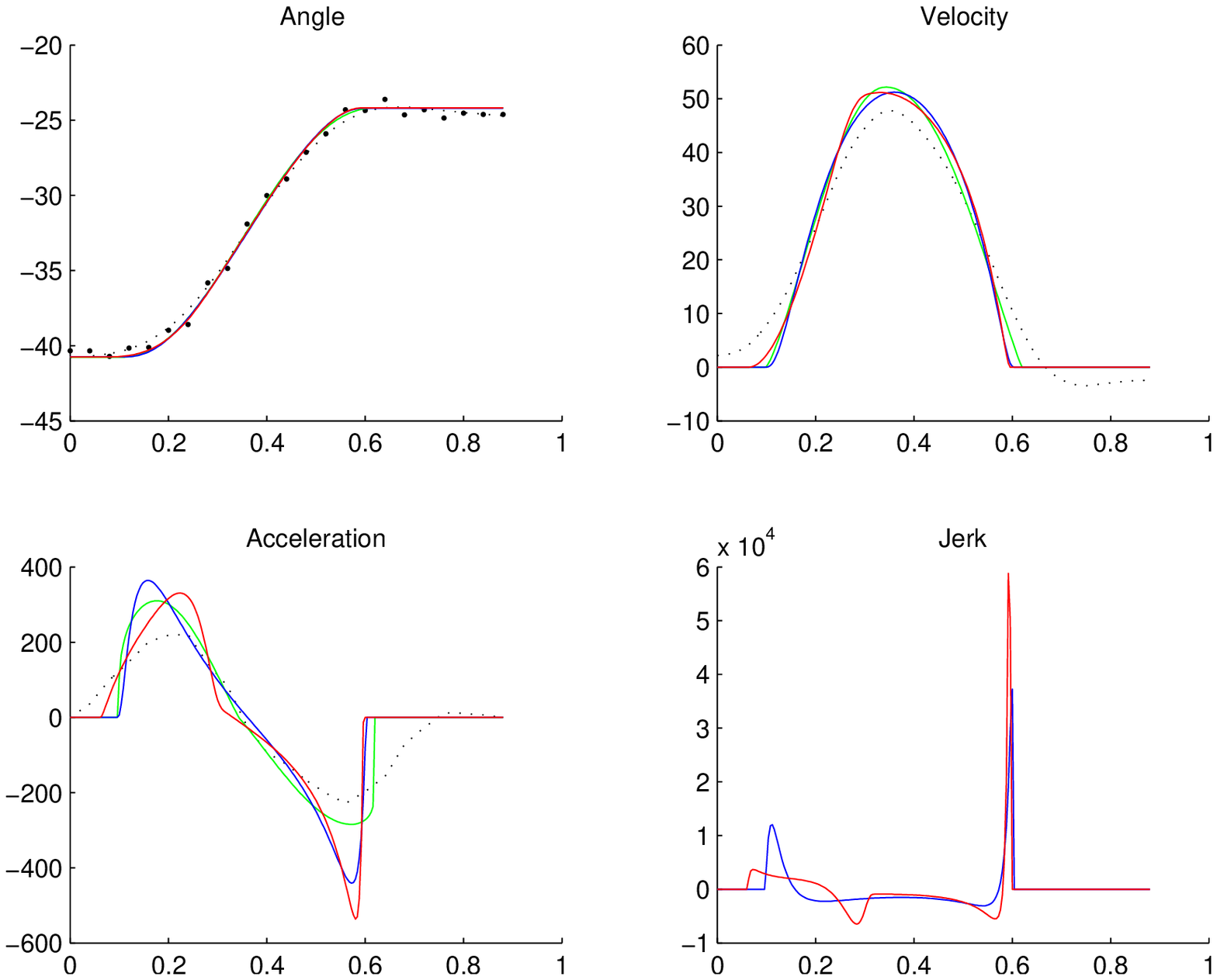, width=6 cm}}
\caption{\label{235tot} Angles (in degrees) and angular velocity, acceleration and jerk, for global optimization.
Derivatives of order 1 to 3 have been calculated by using analytical expression of displacements expressed with sigmoid laws.
Experimental data $\theta_j^i $ are plotted by black points,
Smoothing data $\widehat \theta_j$ are plotted by dashed black line, 
INVEXP sigmoid model is  plotted in red continuous line, 
NORM sigmoid model is plotted in blue continuous line, 
SYM sigmoid model is plotted in green continuous line.}
\end{figure}

\begin{figure}[ht]
\begin{center}
\epsfig{file=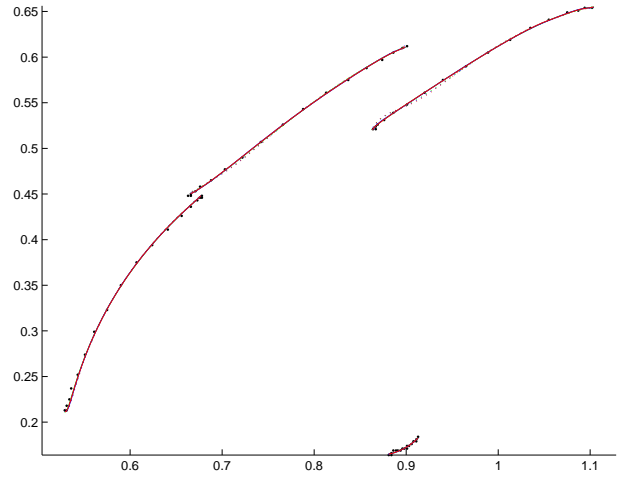, width=8 cm}
\end{center}
\caption{\label{240tot}%
Trajectories of points $A_1$, $A_2$, $A_3$ and $A_4$.
INVEXP sigmoid model is  plotted in red continuous line (for global optimization) and red dashed line (for optimization for each angle).
NORM sigmoid model is  plotted in blue continuous line (for global optimization) and blue  dashed line (for optimization for each angle).
SYM sigmoid model is  plotted in green continuous line (for global optimization) and green dashed line (for optimization for each angle).}
\end{figure}

The results obtained from the \nombrettp\ pointing tasks and 120 squat-jumps are summarized in the tables 
\ref{tab01tp} and \ref{tab10tp} for pointing task
and \ref{tab01sj} and \ref{tab10sj}    for squat Jumps
given in the appendix \ref{ensemble_tableau}.
Examples of experimental and modeled data are provided for the pointing tasks and squat-jumps in the figures \ref{220tot} to \ref{240tot} and \ref{SJFig1} to \ref{SJFig4} respectively. Plots of velocities, accelerations and jerks of modeled data were obtained from analytical derivation of sigmoid models with respect to time.

%%%%%%%%%%%%%%%%%%%%%%%%%%%%%%%%%%%%%%%%%%%%%%%%%%%%%%%%%%%%%%%%%%%%%%%%%%%
%%%%%%%%%%%%%%%%%%%%%%%%%%%%%%%%%%%%%%%%%%%%%%%%%%%%%%%%%%%%%%%%%%%%%%%%%%%
%%%%% DEBUT INSERTION fichier crée par Sweave sur  meilleurs_resultats
%\input{meilleurs_resultats} 
% ATTENTION,% ATTENTION,% ATTENTION,% ATTENTION,
% ATTENTION, pour que ce fichier rnw fonctionne avec Sweave, il est nécessaire
% de faire Sweave sur statistique_finale juste avant, sans taper 
% rm(list=ls()), pour que les data frame dataframeSJ et dataframeTP soient connus !
% taper donc pour une compilation globale des trois fichiers rnw 
%  rm(list=ls());Sweave("statistique_finale.rnw");Sweave("ensemble_tableaux_stats.rnw");Sweave("meilleurs_resultats.rnw");

The obtained results are very accurate from  numerical viewpoint.
Indeed, for pointing task, in 
95\% of cases, for the three models,
the maximal error (defined by \eqref{erreurcoordmax}, \textit{i.e.} the difference between the calculated trajectories and the experimental trajectories for local 
method and  is smaller than 
\begin{subequations}
\label{meilleuresultp}
\begin{align}
&\varepsilon_{\max}=
2.894 \text{ cm},
\intertext{and for semi-gobal or global error,
and it is 
smaller than}
&\varepsilon_{\max}=
1.86\text{ cm}.
\intertext{The mean error  (defined by \eqref{erreurcoordmean})  for the three methods is smaller than}
&\varepsilon_{\text{mean}}=
0.824 \text{ cm}.
\end{align}
\end{subequations}
For for squat jumps, 
the maximal error  for the best method, \textit{i.e.} the kinematic one 
is smaller than 
\begin{subequations}
\label{meilleuresultsj}
\begin{align}
&\varepsilon_{\max}=
8.492 \text{ cm},
\intertext{and the mean error for the kinematic method is smaller than}
&\varepsilon_{\text{mean}}=
2.882\text{ cm}.
\end{align}
\end{subequations}

%%%%% FIN INSERTION fichier crée par Sweave sur  meilleurs_resultats
%%%%%%%%%%%%%%%%%%%%%%%%%%%%%%%%%%%%%%%%%%%%%%%%%%%%%%%%%%%%%%%%%%%%%%%%%%%
%%%%%%%%%%%%%%%%%%%%%%%%%%%%%%%%%%%%%%%%%%%%%%%%%%%%%%%%%%%%%%%%%%%%%%%%%%%

%%%%%%%%%%%%%%%%%%%%%%%%%%%%%%%%%%%%%%%%%%%%%%%%%%%%%%%%%%%%%%%%%%%%%%%%%%%
%%%%%%%%%%%%%%%%%%%%%%%%%%%%%%%%%%%%%%%%%%%%%%%%%%%%%%%%%%%%%%%%%%%%%%%%%%%
%%%%% DEBUT INSERTION fichier crée par Sweave sur statistique_final
%\input{statistique_finale} 

% pour sweave  : 
\setkeys{Gin}{width=0.55\textwidth}

% ATTENTION% ATTENTION% ATTENTION
% ATTENTION, lignes différentes selon l'environnement de Jérôme ou de Thomas !!
% chemin relatif des fonctions R prepa_facteur1.R et signif.codes.R, données SJ et des données TP

\begin{figure}[ht]
\begin{center}
%%%% A COMMENTER POUR arXiv
%\includegraphics{dessins/statistique-fig01tpp}
%%%% A DECOMMENTER POUR arXiv
\includegraphics{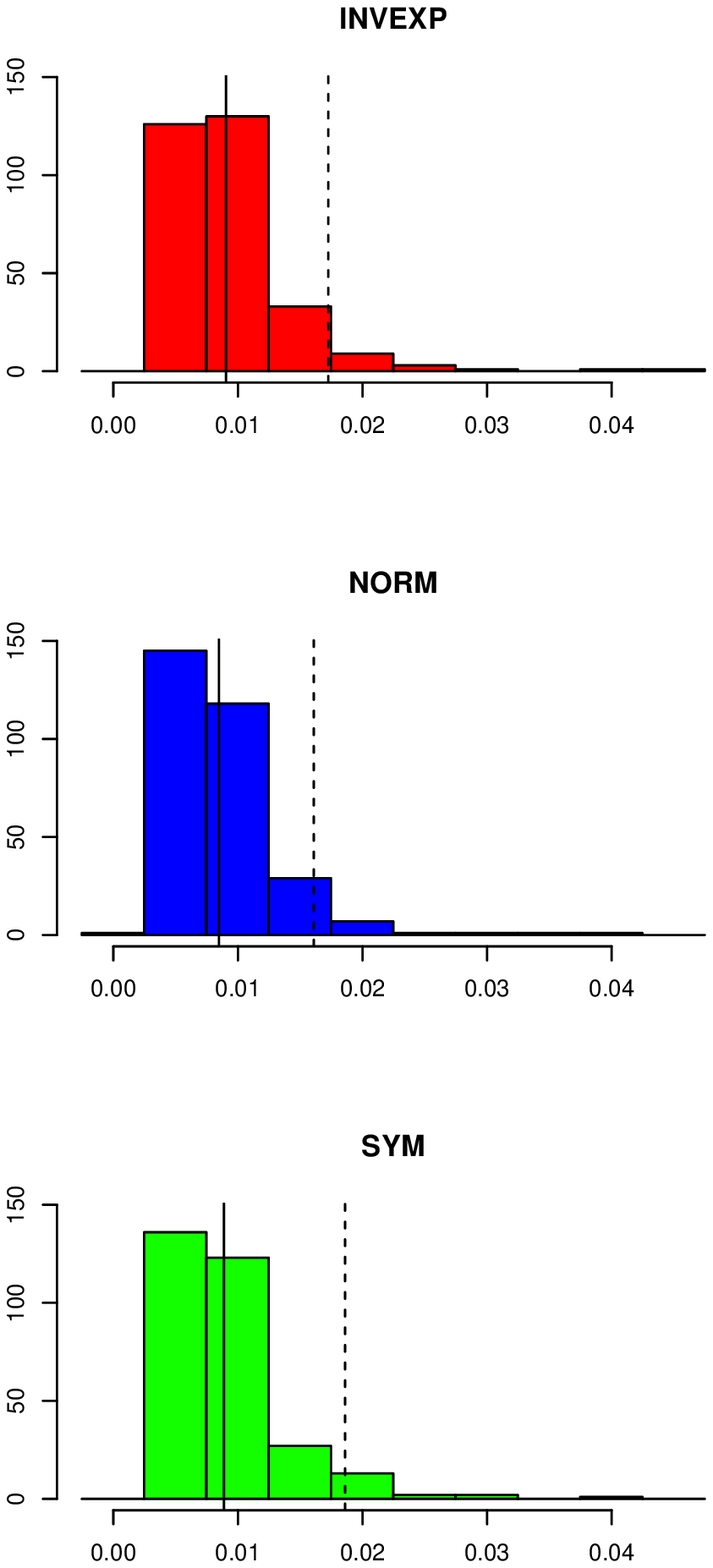}
\end{center}
\caption{\label{fig01statstpp} Histograms of $\varepsilon_{\max}$ for INVEXP, NORM and SYM sigmoid models, for semi-global  optimization 
Continuous and dashed lines indicate mean and 95 \% quantile respectively.}
\end{figure}

\begin{figure}[ht]
\begin{center}
%%%% A COMMENTER POUR arXiv
%\includegraphics{dessins/statistique-fig10tpp}
%%%% A DECOMMENTER POUR arXiv
\includegraphics{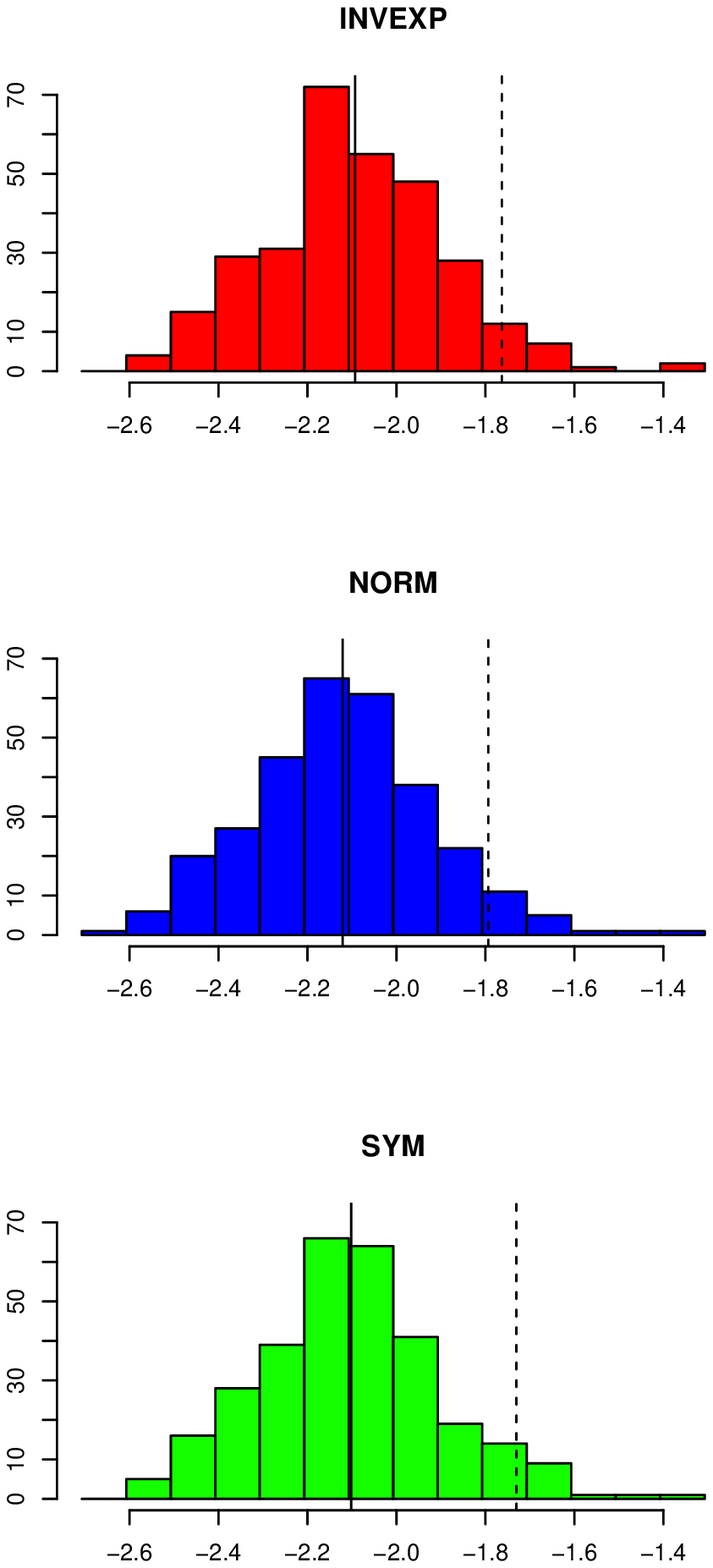}
\end{center}
\caption{\label{fig10statstpp} Histograms of $\log_{10}(\varepsilon_{\max})$ for INVEXP, NORM and SYM sigmoid models, for semi-global  optimization 
Continuous and dashed lines indicate mean and 95 \% quantile respectively.}
\end{figure}

%For both maximal and mean errors given in \eqref{erreurcoordmax} et \eqref{erreurcoordmean}, the Shapiro-Wilk test reported unnormal distributions.
%Thus, statistical tests were realized on normally distributed $\log_{10}$ of observations (i.e., errors and computation time).
%Le test de Shapiro-Wilk nous montre que les erreurs moyennes et maximales, données par \eqref{erreurcoordmax} et \eqref{erreurcoordmean}
%ne sont pas normales, contrairement au logarithme 10 de ces erreurs.
%Comparer, par exemple, pour le squat jump, les figures \ref{fig01statstpp} et \ref{fig10statstpp}.
%Il en est de même pour le temps CPU de calcul.
%Aini, sauf mention contraire, toutes les statistiques sont faites sur les $\log_{10}$ des erreurs et des temps CPU.

%%%%%%%%%%%%%%%%%%%%%%%%%%%%%%%%%%%%%%%%%%%%%%%%%%%%%%%%%%%%%%%%%%%%%%%%%%%%%%%%%%%%%%%%%
\subsection{Pointing task}

Basic descriptive statistics of measured values are given in tables \ref{tab01tp} and \ref{tab10tp}.
%Les statistiques élémentaires (moyennes et écart-types) des $\log_{10}$ des erreurs et des temps CPU sont donnés dans le tableau \ref{tab01tp}.
%Voir aussi le quantile à 95 \% des erreurs dans le tableau \ref{tab10tp}.

%Remark that, if $Q=304$ is the number of measures of error ${\varepsilon}_i$, we have 
%by setting ${\eta}_i=\log_{10}\left({\varepsilon}_i\right)$
%\begin{equation*}
%{\left(\prod_i {\varepsilon}_i\right)}^{1/Q}=10^{\left(\frac{1}{Q}\sum_i {\eta}_i\right)}=10^{\overline{\log_{10}(\varepsilon)}},
%\end{equation*}
%which means that the geometric mean 
%of error is equal to $10^m$ where $m$ is the mean of $\log_{10}$ of error: see table \ref{tab20tp}.

Recall that for $p\in [0,1]$
\begin{itemize}
\item
'***' means $p<0.001$;
\item
'**' means $p<0.01$;
\item
'*' means $p<0.05$;
\item
'.' means $p<.1$.
\end{itemize}

Among the three anovas performed for computation time, maximal and mean errors, the highest p-value was equal to
$1.48e-119$ (***) suggesting that both optimization methods and sigmoid models are associated to significantly different results.
%On étudie la dépendance entre l'erreur et le facteur $'XY'$  où X est dans \{"loc","semi","glob"\}
%et Y dans \{"sym","norm","invexp"\}.
%On fait trois 
%anova pour des  mesure répétées
%pour le logarithme $10$ du temps CPU, de l'erreur maximale et de l'erreur moyenne; le maximum des trois 
%probabilités critiques est égal à 
%$1.48e-119$ (***). 
%Les trois  méthodes d'optimisation et les trois modèles sont donc significativement différentes, aussi bien en terme de temps de calcul que d'erreur.

%Quelque soit le modèle, 
%la méthode d'optimisation semi est meilleure que la méthode locale  en terme de temps que d'erreur :
%On a fait un test unilatéral pour tester les hypothèses $H_1$:
%"semisym-locsym $\leq 0$",
%"seminorm-locnorm $\leq 0$"
%ou 
%"semiinvexp-locinvexp $\leq 0$".
%On trouve en effet une probabilité 
%égale à 
%\begin{itemize}
%\item
%0 pour l'erreur maximale;
%\item
%0 pour l'erreur moyenne.
%\end{itemize}

Considering both mean and maximal errors, post-hoc tests revealed greater adequation of original data with semi-global optimization procedure than with local one 
($p=0$) for each of the three sigmoid models.

%Pour les temps de calcul, la méthode semi-globale est plus longue  que la méthode locale pour les modèles sym et invexp, mais pas pour le modèle norm:
%on teste l'hypothèse $H_1$
%"semi-loc $\geq 0$" pour les temps cpu et 
%on trouve des probabilités critiques égales à 
%\begin{itemize}
%\item
%0 pour le modèle sym;
%\item
%0 pour le modèle invexp;
%\item
%1 pour le modèle norm.
%\end{itemize}

Computation time reported for semi-global optimization was significantly higher than durations obtained with local method for SYM and INVEXP models 
($p=0$) but not for NORM one 
($p=1$).

%Nous faisons maintenant des Tests  unilatéraux de Tukey post hoc. 

%\textbf{Attention, il faut diviser partout les proba critiques obtenues par le nombre de comparaisons faites. mais elles sont toutes tellement petites qu'on s'en fiche !!
%\'Evoquer correction de bonferroni ?}

Comparing semi-global and global methods revealed no significant difference for computation time 
($p=1$), 
maximal ($p\geq$ 0.3291) and mean ($p\geq$ 0.6166) errors.

%En terme d'erreur, il n'y a pas d'amélioration significative en passant de la méthode semi globale à la méthode globale, et ce pour les trois modèles:
%les trois probabilités critiques obtenues sont toutes plus grandes que
%0.3291 pour le modèle l'erreur maximale et 
%0.6166 pour le modèle l'erreur moyenne.
%%% debut ajout 
%En terme de temps, 
%on teste l'hypothèse $H_1$
%"glob-semi$\geq 0$" pour les temps cpu et 
%on trouve des probabilités critiques égales à 
%\begin{itemize}
%\item
%1 pour le modèle sym;
%\item
%1 pour le modèle invexp;
%\item
%1 pour le modèle norm.
%\end{itemize}
%%% fin  ajout 

%Il peut être intéressant de comparer entre eux les trois modèles pour la méthode loc ; un test post hoc de Tukey 
%bilatéral cette fois nous montre qu'il n'y a pas de différence significative entre eux, aussi bien pour l'erreur maximale 
%($p\geq 0.7999$) que l'erreur moyenne
%($p\geq 0.7737$).

%Il en est de même pour les méthode semi et globale
%($p\geq 0.1315$).

For the local optimization case, no significant difference was found between the three sigmoid models when comparing maximal ($p\geq 0.7999$) and mean ($p\geq 0.7737$) errors.
Similar results are obtained for global (maximal: $p\geq 0.1315$, mean: $p\geq 0.1957$) and semi-global
(maximal: $p\geq 0.157$, mean: $p\geq 0.1877$) cases.

%%%%%%%%%%%%%%%%%%%%%%%%%%%%%%%%%%%%%%%%%%%%%%%%%%%%%%%%%%%%%%%%%%%%%%%%%%%%%%%%%%%%%%%%%
\subsection{Squat Jump}

\begin{figure}[!tp]
\begin{center}$
\begin{array}{ccc}
\includegraphics[width=1.7in]{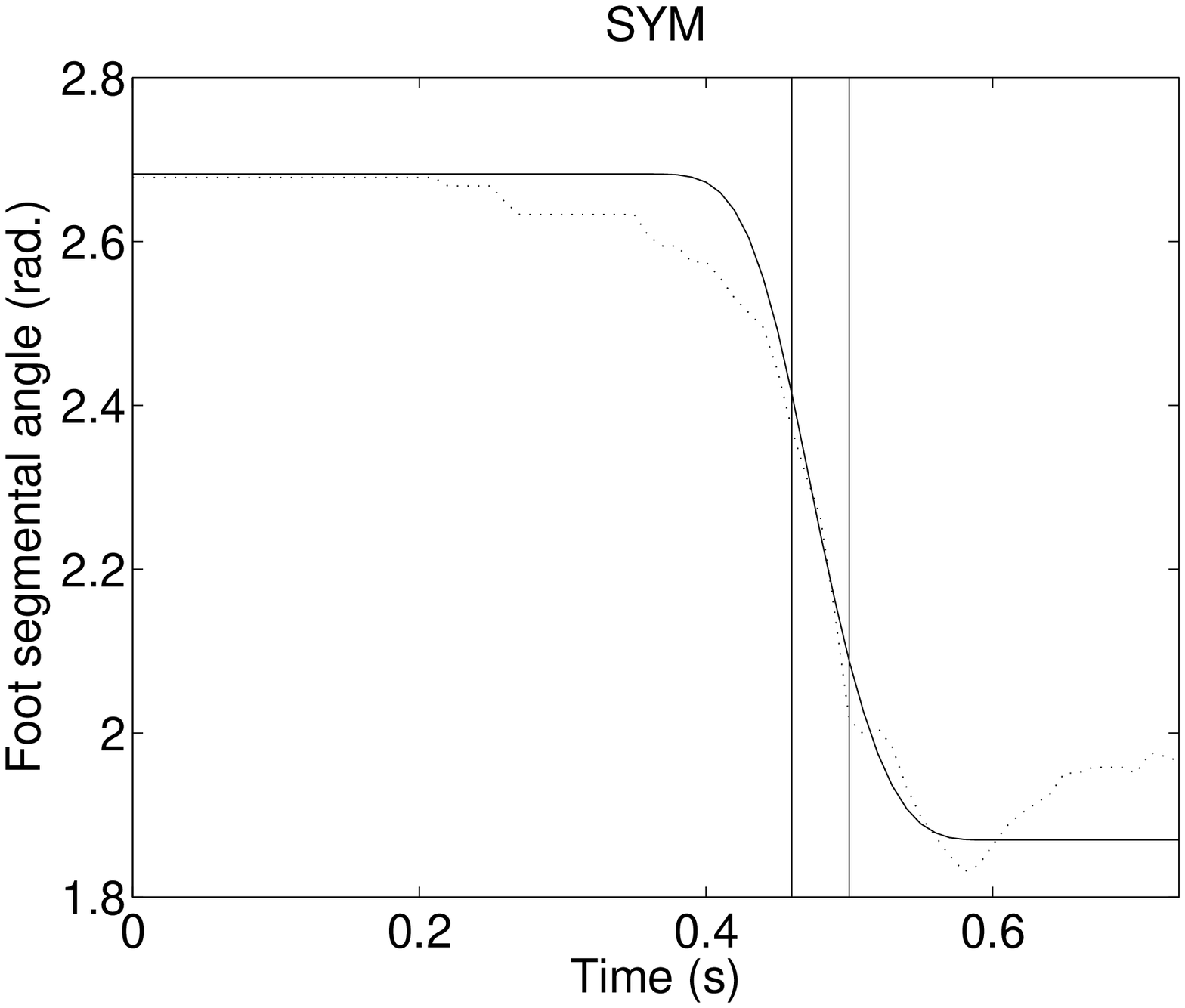} & \includegraphics[width=1.7in]{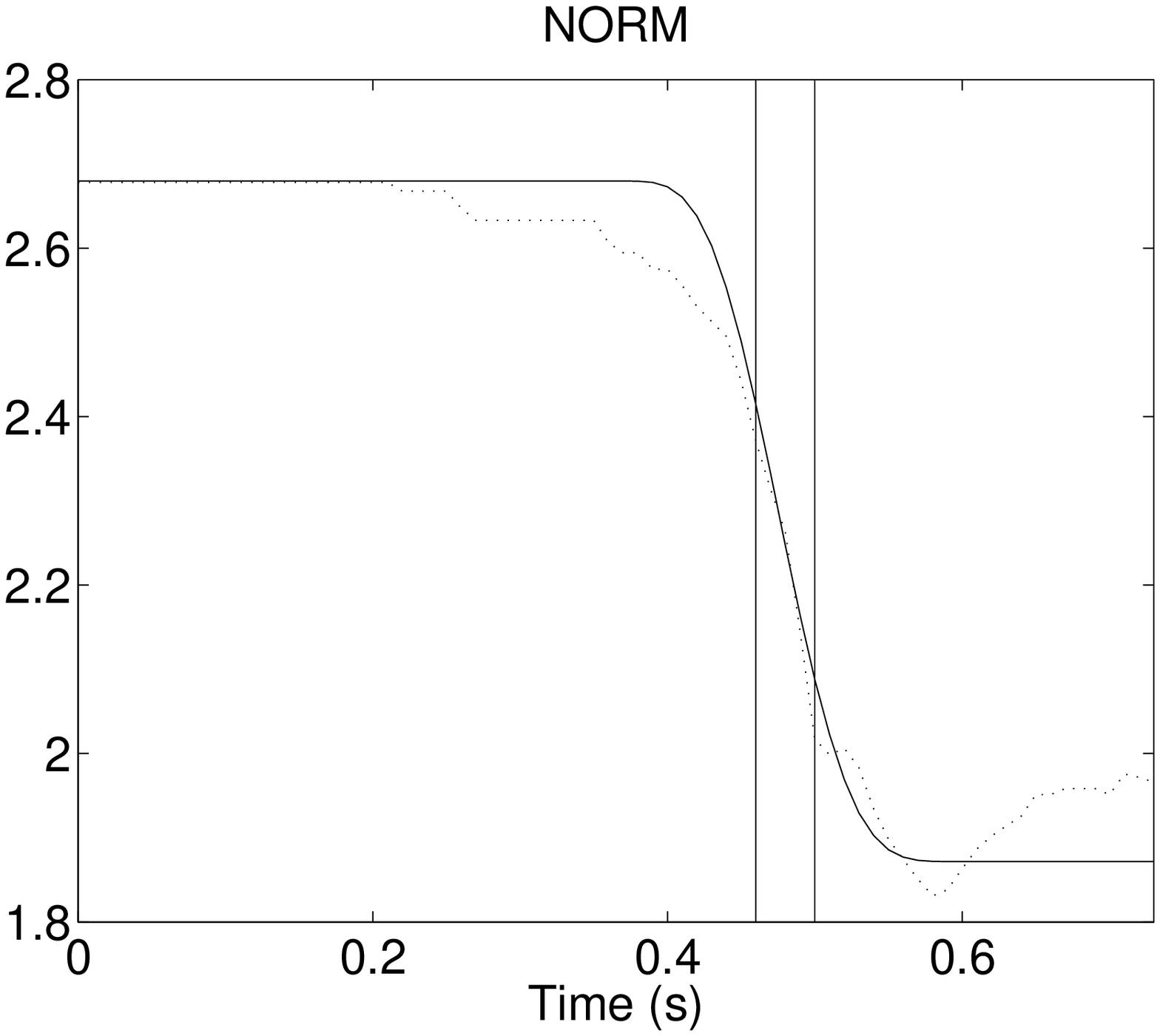} & \includegraphics[width=1.7in]{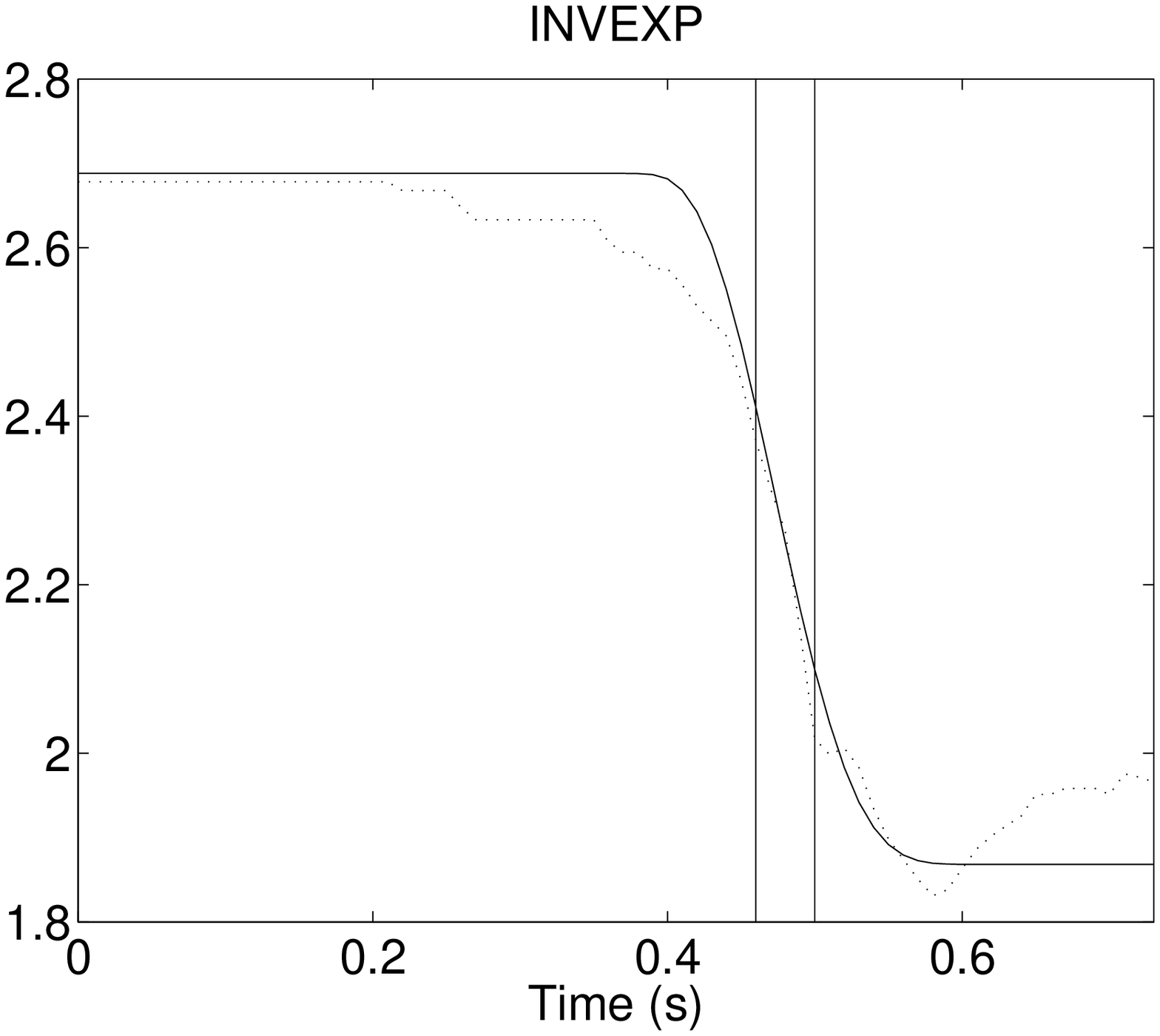} \\
\includegraphics[width=1.7in]{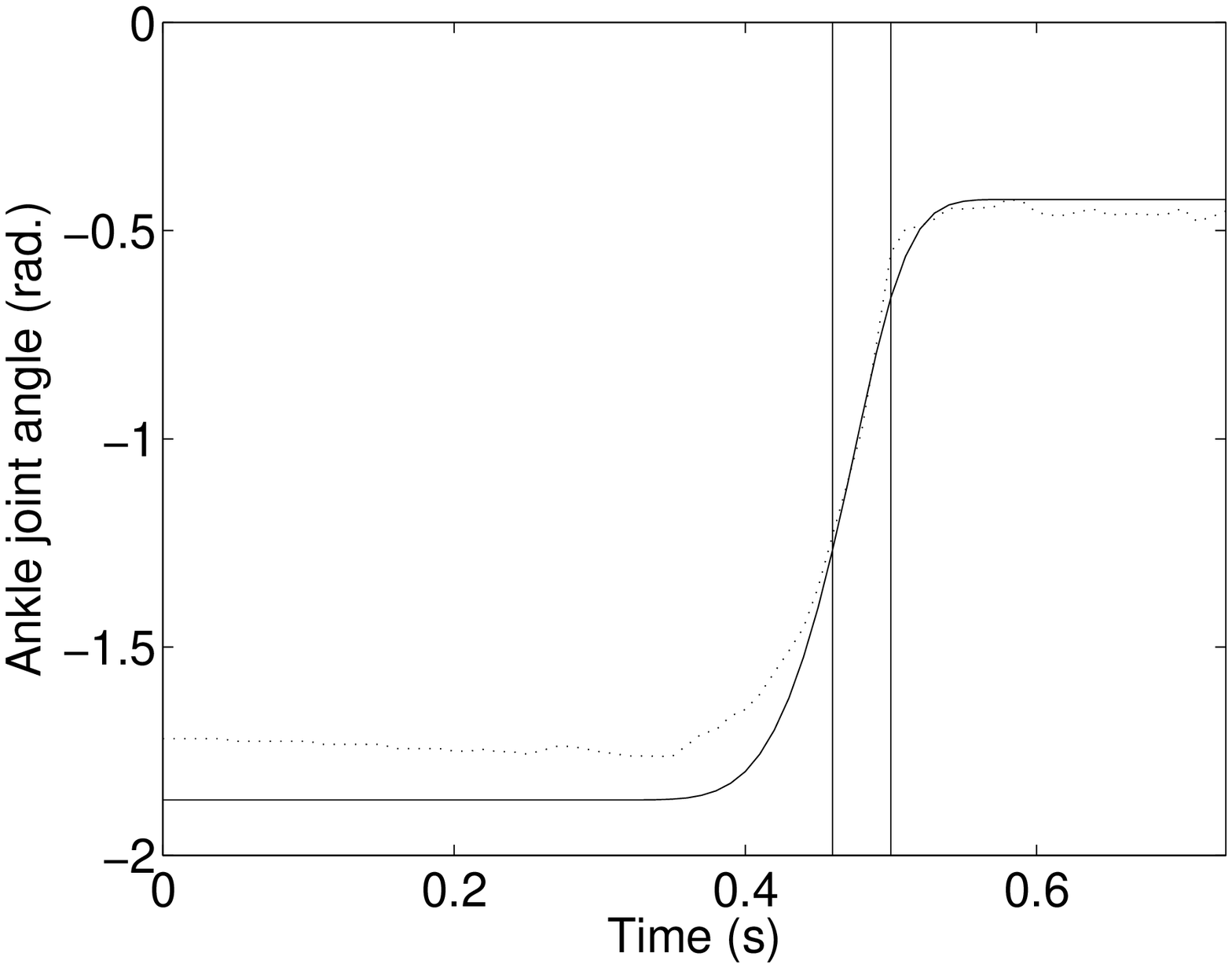} & \includegraphics[width=1.7in]{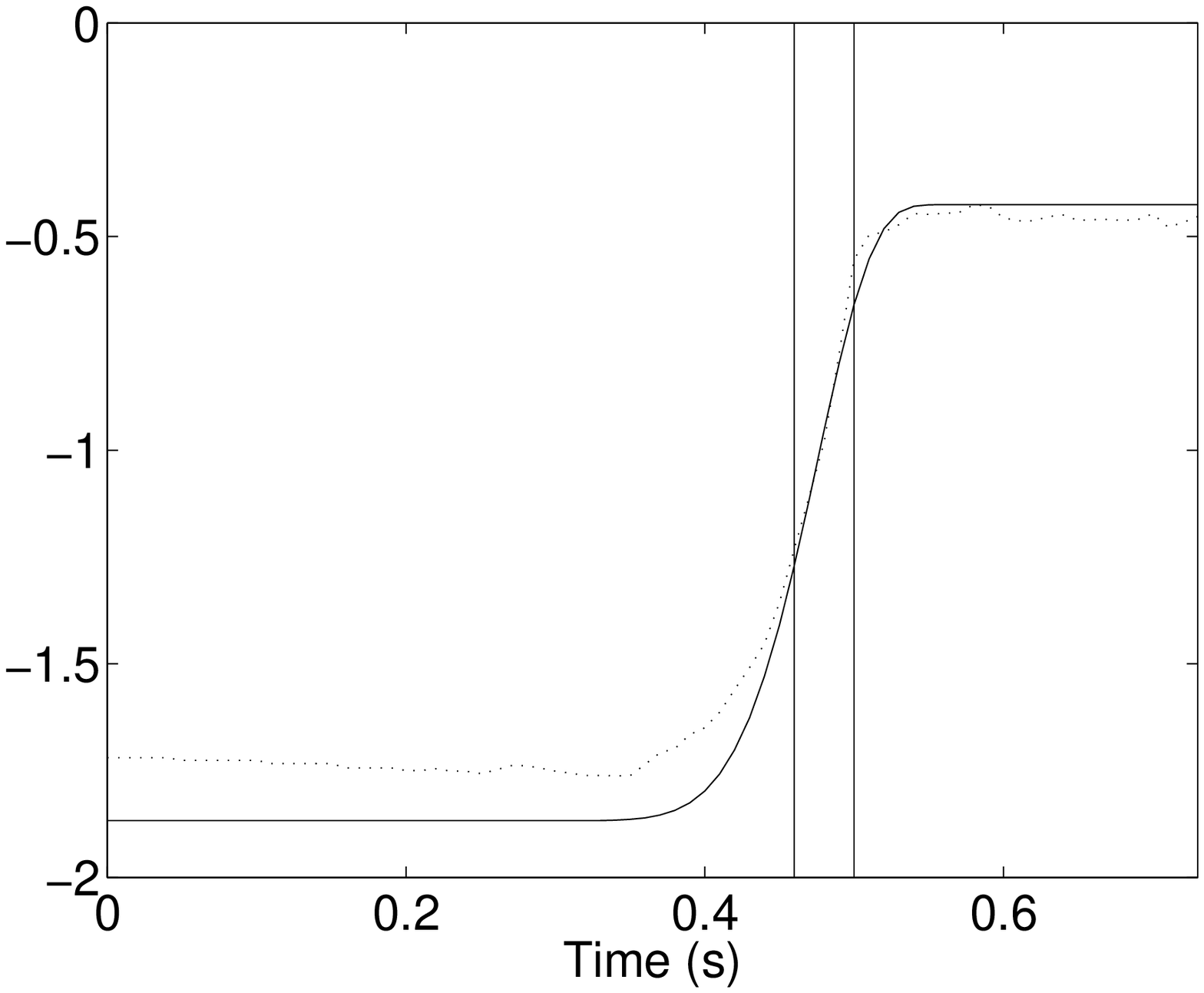} & \includegraphics[width=1.7in]{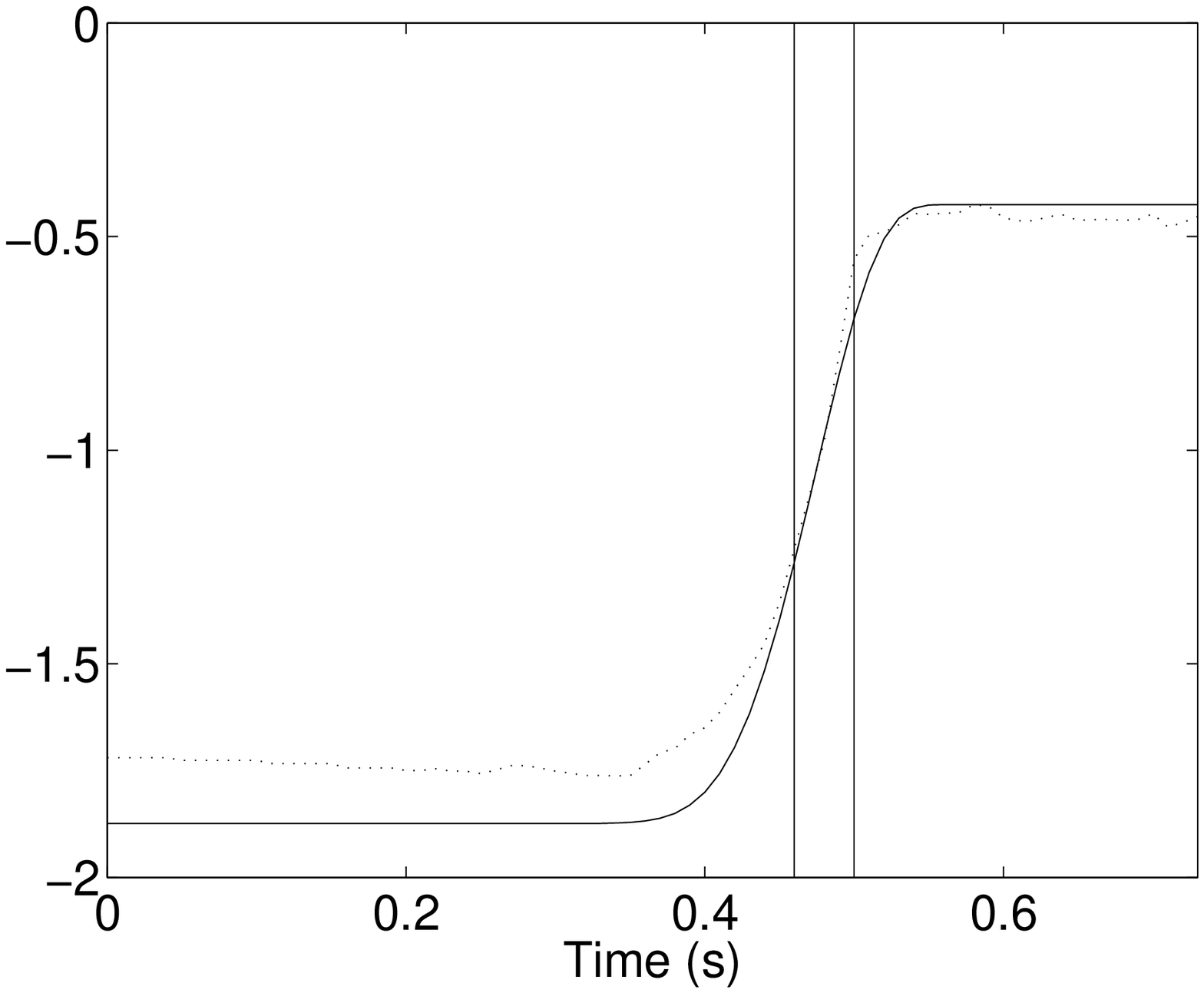} \\
\includegraphics[width=1.7in]{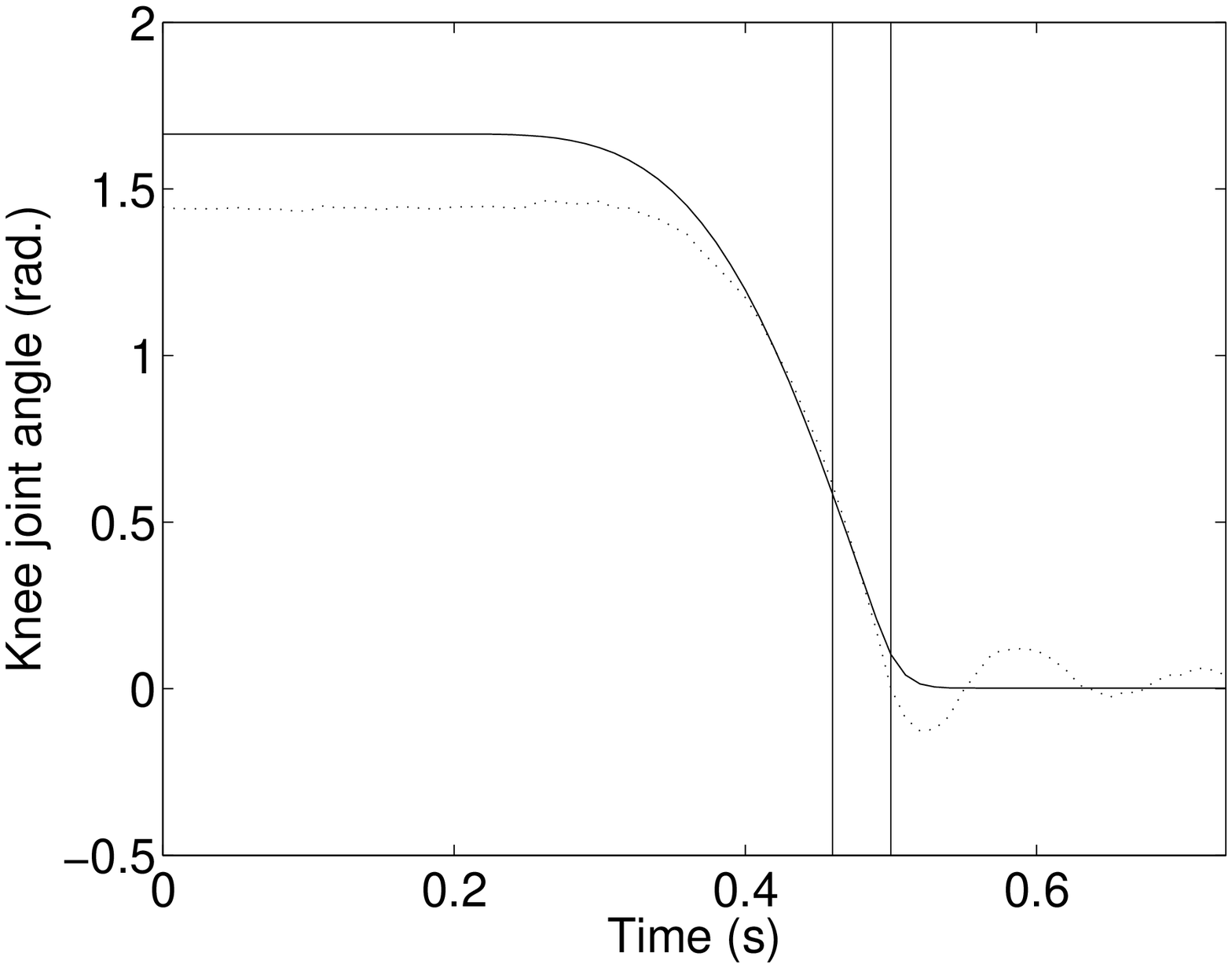} & \includegraphics[width=1.7in]{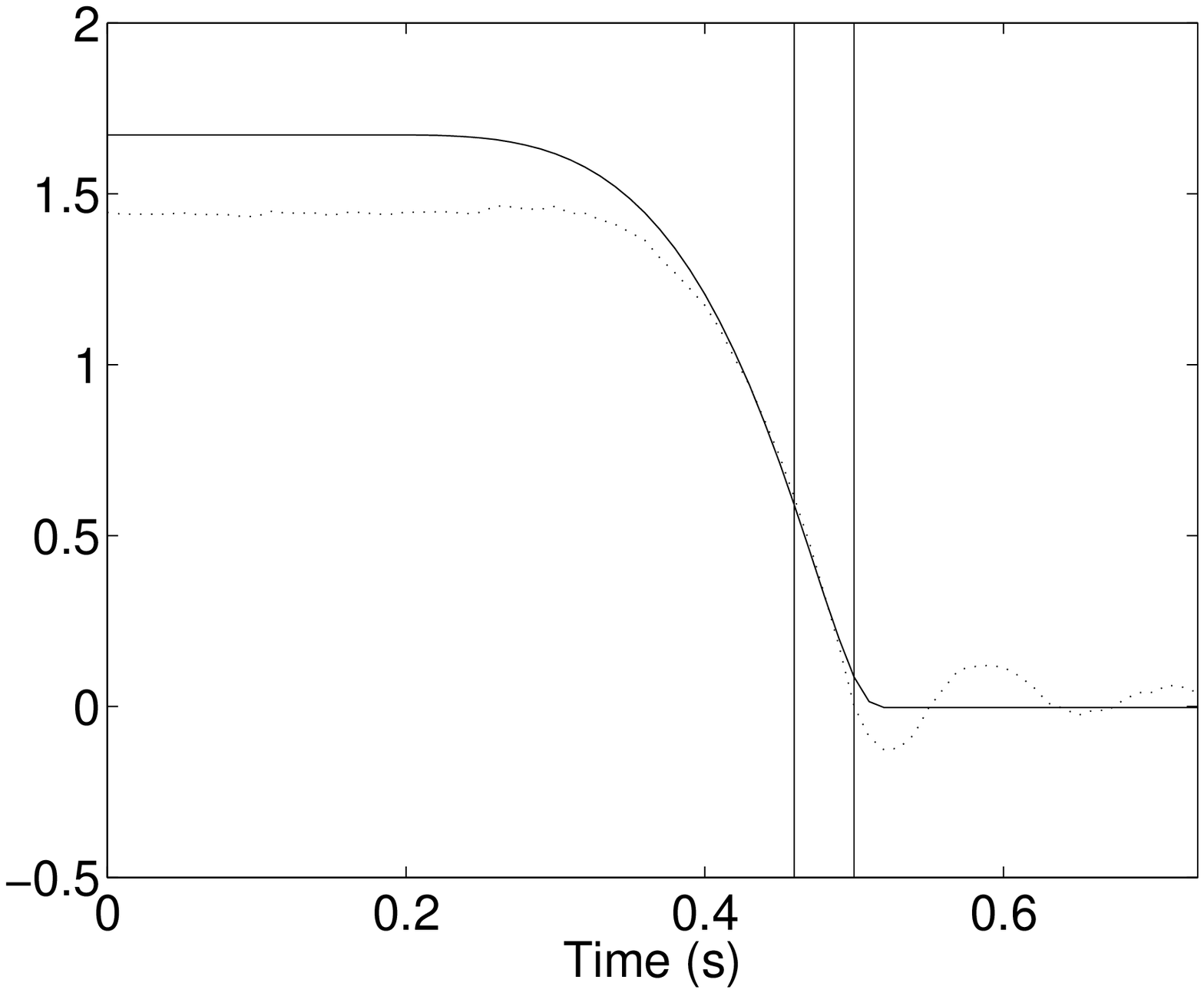} & \includegraphics[width=1.7in]{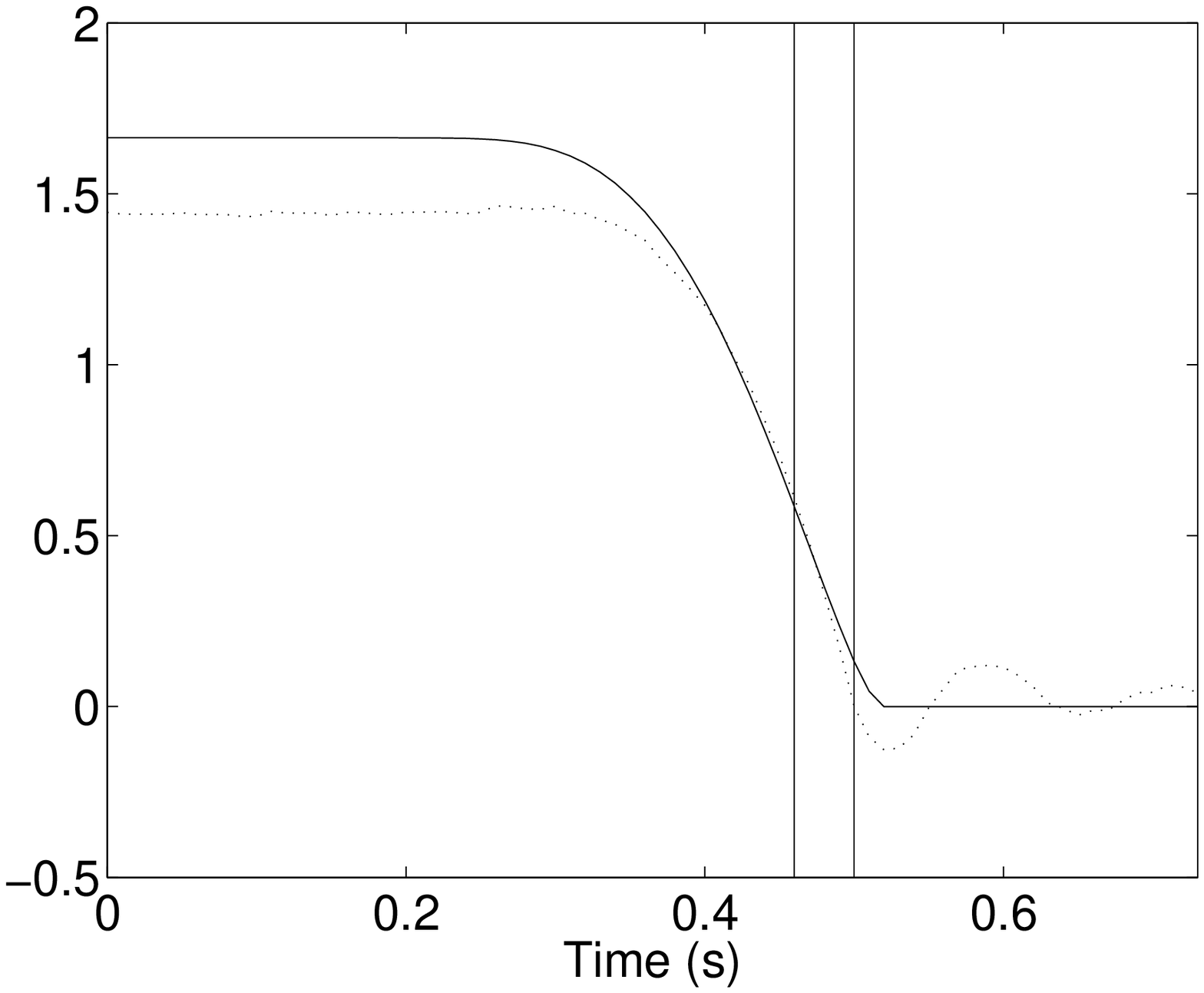} \\
\includegraphics[width=1.7in]{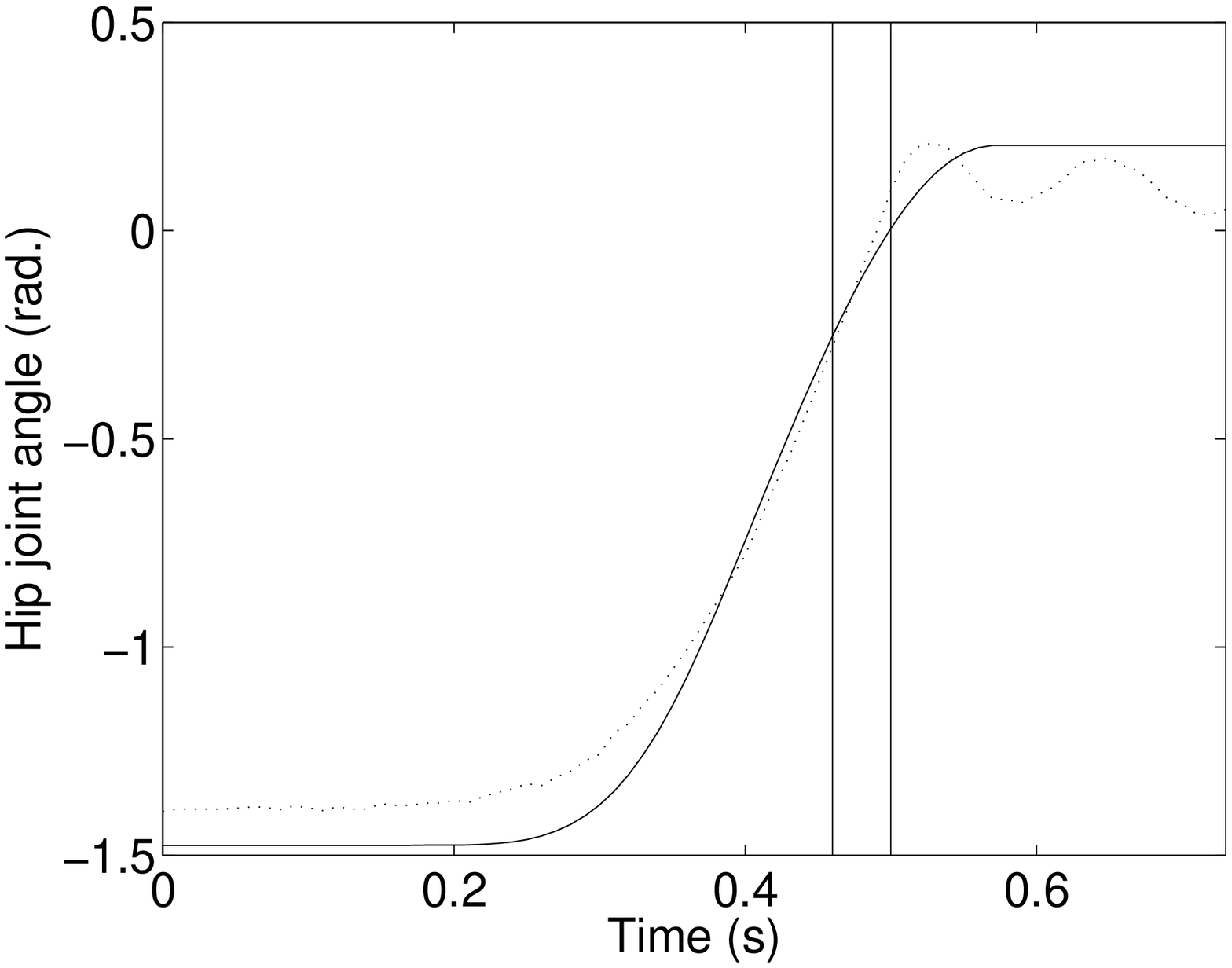} & \includegraphics[width=1.7in]{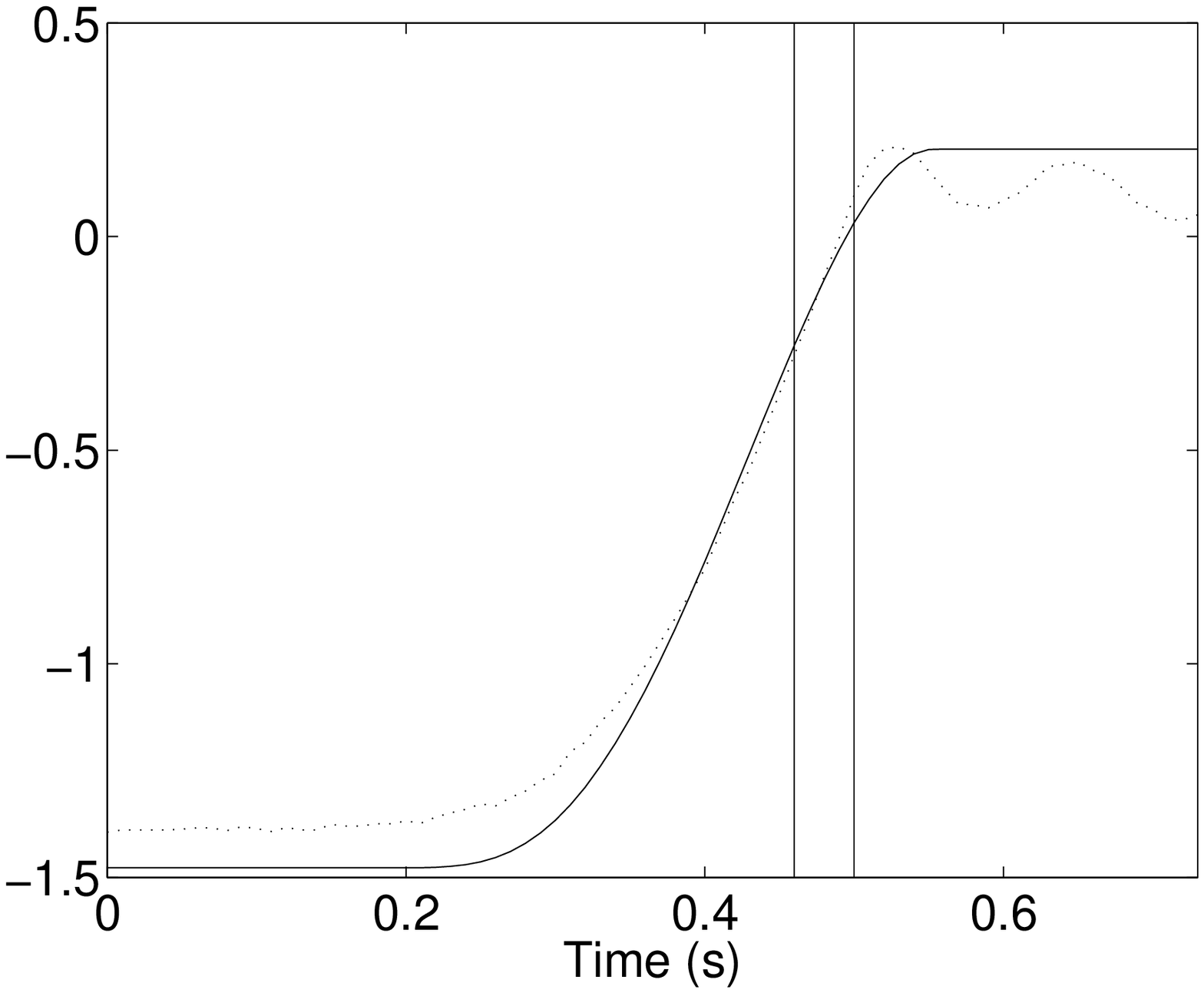} & \includegraphics[width=1.7in]{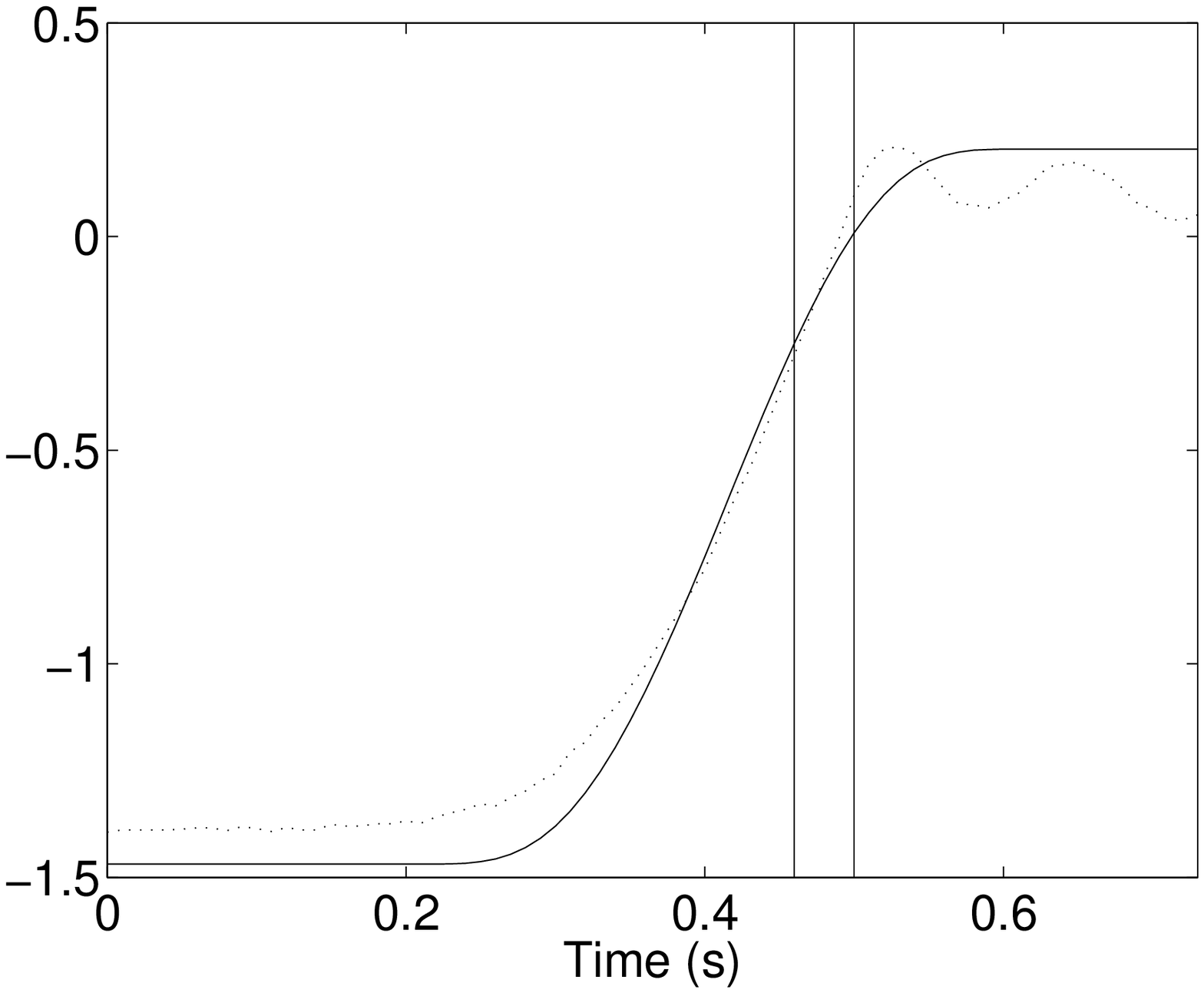}
\end{array}$
\end{center}
\caption{\label{SJFig1}Time histories of joint angles. Dotted and plain curves correspond to experimental and kinematic stage modeled data respectively. Vertical lines indicate $t_1$ and take-off instants.}
\end{figure}

\begin{figure}[!tp]
\begin{center}$
\begin{array}{ccc}
\includegraphics[width=1.7in]{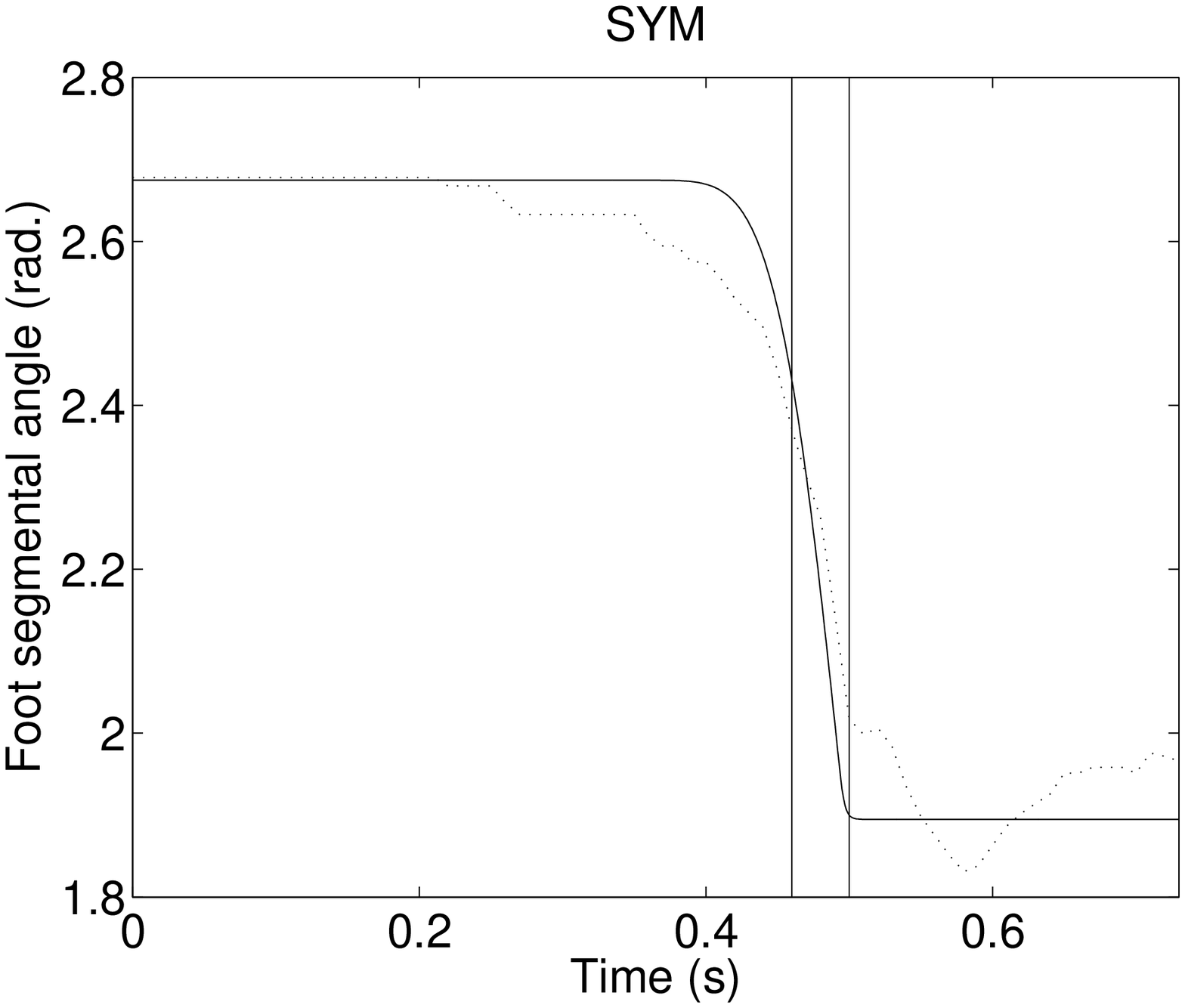} & \includegraphics[width=1.7in]{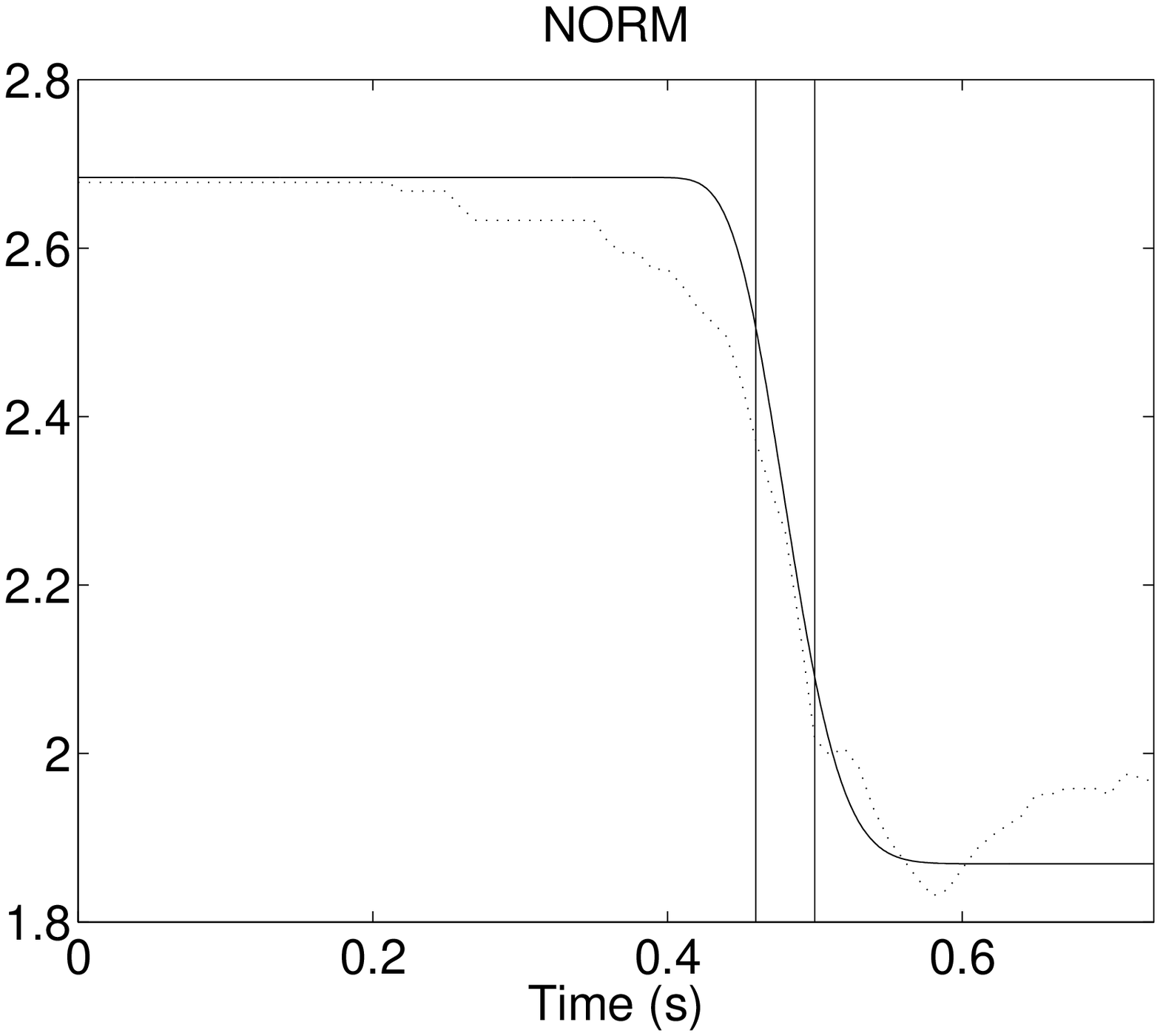} & \includegraphics[width=1.7in]{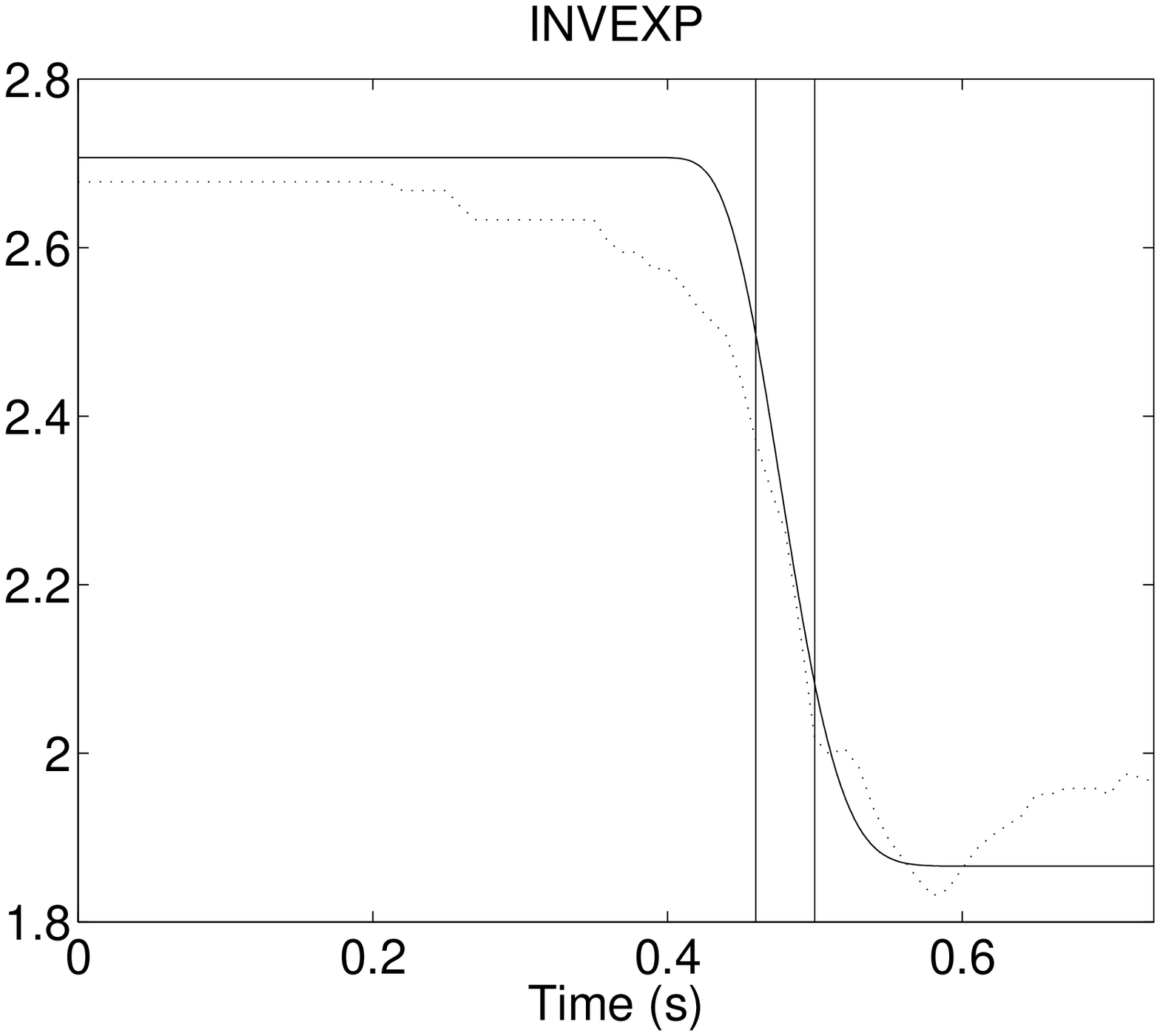} \\
\includegraphics[width=1.7in]{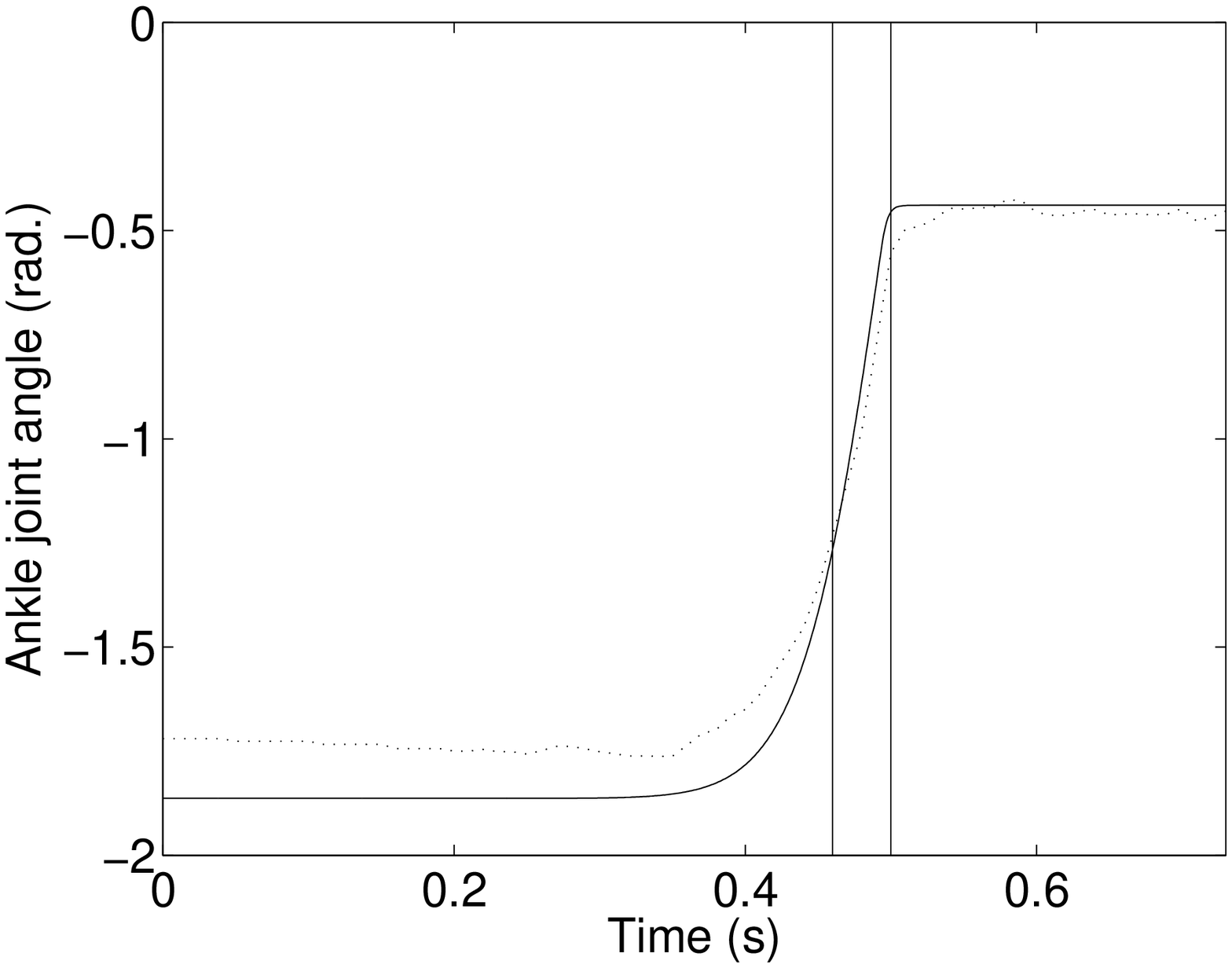} & \includegraphics[width=1.7in]{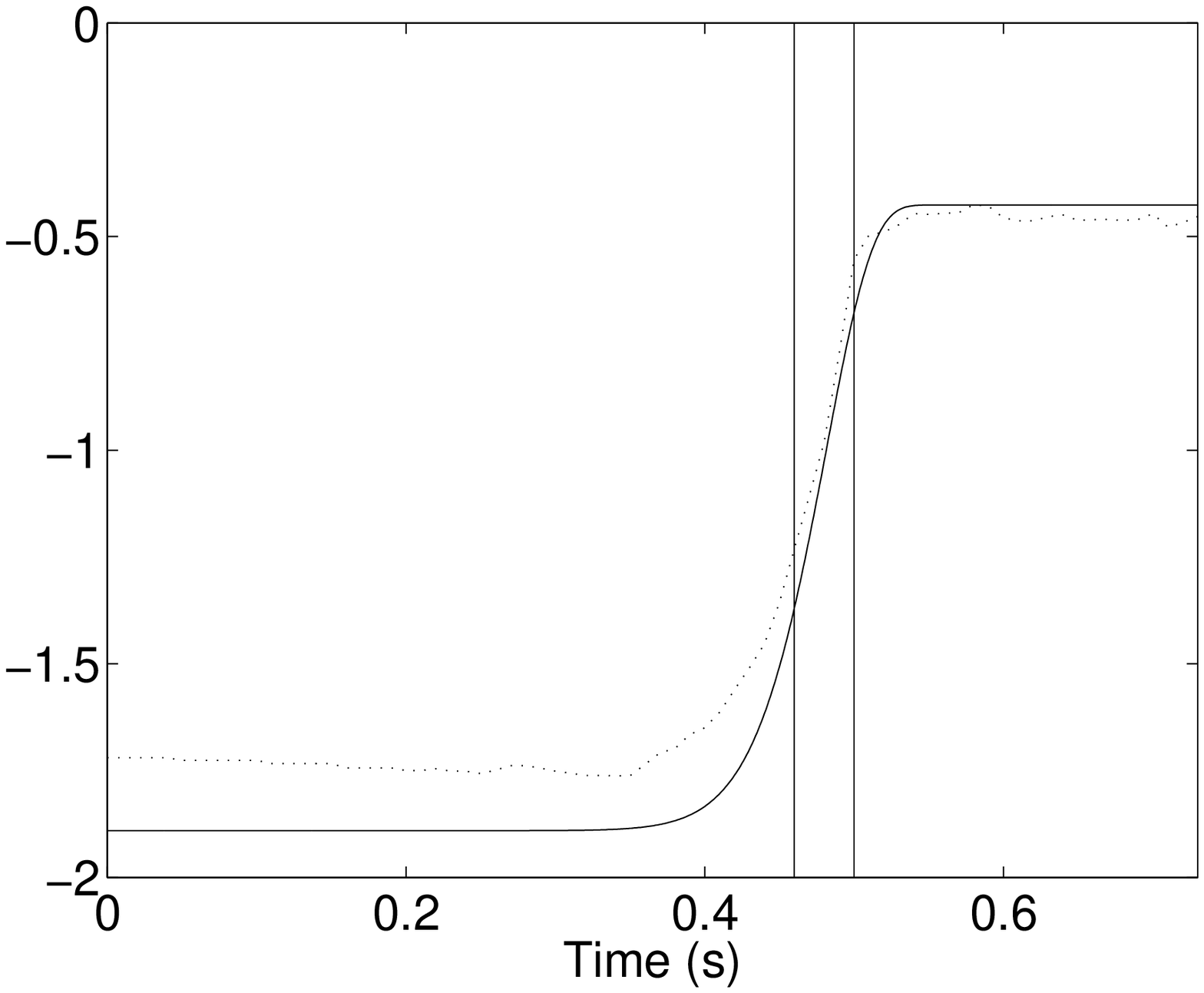} & \includegraphics[width=1.7in]{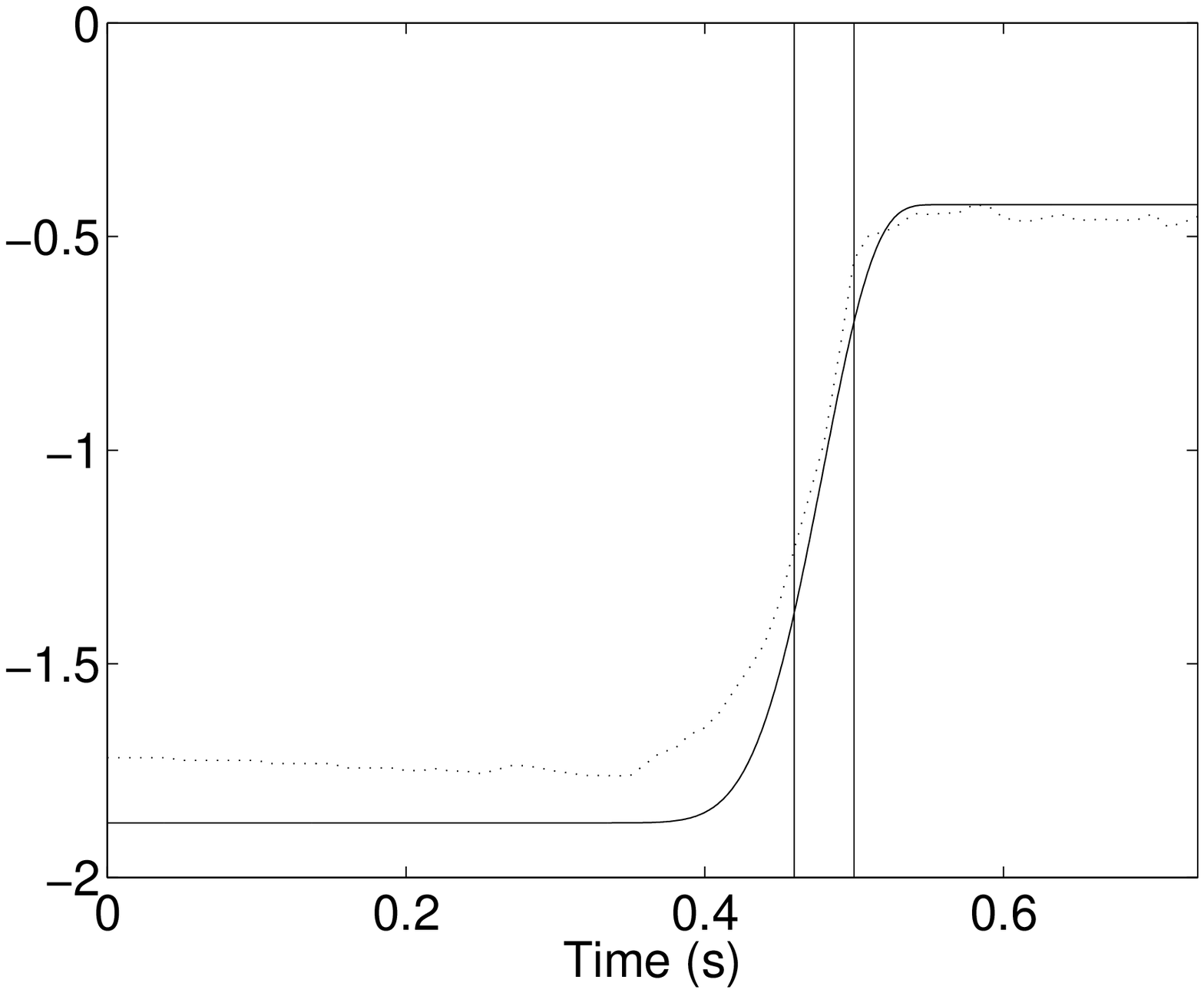} \\
\includegraphics[width=1.7in]{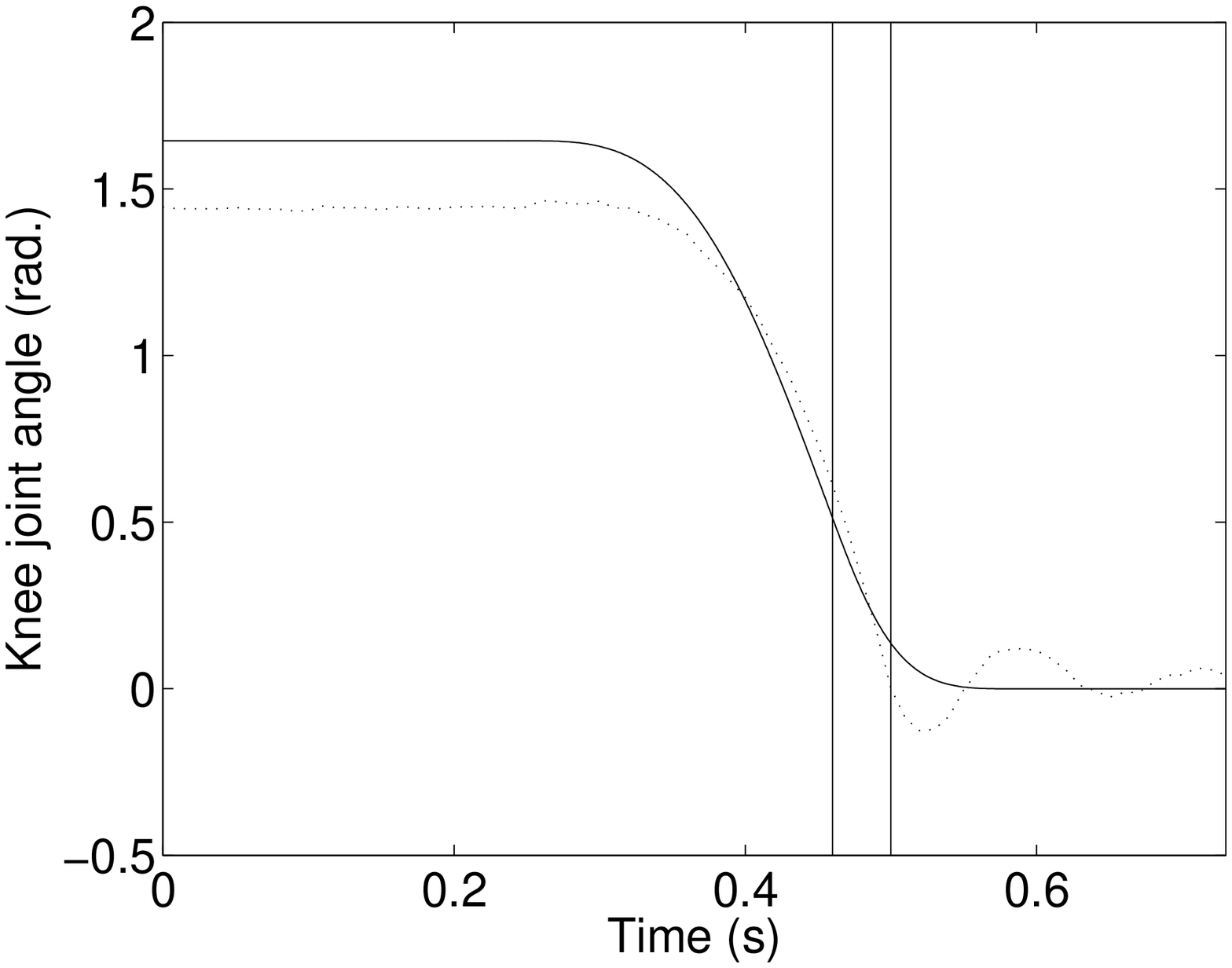} & \includegraphics[width=1.7in]{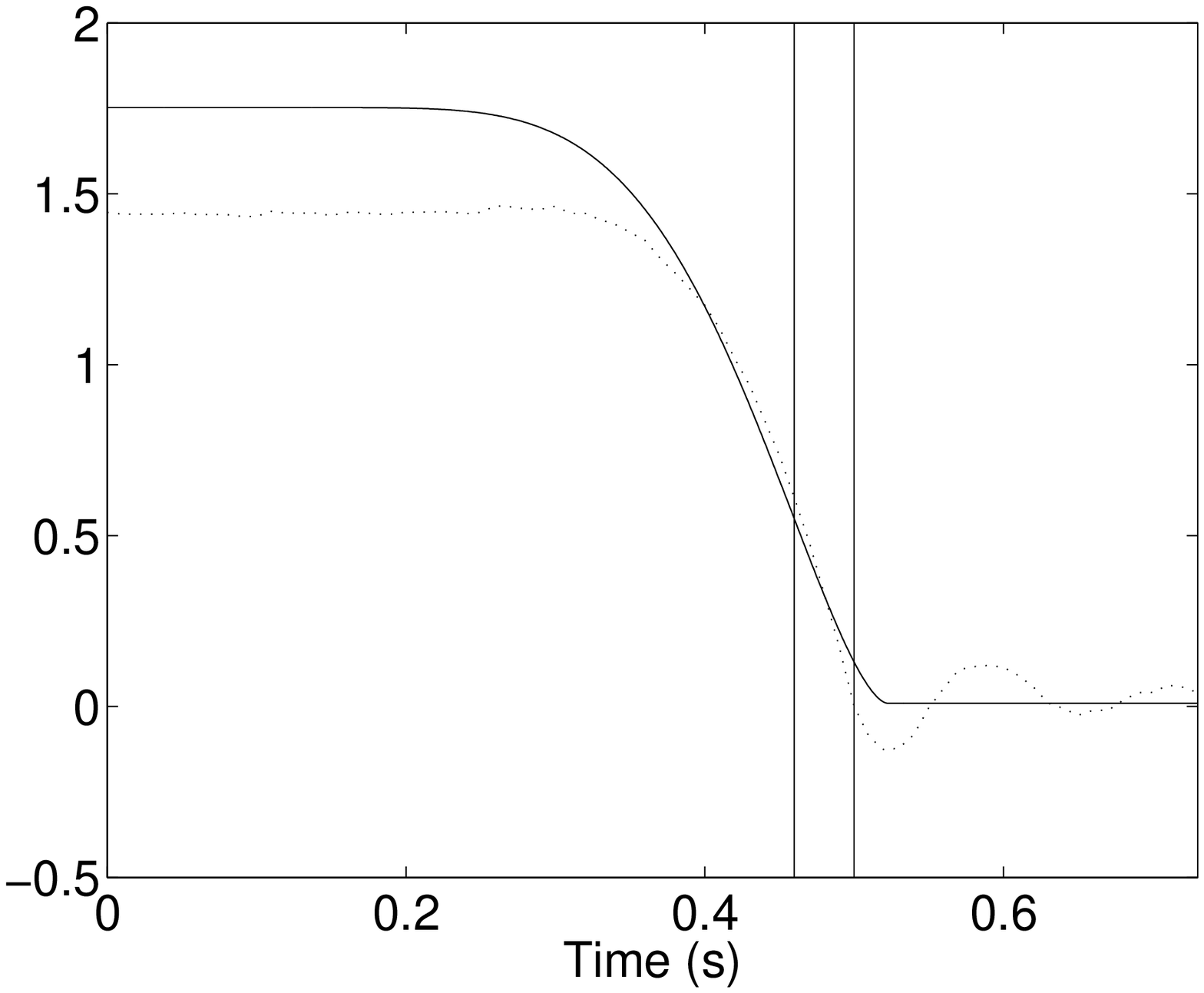} & \includegraphics[width=1.7in]{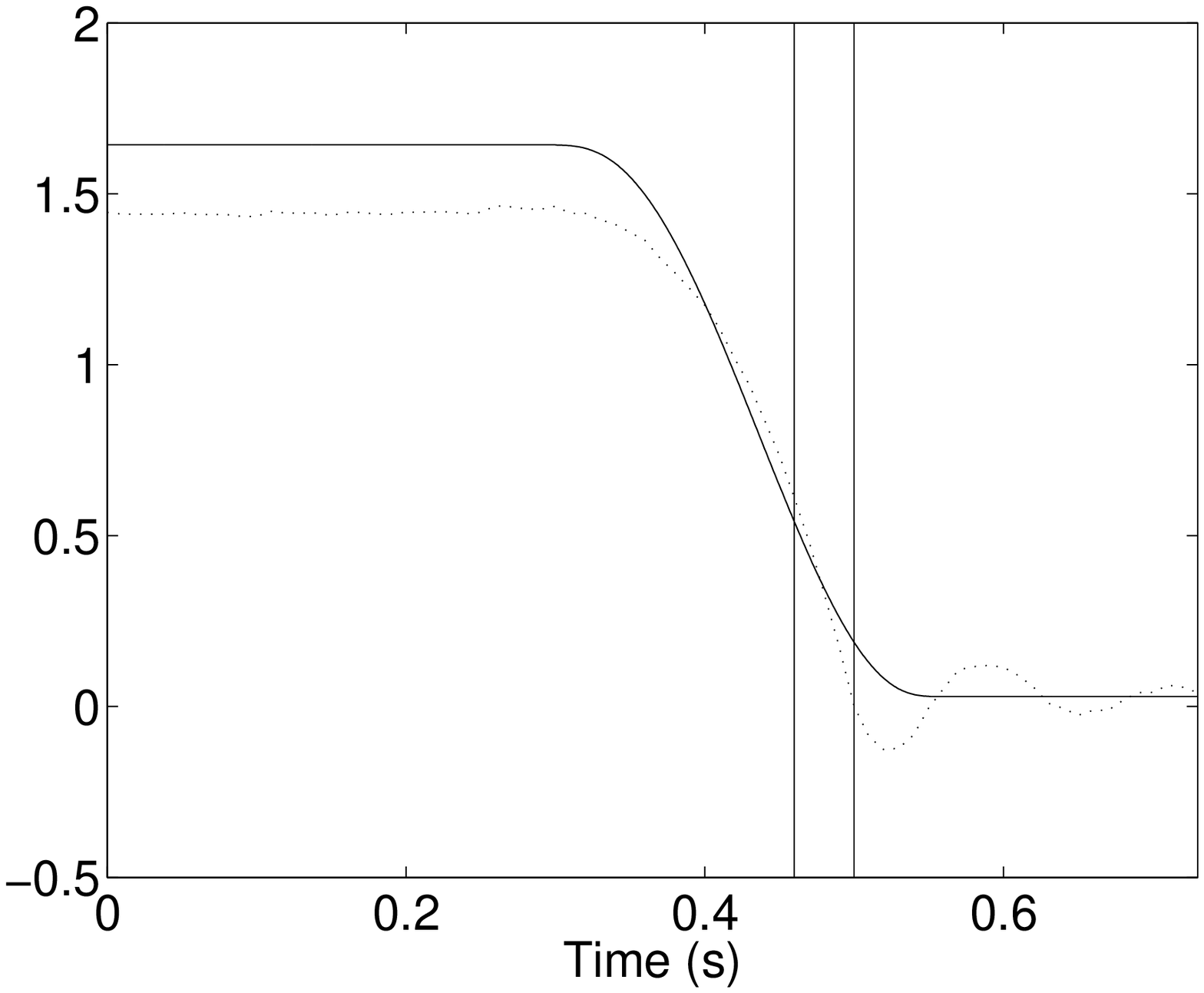} \\
\includegraphics[width=1.7in]{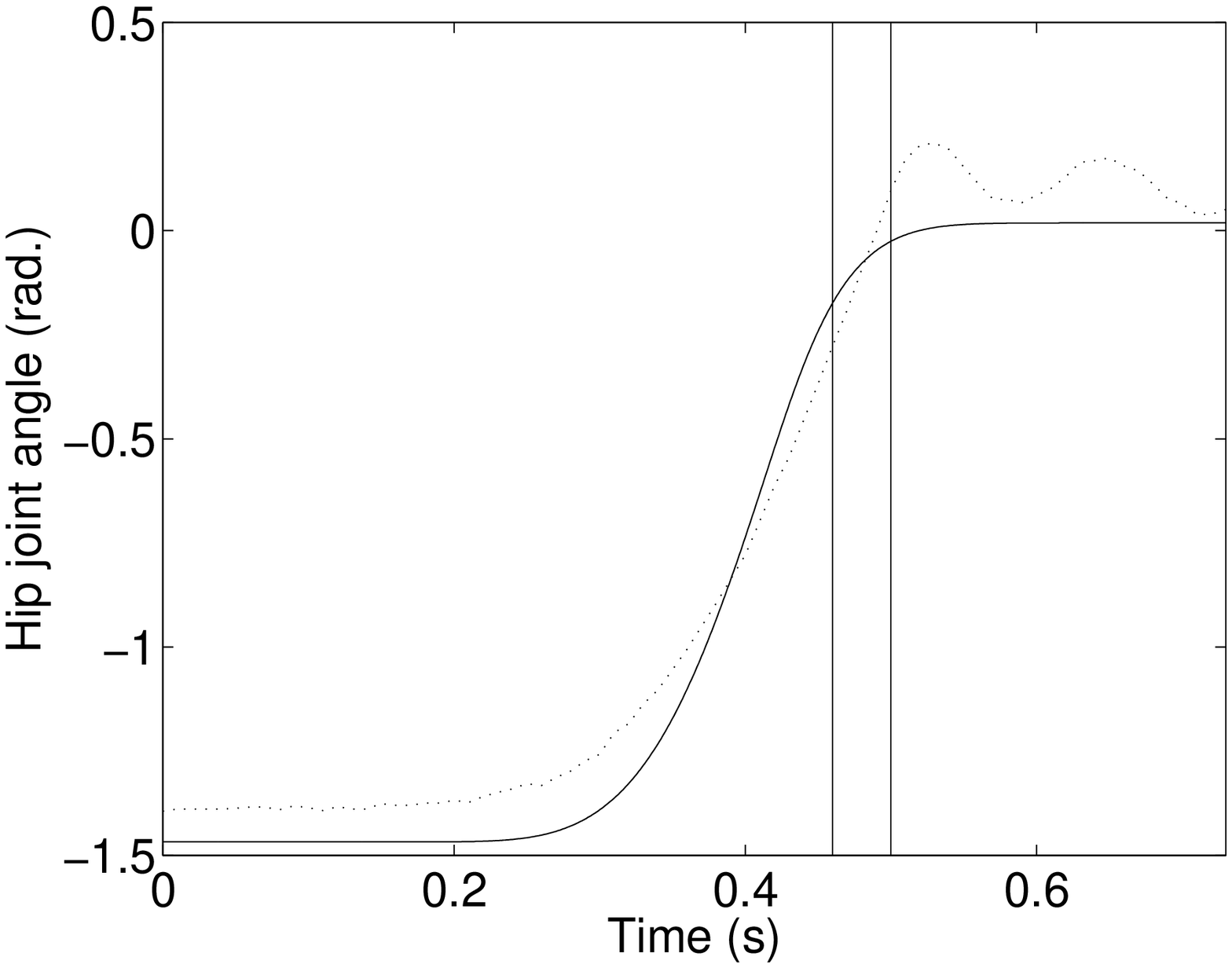} & \includegraphics[width=1.7in]{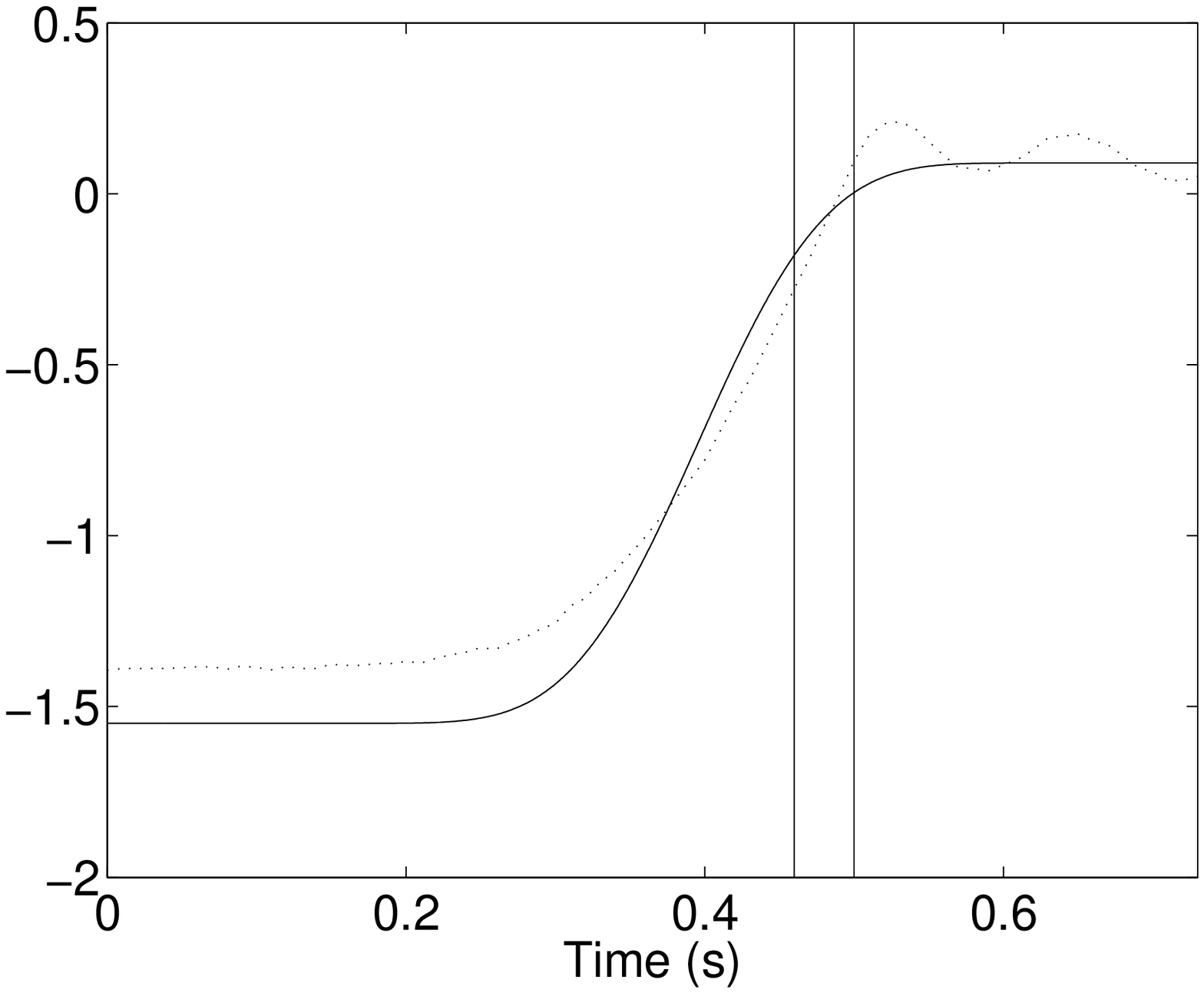} & \includegraphics[width=1.7in]{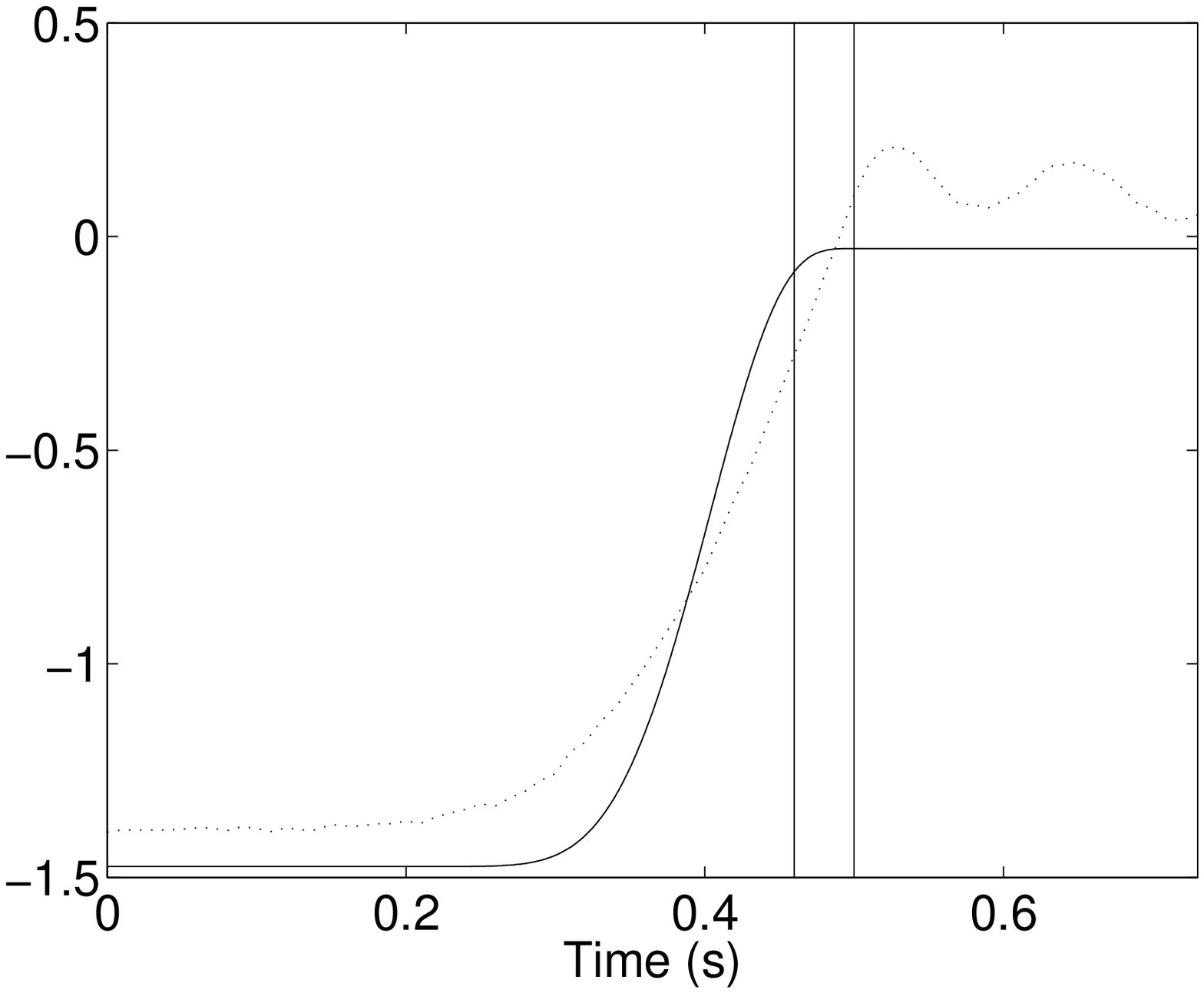}
\end{array}$
\end{center}
\caption{\label{SJFig2}Time histories of joint angles. Dotted and plain curves correspond to experimental and dynamic stage modeled data respectively. Vertical lines indicate $t_1$ and take-off instants.}
\end{figure}

\begin{figure}[!tp]
\begin{center}$
\begin{array}{ccc}
\includegraphics[width=1.7in,height=0.8in]{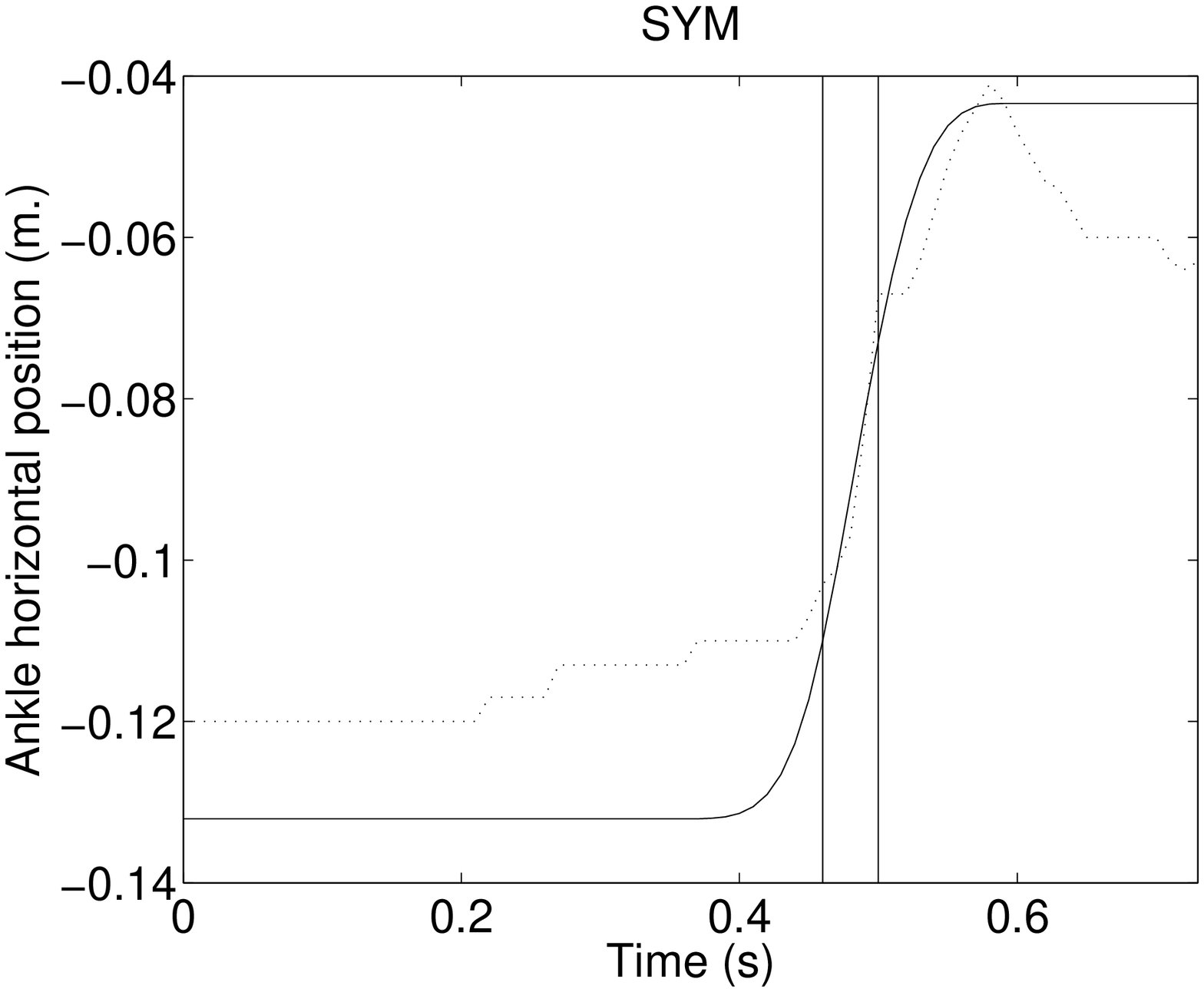} & \includegraphics[width=1.7in,height=0.8in]{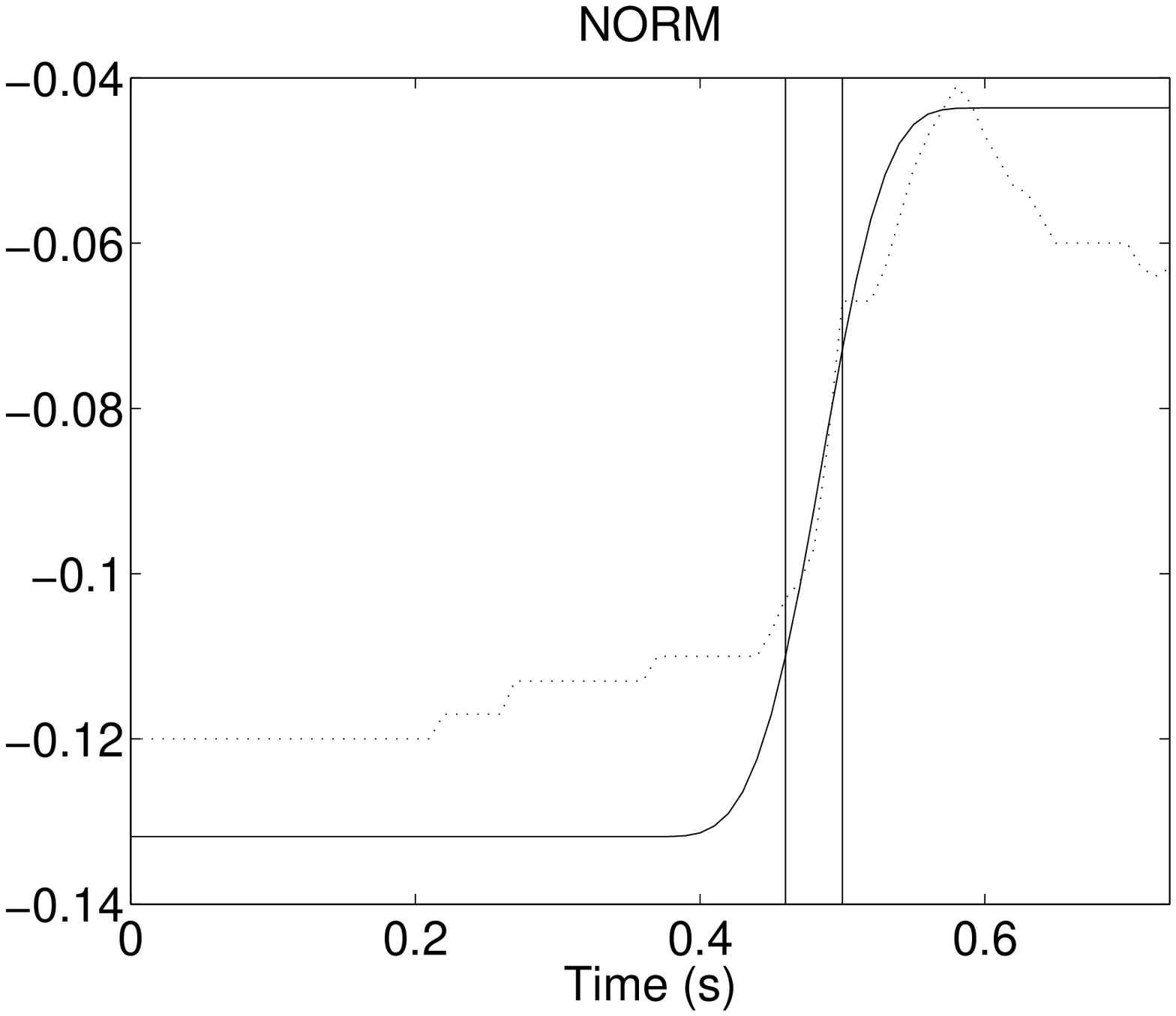} & \includegraphics[width=1.7in,height=0.8in]{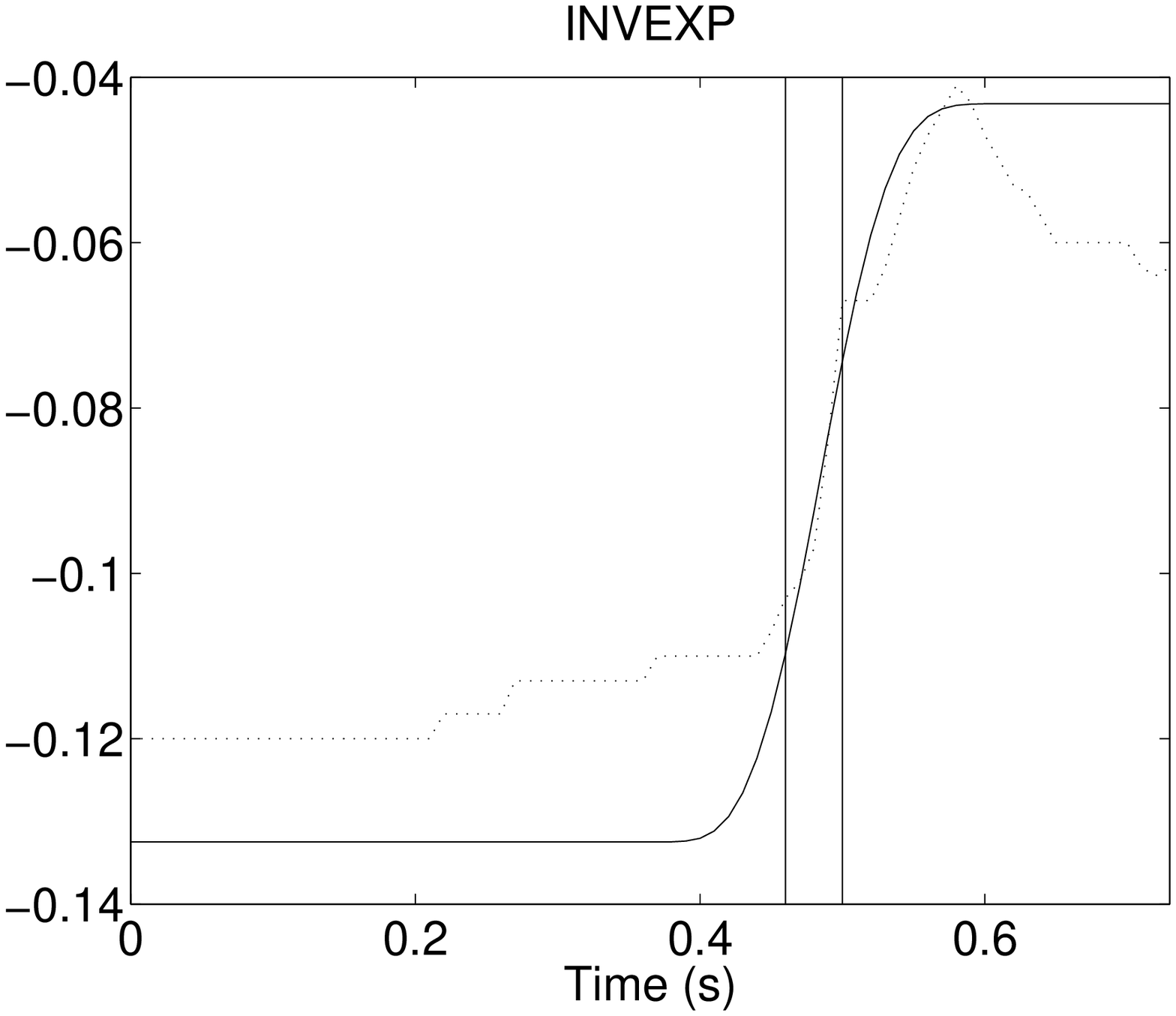} \\
\includegraphics[width=1.7in,height=0.8in]{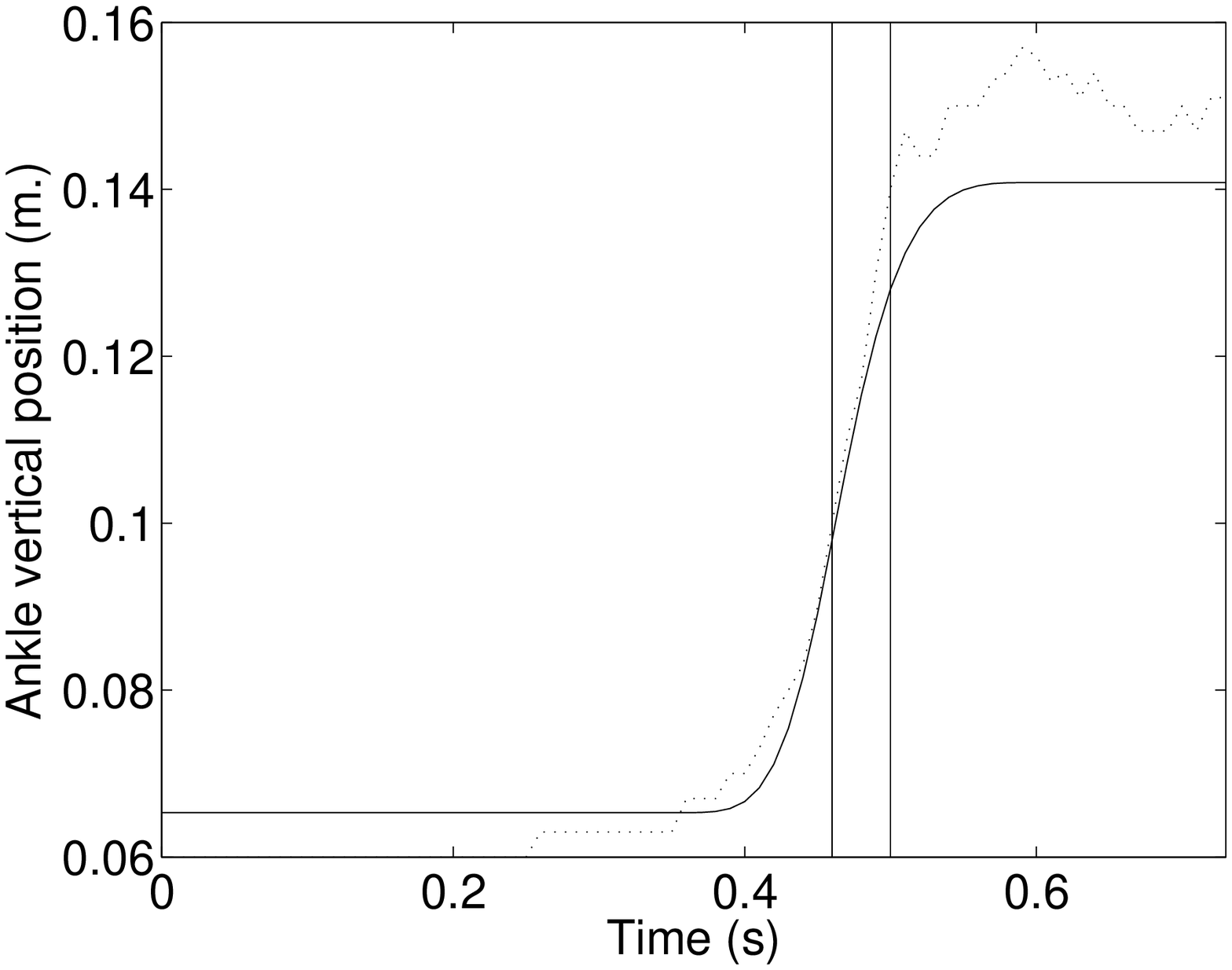} & \includegraphics[width=1.7in,height=0.8in]{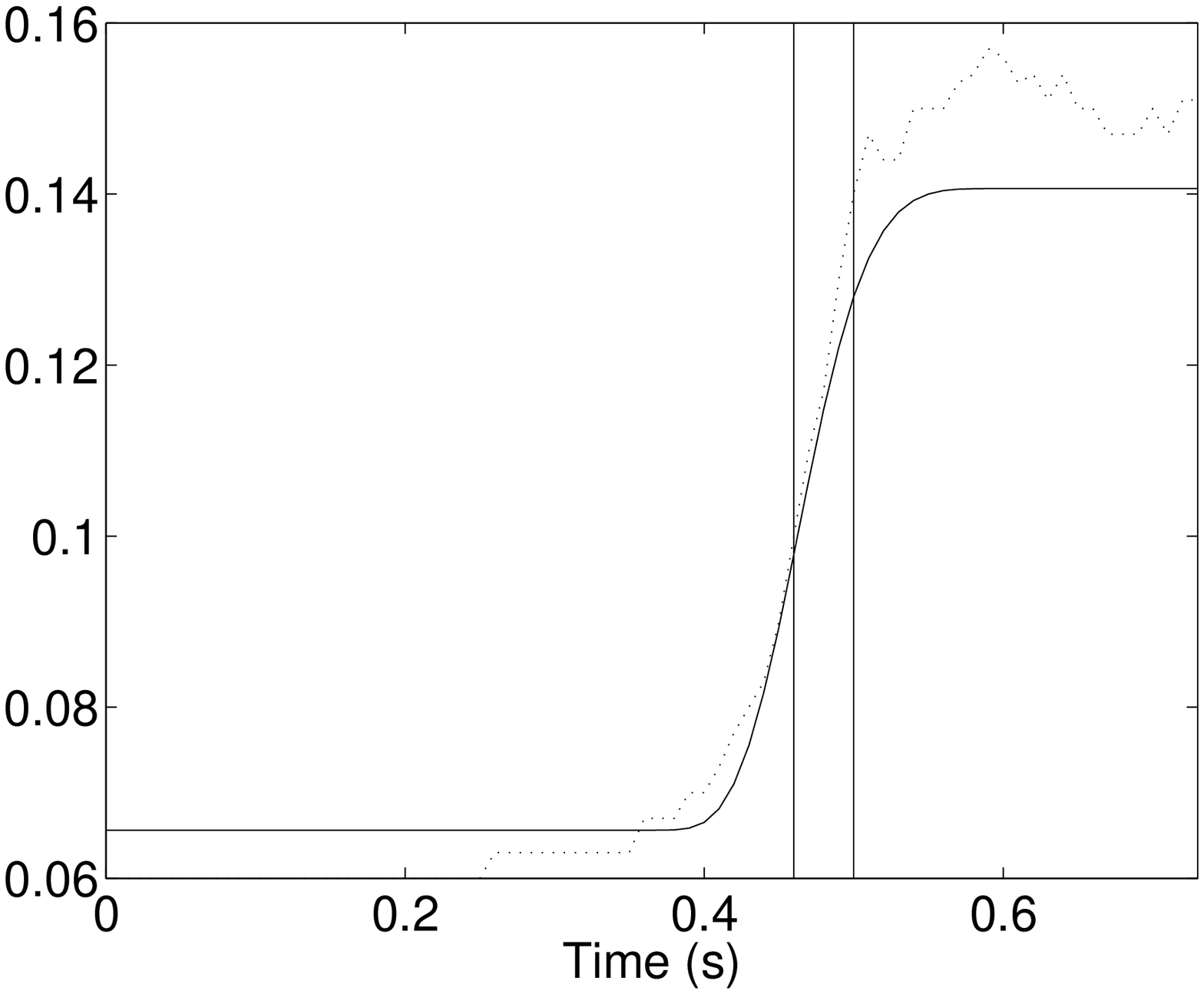} & \includegraphics[width=1.7in,height=0.8in]{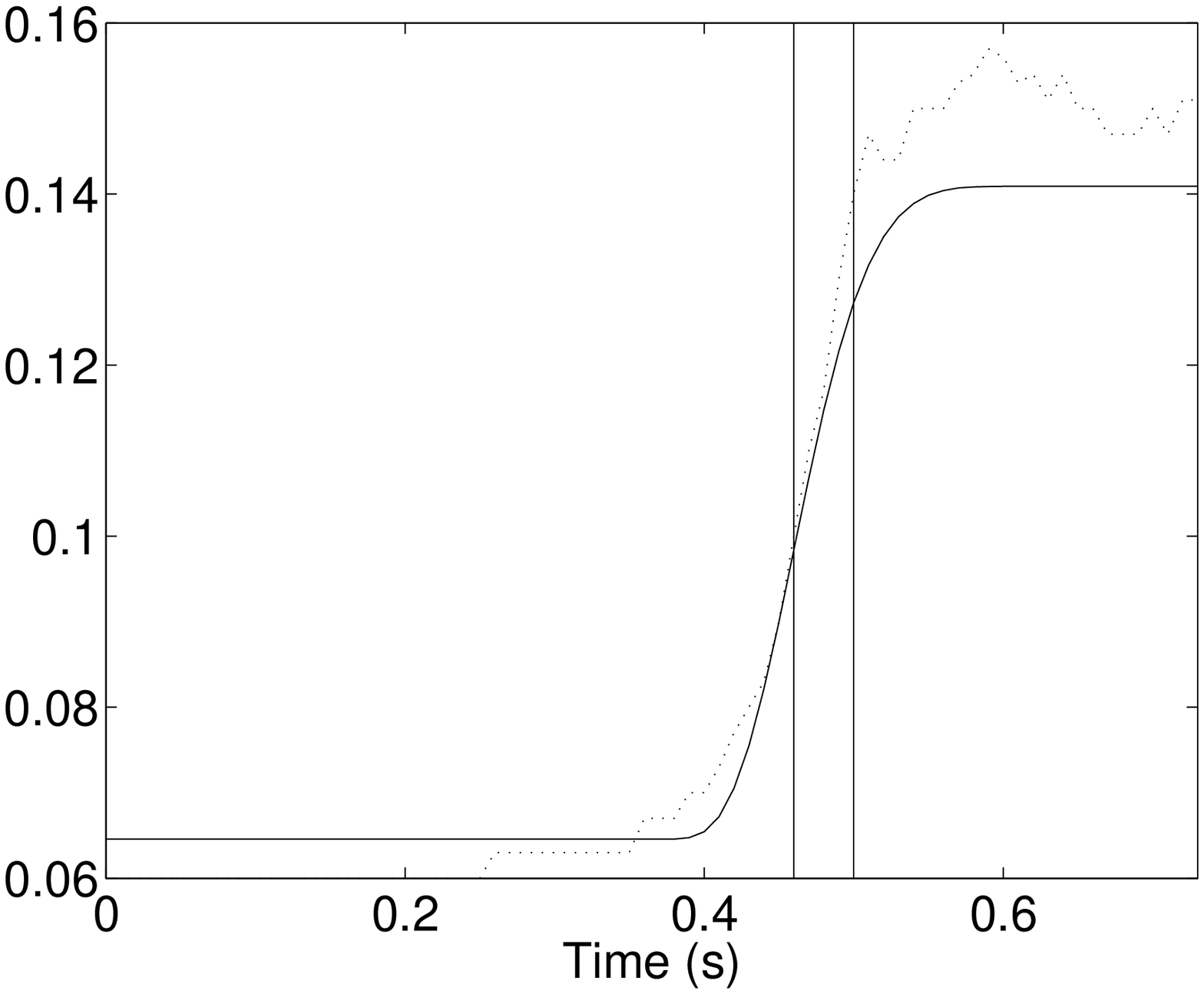} \\
\includegraphics[width=1.7in,height=0.8in]{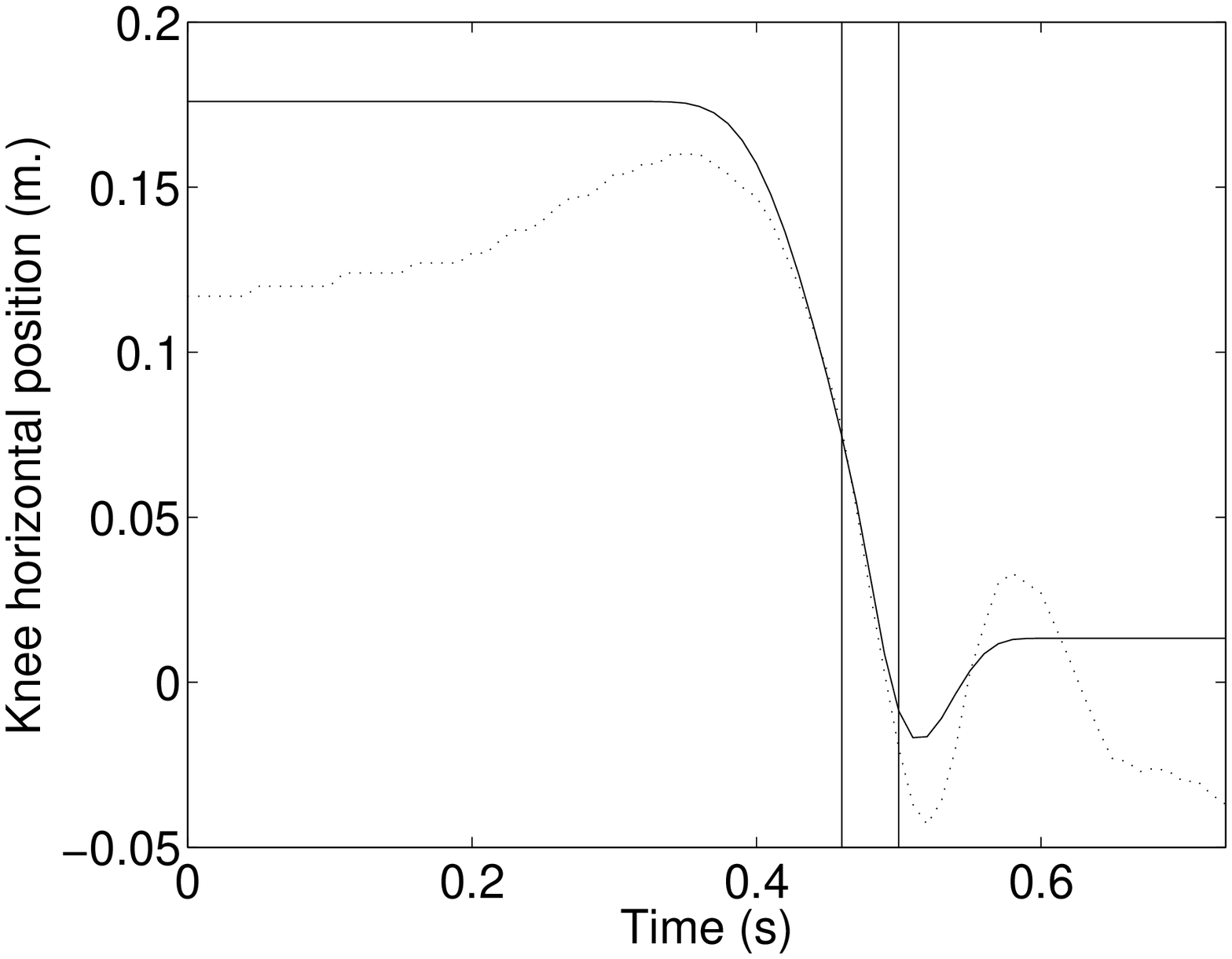} & \includegraphics[width=1.7in,height=0.8in]{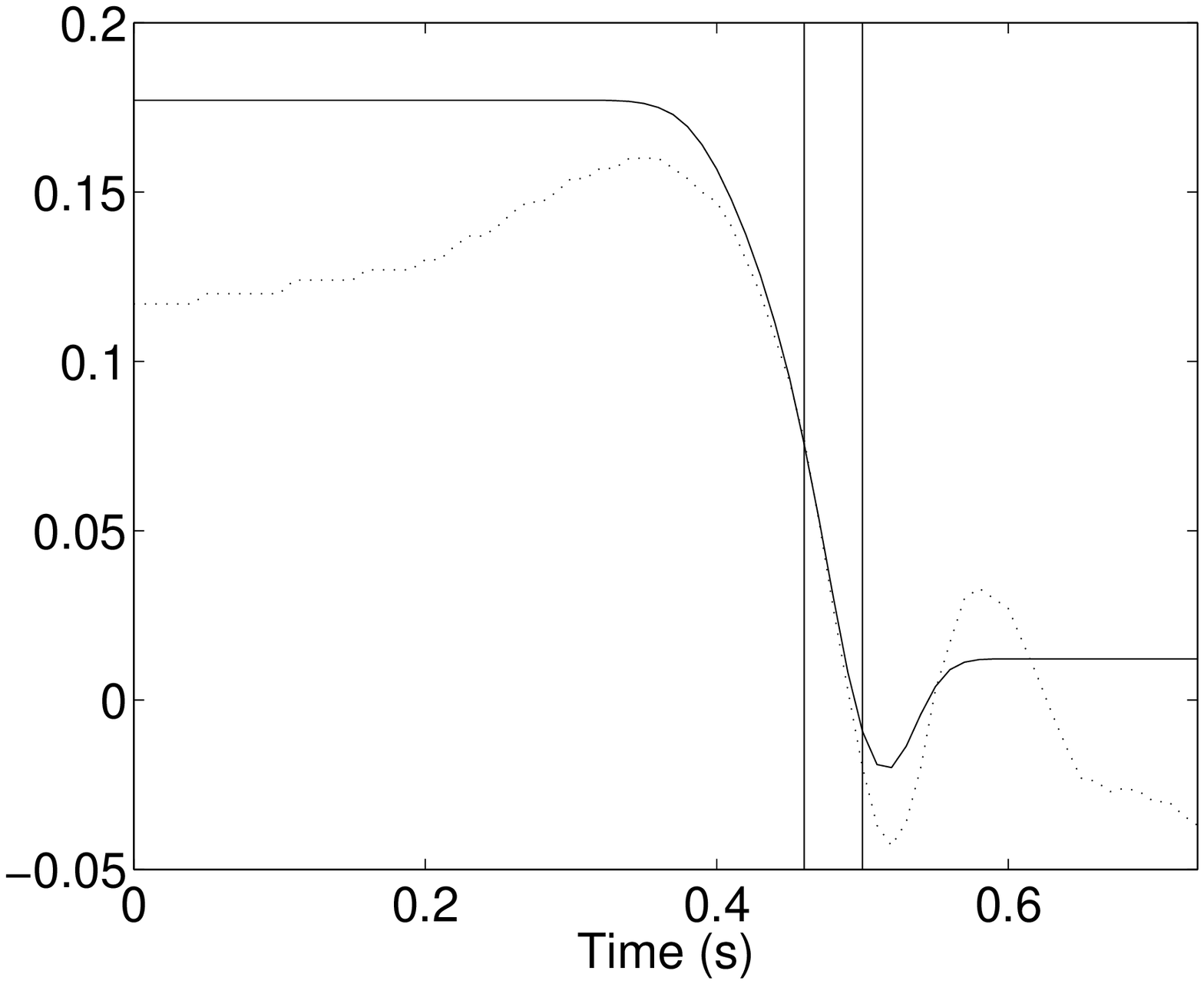} & \includegraphics[width=1.7in,height=0.8in]{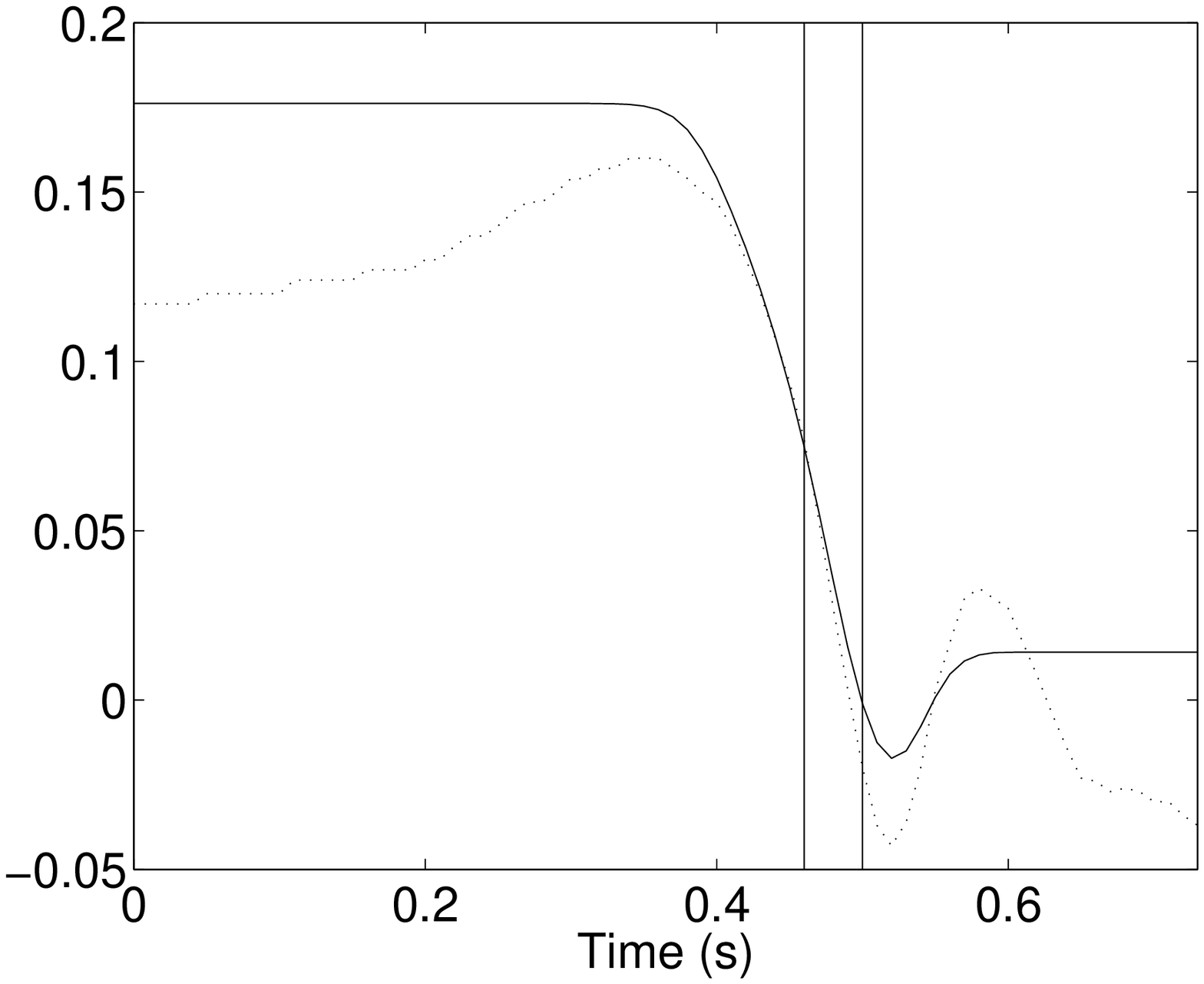} \\
\includegraphics[width=1.7in,height=0.8in]{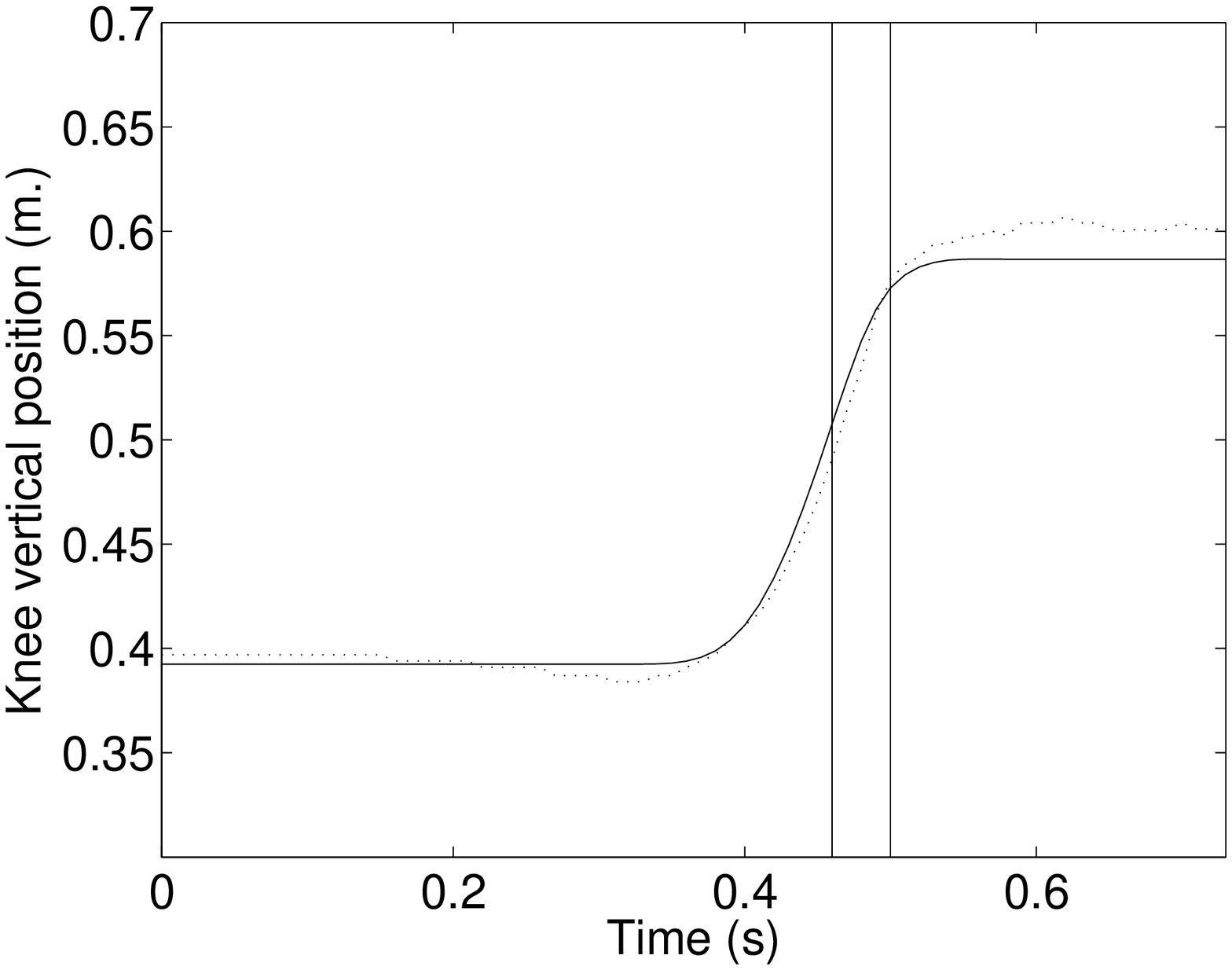} & \includegraphics[width=1.7in,height=0.8in]{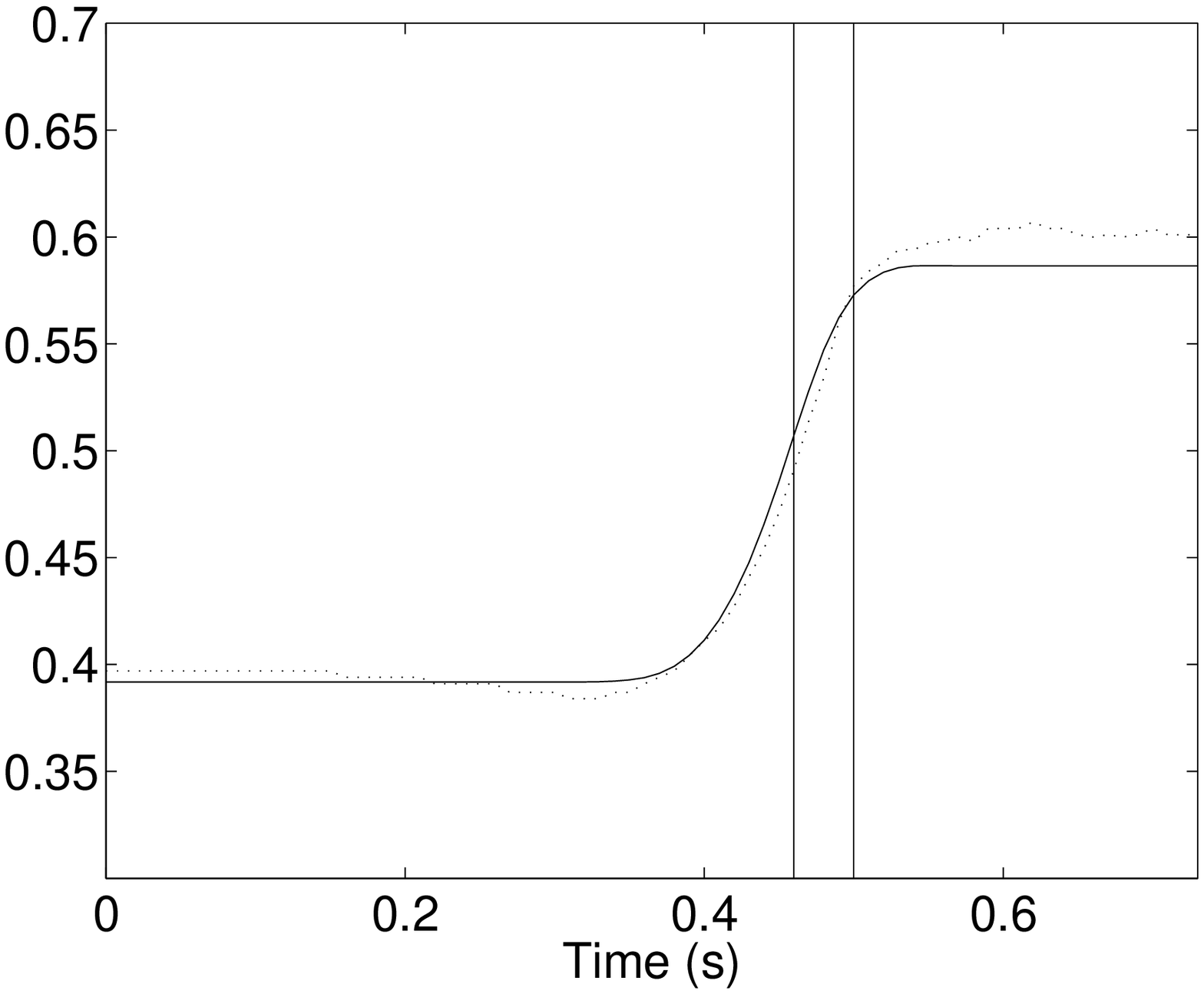} & \includegraphics[width=1.7in,height=0.8in]{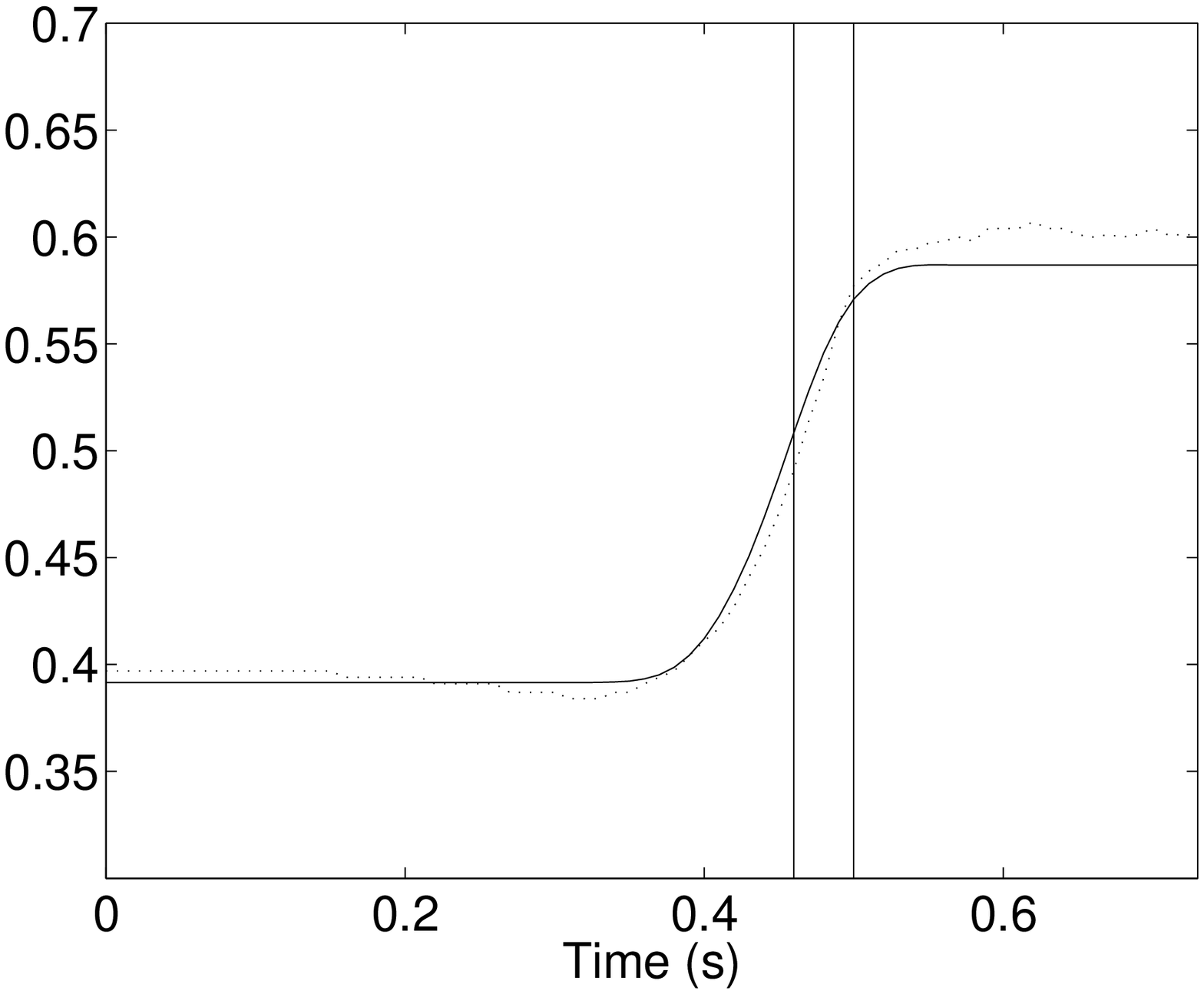} \\
\includegraphics[width=1.7in,height=0.8in]{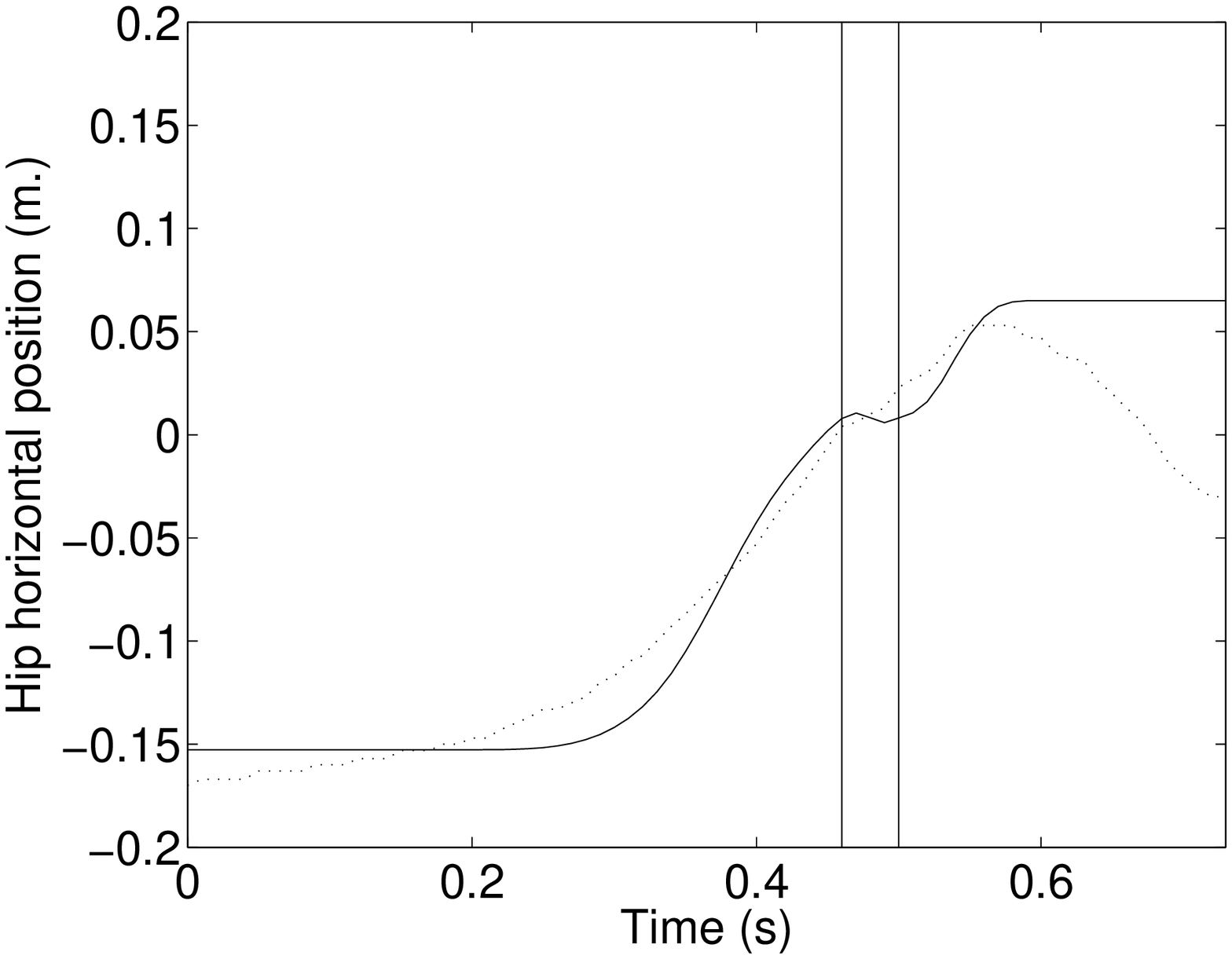} & \includegraphics[width=1.7in,height=0.8in]{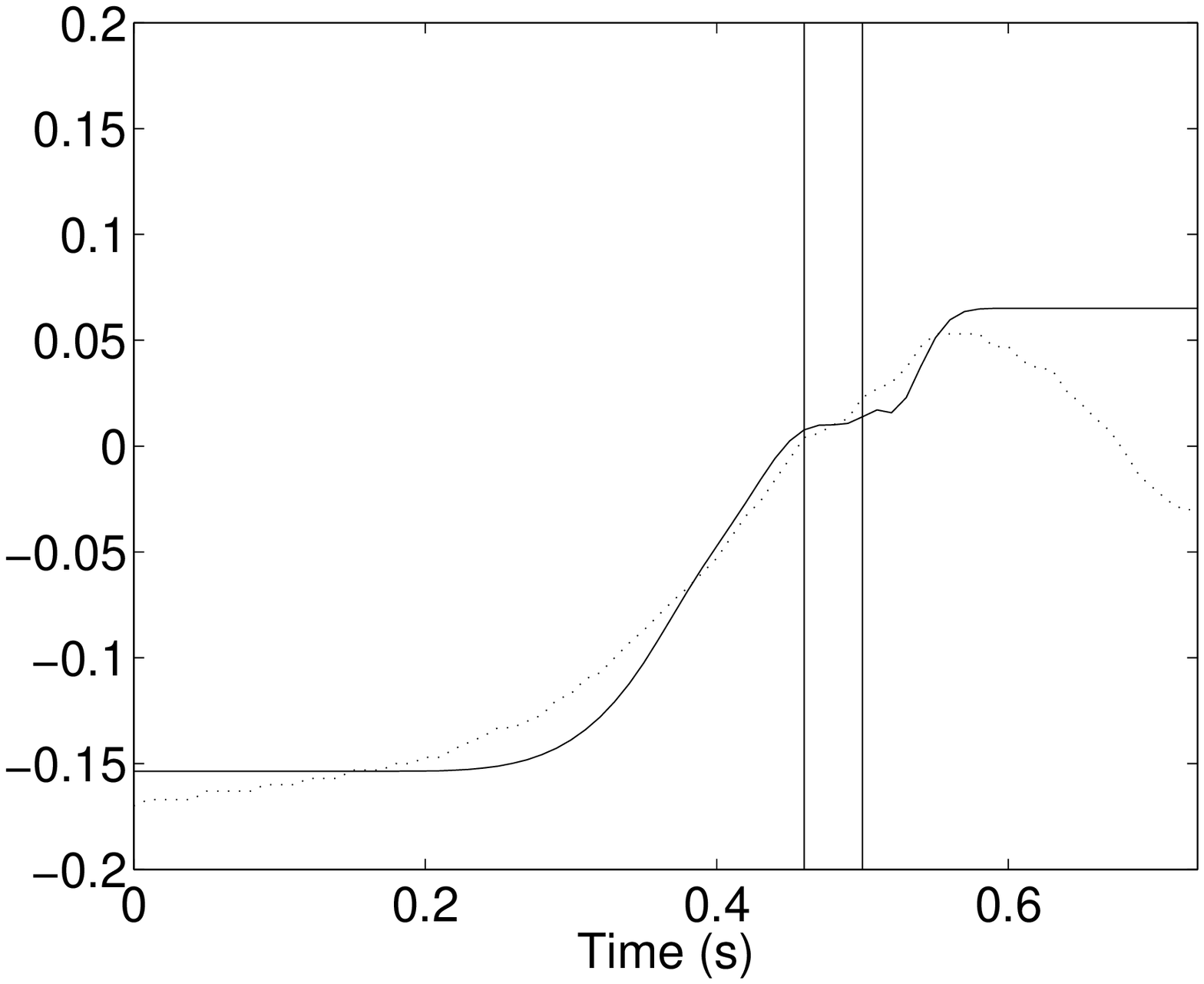} & \includegraphics[width=1.7in,height=0.8in]{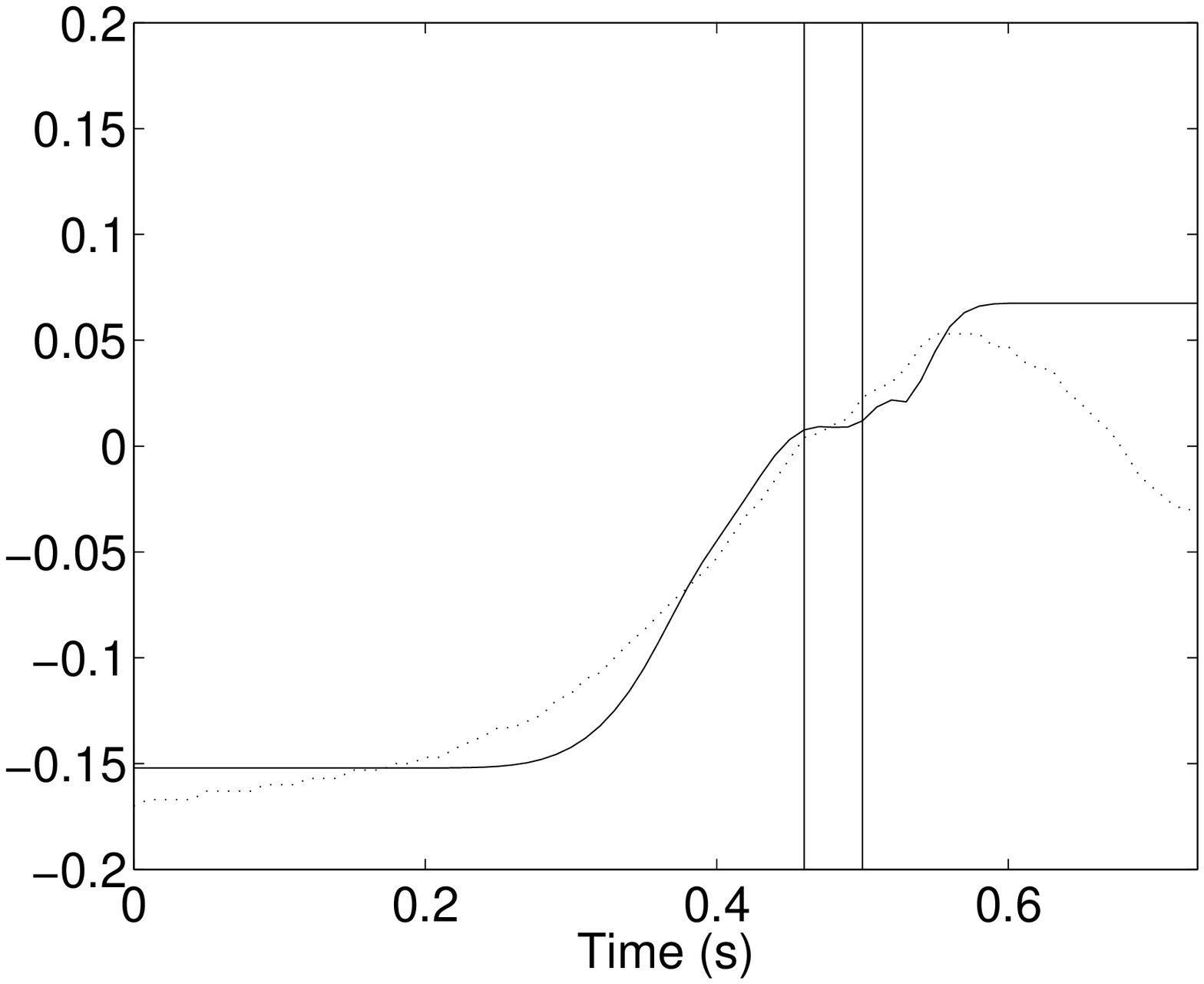} \\
\includegraphics[width=1.7in,height=0.8in]{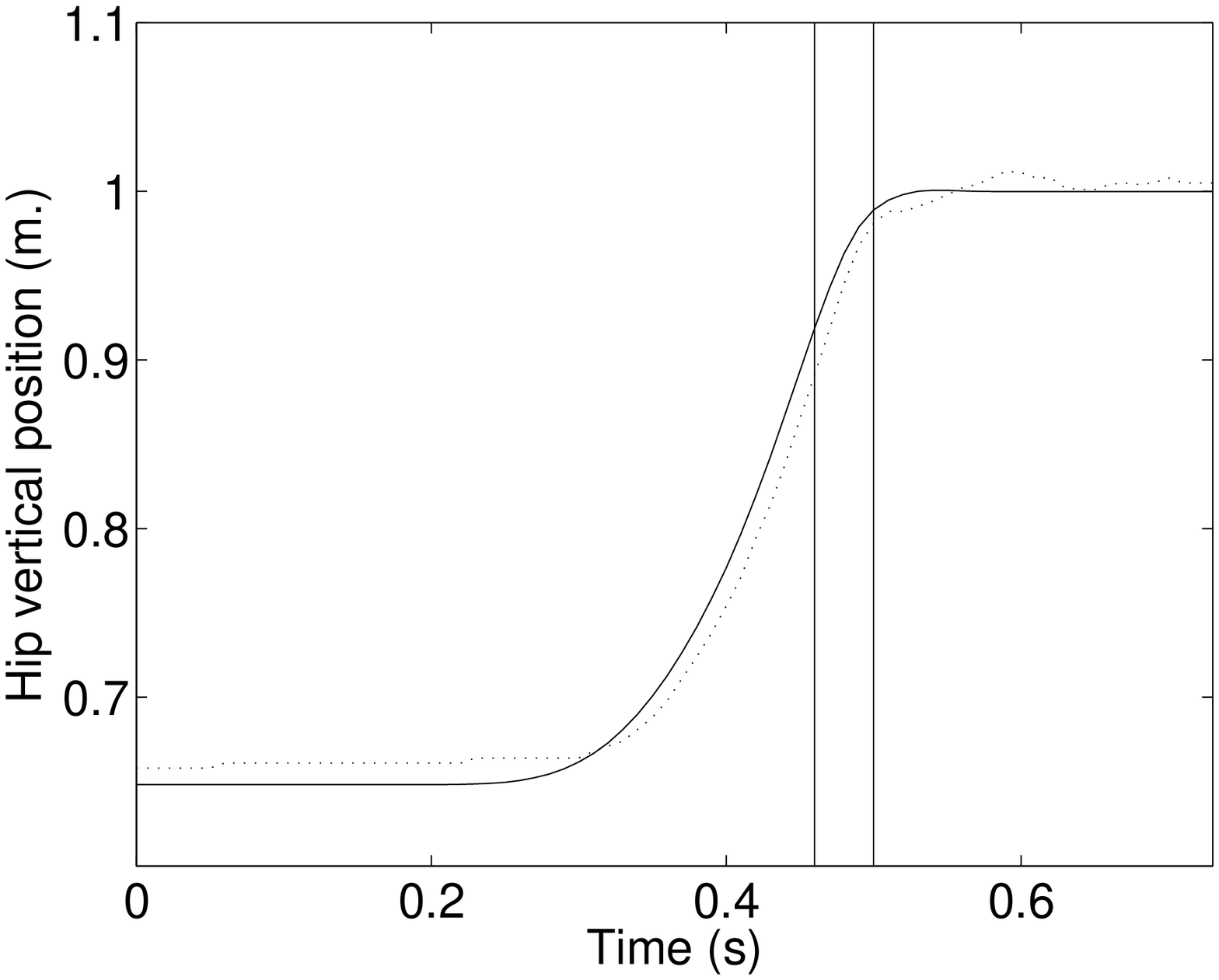} & \includegraphics[width=1.7in,height=0.8in]{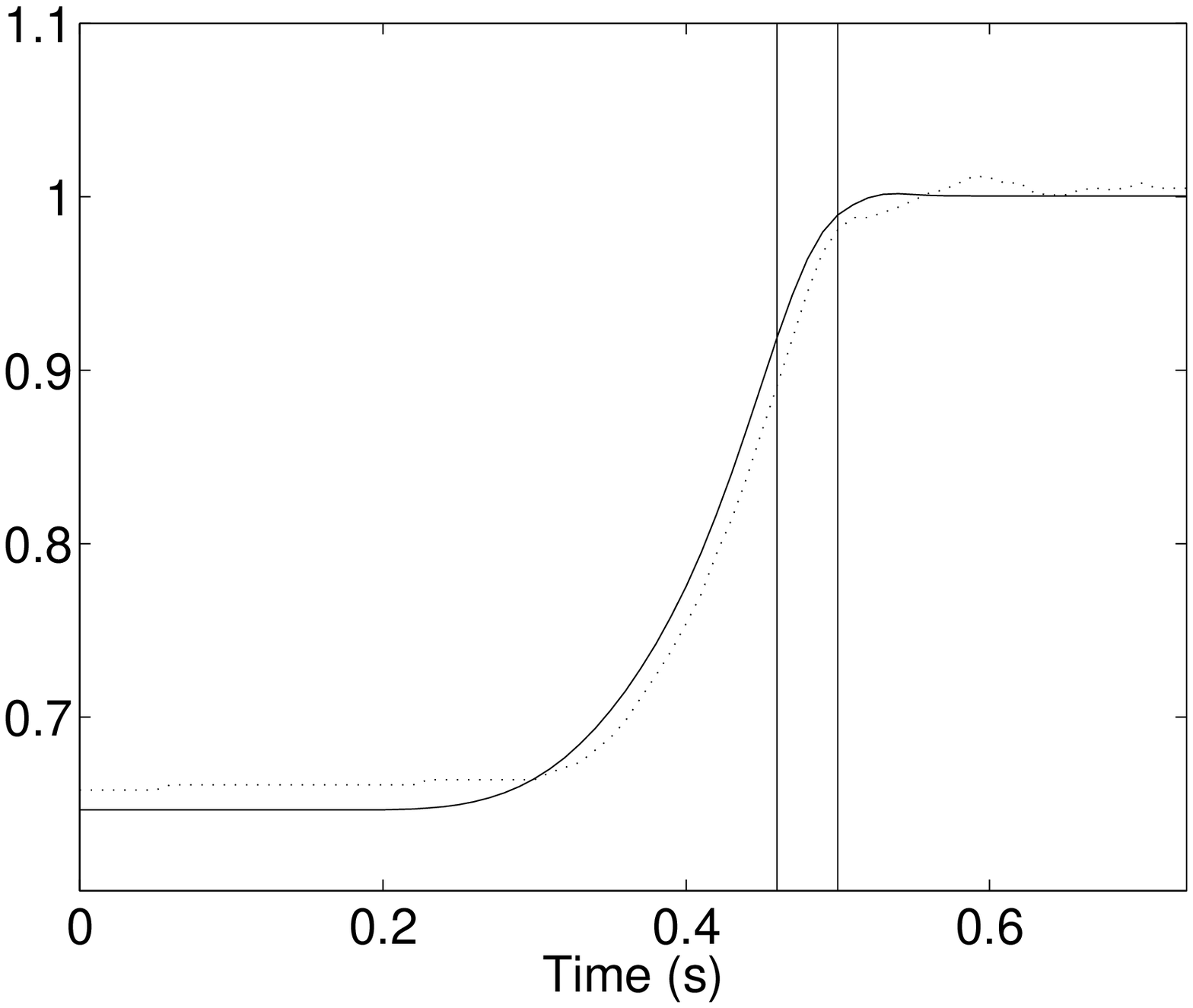} & \includegraphics[width=1.7in,height=0.8in]{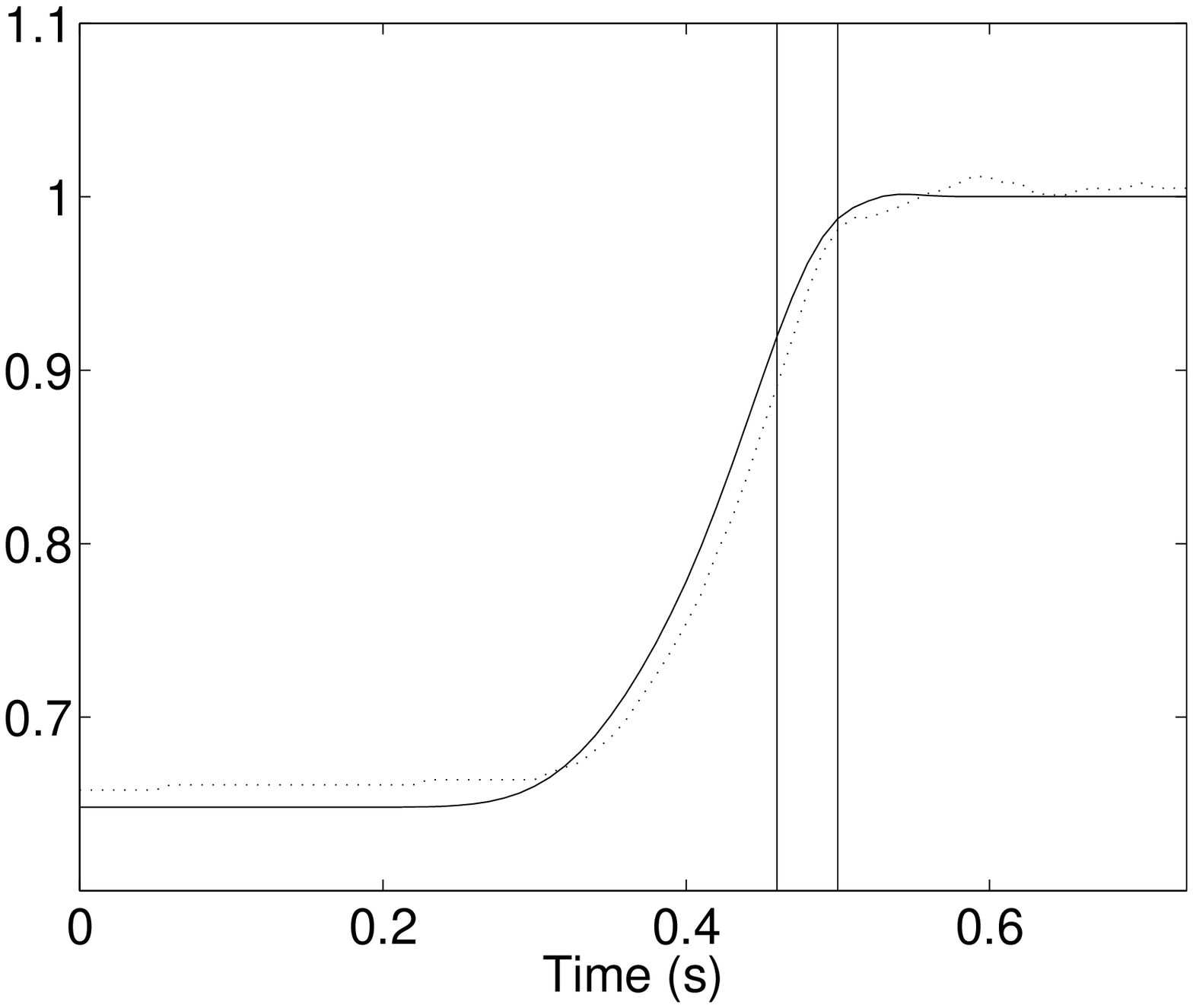} \\
\includegraphics[width=1.7in,height=0.8in]{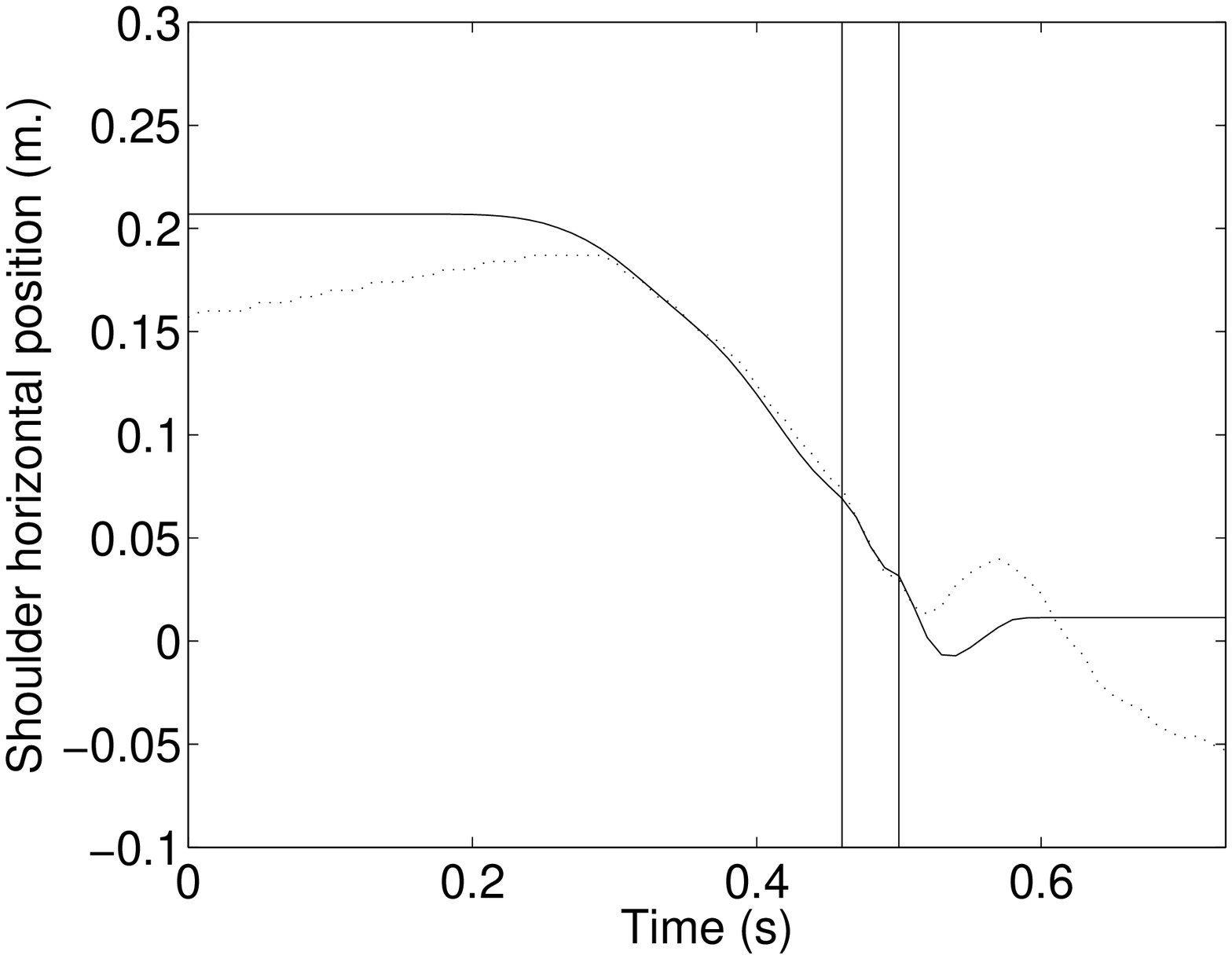} & \includegraphics[width=1.7in,height=0.8in]{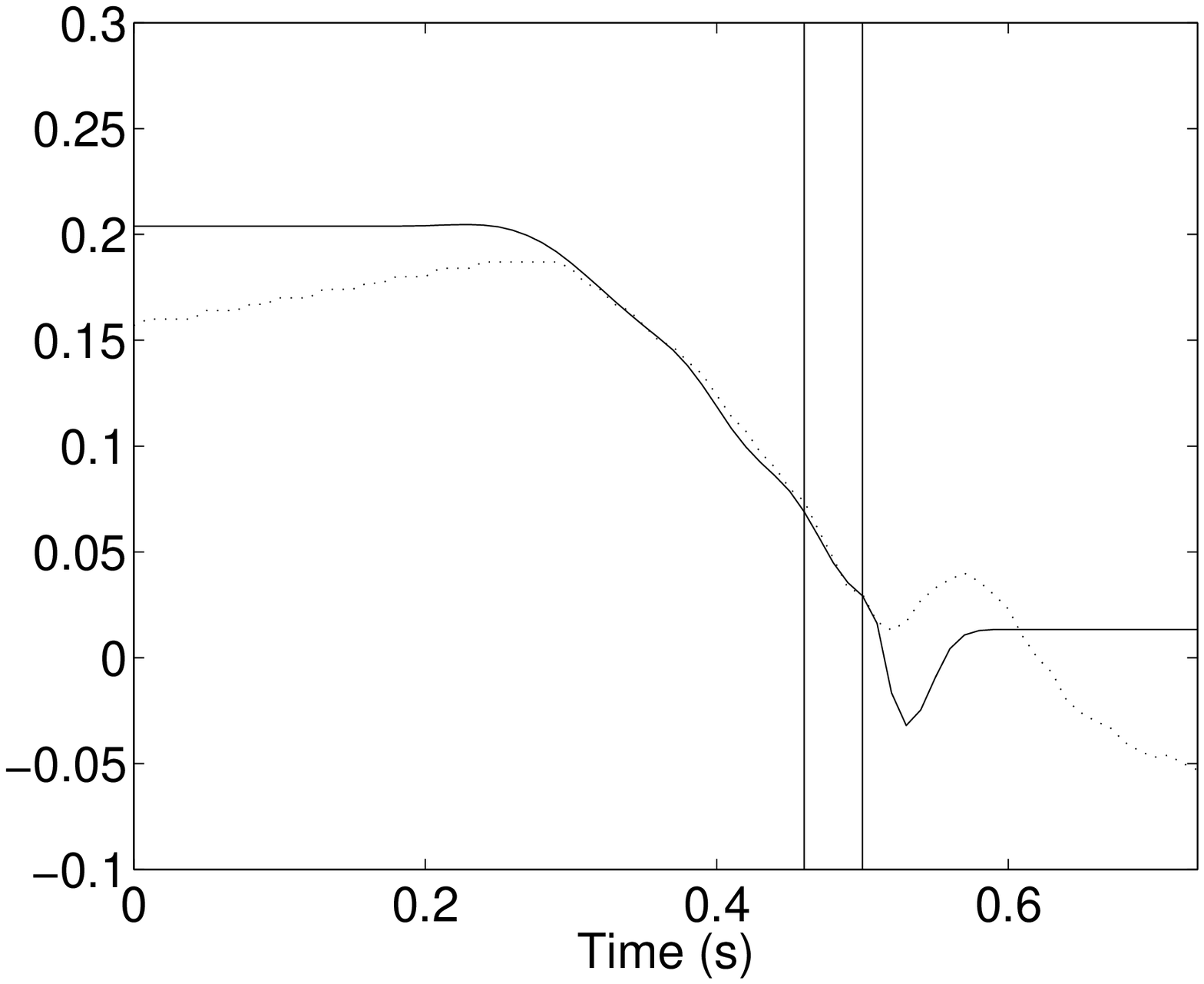} & \includegraphics[width=1.7in,height=0.8in]{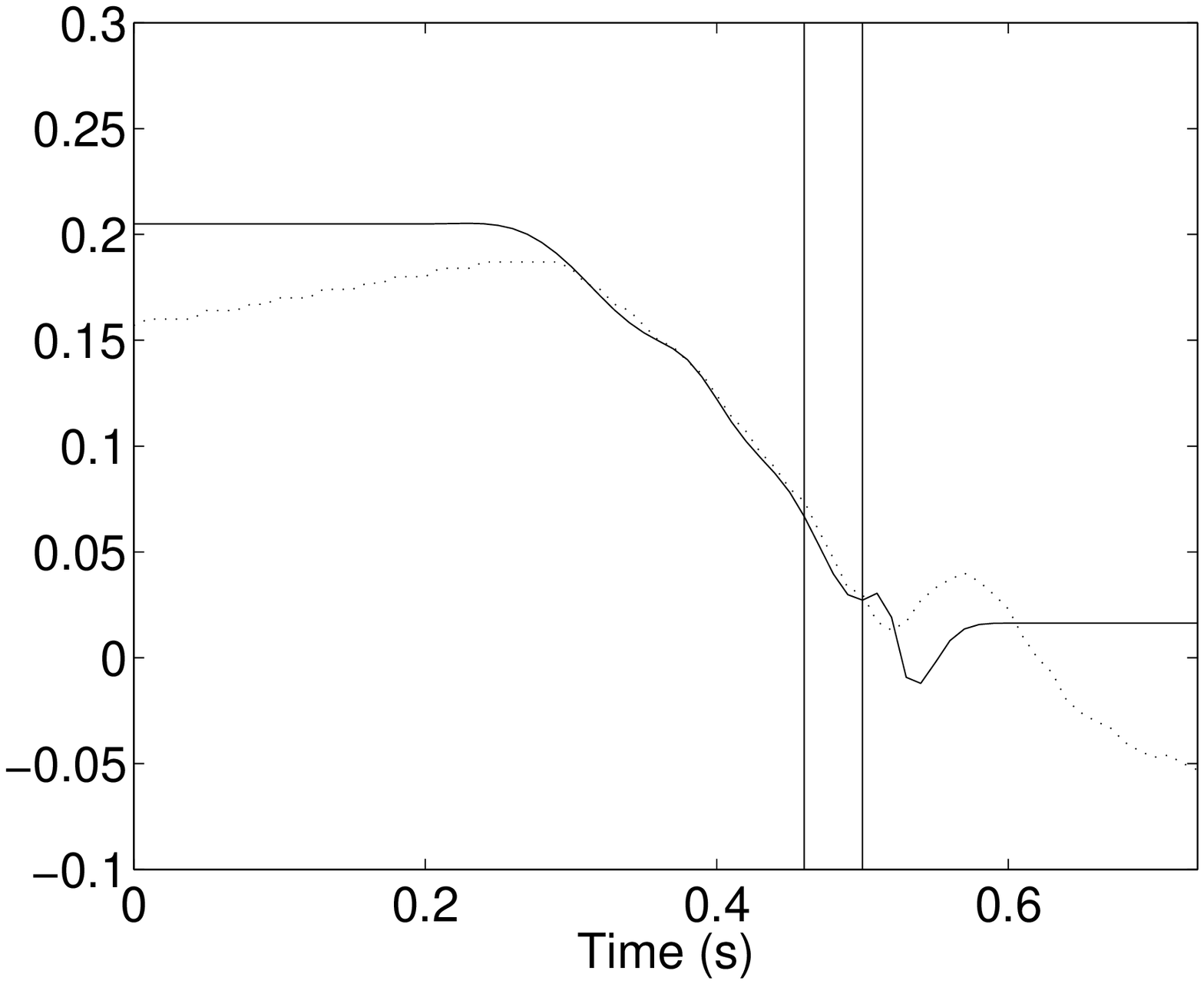} \\
\includegraphics[width=1.7in,height=0.8in]{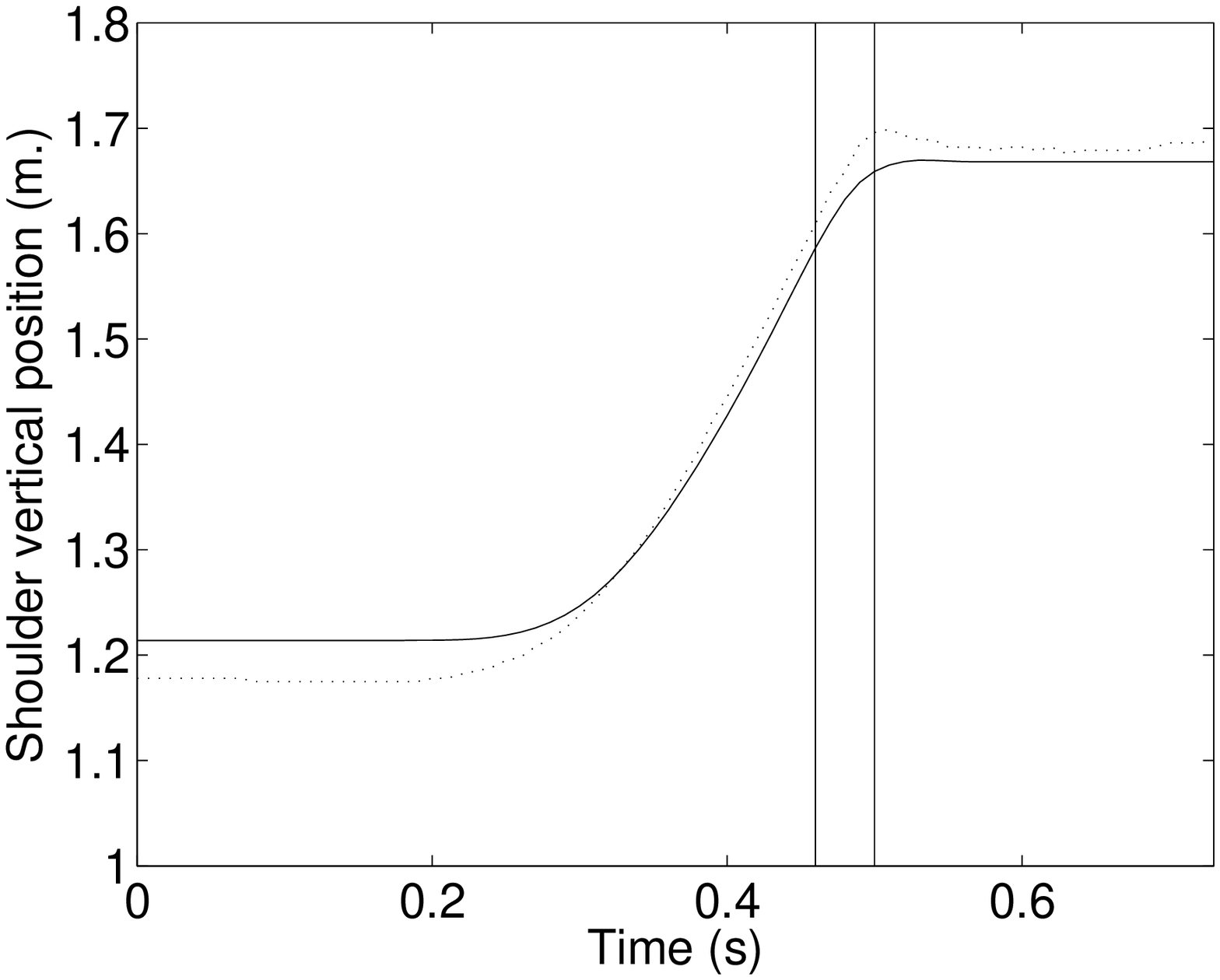} & \includegraphics[width=1.7in,height=0.8in]{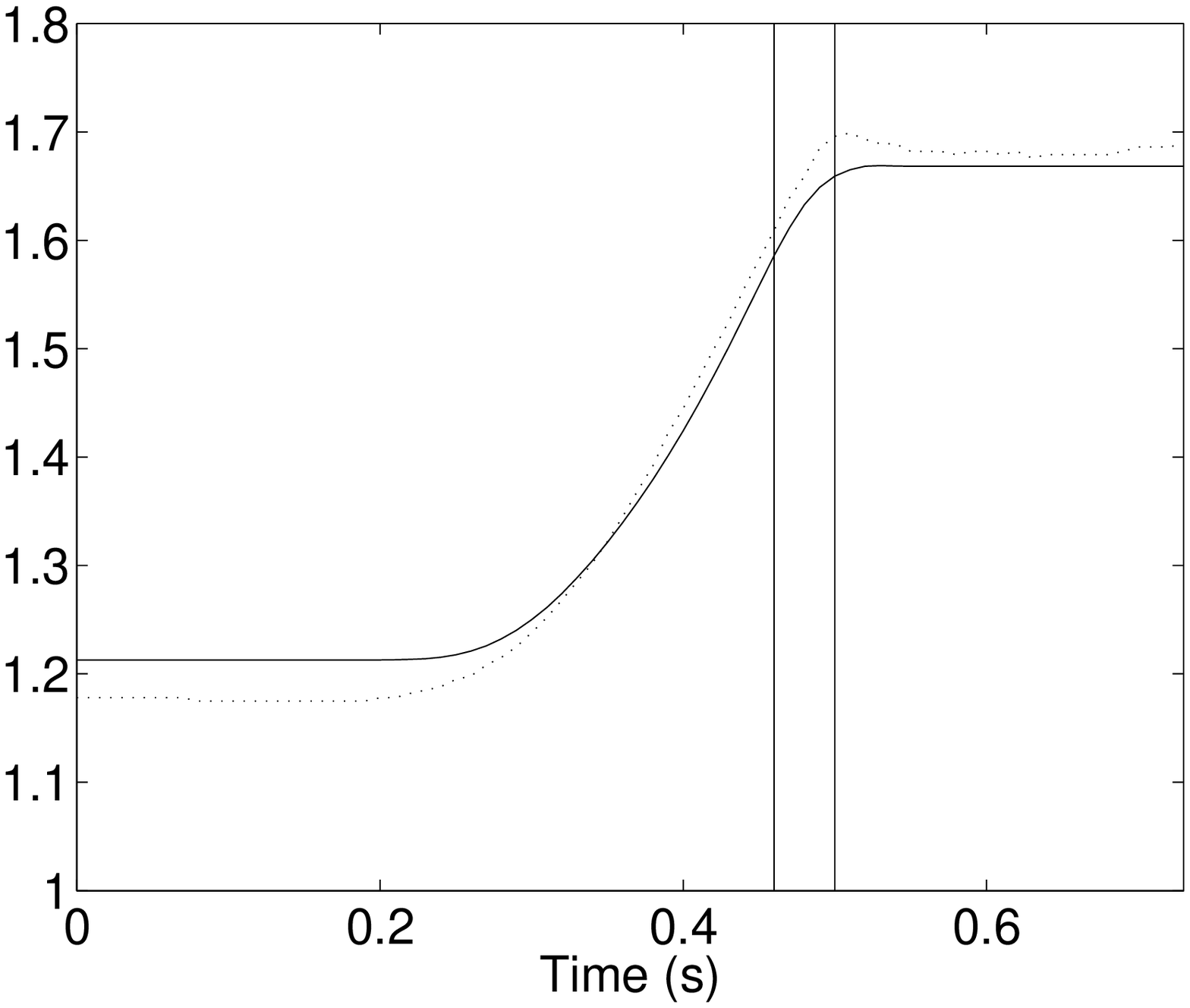} & \includegraphics[width=1.7in,height=0.8in]{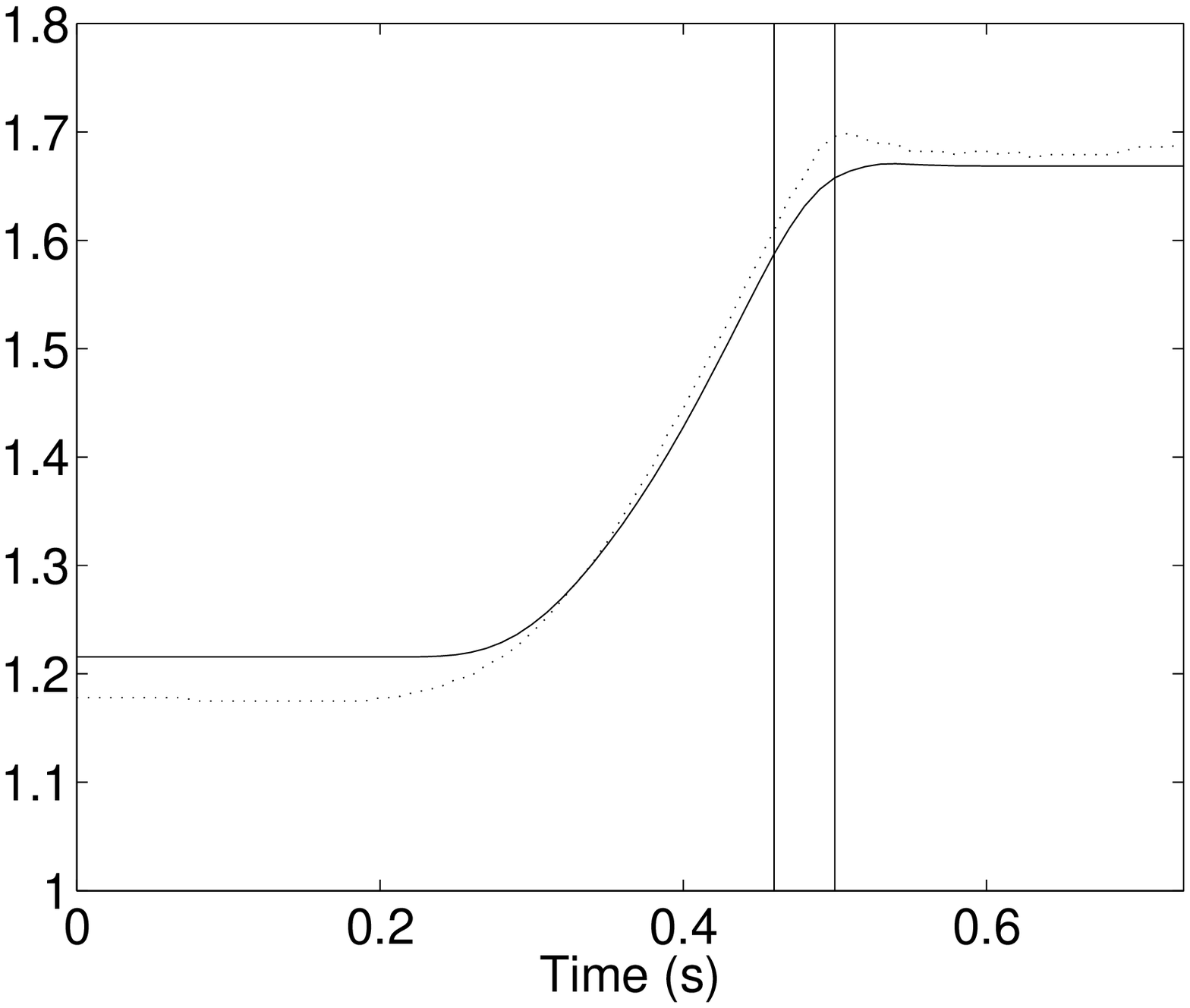}
\end{array}$
\end{center}
\caption{\label{SJFig3}Time histories of joints relative positions. Dotted and plain curves correspond to experimental and kinematic stage modeled data respectively. Vertical lines indicate $t_1$ and take-off instants.}
\end{figure}

\begin{figure}[!tp]
\begin{center}$
\begin{array}{ccc}
\includegraphics[width=1.7in,height=0.8in]{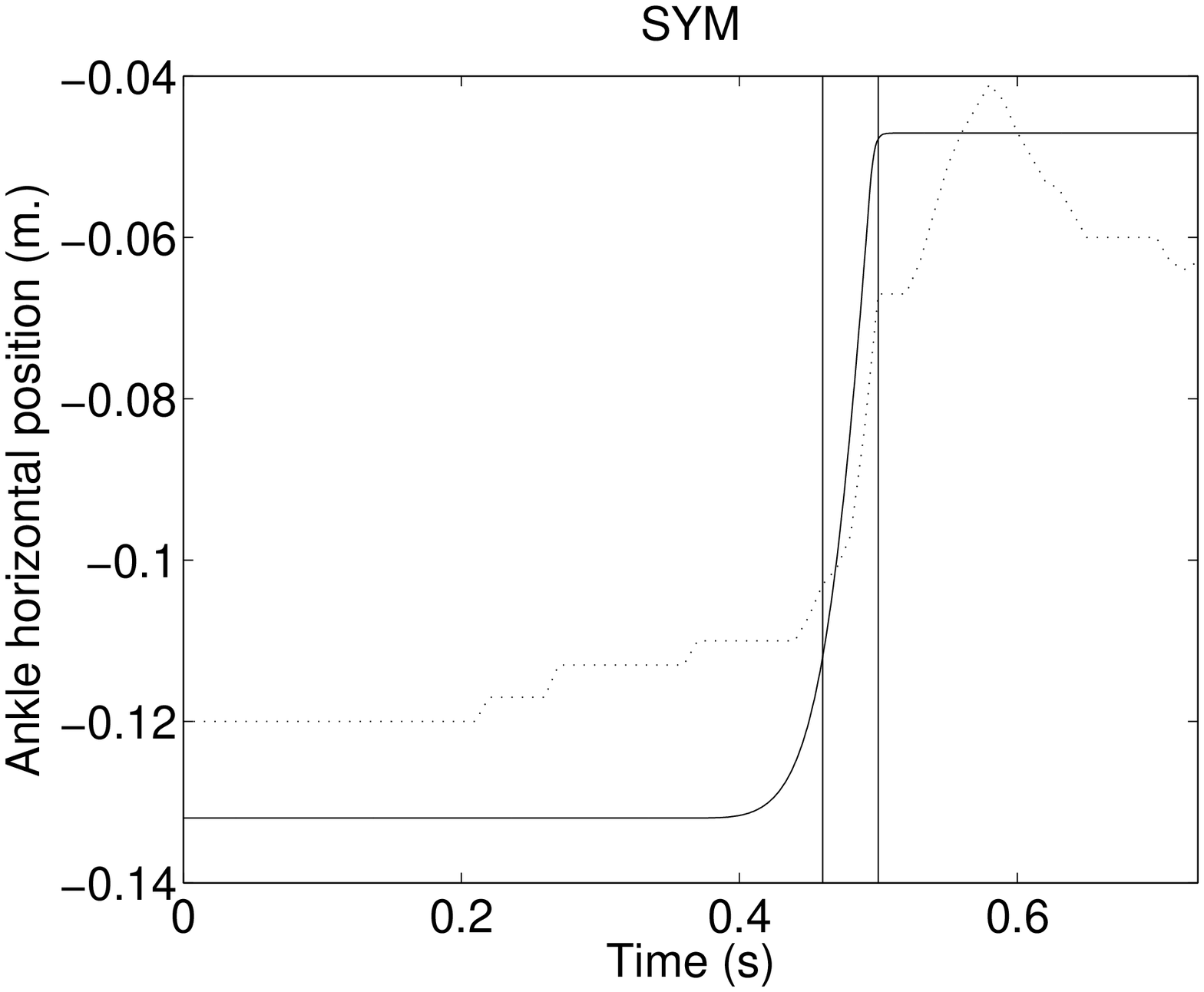} & \includegraphics[width=1.7in,height=0.8in]{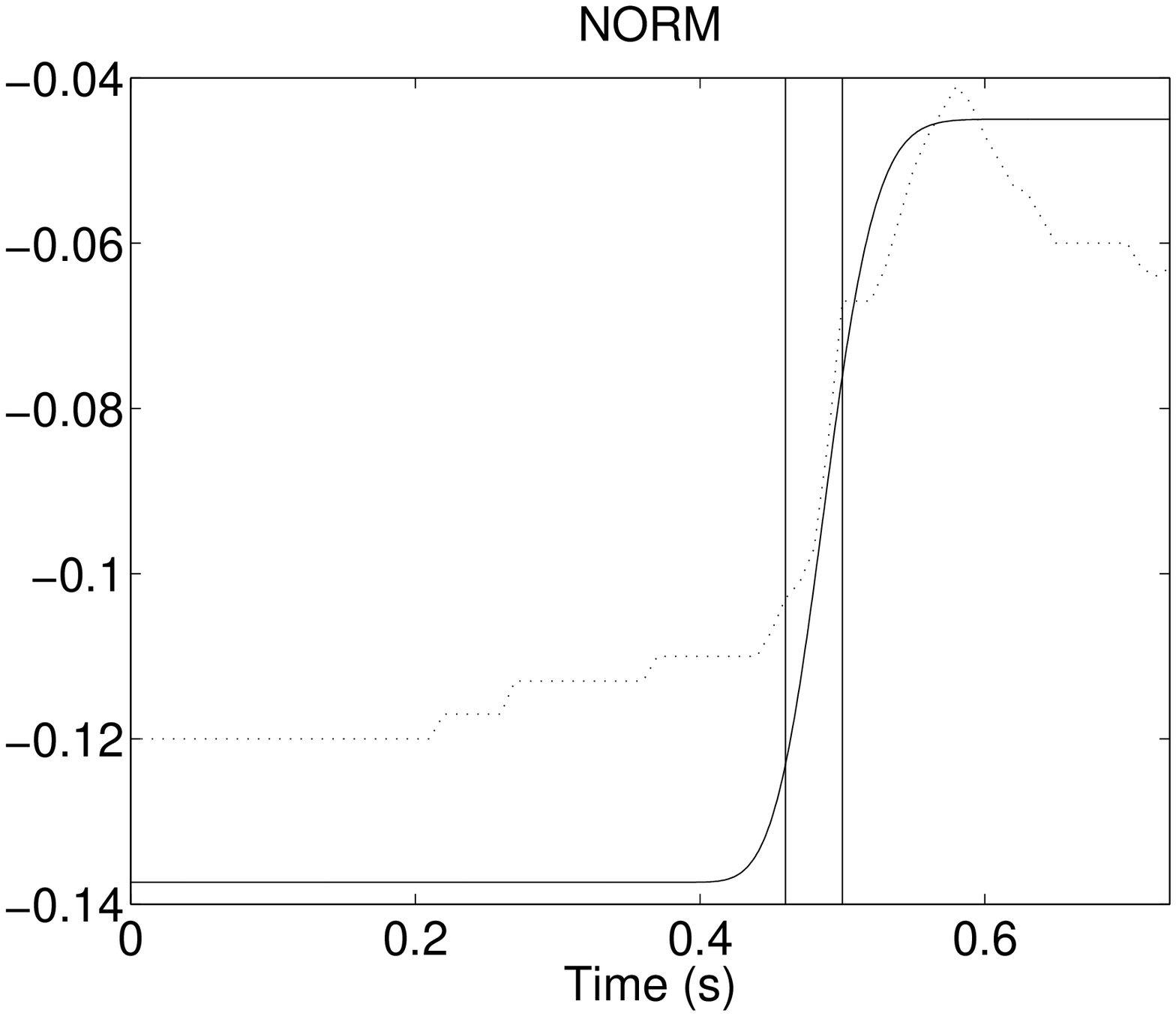} & \includegraphics[width=1.7in,height=0.8in]{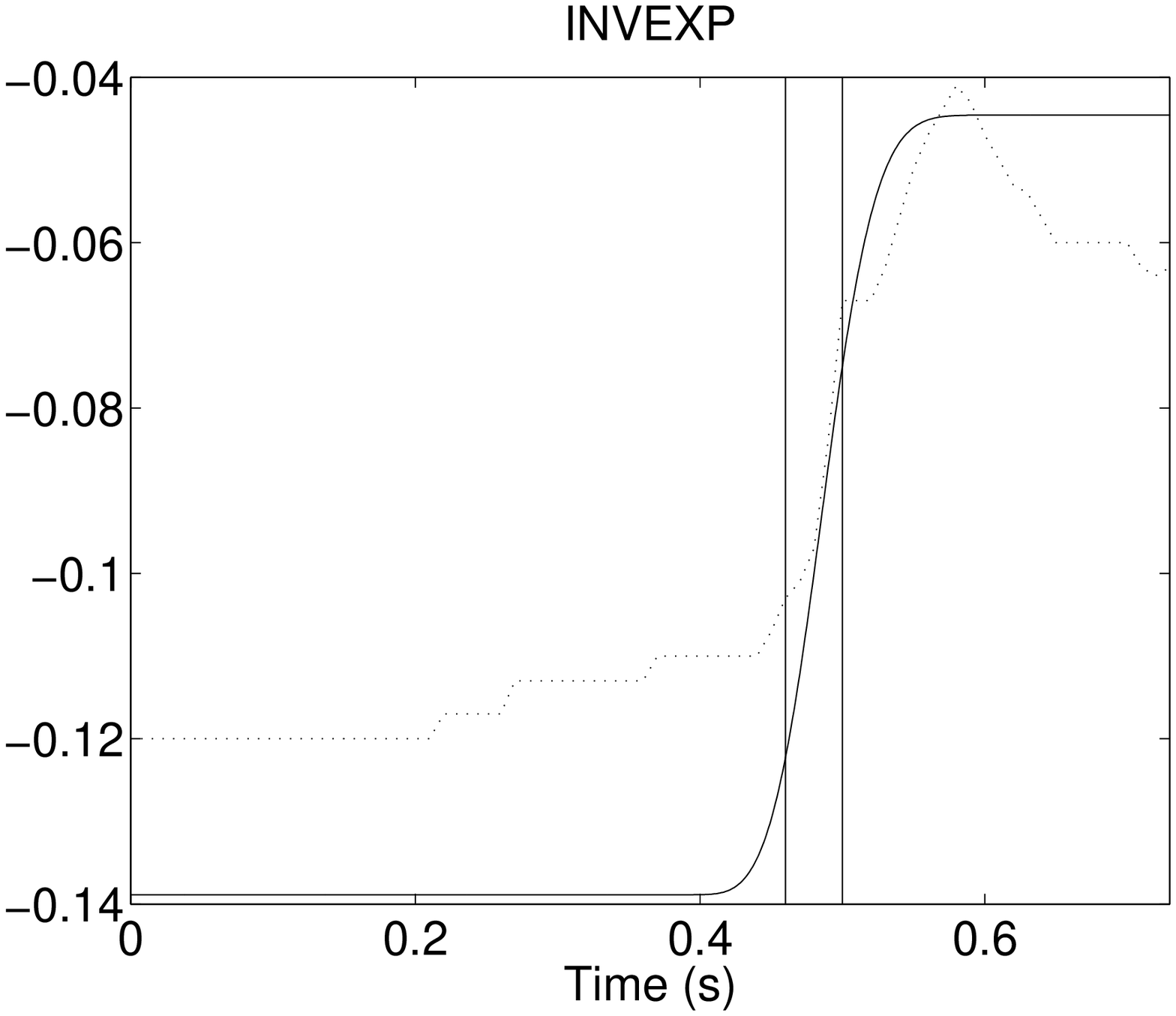} \\
\includegraphics[width=1.7in,height=0.8in]{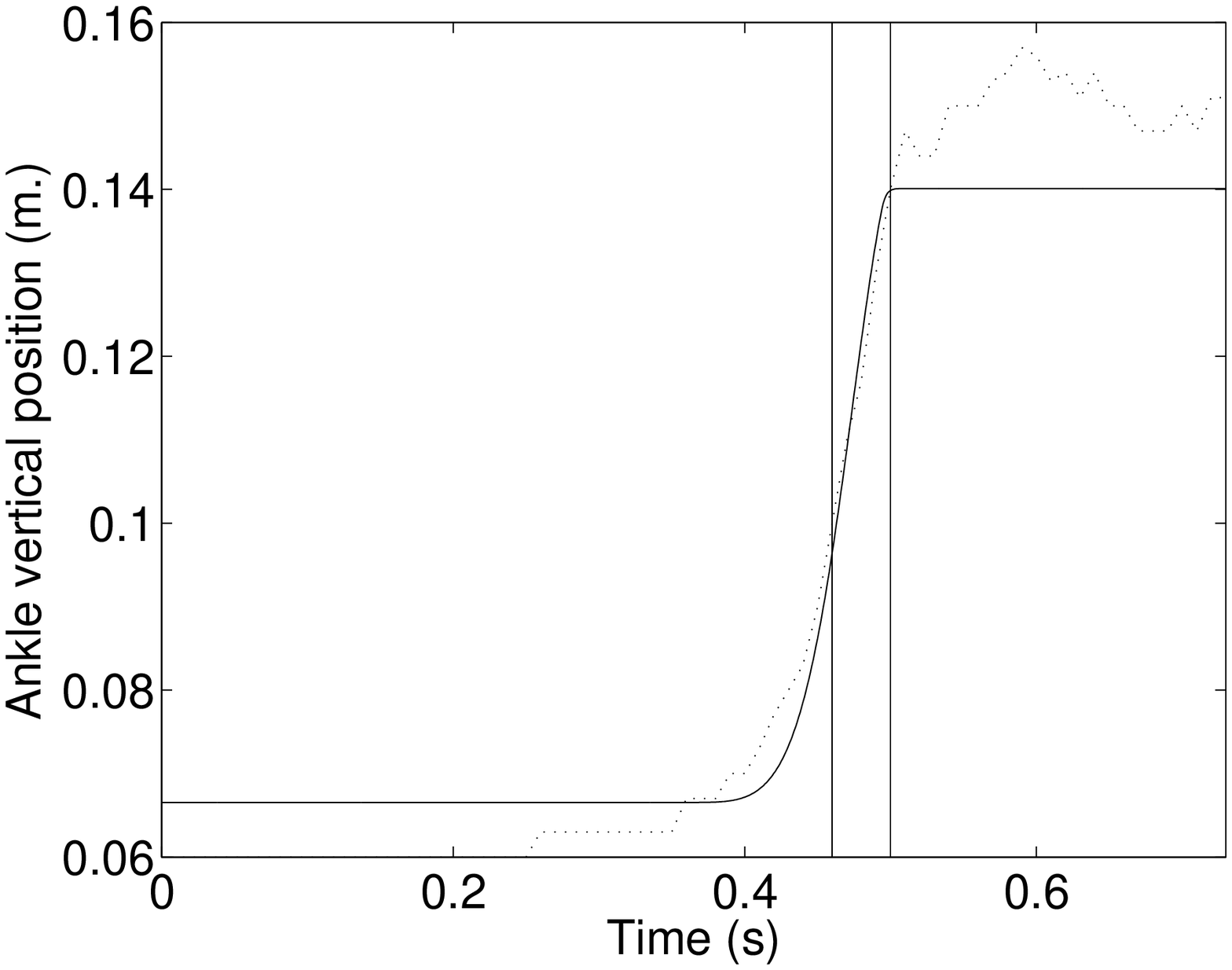} & \includegraphics[width=1.7in,height=0.8in]{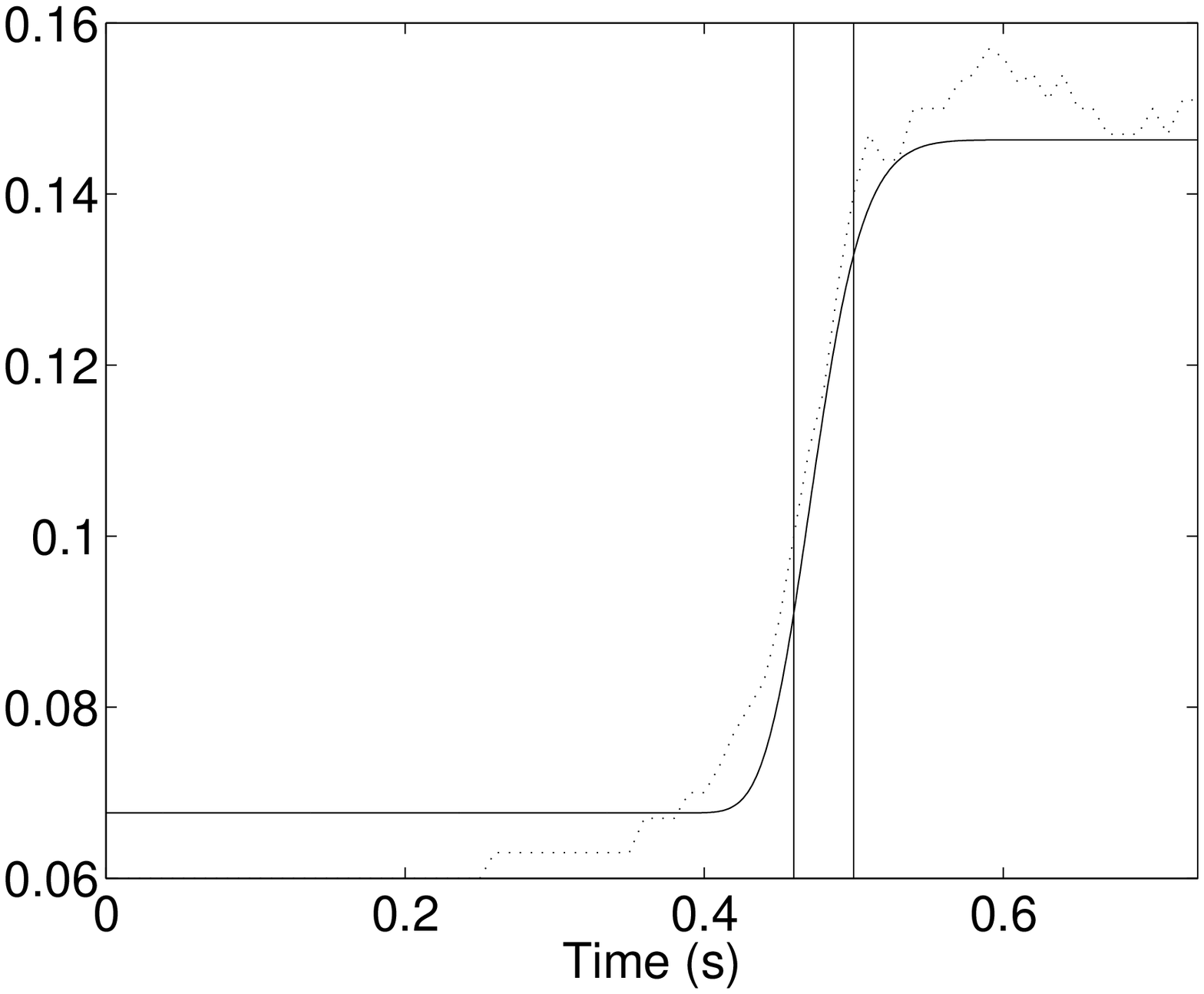} & \includegraphics[width=1.7in,height=0.8in]{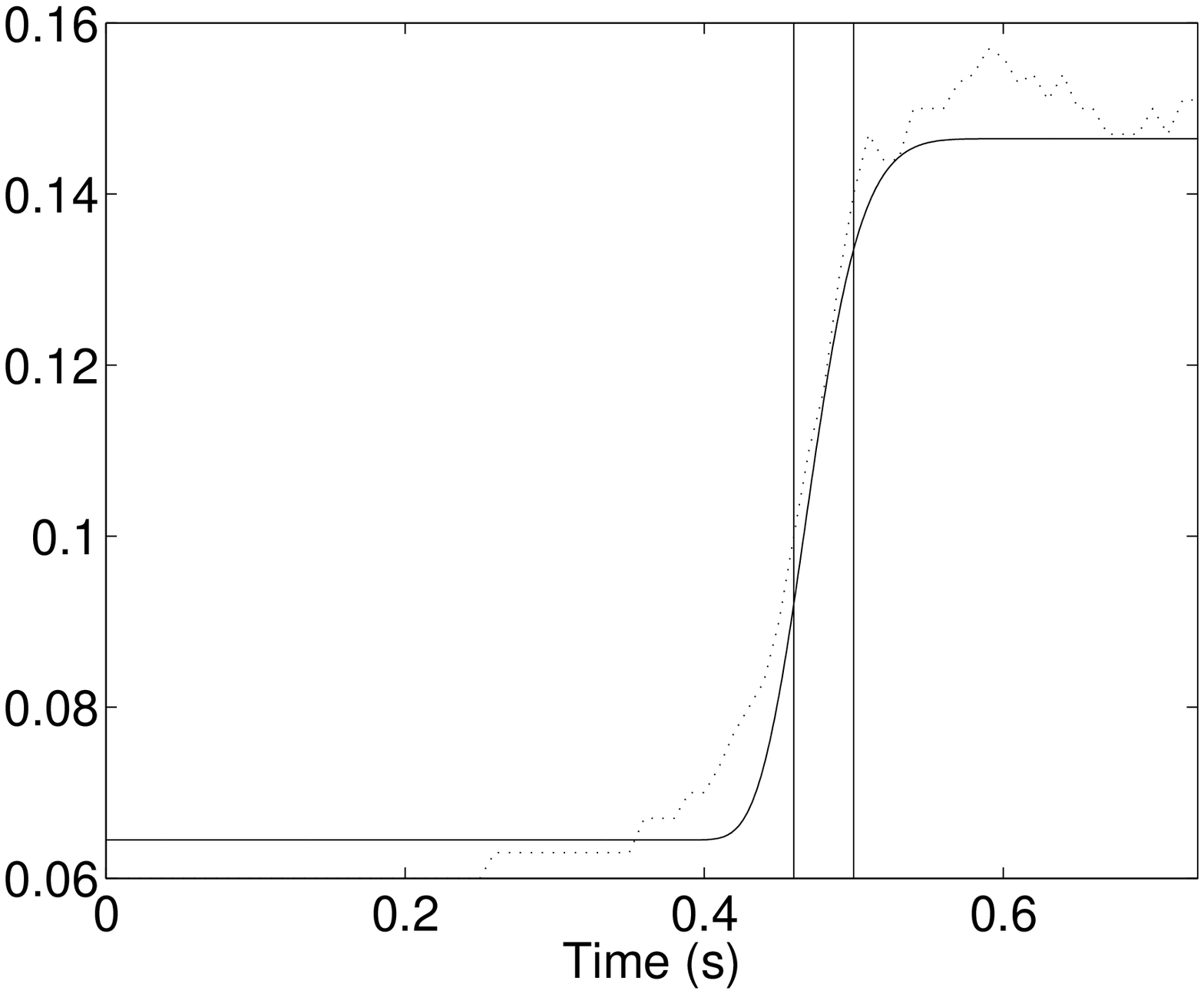} \\
\includegraphics[width=1.7in,height=0.8in]{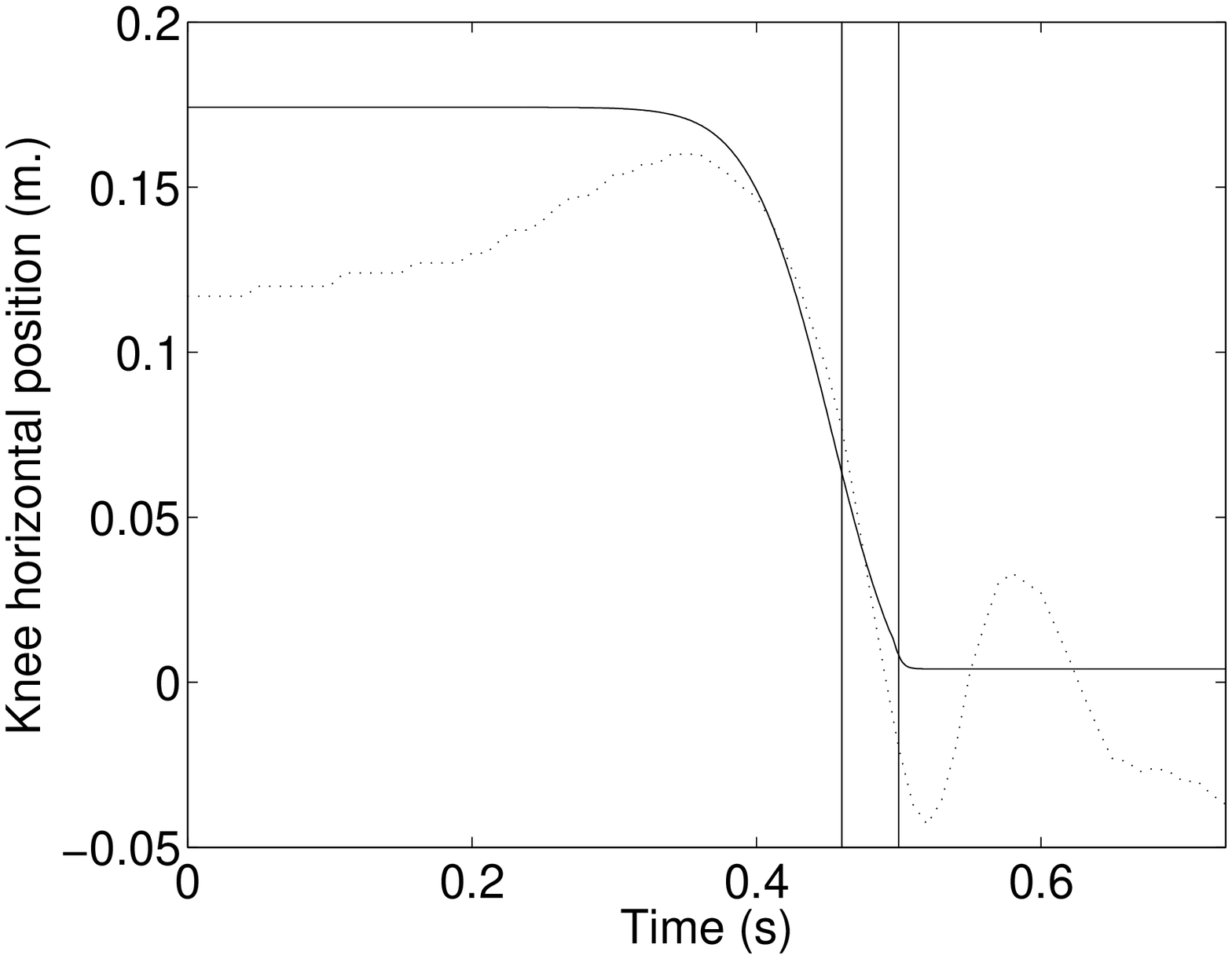} & \includegraphics[width=1.7in,height=0.8in]{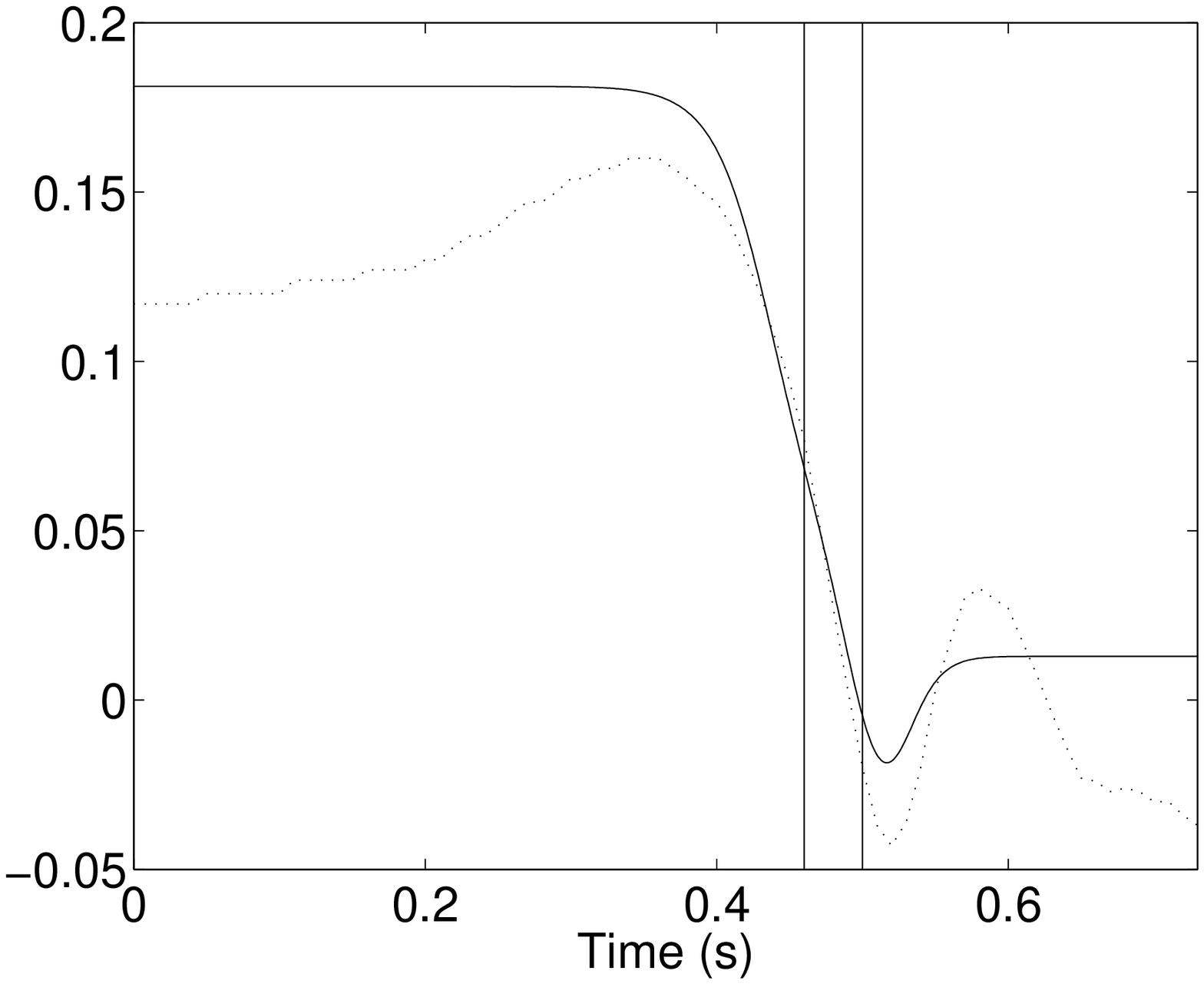} & \includegraphics[width=1.7in,height=0.8in]{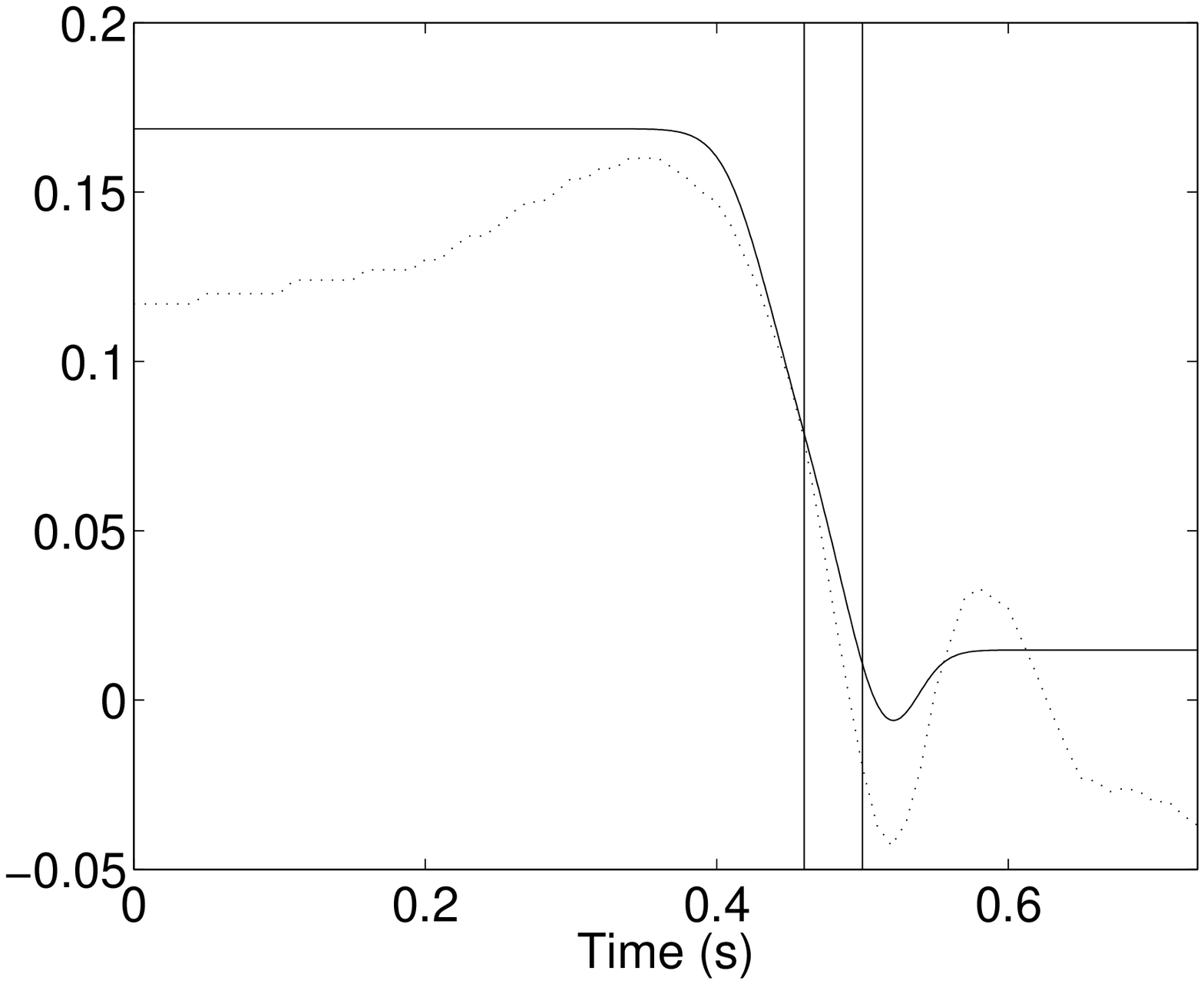} \\
\includegraphics[width=1.7in,height=0.8in]{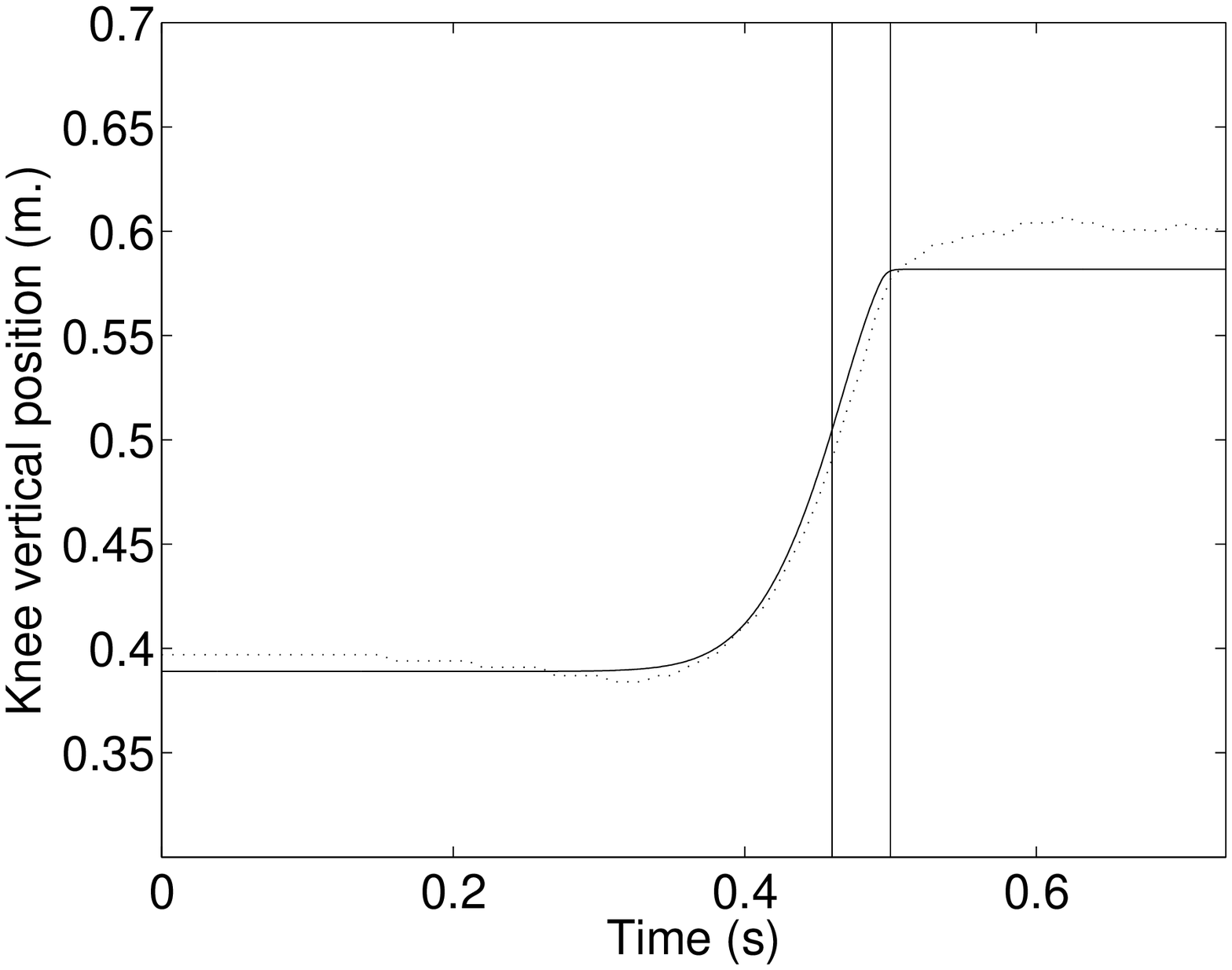} & \includegraphics[width=1.7in,height=0.8in]{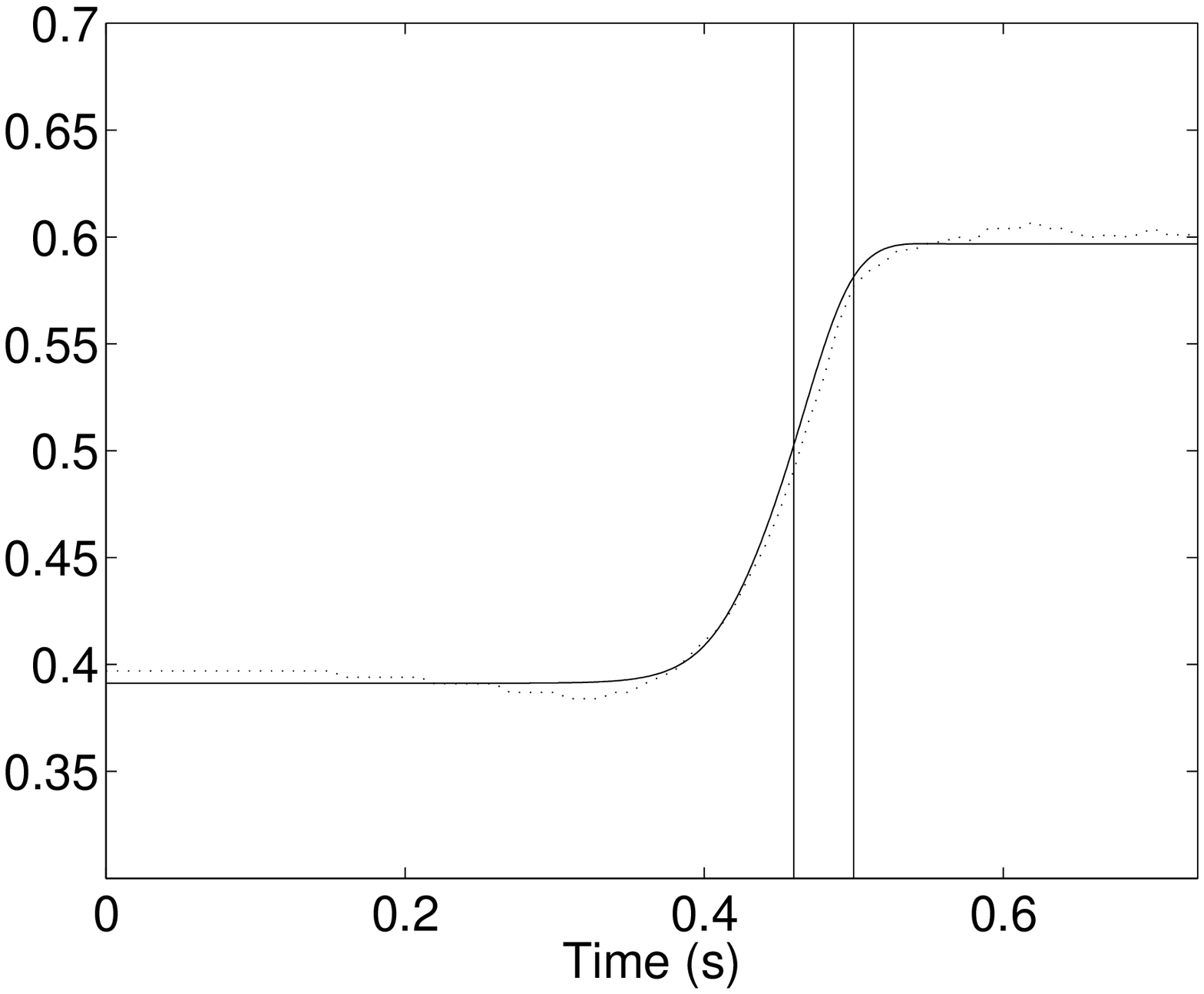} & \includegraphics[width=1.7in,height=0.8in]{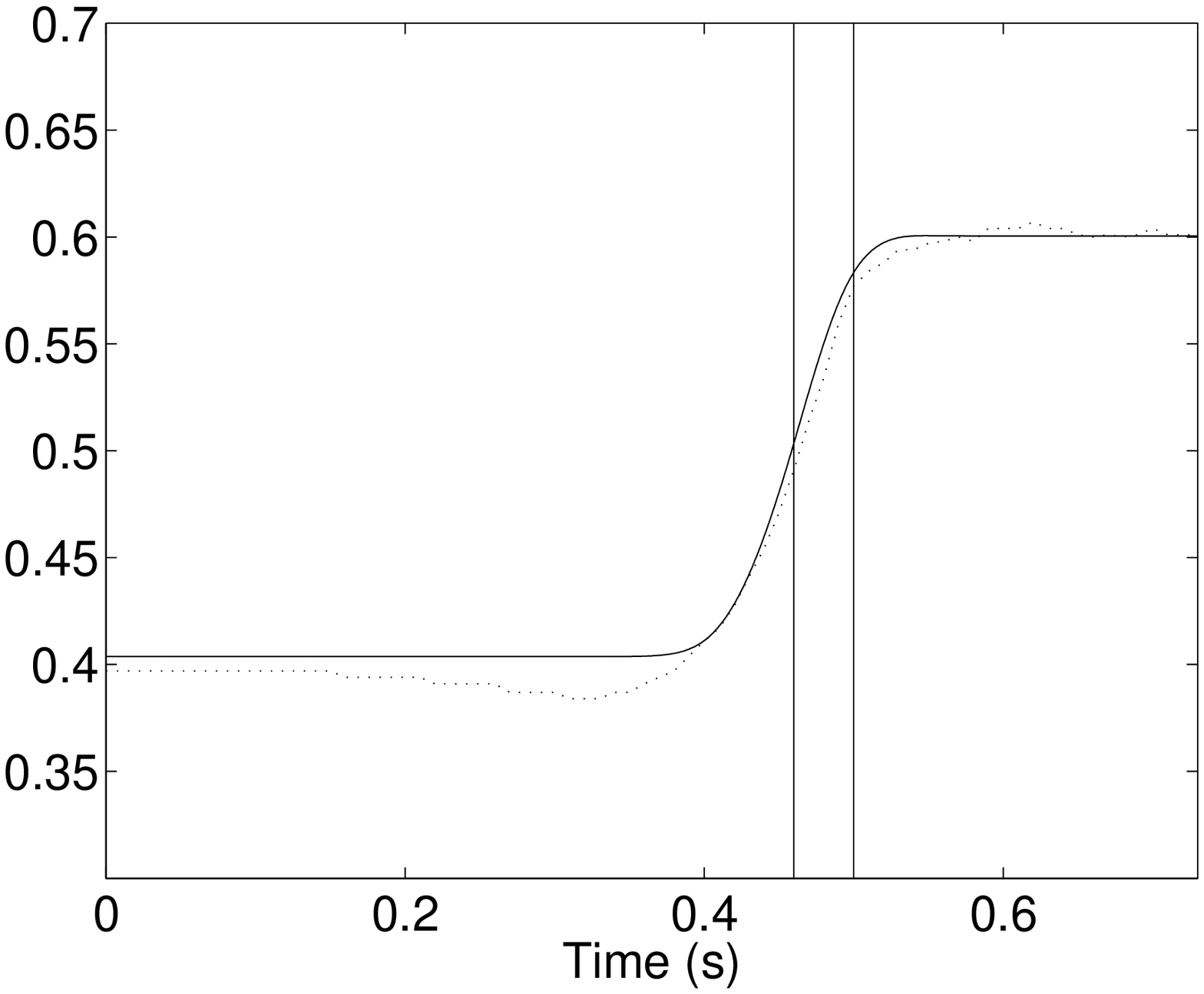} \\
\includegraphics[width=1.7in,height=0.8in]{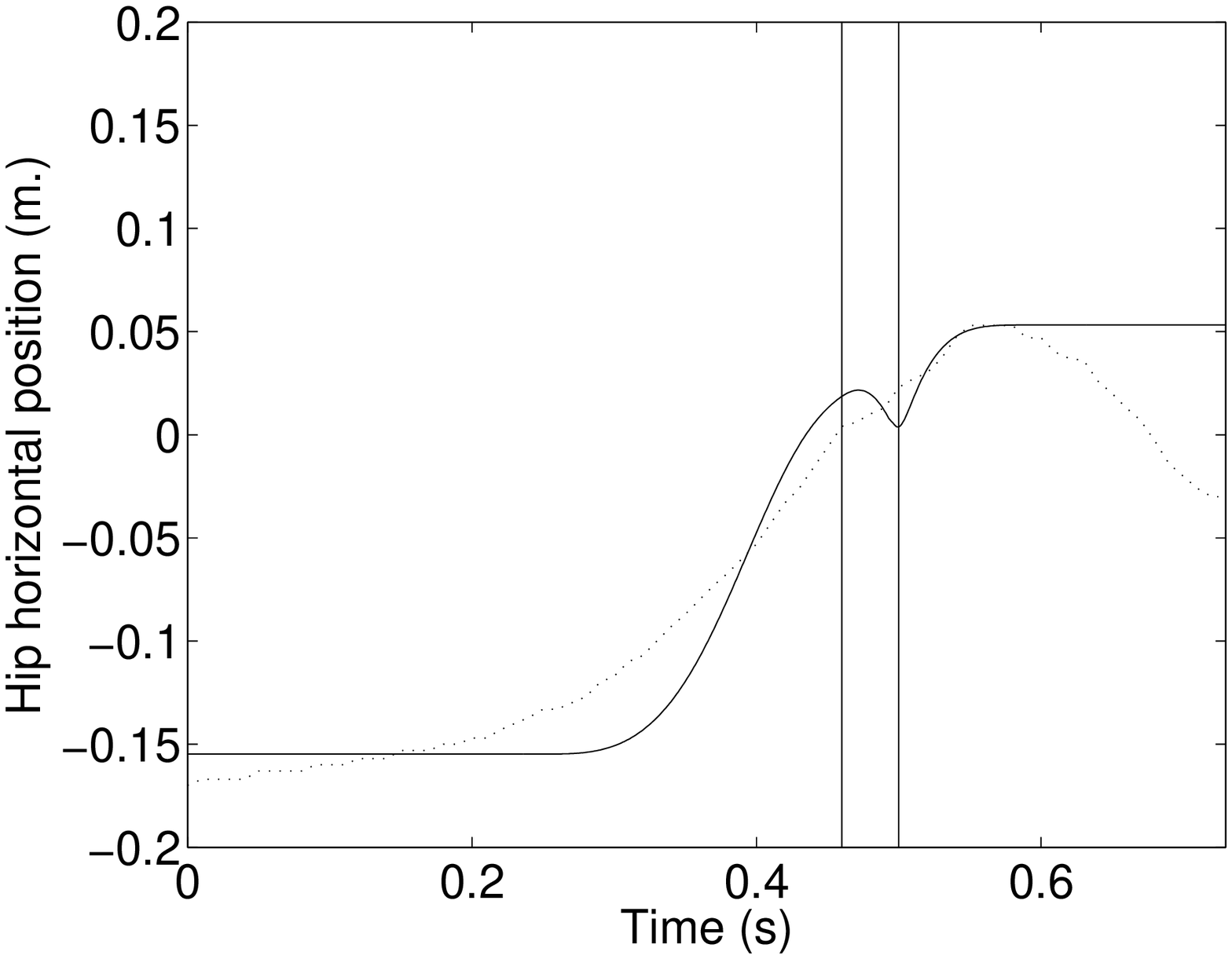} & \includegraphics[width=1.7in,height=0.8in]{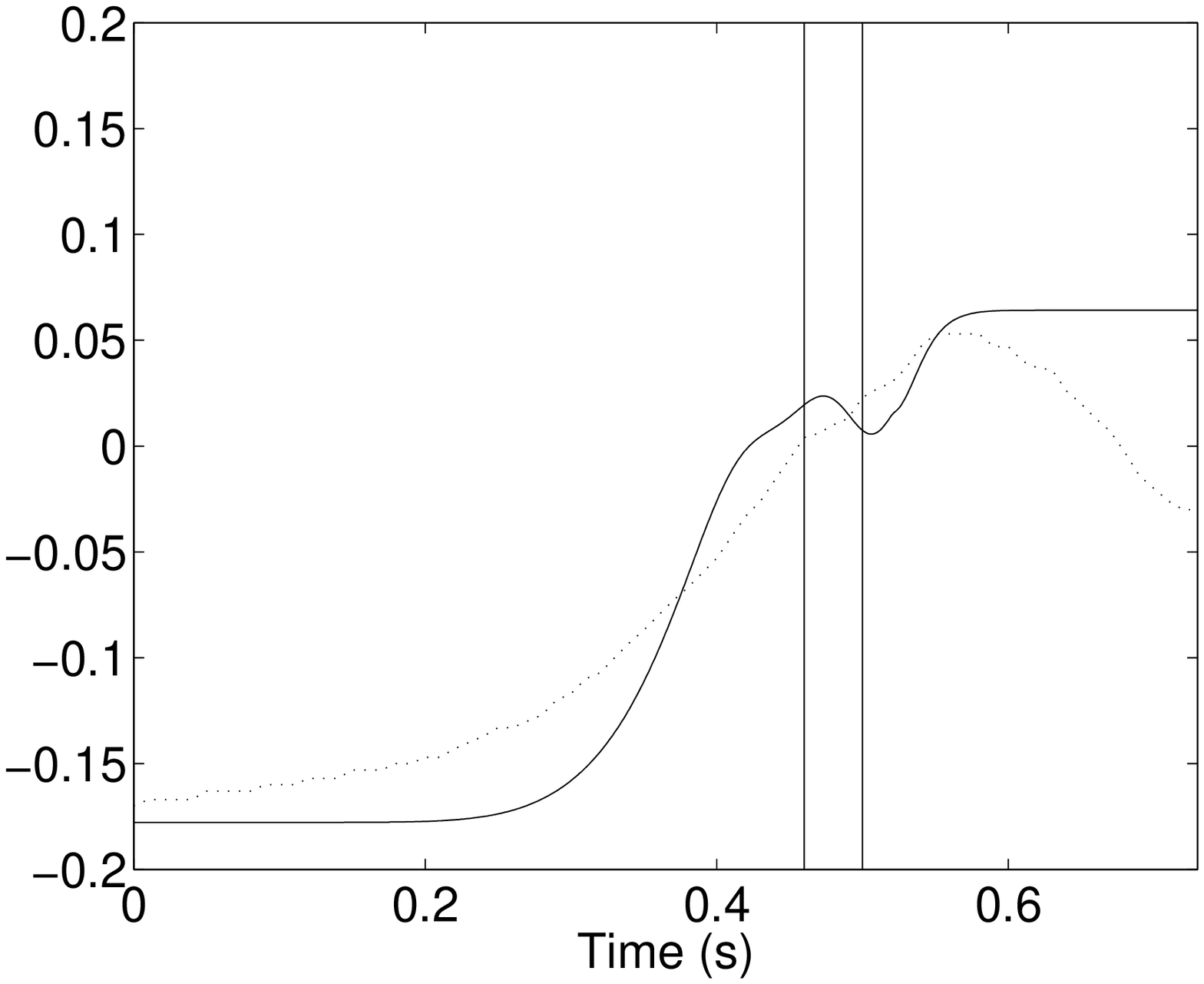} & \includegraphics[width=1.7in,height=0.8in]{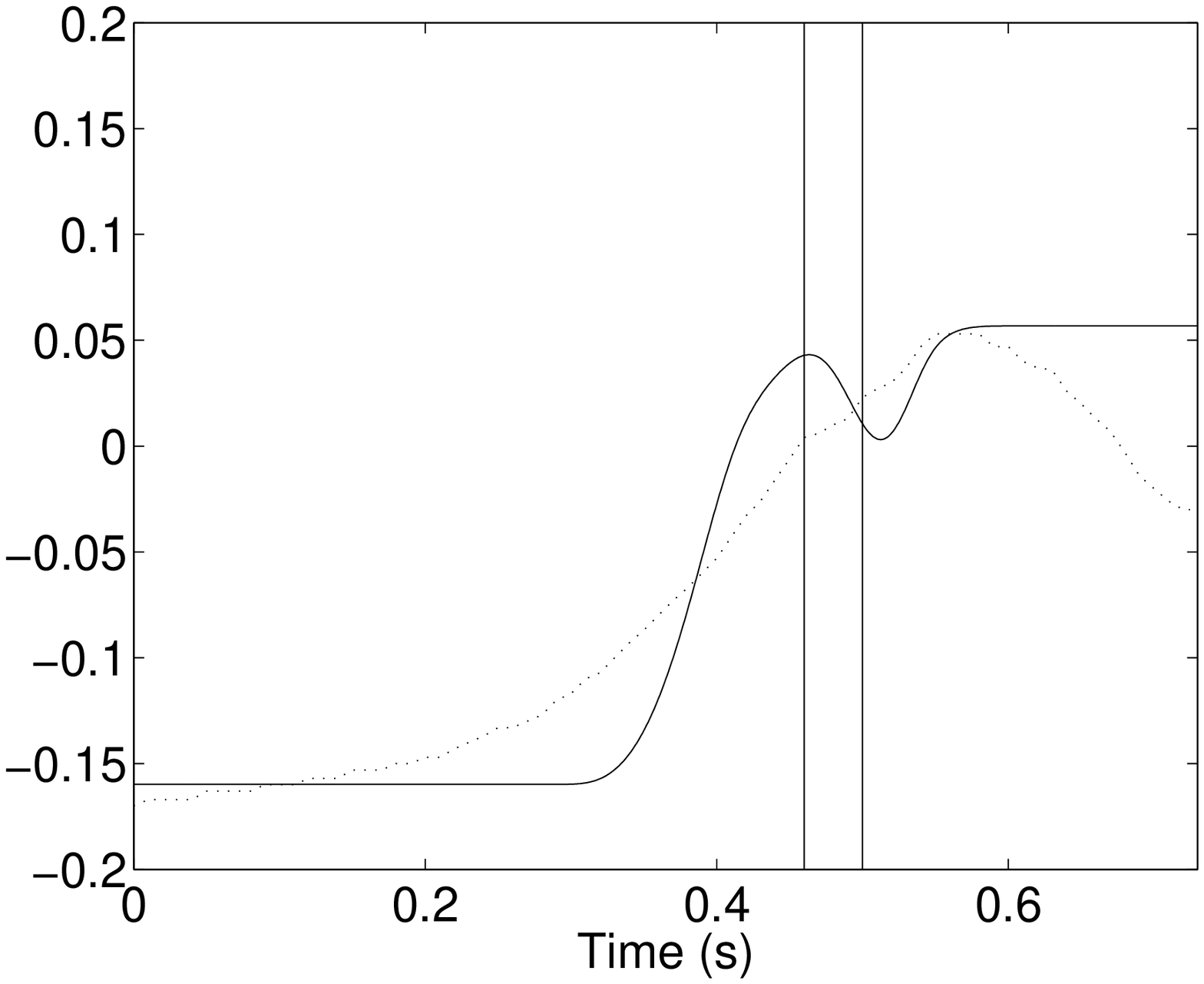} \\
\includegraphics[width=1.7in,height=0.8in]{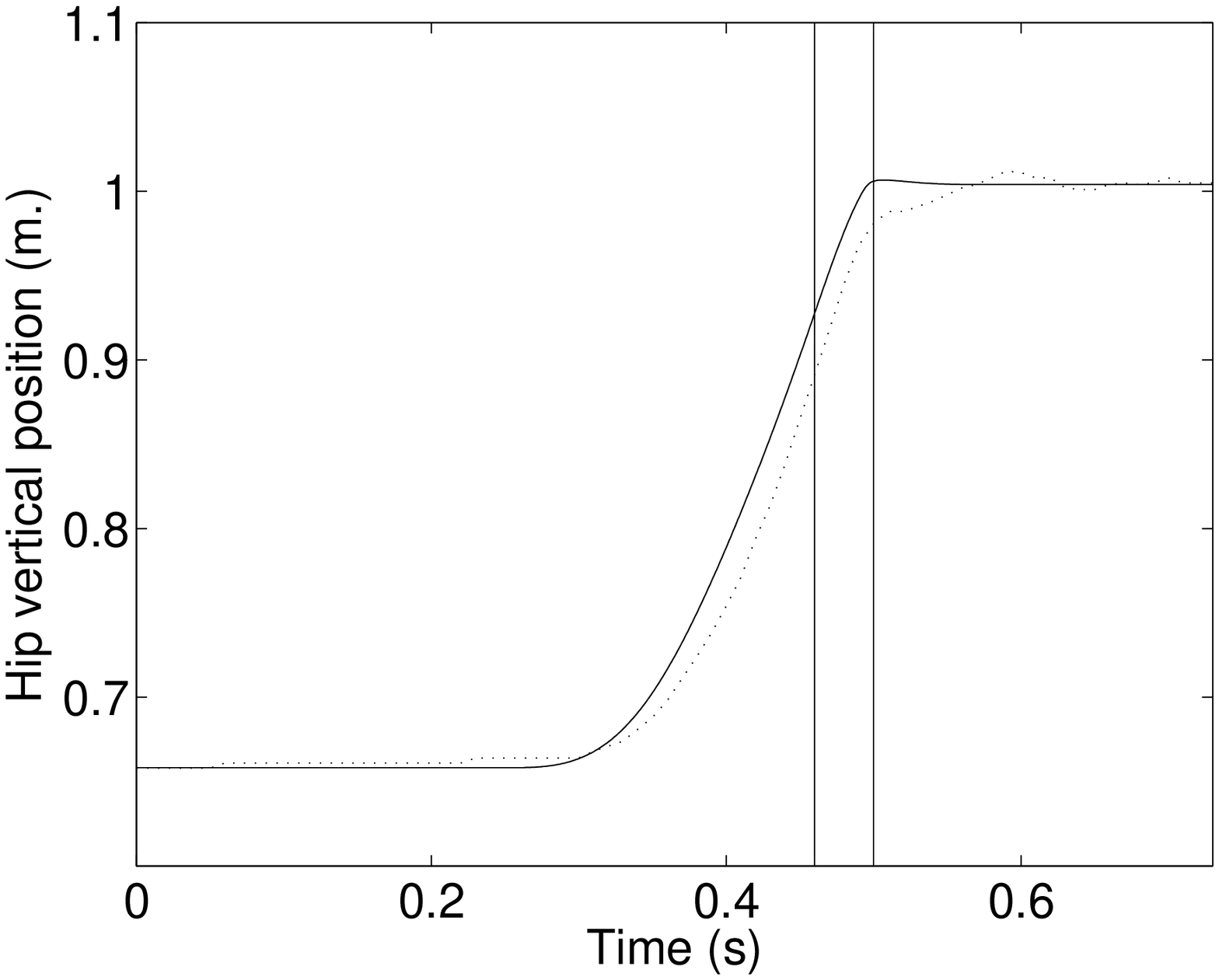} & \includegraphics[width=1.7in,height=0.8in]{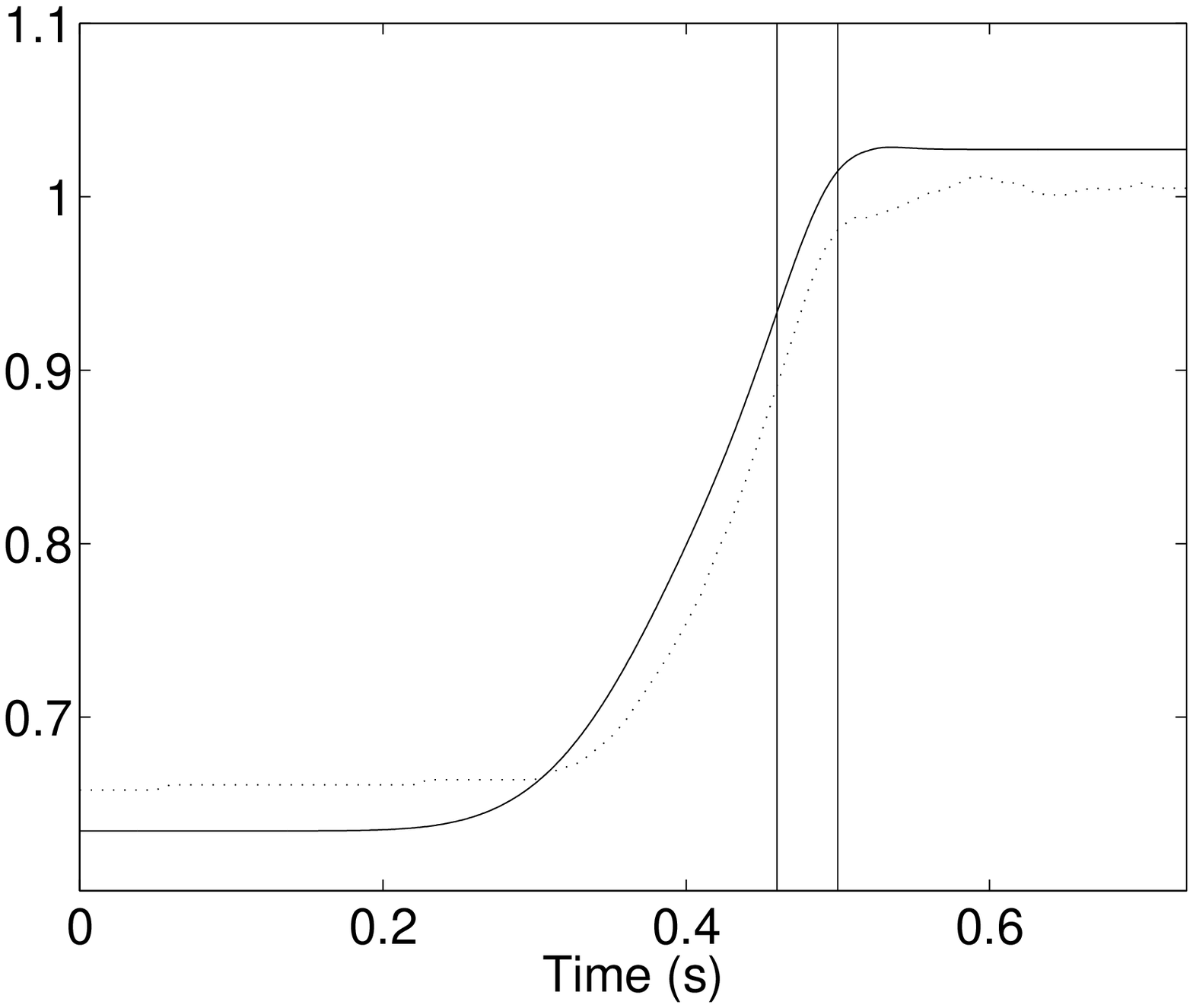} & \includegraphics[width=1.7in,height=0.8in]{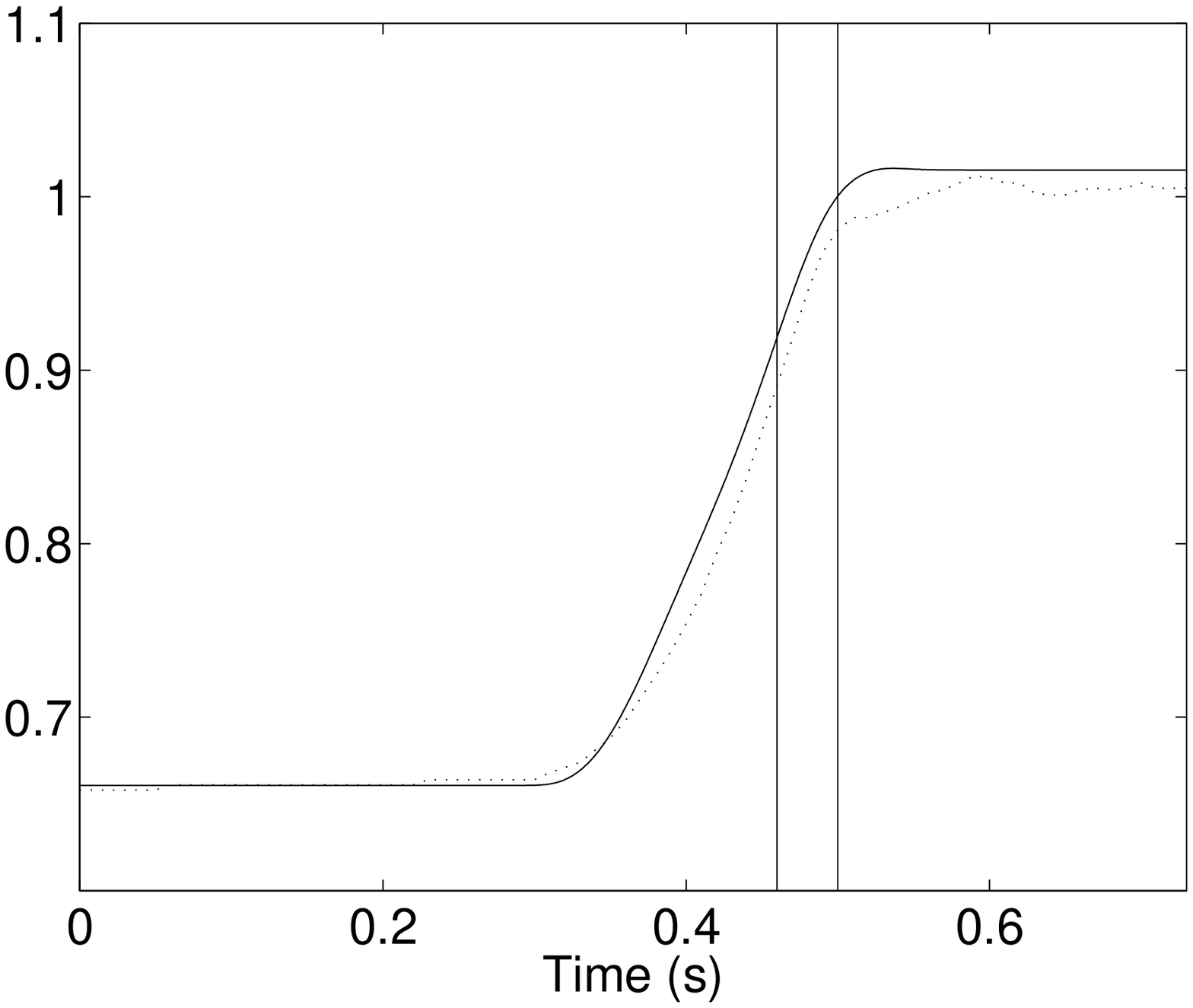} \\
\includegraphics[width=1.7in,height=0.8in]{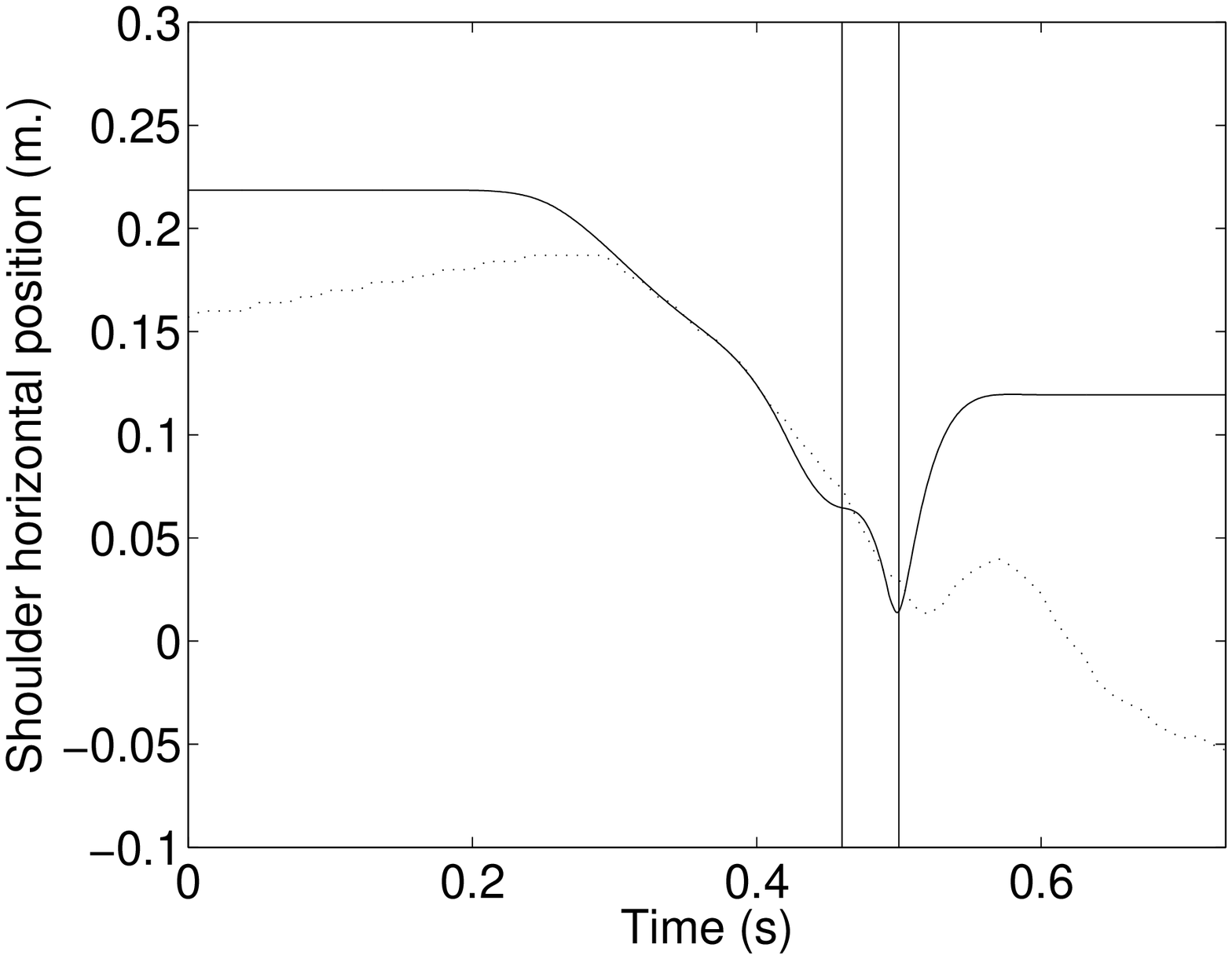} & \includegraphics[width=1.7in,height=0.8in]{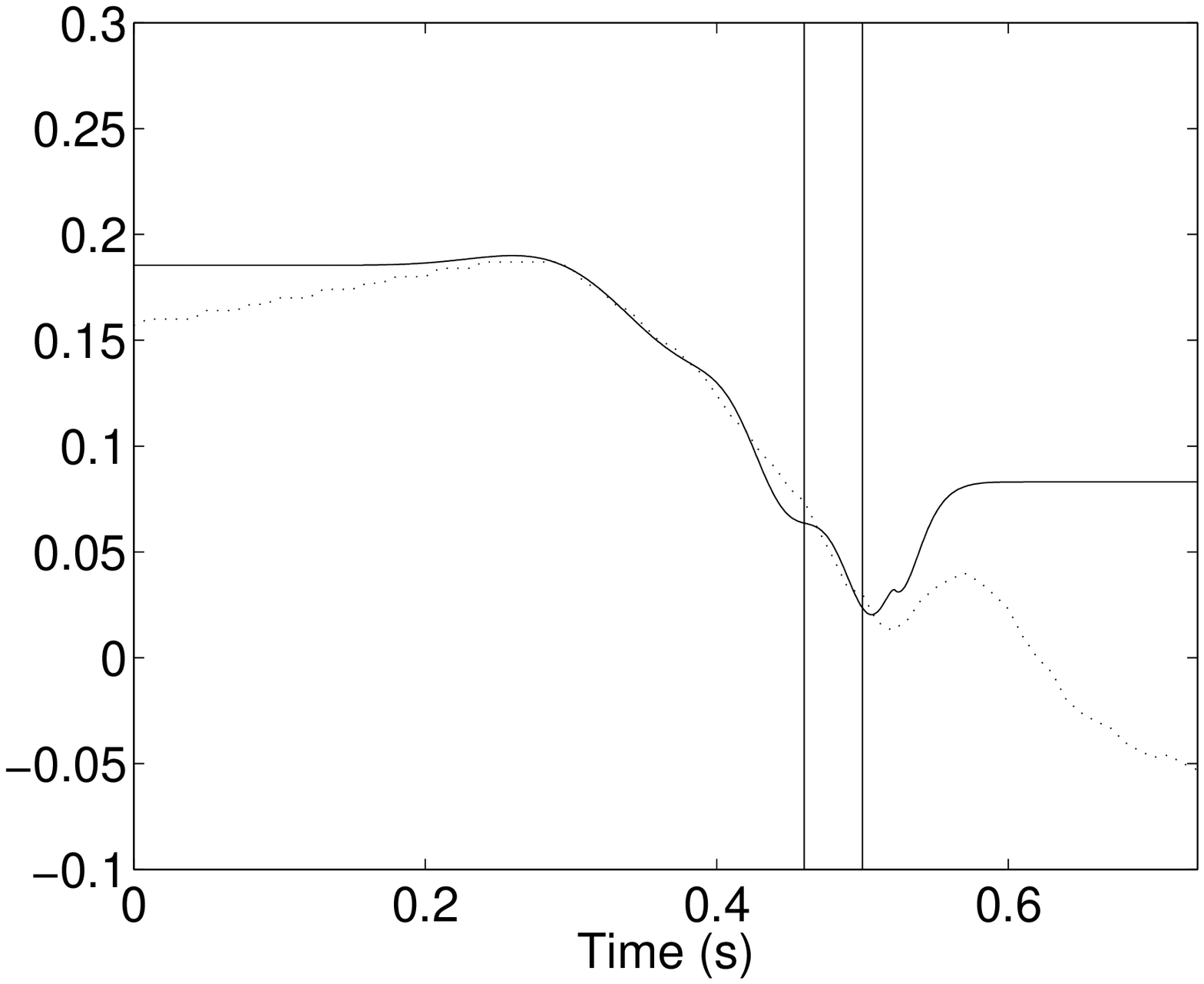} & \includegraphics[width=1.7in,height=0.8in]{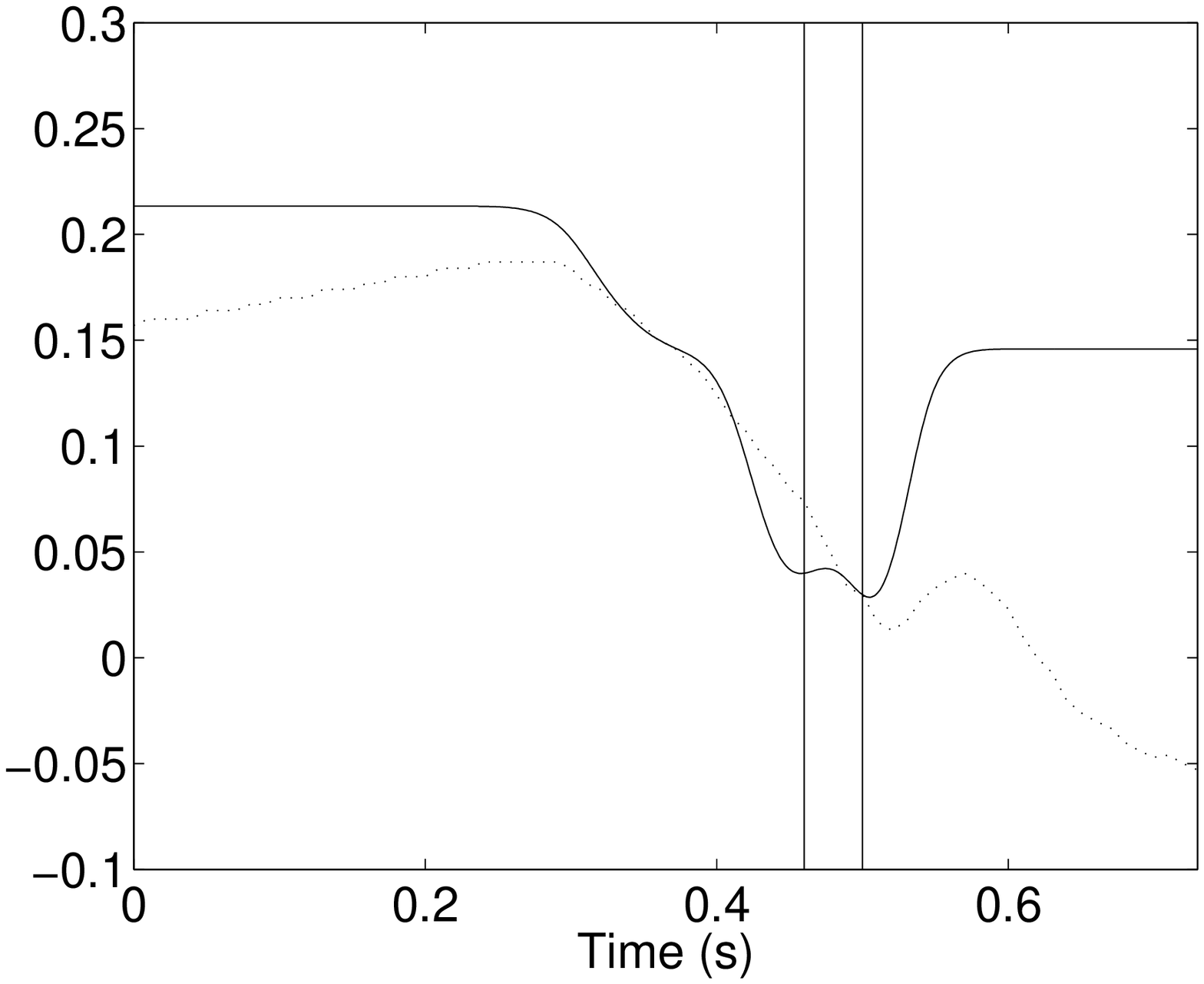} \\
\includegraphics[width=1.7in,height=0.8in]{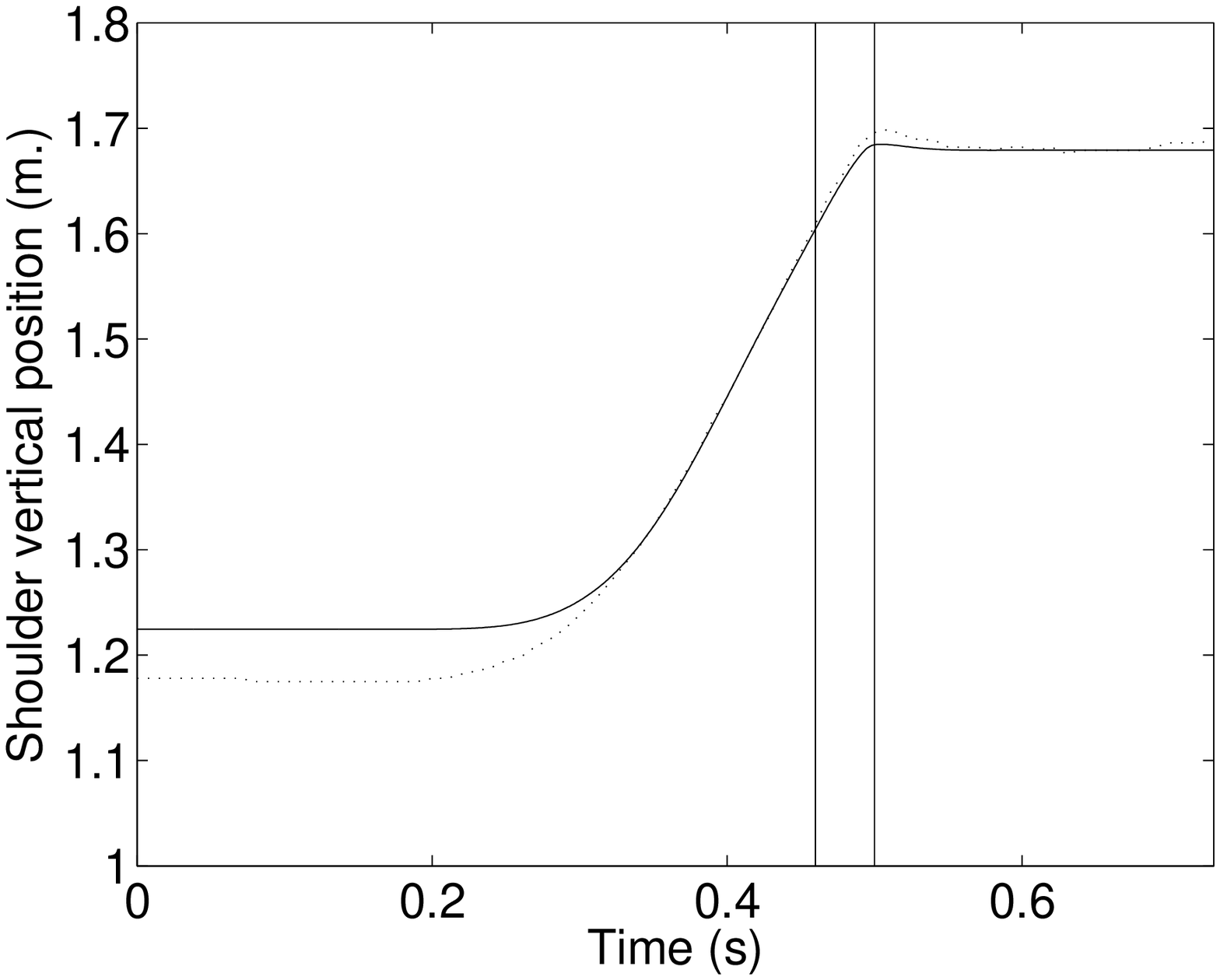} & \includegraphics[width=1.7in,height=0.8in]{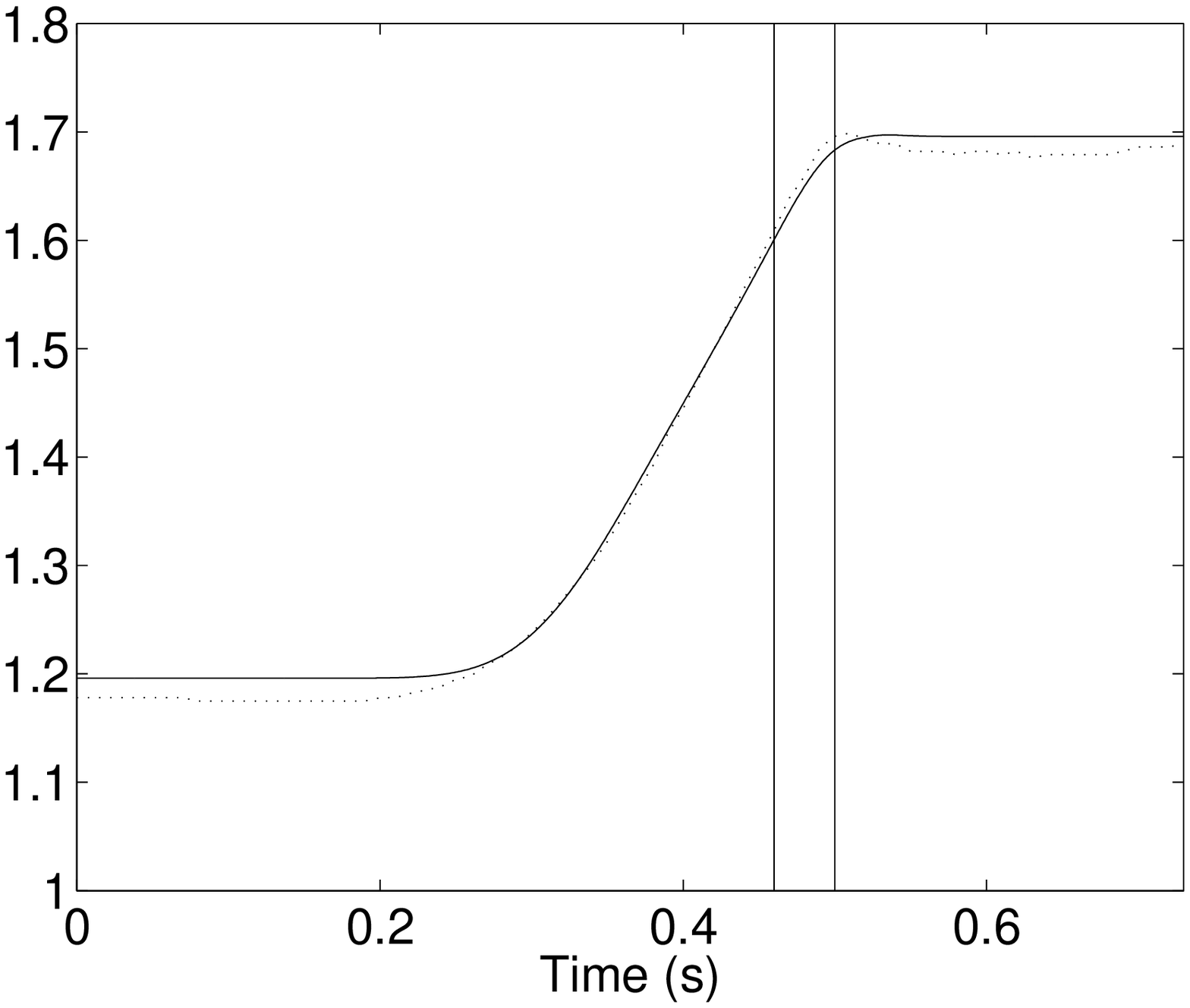} & \includegraphics[width=1.7in,height=0.8in]{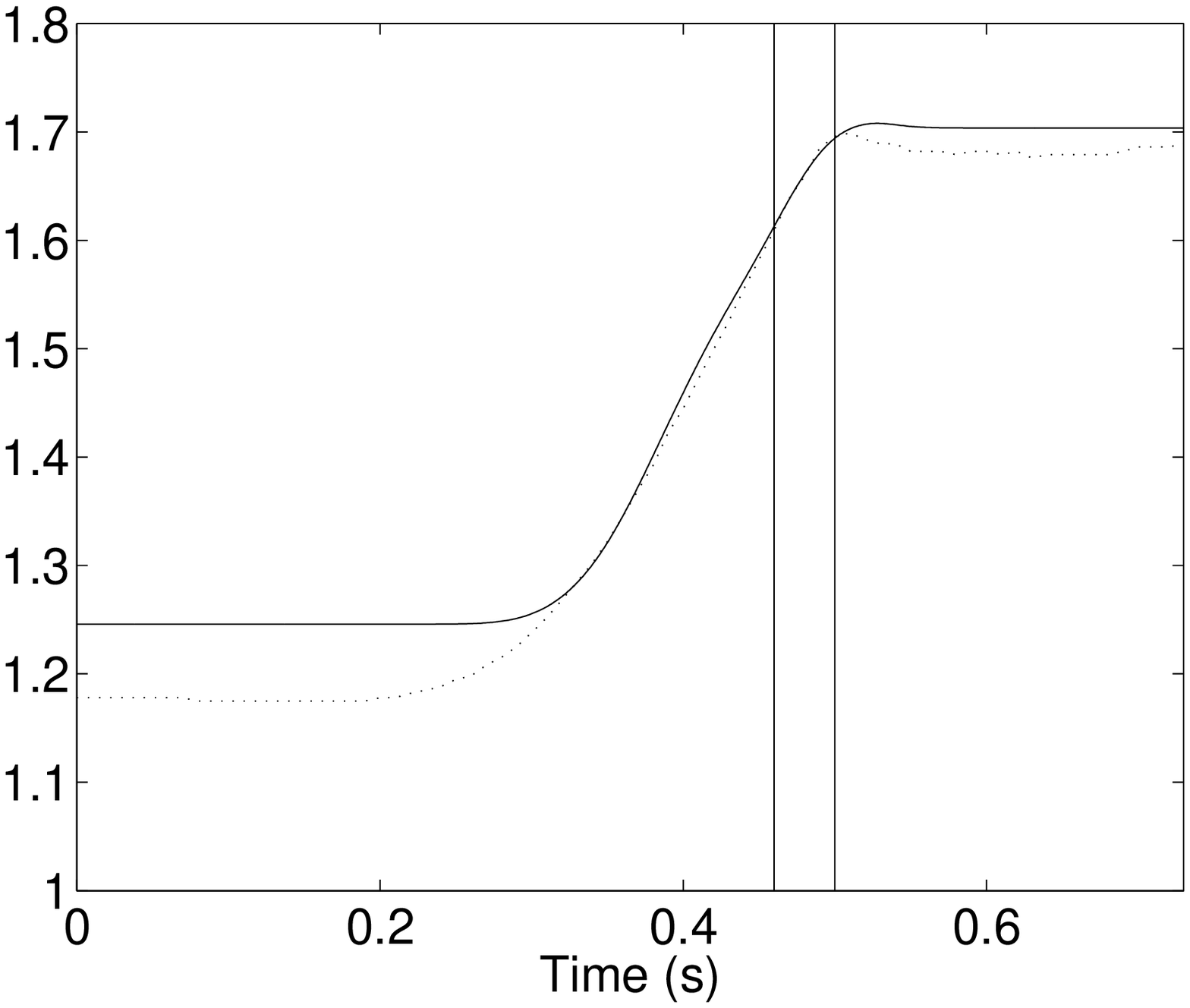}
\end{array}$
\end{center}
\caption{\label{SJFig4}Time histories of joints relative positions. Dotted and plain curves correspond to experimental and dynamic stage modeled data respectively. Vertical lines indicate $t_1$ and take-off instants.}
\end{figure}

% ATTENTION
% c'est presque un copier coller de la première sous section !!
% sauf le chargement des données !!

Basic descriptive statistics of measured values are given in tables \ref{tab01sj} and \ref{tab10sj}.
Examples of experimental and modeled joint angles time histories are presented for both kinematic and dynamic optimization methods in the figures \ref{SJFig1} and \ref{SJFig2} respectively.
Time histories of relative joints position are presented in the figure \ref{SJFig3}.

%Les statistiques élémentaires (moyennes et écart-types) des $\log_{10}$ des erreurs et des temps CPU sont donnés dans le tableau \ref{tab01sj}.
%Voir aussi le quantile à 95 \% des erreurs dans le tableau \ref{tab10sj}. 

Among the three anovas performed for computation time, maximal and mean errors, the highest p-value was equal to
$1.856e-136$ (***) suggesting that both optimization methods and sigmoid models are associated to significantly different results.

%On étudie la dépendance entre l'erreur et le facteur $'XY'$  où X est dans \{"Cin","Dyn"\}
%et Y dans \{"sym","norm","invexp"\}.
%On fait trois 
%anova pour des  mesure répétées
%pour le logarithme $10$ du temps CPU, de l'erreur maximale et de l'erreur moyenne; le maximum des trois 
%probabilités critiques est égal à 
%$1.856e-136$ (***). 
%Les deux méthodes d'optimisation et les trois modèles sont donc significativement différentes, aussi bien en terme de temps de calcul que d'erreur.

%De même, on observe que la méthode Cin est plus rapide que la méthode Dyn, quelque soit les trois modèles
%($p\leq 0$).

Compared to constrained optimization, unconstrain\-ed method executed faster 
($p=0$) and fitted better kinetic original data for both maximal and mean errors 
($p=0$) whatever the sigmoid model.

%Quelque soit le modèle, 
%la méthode d'optimisation Cin  est meilleure que la méthode Dyn   en terme de temps que d'erreur :
%On a fait un test unilatéral de Tukey post hoc pour tester les hypothèses $H_1$:
%"Cinsym-Dynlocsym $\leq 0$",
%"Cinsym-Dynlocsym $\leq 0$"
%ou 
%"Cinsym-Dynlocsym $\leq 0$".
%On trouve en effet une probabilité 
%égale à 
%\begin{itemize}
%\item
%0 pour l'erreur maximale;
%\item
%0 pour l'erreur moyenne.
%\end{itemize}

Considering the unconstrained optimization, no significant difference was found between the sigmoid models for maximal ($p\geq 0.06552$) and mean ($p\geq 0.06285$) errors.

%Comparons  entre eux les trois modèles pour la méthode Cin ; un test post hoc de Tukey 
%bilatéral  nous montre qu'il n'y a pas de différence significative entre eux, aussi bien pour l'erreur maximale 
%($p\geq 0.06552$) que l'erreur moyenne
%($p\geq 0.06285$).

%%% debut ajout
%Il n'en est pas de même pour la  méthode Dyn : 
%on obtient pour les erreurs maximales
%\begin{itemize}
%\item
%$p=4.248e-06$ (sym vs nrom)
%\item
%$p=0.07643$ (sym vs exp)
%\item
%$p=0.02131$ (norm vs exp)
%\end{itemize}
%et pour les erreurs moyennes :
%\begin{itemize}
%\item
%$p=1.584e-10$ (sym vs nrom)
%\item
%$p=4.901e-06$ (sym vs exp)
%\item
%$p=0.1966$ (norm vs exp)
%\end{itemize}
%%% fin ajout 

Post-hoc tests revealed significant differences between the sigmoid models for the constrained optimization method.
Maximal errors mesaured for NORM mo\-del were higher than those obtained from SYM ($p=4.248e-06$ (***)) and 
INVEXP ($p=0.02131$ (*)).
Considering mean errors, SYM model fitted best the original data compared to NORM 
($p=1.584e-10$ (***)) and INVEXP ($p=4.901e-06$ (***)).

%\textbf{XXXX conclusion : en terme d'erreur, amélioration pour les trois modèles en passant de Dyn à Cin.
%De plus, plus rapide.
%En terme de modèles, ils sont équivalents pour la méthode Cin.
%Tableau synthétique ?}

%%%%% FIN INSERTION fichier crée par Sweave sur statistique_final
%%%%%%%%%%%%%%%%%%%%%%%%%%%%%%%%%%%%%%%%%%%%%%%%%%%%%%%%%%%%%%%%%%%%%%%%%%%
%%%%%%%%%%%%%%%%%%%%%%%%%%%%%%%%%%%%%%%%%%%%%%%%%%%%%%%%%%%%%%%%%%%%%%%%%%%

%%%%%%%%%%%%%%%%%%%%%%%%%%%%%%%%%%%%%%%%%%%%%%%%%%%%%%%%%%%%%%%%%%%%
%%%%%%%%%%%%%%%%%%%%%%%%%%%%%%%%%%%%%%%%%%%%%%%%%%%%%%%%%%%%%%%%%%%%
\section{Discussion}
\label{discussion}

This study evaluated different optimization methods to fit joint trajectories produced during pointing tasks and squat jumps.
The evolution of joint angles during the movements was modeled using three sigmoid shaped functions.
Assuming a constant length of the limbs, the whole movement was reconstructed from the sigmoid models parameters.
For each movement type (i.e. pointing tasks and squat jumps) and sigmoid model, different optimization methods were investigated.

In the literature, only Plamondon used a similar approach. However among the published articles, experimental data were presented only in 
\cite{plamondon98}.
Furthermore, no quantitative results were provided and the data was presented for a single subject. This does not allow to compare the present models with Plamondon's one. However, as mentioned earlier, the models used in the present study are defined on a bounded time interval contrarily to the log-normal models for which the end of the movement is not clearly defined.

\subsection{Rigid bodies assumption}
Differences between original and reconstructed data we\-re lower for pointing tasks than for squat-jumps. This result could be explained by the relatively greater amplitude of the joint trajectories during the jumping movement. Moreover, the modeling of the skeleton assumes rigid bodies between the joints. Considering the pointing tasks, it can be supposed that the length of the modeled limbs is quite constant. This assumption is supported by the similarity of the errors observed for global and semi-global methods. The rigid bodies assumption would be less true for squat-jump, especially for the trunk limb. Indeed, the spine is composed of many joints which allow bending of the trunk and thus, the trunk may be divided into two \cite{Kingma1996,deLooze1992,Plamondon1996} or three \cite{deLeva1993} segments to ensure that the rigid bodies model is close enough to the reality of the movement.

\subsection{Planar movement}
The modeling methods proposed in this study deal with planar movements. The higher errors obtained with modeling of squat jumps may be explained by the movement of the joints along the transverse axis, especially for the knee. In comparison, pointing tasks would be closer to a real planar movement since the movement is performed on a planar surface.

\subsection{Optimization methods computing velocity}

Considering the pointing tasks, computation lasted lon\-ger for semi-global method than for local one.
Global optimization executed with similar velocity compared to semi-global method. Thus, global optimization should be used unless specific purposes are researched.
For the squat-jumps, the present results show that unsurprisingly, using the constrained method is much more longer than the unconstrained optimization.

\subsection{Optimization methods accuracy}
For pointing tasks, accuracy of the model was higher for semi-global optimization than for local one.
This suggest that modeling should consider the joints movements together to achieve better fitting of original data.
It should be noticed that local optimization could have considered the dependence of the distal joints trajectories to the proximal ones.
Global optimization did not lead to better results compared to semi-global method. This result is consistent with both the planarity of movement and rigid bodies assumptions.

\subsection{Sigmoid models accuracy}

%{\small 
\begin{table} [ht]
\begin{center}
\begin{tabular}{|l|c|c|c|}
        \hline
& {{SYM}} & {{NORM}} & {{INVEXP}}\\
        \hline
{{Computation time}}  & {{+}} & {{+++}} & {{+}}\\
        \hline
{{Definition space}} & {{++}} & {{+}} & {{+++}}\\
        \hline
{{Accuracy}} & {{0}} & {{0}} & {{0}}\\
        \hline
{{Mathematical regularity}} & {{+}} & {{+++}} & {{+++}}\\
        \hline
\end{tabular}
\vspace{0.5 cm}
\caption{\label{tabrecapconc}Summary of the results obtained for the sigmoid models. + and 0 indicate the existence or absence of advantages respectively}
\end{center}
\end{table}

For both pointing tasks and squat jumps, similar accuracy was obtained with the three models of sigmoids. Among the two movements and the optimization methods, it appears that the NORM model allows fastest computation. Considering SYM and INVEXP models, their relatively slower execution can be explained by the non-linear equation solving and the numerical integration respectively. NORM model formulation takes advantage of the native implementation of the erf function in Matlab software thus ensuring fast computation.

\subsection{Practical considerations}
The present results show that joint trajectories during planar movements such as pointing tasks or squat-jumps can be modeled using meaningful kinematic parameters.
Table \ref{tabrecapconc} presents a summary of the results obtained with the different optimization methods and sigmoid models.

Among the three sigmoid models tested in this study, it appears that the NORM model is computed faster and allows better data fitting of the pointing tasks than other models.
On the contrary, for squat-jumps, INVEXP and SYM models fitted better original data.
From these results, it can be suggested that INVEXP and NORM models should be used preferentially.
Indeed, the INVEXP model did not lead to better results and needs substantial computation time compared to other models.
Despite the important computation time, INVEXP model may be useful for modeling specific movements, especially fast movements, which may not allow a good fitting with NORM model due to the relatively small definition domain of this model in the $\alpha,\beta,\kappa$ space.
For relatively slow and smooth movements, NORM model should be primarily used.

Considering the class of the three models, INVEXP or NORM models should be used when the jerk has to be computed, since it can be analytically determined from the models formulation. If the jerk is not considered as a relevant parameter, both velocities and accelerations can be obtained analytically whatever the used model. Furthermore, slow data acquisition rates should not affect much the quality of the fits since only three points are needed to compute the shape parameters of the three models.

%%%%%%%%%%%%%%%%%%%%%%%%%%%%%%%%%%%%%%%%%%%%%%%%%%%%%%%%%%%%%%%%%%%%%%%%%%%
%%%%%%%%%%%%%%%%%%%%%%%%%%%%%%%%%%%%%%%%%%%%%%%%%%%%%%%%%%%%%%%%%%%%%%%%%%%
%%%%% DEBUT INSERTION fichier crée par Sweave sur  conclusion 
%\input{conclusion} 

% ATTENTION,% ATTENTION,% ATTENTION,% ATTENTION,
% ATTENTION, pour que ce fichier rnw fonctionne avec Sweave, il est nécessaire
% de faire Sweave sur statistique_finale juste avant, sans taper 
% rm(list=ls()), pour que les data frame dataframeSJ et dataframeTP soient connus !
% taper donc pour une compilation globale des quatre fichiers rnw 
%  rm(list=ls());Sweave("statistique_finale.rnw");Sweave("ensemble_tableaux_stats.rnw");Sweave("meilleurs_resultats.rnw");Sweave("conclusion.rnw");

%%%%%%%%%%%%%%%%%%%%%%%%%%%%%%%%%%%%%%%%%%%%%%%%%%%%%%%%%%%%%%%%%%%%
%%%%%%%%%%%%%%%%%%%%%%%%%%%%%%%%%%%%%%%%%%%%%%%%%%%%%%%%%%%%%%%%%%%%
\section{Conclusion}
\label{conclusion}

This study shows that complex planar movements can be modeled by using a small set of meaningful kinematic parameters defining the time history of joint angles with high accuracy 
(in 95\% of cases, mean errors obtained from 
pointing task and squat jump
were respectively inferior to 
1
and 
3
centimeters).
This approach can provide a continuous description of the movement and thus may be used to analyze the evolution throughout the movement of parameters which need differentiation of raw data with respect to time without performing numerical computations.
Especially, this could avoid well known magnification of error resulting from such procedure.
Furthermore, the modeling procedure can be applied for fast movements, as well as when acquisition rate is slow, as only 3 points are required to get the sigmoid parameters.
Moreover, the flexibility of the new sigmoid models should lead to increased realism of movements obtained from procedural animation.
Further researches should assess the relevance of such modeling strategy for three dimensional movements and the relation between the model parameters and the central nervous system processes implied in motor control.

%%%%% FIN INSERTION fichier crée par Sweave sur  conclusion
%%%%%%%%%%%%%%%%%%%%%%%%%%%%%%%%%%%%%%%%%%%%%%%%%%%%%%%%%%%%%%%%%%%%%%%%%%%
%%%%%%%%%%%%%%%%%%%%%%%%%%%%%%%%%%%%%%%%%%%%%%%%%%%%%%%%%%%%%%%%%%%%%%%%%%%

\appendix

%%%%%%%%%%%%%%%%%%%%%%%%%%%%%%%%%%%%%%%%%%%%%%%%%%%%%%%%%%%%%%%%%%%%%%%%%%%
%%%%%%%%%%%%%%%%%%%%%%%%%%%%%%%%%%%%%%%%%%%%%%%%%%%%%%%%%%%%%%%%%%%%%%%%%%%
%%%%% DEBUT INSERTION fichier crée par Sweave sur ensemble_tableaux_stats
%\input{ensemble_tableaux_stats} 

% ATTENTION,% ATTENTION,% ATTENTION,% ATTENTION,
% ATTENTION, pour que ce fichier rnw fonctionne avec Sweave, il est nécessaire
% de faire Sweave sur statistique_finale juste avant, sans taper 
% rm(list=ls()), pour que les data frame dataframeSJ et dataframeTP soient connus !
% taper donc pour une compilation globale des trois fichiers rnw 
%  rm(list=ls());Sweave("statistique_finale.rnw");Sweave("ensemble_tableaux_stats.rnw");Sweave("meilleurs_resultats.rnw");

\clearpage

%%%%%%%%%%%%%%%%%%%%%%%%%%%%%%%%%%%%%%%%%%%%%%%%%%%%%%%%%%%%%%%%%%%%%%%%%%%%%%%%%%%%%%%%%%%%
\section{Set of tables of statistical results}
\label{ensemble_tableau}

%%%%%%%%%%%%%%%%%%%%%%%%%%%%%%%%%%%%%%%%%%%%%%%%%%%%%%%%%%%%%%%%%%%%%%%%%%%%%%%%%%%%%%%%%
\subsection{Pointing task}

\begin{table} [ht]
\begin{center}
\begin{tabular} {|l|l|l|l|l|}
\hline 
data & method $\setminus$ model &  SYM & NORM & INVEXP \\
\hline  \hline 
computation time & 
local
& $0.776 \pm 0.24$
& $0.471 \pm 0.23$
& $1.41 \pm 0.17$
\\ 
%\hline 
\cline{2-5}
&
semi-global
& $1.361 \pm 0.33$
& $0.309 \pm 0.32$
& $1.836 \pm 0.37$
\\ 
%\hline
\cline{2-5}
&
global
& $1.248 \pm 0.45$
& $0.079 \pm 0.41$
& $1.66 \pm 0.5$
\\ \hline\hline 
maximal error  & 
local
& $-1.915 \pm 0.22$
& $-1.917 \pm 0.22$
& $-1.908 \pm 0.21$
\\ 
%\hline 
\cline{2-5}
&
semi-global
& $-2.102 \pm 0.2$
& $-2.121 \pm 0.2$
& $-2.093 \pm 0.2$
\\ 
%\hline
\cline{2-5}
&
global
& $-2.118 \pm 0.2$
& $-2.139 \pm 0.2$
& $-2.109 \pm 0.2$
\\ \hline \hline
mean error  & 
local
& $-2.415 \pm 0.22$
& $-2.422 \pm 0.21$
& $-2.412 \pm 0.21$
\\ 
%\hline 
\cline{2-5}
&
semi-global
& $-2.591 \pm 0.18$
& $-2.612 \pm 0.18$
& $-2.586 \pm 0.18$
\\ 
%\hline
\cline{2-5}
&
global
& $-2.6 \pm 0.18$
& $-2.62 \pm 0.18$
& $-2.594 \pm 0.2$
\\ \hline 
\end{tabular}
\vspace{0.5 cm}
\caption{\label{tab01tp}Elementary statistics on  $\log_{10}$ (mean $\pm$ standard deviation).}
\end{center}
\end{table}

\begin{table} [ht]
\begin{center}
\begin{tabular} {|l|l|l|l|l|}
\hline 
data & method $\setminus$ model &  SYM & NORM & INVEXP \\
\hline  \hline 
maximal error  & 
local
& $2.768$
& $2.894$
& $2.781$
\\ 
%\hline 
\cline{2-5}
&
semi-global
& $1.86$
& $1.609$
& $1.725$
\\ 
%\hline
\cline{2-5}
&
global
& $1.645$
& $1.486$
& $1.602$
\\ \hline \hline
mean error & 
local
& $0.824$
& $0.824$
& $0.812$
\\ 
%\hline 
\cline{2-5}
&
semi-global
& $0.521$
& $0.451$
& $0.484$
\\ 
%\hline
\cline{2-5}
&
global
& $0.493$
& $0.447$
& $0.47$
\\ \hline 
\end{tabular}
\vspace{0.5 cm}
\caption{\label{tab10tp}  95 \% cases error in centimeter.}
\end{center}
\end{table}

%\clearpage

%%%%%%%%%%%%%%%%%%%%%%%%%%%%%%%%%%%%%%%%%%%%%%%%%%%%%%%%%%%%%%%%%%%%%%%%%%%%%%%%%%%%%%%%%
\subsection{Squat Jumps}

% ATTENTION
% c'est presque un copier coller de la première sous section !!

\begin{table} [ht]
\begin{center}
\begin{tabular} {|l|l|l|l|l|}
\hline 
data & method $\setminus$ model &  SYM & NORM & INVEXP \\
\hline  \hline 
computation time & 
kinematic
& $1.334 \pm 0.27$
& $0.26 \pm 0.24$
& $0.978 \pm 0.19$
\\ 
%\hline 
\cline{2-5}
&
dynamic
& $2.114 \pm 0.28$
& $1.311 \pm 0.36$
& $2.096 \pm 0.24$
\\ \hline\hline 
maximal error & 
kinematic
& $-1.311 \pm 0.16$
& $-1.33 \pm 0.17$
& $-1.297 \pm 0.16$
\\ 
%\hline 
\cline{2-5}
&
dynamic
& $-1.033 \pm 0.21$
& $-0.962 \pm 0.2$
& $-1.001 \pm 0.19$
\\ \hline\hline 
mean error & 
kinematic
& $-1.771 \pm 0.13$
& $-1.784 \pm 0.14$
& $-1.76 \pm 0.13$
\\ 
%\hline 
\cline{2-5}
&
dynamic
& $-1.606 \pm 0.16$
& $-1.535 \pm 0.18$
& $-1.554 \pm 0.14$
\\ \hline 
\end{tabular}
\vspace{0.5 cm}
\caption{\label{tab01sj}  Elementary statistics on  $\log_{10}$ (mean $\pm$ standard deviation).}
\end{center}
\end{table}

\begin{table} [ht]
\begin{center}
\begin{tabular} {|l|l|l|l|l|}
\hline 
data & method $\setminus$ model &  SYM & NORM & INVEXP \\
\hline  \hline 
maximal error  & 
kinematic
& $8.392$
& $8.492$
& $8.492$
\\ 
%\hline 
\cline{2-5}
&
dynamic
& $21.134$
& $26.257$
& $21.213$
\\ \hline \hline
mean error & 
kinematic
& $2.688$
& $2.688$
& $2.882$
\\ 
%\hline 
\cline{2-5}
&
dynamic
& $4.935$
& $5.842$
& $5.034$
\\ \hline 
\end{tabular}
\vspace{0.5 cm}
\caption{\label{tab10sj}  95 \% cases error in centimeter.}
\end{center}
\end{table}

%%%%% FIN INSERTION fichier crée par Sweave sur  ensemble_tableaux_stats
%%%%%%%%%%%%%%%%%%%%%%%%%%%%%%%%%%%%%%%%%%%%%%%%%%%%%%%%%%%%%%%%%%%%%%%%%%%
%%%%%%%%%%%%%%%%%%%%%%%%%%%%%%%%%%%%%%%%%%%%%%%%%%%%%%%%%%%%%%%%%%%%%%%%%%%

%%%%%%%%%%%%%%%%%%%%%%%%%%%%%%%%%%%%%%%%%%%%%%%%%%%%%%%
%          end of text
%%%%%%%%%%%%%%%%%%%%%%%%%%%%%%%%%%%%%%%%%%%%%%%%%%%%%%%

%%%%%%%%%%%%%%%%%%%%%%%%%%%%%%%%%%%%%%%%%%%%%%%%%%%%%%%%%%%%%
%%%%%%%%%%%%%%%%%%%%%%%%%%%%%%%%%%%%%%%%%%%%%%%%%%%%%%%%%%%%%
% si Bibliograhie  faite avec bibtex

\bibliographystyle{alpha}
\bibliography{sigmo_TCJBCVPL_arXiv_2012}

\newcommand{\etalchar}[1]{$^{#1}$}
\begin{thebibliography}{MMMR96}

\bibitem[ARW01]{Ahnetali01}
Sung~Joon Ahn, Wolfgang Rauh, and Hans-Jürgen Warnecke.
\newblock Least-squares orthogonal distances fitting of circle, sphere,
  ellipse, hyperbola, and parabola.
\newblock {\em Pattern Recognition}, 34(12):2283--2303, 2001.

\bibitem[BBM12]{SJ_JBYBLM_12}
Jérôme Bastien, Yoann Blache, and Karine Monteil.
\newblock Estimation of anthropometrical and inertial body parameters using
  double integration of residual torques and forces during squat jump.
\newblock \iflanguage{french}{Soumis au \emph{Journal of Biomechanical
  Engineering}. Disponible sur
  {\footnotesize{\url{http://utbmjb.chez-alice.fr/recherche/articles_provisoires/squatJump_JBYBKM_2013.pdf}}}}%
  {Submitted to \emph{Journal of Biomechanical Engineering}. Aivalable on
  {\footnotesize{\url{http://utbmjb.chez-alice.fr/recherche/articles_provisoires/squatJump_JBYBKM_2013.pdf}}}},
  2012.

\bibitem[BC13]{bastiencreveauxencours12}
J.~Bastien and T.~Creveaux.
\newblock Modelling of joint displacement by sigmoid function. {M}athematical
  formalization.
\newblock {I}n preparation, 2013.

\bibitem[BLM10]{jbplkm2008}
Jérôme Bastien, Pierre Legreneur, and Karine Monteil.
\newblock A geometrical alternative to jacobian rank deficiency method for
  planar workspace characterisation.
\newblock {\em Mechanism and Machine Theory}, 45:335--348, 2010.

\bibitem[Bra70]{bradbury1970}
Michael~W.B. Bradbury.
\newblock The effect of rubidium on the distribution and movement of potassium
  between blood, brain and cerebrospinal fluid in the rabbit.
\newblock {\em Brain Research}, 24(2):311--312, 1970.

\bibitem[CBL09]{TCJBPLACAPS09ACAPS}
Thomas Creveaux, Jérôme Bastien, and Pierre Legreneur.
\newblock Model of joint angle displacement: application to vertical jumping.
\newblock In {\em 13 ième congrès international de l'ACAPS}, Approche
  {P}luridisciplinaire de la {M}otrocité {H}umaine, pages 49--50, Lyon, October
  2009.

\bibitem[CdB81]{boor}
D.~Conte and C.~de~Boor.
\newblock {\em Elementary numerical analysis. {A}n algorithmic approach}.
\newblock Mc Graw-Hill, 1981.

\bibitem[CJ89]{MR1011392}
Maurice~G. Cox and Helen~M. Jones.
\newblock An algorithm for least-squares circle fitting to data with specified
  uncertainty ellipses.
\newblock {\em IMA J. Numer. Anal.}, 9(3):285--298, 1989.

\bibitem[Cre09]{creveaux09}
Thomas Creveaux.
\newblock {\em Des données expérimentales à la modélisation d'un mouvement
  dynamique : cas du squat-jump}.
\newblock PhD thesis, Université Claude Bernard Lyon 1, 2009.

\bibitem[Deb79]{debouche79}
C.~Debouche.
\newblock Présentation coordonnée de différents modèles de croissance.
\newblock {\em Revue de statistique appliquée}, 27(4):5--22, 1979.

\bibitem[DG06]{MR2212485}
J{\'o}zsef Dombi and Norbert Gy{\H{o}}rb{\'{\i}}r{\'o}.
\newblock Addition of sigmoid-shaped fuzzy intervals using the {D}ombi operator
  and infinite sum theorems.
\newblock {\em Fuzzy Sets and Systems}, 157(7):952--963, 2006.

\bibitem[dL93]{deLeva1993}
P.~de~Leva.
\newblock Validity and accuracy of four methods for locating the center of mass
  of young male and female athletes.deleva1993.
\newblock In {\em Proceedings of the XIVth Congress of the International
  Society of Biomechanics}, pages 318--319, Paris, France, 1993. Imprimerie
  Laballery.

\bibitem[dLKBT92]{deLooze1992}
M.~P. de~Looze, I.~Kingma, J.~B.~J. Bussmann, and H.~M. Toussaint.
\newblock Validation of a dynamic linked segment model to calculate joint
  moments in lifting.
\newblock {\em Clinical Biomechanics}, 7:161--169, 1992.

\bibitem[Dra95]{MR1365401}
John~A. Drakopoulos.
\newblock Sigmoidal theory.
\newblock {\em Fuzzy Sets and Systems}, 76(3):349--363, 1995.

\bibitem[EMM07]{Endoetali07}
H.~Endo, T.~Murahashi, and E.~Marui.
\newblock Accuracy estimation of drilles holes with small diameter and
  influence of drill parameter on the machining accuracy when drillin in mild
  steel shett.
\newblock {\em Machine Tools and manufacture}, 47:175--181, 2007.

\bibitem[Fan90]{MR1132365}
De~Liang Fan.
\newblock On formulas for calculating parameters of least square circles.
\newblock {\em J. Southeast Univ.}, 20(6):96--101, 1990.

\bibitem[Fin52]{MR0048770}
D.~J. Finney.
\newblock {\em Probit analysis. {A} statistical treatment of the sigmoid
  response curve}.
\newblock Cambridge, at the University Press, 2 edition, 1952.

\bibitem[IS89]{vanIngenShenau1989}
van~G.J. Ingen~Shenau.
\newblock From rotation to translation: constraints on multi-joint movements
  and the unique action of bi-articular muscles.
\newblock {\em Hum Mov Sc}, 8:301--377, 1989.

\bibitem[KdLT{\etalchar{+}}96]{Kingma1996}
I.~Kingma, M.~P. de~Looze, H.~M. Toussaint, H.~G. Klijnsma, and T.~B.~M.
  Bruijnen.
\newblock Validation of a full body 3-d dynamic linked segment model.
\newblock {\em Human Movement Science}, 15:833--860, 1996.

\bibitem[KS96]{MR1400774}
Joe Kilian and Hava~T. Siegelmann.
\newblock The dynamic universality of sigmoidal neural networks.
\newblock {\em Inform. and Comput.}, 128(1):48--56, 1996.

\bibitem[Kum00]{kumazawa00}
Itsuo Kumazawa.
\newblock Compact and parametric shape representation by a tree of sigmoid
  functions for automatic shape modeling.
\newblock {\em Pattern Recognition Letters}, 21(6-7):651--660, 2000.

\bibitem[LCB11]{legreneuretali11}
Pierre Legreneur, Thomas Creveaux, and Vincent Bels.
\newblock Control of poly-articular chain trajectory using temporal sequence of
  its joints displacements.
\newblock {\em Intelligent Control and Automation}, 2(1):38--46, 2011.

\bibitem[LG63]{lindenmann1963}
J~Lindenmann and G.E. Gifford.
\newblock Studies on vaccinia virus plaque formation and its inhibition by
  interferon {I}. {D}ynamics of plaque formation by vaccinia virus.
\newblock {\em Virology}, 19:283--293, 1963.

\bibitem[MF54]{Fitts54}
Paul~M. M.~Fitts.
\newblock The information capacity of the human motor system in controlling the
  amplitude of movement.
\newblock {\em Journal of Experimental Psychology}, 6:381--391, 1954.
\newblock (Reprinted in Journal of Experimental Psychology: General,
  121(3):262-269, 1992).

\bibitem[MFP64]{FittsPeterson64}
Paul~M. M.~Fitts and James~R. Peterson.
\newblock Information capacity of discrete motor responses.
\newblock {\em Journal of Experimental Psychology}, 2:103--112, 1964.

\bibitem[MK91]{MR1099915}
L.~Moura and R.~Kitney.
\newblock A direct method for least-squares circle fitting.
\newblock {\em Comput. Phys. Comm.}, 64(1):57--63, 1991.

\bibitem[MMMR96]{menon96}
Anil Menon, Kishan Mehrotra, Chilukuri~K. Mohan, and Sanjay Ranka.
\newblock Characterization of a class of sigmoid functions with applications to
  neural networks.
\newblock {\em Neural Networks}, 9(5):819--835, 1996.

\bibitem[Nar97]{MR1440301}
Sridhar Narayan.
\newblock The generalized sigmoid activation function: competitive supervised
  learning.
\newblock {\em Inform. Sci.}, 99(1-2):69--82, 1997.

\bibitem[PCF03]{plamondonchunchua03}
R.~Plamondon and A.W. Chunhua~Feng.
\newblock A kinematic theory of rapid human movements. {P}art {IV}. a formal
  mathematical proof and new insight.
\newblock {\em Biol. Cybern.}, 89:126--138, 2003.

\bibitem[Pea91]{pearson1991}
David~E. Pearson.
\newblock Probability analysis of blended coking coals.
\newblock {\em International Journal of Coal Geology}, 19(1-4):109--119, 1991.

\bibitem[PGD96]{Plamondon1996}
A.~Plamondon, M.~Gagnon, and P.~Desjardins.
\newblock Validation of two 3-d segments models to calculate the net reaction
  forces and moments at the l5/s1 joint in lifting.
\newblock {\em Clinical Biomechanics}, 11:101--110, 1996.

\bibitem[Pla95a]{plamondon95a}
R.~Plamondon.
\newblock A kinematic theory of rapid human movements. {P}art {I}. {M}ovement
  representation and generation.
\newblock {\em Biol. Cybern.}, 72(4):295--307, 1995.

\bibitem[Pla95b]{plamondon95b}
R.~Plamondon.
\newblock A kinematic theory of rapid human movements. {P}art {II}. {M}ovement
  time and control.
\newblock {\em Biol. Cybern.}, 72(4):309--320, 1995.

\bibitem[Pla98]{plamondon98}
R.~Plamondon.
\newblock A kinematic theory of rapid human movements. {P}art {III}. {K}inetic
  outcomes.
\newblock {\em Biol. Cybern.}, 78:133--145, 1998.

\bibitem[{R D}11]{R}
{R Development Core Team}.
\newblock {\em R: A Language and Environment for Statistical Computing}.
\newblock R Foundation for Statistical Computing, Vienna, Austria, 2011.
\newblock {ISBN} 3-900051-07-0.

\bibitem[Raz97]{razet97}
A.~Razet.
\newblock Résolution analytique d'un cercle de moindres carrés pour une
  utilisation en interferométrie.
\newblock {\em Bulletin du Bureau National de Métrologie}, 108:39--48, 4 1997.
\newblock {B}ureau {N}ational de {M}étrologie, {C}onservatoire National des
  Arts et Métier, 292 rue Saint-Martin, 75141 Paris Cedex 03, France.

\bibitem[Raz98]{razet98}
A.~Razet.
\newblock Analytical resolution of least-square applications for the circle in
  interferometry and radiometry.
\newblock {\em Metrologia}, 35:143--149, 1998.

\bibitem[SC03]{MR2004649}
Yogesh Singh and Pravin Chandra.
\newblock A class of {$+1$} sigmoidal activation functions for {FFANN}s.
\newblock {\em J. Econom. Dynam. Control}, 28(1):183--187, 2003.

\bibitem[SL81]{SoechtingLacquantini1981}
J.F Soechting and F.~Laquantini.
\newblock Invariant characteristics of a pointing movement in man.
\newblock {\em J Neurosci}, 1:710--720, 1981.

\bibitem[VBML08]{cvjbplkm2008ISCA}
Clément Villars, Jérôme Bastien, Karine Monteil, and Pierre Legreneur.
\newblock Kimatic modelisation of joint displacement: validation in human
  pointing task.
\newblock In {\em Industrial Simulation Conference (ISC 08)}, CESH, Lyon,
  France, June 2008.

\bibitem[Vil08]{villars08}
Clément Villars.
\newblock Les tâches de pointages : approches expérimentale et théorique.
\newblock Master's thesis, Université Claude Bernard Lyon 1, 2008.

\bibitem[Win90]{Winter1990}
D.~Winter.
\newblock {\em Biomechanics and motor control of human movement}.
\newblock Wiley-Interscience, 1990.

\bibitem[YK03]{MR1976213}
Beong~In Yun and Philsu Kim.
\newblock A new sigmoidal transformation for weakly singular integrals in the
  boundary element method.
\newblock {\em SIAM J. Sci. Comput.}, 24(4):1203--1217 (electronic), 2003.

\bibitem[ZSG86]{Zelaznik1986}
H.~N. Zelaznik, R.~A. Schmidt, and S.~C. Gielen.
\newblock Kinematics properties of rapid aimed hand movements.
\newblock {\em J Mot Behav}, 18(4):353--372, 12 1986.

\end{thebibliography}

\end{document}